%% file: simmpc.tex
\documentclass[acmsmall,screen,nonacm]{acmart}

\AtBeginDocument{%
  \providecommand\BibTeX{{%
    \normalfont B\kern-0.5em{\scshape i\kern-0.25em b}\kern-0.8em\TeX}}}

\usepackage{subfig}
\usepackage{enumitem}
\usepackage[normalem]{ulem}
\usepackage{multirow}
\usepackage{enumitem}
\usepackage{tikz}
\usetikzlibrary{shapes,arrows}
\usepackage{nicefrac}
\usepackage{color,soul}
\usepackage[most]{tcolorbox}
\usepackage{adjustbox} %

\usepackage{collcell,xcolor,xfp}
\usepackage{xparse}

\NewDocumentCommand\firstitem{>{\SplitList{,}}m}
{%
	{\ProcessList{#1}{#1}}%
}

\newcommand{\fmtnum}[1]{%
	\ifnum\fpeval{\firstitem{#1} < 0} = 1
	\cellcolor{red}{$#1$}%
	\else
	\ifnum\fpeval{\firstitem{#1} < 1} = 1
	\cellcolor{green}{$#1$}%
	\else
	\cellcolor{blue}{$#1$}%
	\fi
	\fi
}

\makeatletter
\newcommand{\hyperrefitem}[2][]{%
	\ifnum\pdfstrcmp{\@currenvir}{enumerate}=0
	\stepcounter{\@enumctr}%
	\item[{\hyperref[#1]{\csname label\@enumctr\endcsname}}]
	\fi
	\ifnum\pdfstrcmp{\@currenvir}{itemize}=0
	\item[{\hyperref[#1]{\csname\@itemitem\endcsname}}]
	\fi
	\hyperref[#1]{#2}%
}
\makeatother

\definecolor{darkpastelgreen}{rgb}{0.01, 0.75, 0.24}
\definecolor{orange-red}{rgb}{1.0, 0.27, 0.0}
\definecolor{model-green}{rgb}{0.47375, 0.99755, 0.34963}
\definecolor{model-blue}{rgb}{0.26967, 0.34878, 0.79631}

\DeclareMathAlphabet{\mathcal}{OMS}{cmsy}{m}{n}

\newcommand{\R}{\mathbb{R}}

\newcommand{\norm}[1]{\left\lVert#1\right\rVert}

\begin{document}

\title{Simulating Interaction Movements via Model Predictive Control}

\author{Markus Klar}
\orcid{0000-0003-2445-152X}

\email{markus.klar@uni-bayreuth.de}

\author{Florian Fischer}
\orcid{0000-0001-7530-6838}
\email{florian.j.fischer@uni-bayreuth.de}

\author{Arthur Fleig}
\orcid{0000-0003-4987-7308}
\email{arthur.fleig@uni-bayreuth.de}

\author{Miroslav Bachinski}
\email{miroslav.bachinski@uni-bayreuth.de}
\orcid{0000-0002-2245-3700}

\author{J{\"o}rg M{\"u}ller}
\email{joerg.mueller@uni-bayreuth.de}
\orcid{0000-0002-4971-9126}
\affiliation{%
	\institution{University of Bayreuth}
	\city{Bayreuth}
	\streetaddress{Universit{\"a}tsstra{\ss}e 30}
	\postcode{95440}
	\country{Germany}}

\renewcommand{\shortauthors}{Klar, et al.}

\begin{abstract}
	
	We present a method to simulate movement in interaction with computers, using Model Predictive Control (MPC). The method starts from understanding interaction from an Optimal Feedback Control (OFC) perspective. We assume that users aim to minimize an internalized cost function, subject to the constraints imposed by the human body and the interactive system. In contrast to previous linear approaches used in HCI, MPC can compute optimal controls for nonlinear systems. This allows us to use state-of-the-art biomechanical models and handle nonlinearities that occur in almost any interactive system. Instead of torque actuation, our model employs second-order muscles acting directly at the joints. We compare three different cost functions and evaluate the simulated trajectories against user movements in a Fitts' Law type pointing study with four different interaction techniques. Our results show that the combination of distance, control, and joint acceleration cost matches individual users’ movements best, and predicts movements with an accuracy that is within the between-user variance. To aid HCI researchers and designers, we introduce CFAT, a novel method to identify maximum voluntary torques in joint-actuated models based on experimental data, and give practical advice on how to simulate human movement for different users, interaction techniques, and tasks.
	
\end{abstract}



\maketitle

\section{Introduction}
Movement during interaction can be understood from an Optimal Feedback Control (OFC) perspective~\cite{fischer2021optimal}:
During interaction, users aim to compute muscle control signals, which control the dynamical systems of the users' bodies, which in turn control interactive systems.
OFC states that users aim to minimize an internal cost function subject to the constraints imposed by the users' bodies and the interactive systems.
They do so by observing the state of the interactive system and \textit{continuously} adjusting their controls to further their goals.

It is this observed continuity and the adjustment of controls that drives the desire to model interaction beyond summary statistics, in order to predict movement along the \textit{entire} interaction loop between human and computer, including, e.g., joint postures or cursor trajectories, \textit{on a moment-by-moment basis}.
Taking the OFC perspective allows us to accomplish these things, by modeling interaction as an optimization problem.
Here, the optimization variable is a continuous muscle signal, i.e., a function of time, that drives the user's movement, which ``controls'' the system.
Thus, this optimization problem %
is usually referred to as an \textit{Optimal Control Problem (OCP)}.
The \textit{feedback} part of OFC is due to solving the OCP in a feedback manner to model the user's ability to adjust their control during interaction, e.g., to react to unforseen circumstances such as perturbations in the cursor movement.

Previous approaches of modeling interaction from the OFC perspective have employed linear optimal control theory, particularly the Linear Quadratic Regulator (LQR)~\cite{Fischer20}, its stochastic extension LQG~\cite{fischer2021optimal}, and intermittent control methods~\cite{Martin21}. %
These approaches considerably simplify the problem of computing the optimal control signals, by using linear approximations to the human-computer system and quadratic cost functions.
However, these limitations lead to unrealistic simplifications of the human-computer system.
Typically, human movements are simulated only with simple point-mass models, since modeling the kinematic chain already leads to nonlinear dynamics.
Other important nonlinear features, such as those of interactive systems (e.g., transfer functions), similarly cannot be modeled by this linear approach.
Further, quadratic cost functions cannot accurately reflect many tasks in Human-Computer Interaction, such as accurately hitting a button with abrupt boundaries. 

In this paper, we extend the OFC approach to Human-Computer Interaction to nonlinear dynamics and non-quadratic cost functions by using Model Predictive Control (MPC)~\cite{GP17}.  
This allows us to investigate the simulation of human movement during interaction with computers using a state-of-the-art nonlinear biomechanical model of the human upper extremity in combination with nonlinear interaction dynamics such as pointer acceleration~\cite{mueller2017}.

MPC as a method has various strengths, such as the easy inclusion of constraints and certain theoretical functionality guarantees to provide trust and reliability, but the main idea behind MPC is \textit{complexity reduction} in time.
It takes the above OCP, which can be computationally hard to solve for the whole interaction/movement duration, and breaks it down into iterative sub-problems of much smaller duration, which are thus considerably easier to solve.
After solving a sub-problem, only the first part of the resulting optimal control sequence is applied to the system, resulting in a new system state.
The horizon is then shifted by one step, i.e., the next sub-problem starts with this new state.
This makes the MPC a closed-loop feedback controller, which is inherently robust against %
perturbations that may occur during interaction.

In summary, the contribution of this work to the field of HCI is fourfold:
\begin{enumerate}%
	\item a nonlinear method combining biomechanical modeling and \textit{Model Predictive Control} to simulate human movement during interaction on a moment-by-moment basis;
	\item a comparison of three different cost functions in their ability to generate biomechanically plausible movements, as observed in a new user study;
	\item an evaluation of our approach using distance, control and joint acceleration costs (\textit{JAC}) for the use case of different mid-air pointing interaction techniques;
    \item practical advice on how to apply our approach, including the generation of customized user models, with individual biomechanical properties and strategies, and \textit{CFAT}, a novel method to infer the maximum voluntary torques used in an interaction task.
\end{enumerate}

The paper is structured as follows.
Related work is discussed in Section~\ref{sec:related-work}.
The core of this paper, our simulation approach using MPC (and practical advice on its use), is presented in Section~\ref{sec:mpc-approach}.
In Section~\ref{sec:cfat} we introduce \textit{CFAT}, a method to compute maximum voluntary torques for joint actuated models such as the one used in this paper. 
Then follows an evaluation of our approach, applied to the use case of ISO pointing in VR, in Sections~\ref{sec:use-case} and~\ref{sec:results}, where we show 
that our simulation is able to predict biomechanically plausible user movements. 
A discussion of the advantages and limitations of MPC ensues in Section~\ref{sec:discussion}.
Section~\ref{sec:conclusion} concludes the paper.

\section{Related Work}\label{sec:related-work}
\subsection{Forward Models of Interaction Movements}\label{ssec:relwork-forward}
Forward models of movement during interaction with computers can predict variables such as movement duration, 
joint angles, or muscle activations.
Depending on what they predict, they can be categorized as \textit{summary statistics} (e.g., movement duration), \textit{end-effector models} (e.g., end-effector position), or \textit{kinematic chain models} (e.g., body joint trajectories).

The most widely used \textit{summary statistics} model of the end-effector is Fitts' Law~\cite{fitts1954information}.
It allows to predict the overall movement time $MT$ from the distance $D$ and width $W$ of the target as $MT = a + b * (D/W + 1)$ (in the Shannon formulation~\cite{mackenzie92}).
It is important to note that Fitts' Law has been developed to describe movement of the human hand.
The fact that the same law can also be used to describe the movement of a virtual end-effector such as the mouse pointer, mediated by input devices and computer programs, is one of the great insights of HCI~\cite{card78}. %
The parameters $a$ and $b$ must be identified for each user and type of movement (e.g., interaction technique) separately.
Recently, more advanced models have been developed to predict, %
e.g., the failure rate and button press timing in moving-target acquisition tasks~\cite{Lee18, Lee21}.

\textit{End-effector models} describe the entire trajectory of the end-effector during the movement.
A classical end-effector model of hand movement is the minimum jerk model~\cite{flash1985coordination}.  
In HCI, only few works investigate the motion of the end-effector, although Bootsma et al.~\cite{behindfitts} demonstrate the importance of understanding movement in HCI beyond summary statistics.
M\"uller et al.~\cite{muller2017control} give an introduction to end-effector models in HCI.
They investigate the kinematics of mouse movements and compare four models from manual control theory.
Quinn and Zhai~\cite{quinn2018} demonstrate how the minimum jerk model can be used to model finger movements during gesture typing.
Jokinen et al.~\cite{Jokinen21} frame touchscreen typing as a visuomotor coordination task and show that optimal supervisory control allows to generate human-like eye-hand movement patterns.
Fischer et al.~\cite{Fischer20, fischer2021optimal} compared the applicability of different optimal control methods to simulate and predict mouse pointing trajectories, and introduced a general optimal control framework for Human-Computer Interaction. 
The focus there lies on controllers, such as the Linear-Quadratic Gaussian Regulator (LQG), which are able to describe mouse pointing while also incorporating signal-dependent noise via a linear-quadratic optimal control problem.
The limitation to linear system dynamics
rules out its application to more complex models of human biomechanics.
Moreover, all of the above works have only analyzed motion in 1D or 2D, although they are in general not limited to 1D or 2D motion.
The only end-effector model that is evaluated with 3D mid-air movements in HCI that we are aware of is the recent work of Bachinski et al.~\cite{bachinski20}, who investigate a 2nd and a 3rd order lag for modeling mid-air movements. %

In this work, we are aiming to model not only end-effector movements in 3D but using a biomechanical model of the human upper body to observe joint angles or velocities, or, even aggregated muscle activation, observed during interaction.
Therefore we consider \textit{kinematic chain models} that, in contrast to pure end-effector models, also make predictions about the underlying causes of the movement %
by modeling the entire kinematic chain.
In particular, this allows to predict ergonomic variables such as joint angles and joint moments.
Most of the previous work on biomechanical models of human movement outside of HCI has concentrated on the substantially simpler 2D case and simple linked-segment models~\cite{harris1998signal, uno1989formation, Takeda19, Li04}.
Linked-segment models use simplified bones as sticks and hinge joints, usually without movement constraints. 
In movement science, the minimum torque change model~\cite{uno1989formation} has been proposed, transferring the idea of the minimum jerk model (i.e., maximization of ``smoothness'') to a simple 2D linked-segment model.
This model requires the exact movement time as well as all joint angles, velocities, and torques of the initial and final postures as input, and yields the kinematics and dynamics of the movement between initial and final state as output.
Li and Todorov~\cite{Li04} present a control method for a 2D linked-segment model using the iterative Linear Quadratic Regulator (iLQR), which minimizes difference between current and target posture plus quadratic control costs.
However, this assumes that the final body posture is known in advance, which is not necessarily the case when the only goal is to move an end-effector (i.e., the fingertip or a virtual cursor) to a target.
Moreover, the model has not yet been extended to the 3D case. %

\subsection{Inverse Biomechanical Simulation in HCI}\label{ssec:relwork-bio}
In contrast to forward models, inverse biomechanical simulation takes as input human movement data and performs inverse estimations of how a specific movement was created. 
The method stems from the fields of biomechanics and rehabilitation and allows to compute accurate physiological indices of movements~\cite{delp2007,rasmussen2003anybody}.
Given motion capture data, it allows to estimate multiple internal variables such as joint angles, joint moments, muscle forces and activation, and neural excitation signals. %
At the core of the biomechanical simulation is a musculoskeletal model, which represents the kinematic, inertial, dynamic, force generation, and neural control properties of the human body~\cite{Saul2014}.
Biomechanical simulation has been introduced and validated for HCI tasks as a method for ergonomic and fatigue evaluation of post-desktop user interfaces~\cite{bachynskyi2015touch,bachynskyi2014}.
It has also been used as a data generation method to develop summarization models of performance and ergonomics for arm movements~\cite{bachynskyi2015}.
Simplified biomechanical models were adapted as components of simulations for fatigue assessment tools~\cite{10.1145/3025453.3025523,10.1145/2556288.2557130}.
Although one current weakness of biomechanical simulations is its necessity for motion capture data collection in user experiments, these simulations have a large potential in the field of HCI, in particular for the analysis and development of AR, VR, and ubicomp user interfaces.

\subsection{Deep Learning and Muscle Control}
The above works from Sections~\ref{ssec:relwork-forward} and~\ref{ssec:relwork-bio} have investigated the control of the human body or a virtual object using either outcome-specific relationships such as Fitts' Law, or %
optimization methods such as iLQR.
Another approach incorporates recent tools and methods from the field of Deep Learning. %
Most notably, Cheema et al.~\cite{Cheema20} recently presented a method to estimate cumulative fatigue during mid-air interaction, in terms of Borg CR10 ratings.
They use a 3D linked-segment arm model and reinforcement learning to learn a control policy for a Fitts' law type task.
They propose a novel \textit{reward function} -- the analogy of a cost function in OCPs --, based on effort estimated through the \emph{Three Compartment Controller}, and show that this generates faster and more ``ergonomic'' movements compared to a baseline reward of summed normalized instantaneous joint torques.
Their model is shown to be able to predict the Borg CR10 ratings of the movements performed in~\cite{10.1145/3025453.3025523} with good accuracy.
However, Cheema et al.~\cite{Cheema20} did not analyze the realism of the movements generated by their approach in terms of end-effector trajectories or joint angles, but rather in terms of predicted cumulative fatigue, averaged over 12 models.

Following the work of Cheema et al., in~\cite{fischer2021reinforcement} a state-of-the-art Reinforcement Learning (RL) algorithm was used
to learn to move the finger to arbitrary targets within reach. %
The resulting end-effector trajectories follow both Fitts' Law~\cite{fitts1954information} and the \nicefrac{2}{3} Power Law~\cite{Lacquaniti83}.
Lately, Hetzel et al.~have extended the model from~\cite{fischer2021reinforcement} to simulate mid-air keyboard typing, using the same RL method~\cite{Hetzel21}.

Beside these works, %
the objective of many research works making use of Deep Learning methods is not to model or understand human motion, but rather to create interesting and realistic animations for movies or computer games. 
We are not aware of any works from this research area that compare the synthesized movements to actual human movements on a biomechanical level. 
Similarly to the works from movement science, most works have controlled the torques at the joints (e.g.,~\cite{peng2018deepmimic,hamalainen2014online}).
Control on a muscle-level has traditionally been considered to be computationally infeasible.
This is due to the fact that the computation time increases exponentially with the dimensionality of the control problem, called the \textit{curse of dimensionality}. 

Recently, however, two approaches to create movements of muscle-actuated characters have been presented.
Lee et al.~\cite{Lee19} propose a two-level \textit{imitation learning} algorithm for musculoskeletal models.
A high-level controller follows a reference motion and generates target joint angles.
A low-level controller then controls the muscles to generate the appropriate forces.
Imitation learning assumes that reference motions from humans are available. %
Whether and how imitation learning approaches can generate novel interaction movements that are not available as recordings will be an important question for future research.

In contrast to imitation learning, \textit{reference-free approaches} can synthesize novel movements based only on the model description and reward function. %
Jiang et al.~\cite{jiang2019synthesis} present an approach to circumvent the muscle control problem by controlling the character in joint space, while determining maximum joint torques and energy costs from a neural network, learned from a realistic model in OpenSim~\cite{delp2007}.
Jiang et al.~demonstrated their technique on a leg model.
Whether and how this approach can work for a significantly more complex arm model, especially taking the shoulder into account, remains open. 

Control of muscles is particularly necessary when the movements are big and cover very different joint angles, as moment arms change significantly during the movement in such cases.
One example used by Jiang et al.~\cite{jiang2019synthesis} is a jump for maximum height, where the joint torque network prevents overbending of the knees as an optimal strategy.
However, during most interactions, the movements are small and the moment arms change only minimally during the execution.
In this case, actuating the joints either based on simplified muscle dynamics (as we do in this work) or direct torque control
can be a good approximation and is substantially simpler to use than complex musculotendon models.
For small movements, passive forces created by ligaments and musculotendon units also play a smaller role.

\section{Modeling Interaction as Model Predictive Control}\label{sec:mpc-approach}
In this section, we describe our approach to model and simulate human movement during interaction with the computer using MPC.
We lay the theoretical foundation and provide concrete but extensible models and practical advice.
A concrete use case is described in detail in Section~\ref{sec:use-case}.

The crucial difference between movement simulation in HCI and related fields such as movement science and character animation is that in HCI, users primarily control a virtual object.
The goal of the user is therefore to move their body such that their virtual representation reaches a desired state, e.g., selecting a button or dragging a virtual object.

For many interaction techniques, however, there exists an infinite number of body movements that can be used to successfully perform a task that is only defined in virtual space. %
This complicates the simulation of movements, as it is unclear how the model should move
in order to plausibly replicate human behavior during interaction. %
This problem becomes even more crucial in interfaces that lack a unique mapping from the physical to the virtual end-effector, which is the case in the majority of interaction techniques.
For example, using the mouse as input device, movements to the left result in the same cursor movement as movements to the top after a clockwise rotation of the mouse by 90 degrees.
Therefore, in order to understand the entire interaction loop on a moment-by-moment basis, the actual state of the user's body \textit{and} the input device needs to be taken into account, which can be achieved by utilizing a unifying and mathematically rigorous optimal control framework of interaction~\cite{fischer2021optimal}. %

\begin{figure}[h!]
	\centering
	\includegraphics[width=\textwidth]{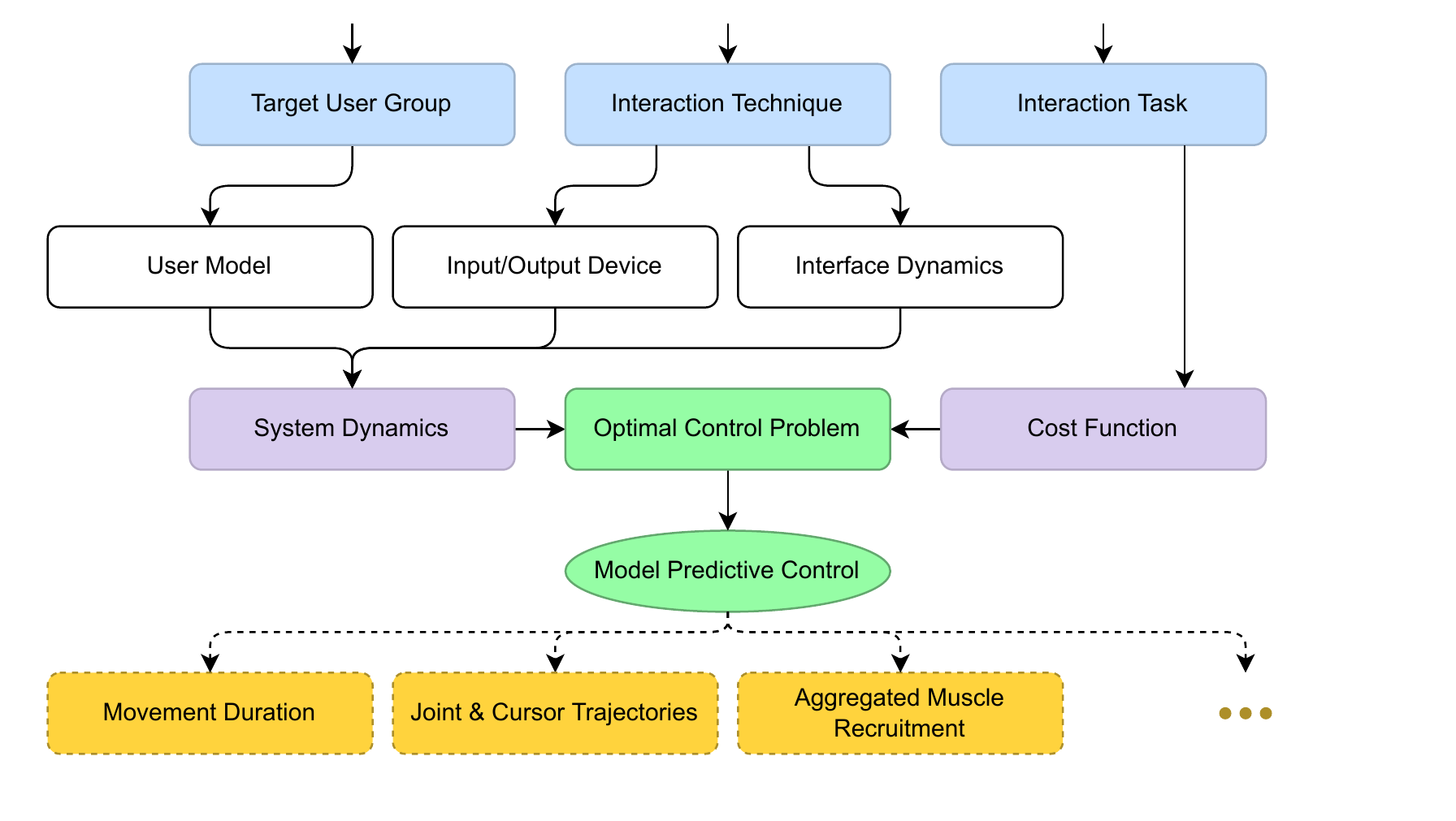}
	\caption{Our proposed simulation framework.
		To run a simulation, one needs to specify the Target User Group, Interaction Task, and Interaction Technique. 
		Based on the Target User Group, a User Model that matches one or more representative user(s) is chosen (or designed). 
		Combined with the Input Device and Interface Dynamics -- both defined by the Interaction Technique -- this results in the System Dynamics.  
		Finally, the Interaction Task imposes a specific Cost Function that the user is assumed to minimize.
		The resulting nonlinear Optimal Control Problem can then be solved with Model Predictive Control, resulting in a simulation of the movement. From this, we can obtain valuable data like movement duration, joint and cursor trajectories, or aggregated muscle recruitment.
	}
	\label{fig:MPC_flow_basic}
\end{figure}

Our framework is depicted in Figure~\ref{fig:MPC_flow_basic}.
Our model takes into account the \textit{target user group}, the \textit{interaction technique}, and the \textit{interaction task}.
The first influence is the \textit{User Model}, see Section~\ref{sec:user_models}, where we match the physical properties of the target users by using state-of-the-art biomechanical models.
The interaction technique consists of two parts in our framework. 
First, modeling of the \textit{Input/Output device} (e.g., motion capture tracking of the index finger, or the combined use of a HTC Vive controller and head-mounted display (HMD)) is described in Section~\ref{sec:inputoutputdevice}.
Second, we define \textit{Interface Dynamics} that determine how the user input is transferred to the virtual system (e.g., to a virtual cursor) in Section~\ref{sec:interfacedynamics}.

Readers that are already familiar with the above concepts and are mostly interested in the core method used to simulate movements may directly skip to Section~\ref{sec:ocp}.
There we introduce the notation of a nonlinear \textit{Optimal Control Problem (OCP)}, which augments the former discussed models of the human body and the interaction technique with a \textit{Cost Function} that formalizes the user-specific objectives for a given interaction task.

In Section~\ref{sec:MPC}, we show how the OCP can be solved via \textit{Model Predictive Control (MPC)}, resulting in a simulation of the complete biomechanical chain during interaction.
We thus obtain not only summary statistics like movement duration or success rates, but also trajectories of joint angles, angular velocities, and accelerations; trajectories of cursor positions, velocities, and accelerations; and biomechanical data such as aggregated muscle recruitment.

\subsection{User Model}\label{sec:user_models}

The biomechanical properties of the user model should match those of the considered users, and the range of possible movements %
must be sufficient to fulfill the task.
To fit in our simulation pipeline (Figure~\ref{fig:MPC_flow_basic}) and act as a part of the system dynamics, the user model should be able to be forward-simulated, i.e., to map the current body state $\mathbf{x}_\text{user}$ (including, e.g., joint angles and angular velocities, and the internal state of the muscles), and the aggregated muscle control signals $u$ to the next (i.e., updated) body state $\mathbf{x}_\text{user}^+$ in a realistic way.\footnote{Here and throughout this work, we denote states that contain a variety of different quantities in bold font, and the individual quantities in regular font.} 
Formally, this mapping can be defined via a function $f_\text{user}$: 
\begin{equation}
	\label{eq:usermodel}
	 \mathbf{x}_\text{user}^+ = f_\text{user}(\mathbf{x}_\text{user},u).
\end{equation}
In this work, a biomechanical, joint-actuated model implemented in a physics engine, coupled with second-order muscle dynamics, will take the role of $f_\text{user}$.
Of course, it is possible to exchange our user model with other models of human motion.

\subsubsection{Upper Extremity Model in MuJoCo}
We make use of the fast physics simulation MuJoCo~\cite{mujoco} to handle the complex biomechanics of human motion.
In~\cite{fischer2021reinforcement}, a MuJoCo model from the state-of-the-art OpenSim~\cite{seth2018opensim} musculoskeletal model from Saul et al.~\cite{Saul2014} was derived. 
We use this MuJoCo model for two reasons: (i) limitations in OpenSim's ability to simulate contacts -- it is very difficult in OpenSim to allow a model to interact with input devices and environmental objects such as a chair or table, while preventing the model from reaching through its torso or legs --; and (ii) computation speed. 

The biomechanical model has seven independent joints\footnote{Some human joints are reflected by multiple model joints.
Therefore, throughout the paper, we use the term \textit{joint} synonymously for a hinge joint in our model.} (i.e., seven DOFs\footnote{degrees of freedom})
and 13 coupled joints, %
representing a shoulder, an elbow, and a wrist. 
The shoulder is the centerpiece of the model and is connected to a torso -- which is made immovable during interaction for simplicity -- 
through a set of three independent and eleven coupled joints. %
The three independent joints set up the angle and extent of the elevation, as well as the rotation of the upper arm.
Ten of the eleven coupled joints are used to accurately describe the motion of clavicle and scapula with respect to the shoulder elevation.
Since the joints build on each other, another coupled joint is used to revert the elevation angle before applying rotation.
The elbow is composed of two independent joints allowing flexion-extension and pronation-supination movements.
For the wrist, we use four joints, two independent and two coupled, which allow accurate flexion-extension and abduction-adduction movements of the hand.
The finger joints are locked in a pointing posture, since they are less important for our example tasks and omitting them considerably simplifies the user model.
The complete model is depicted in Figure~\ref{fig:mujocomodel}; joint angle ranges can be found in the Appendix~\ref{tab:joint-limits}.

To avoid the \textit{curse of dimensionality}, i.e., the exponential growth of computation time with the number of variables to be optimized, we refrain from including muscles in our MuJoCo model.
Instead, we implement simplified muscles that directly act on the joints as follows.
We place a torque actuator that can produce positive and negative torque around the axis of each of the seven independent joints.
At any given time step~$n$, for $i\in\{1, \dots, 7\}$ the applied torque~$\tau^{i}(n)$ of each actuator depends on its current activation~$x_{\sigma}^{i}(n)$, scaled by the maximum voluntary torque~$g^{i}$ for the respective joint: %
\begin{equation}\label{eq:activation2torque_DOF}
	\tau^{i}(n) = g^{i} x_{\sigma}^{i}(n).
\end{equation}
The current activation~$x_{\sigma}^{i}(n)$ of each torque actuator is obtained through a simplified second-order muscle model, which is explained in detail below. %

For practical applications, one challenge with this simplified muscle model is to determine the maximum voluntary torques~$g=(g^{1}, \dots, g^{7})^{\top}\in \R^7$ for each independent joint, as to prevent unrealistic movements. 
To this end, we propose \textit{CFAT}, a tool described in Section~\ref{sec:cfat} to obtain better matching torques.

\begin{figure}
	\centering
	\includegraphics[width=.5\linewidth,trim=0 10cm 0 5cm,clip]{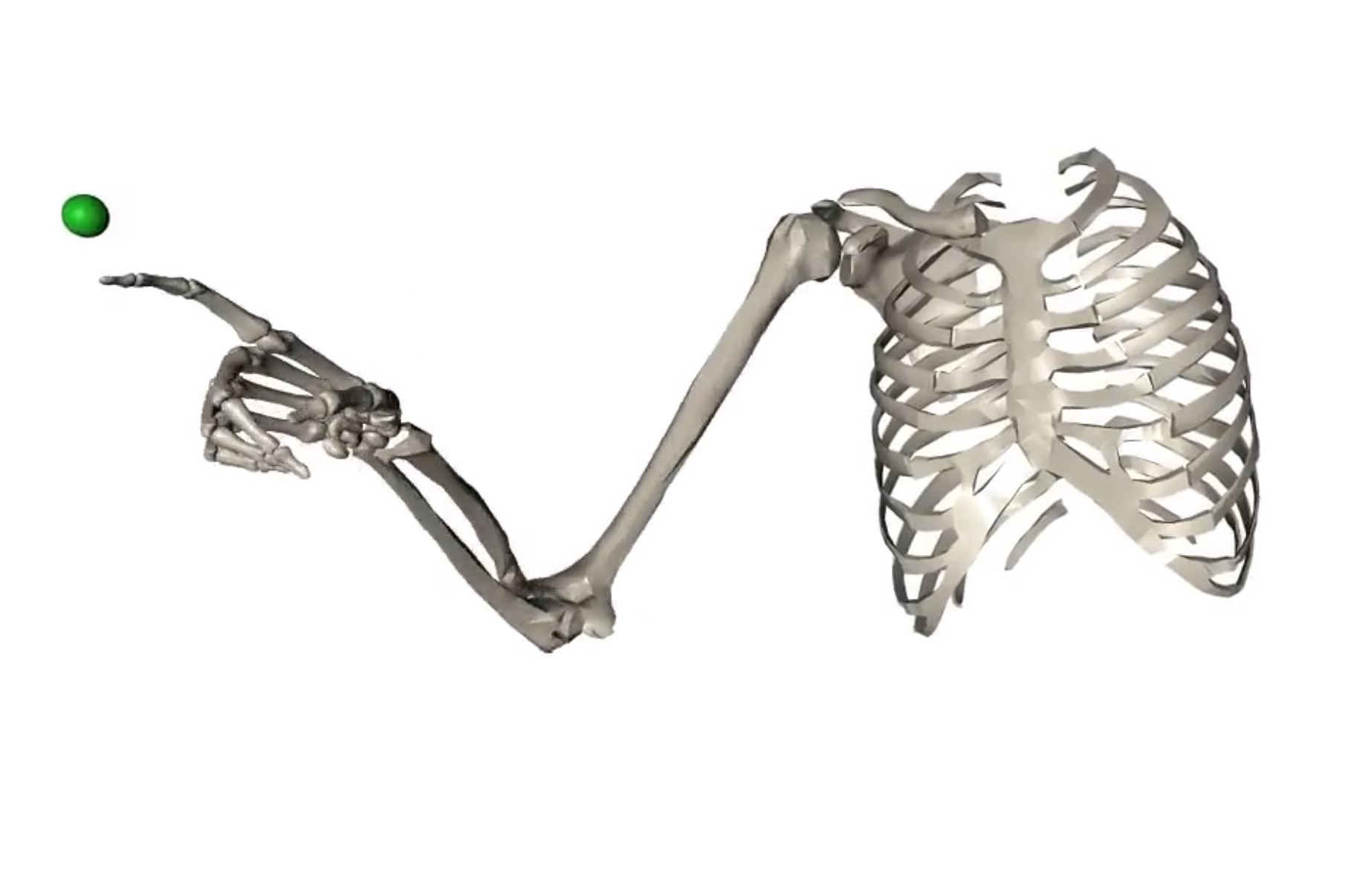}
 	\caption{Visualization of our MuJoCo user model. The green sphere models the motion tracking marker for the physical end-effector.}
	\label{fig:mujocomodel}
\end{figure}

\subsubsection{Second-Order Muscle Dynamics}\label{sec:musclemodel}

Modeling and simulating human muscles has proven to be challenging.
This is not only because of the sheer amount of muscles -- in the original OpenSim model~\cite{Saul2014}, the shoulder and arm alone are moved by a total of 31 muscles -- but also because of the complex interaction of force generation, tendon lengths, tendon positioning, etc.
Optimizing for each muscle activation simultaneously is a challenging problem, which so far has only become feasible through techniques like hierarchical optimization~\cite{liu2009hierarchical} or through aggregation.
We follow the approach by van der Helm et al.~\cite{van2000musculoskeletal}, who aggregate muscles for each DOF using second-order dynamics.
We discretize these muscle dynamics using the forward Euler method~\cite{butcher2016numerical}.
The vector~$x_{\sigma}=(x_{\sigma}^{1}, \dots, x_{\sigma}^{7})^{\top} \in [-1,1]^7$ contains the activation for all seven DOFs. %
The vector of activation derivatives %
is denoted by~$x_{\dot{\sigma}}=(x_{\dot{\sigma}}^{1}, \dots, x_{\dot{\sigma}}^{7})^{\top} \in \R^7$, and is affected by the vector of applied controls denoted by~$u=(u^{1},\dots,u^{7})^{\top}$. 
In formulas, the discrete-time dynamics for each DOF $i\in\{1, \dots, 7\}$ can be described as follows, where $n$ is the current time step and $n+1$ the next one: 

\begin{equation}\label{eq:activationmodel}
	\begin{bmatrix}
		x_{\sigma}^{i}(n+1) \\ 
		x_{\dot{\sigma}}^{i}(n+1)
	\end{bmatrix} =
	\begin{bmatrix}
		1 & \Delta t \\
		\frac{-\Delta t}{(t_e t_a)} & 1 - \Delta t \frac{t_e + t_a}{t_e t_a}
	\end{bmatrix} 
	\begin{bmatrix}
		x_{\sigma}^{i}(n) \\ 
		x_{\dot{\sigma}}^{i}(n)
	\end{bmatrix}
	+ \begin{bmatrix}
		0 \\ \frac{\Delta t}{t_e t_a}
	\end{bmatrix} u^{i}(n),
\end{equation}
with initial constraints
\begin{equation}
	x_{\sigma}(0) = \sigma_0, \; x_{\dot{\sigma}}(0)=\dot{\sigma}_0.
\end{equation}
Here, $\Delta t=~$2~ms is the update interval, %
$t_{e}=30$~ms and $t_{a}=40$~ms are the fixed excitation and activation time constants, respectively, which are taken from van der Helm et al.~\cite{van2000musculoskeletal}, and the $\sigma_0$ and $\dot{\sigma}_0$ are initial values for the activation and its derivative\footnote{An idea about the magnitudes of these initial values is obtained through practical experiments, where these initial values are obtained from data for each trial, see Section~\ref{sec:study-based-simulation}.}.

\subsubsection{Necessary adjustments for other use cases}
Since we focus on the simulation of (right-handed) mid-air pointing movements of adults, we only model the upper extremity of an adult, i.e., the right arm and shoulder.
However, by adjusting the MuJoCo model (which comes down to editing an XML file), e.g., to size or physique, different target user groups can be considered.

We note that, although we use a state-of-the-art biomechanical model, we do not model the complete human body. 
Because of the current model limitations, the interaction technique and interaction task have an influence on the user model.
For example, if the input device is a handheld controller, the hand must be able to hold the controller.
In particular, the (rigid) hand must be re-arranged to match the position of holding the considered device.
More extensive models may contain palm and finger joints to enable fine movements.
Similarly, changing the interaction task may require adjustments in the user model.
If, for example, the task is to grasp and move and some virtual object, the MuJoCo model would require a biomechanically more accurate model of the hand. %

Major changes to the user model may affect the maximum voluntary torques, which is why, in this case, we recommend reapplying the CFAT tool described in Section~\ref{sec:cfat}.

\subsection{Input/Output Device}
\label{sec:inputoutputdevice}
In addition to the biomechanics, we implement several mid-air interaction techniques in MuJoCo. 
Following the scheme from Figure~\ref{fig:MPC_flow_basic}, 
we divide interaction techniques into their (physical) input devices (e.g., a joystick, touch screen, or motion capture system) and output devices (e.g., a monitor or HMD), and the mapping from the information that the computer receives to the virtual state that it displays.

The model of the input device should be able to realistically capture the same data from the user model as the input device captures from the real user.
If, for example, a joystick senses angular movement in two axes, the model of the joystick should be able to obtain the same information. %
Formally, we understand an input device as a function $f_\text{dev}$ that maps the user's current state $\mathbf{x}_\text{user}$ (e.g., body posture) and the current device state $\mathbf{x}_\text{dev}$ (e.g., joystick angle and/or motion capture marker position) to the updated device state $\mathbf{x}_\text{dev}^+$, i.e., %
\begin{equation}\label{eq:xdev}
	\mathbf{x}_\text{dev}^+ = f_\text{dev}(\mathbf{x}_\text{dev}, \mathbf{x}_\text{user}).
\end{equation}

In the considered use case of mid-air pointing without any handheld device, the input device corresponds to the motion capture system PhaseSpace\footnote{https://www.phasespace.com/x2e-motion-capture/}, which allows to continuously track the movement of the user. 
An LED marker is placed at the tip of the right index finger, whose position is used to determine the motion of a virtual cursor.
To model this input device, we use a virtual marker on our MuJoCo user model's index finger to track its position, which we denote as $x_\text{ee}$.
In this particular use case, we thus have 
\begin{equation*}
	\mathbf{x}_\text{dev}=x_\text{ee}\text{.}
\end{equation*}
Since we can obtain this data directly from MuJoCo, we do not need to implement any additional dynamics here, which would be necessary when modeling, e.g., a joystick. %
Therefore, the device dynamics in our case are given by the very simple mapping
\begin{equation}
	f_\text{dev}(\mathbf{x}_\text{dev}, \mathbf{x}_\text{user}) = x_\text{ee}.
\end{equation}
Output modalities may also differ between interaction techniques.
Most commonly, users get a visual feedback via a screen that shows how the virtual environment reacts to their input.
With this information, users can evaluate their actions, e.g., through the position of a virtual cursor, and possibly change their strategy to fulfill the given task, e.g., pointing towards a virtual target.
In this paper, we demonstrate the simulation of mid-air pointing in VR by assuming perfect observation.
This particularly implies that users always see the exact cursor position.

\subsubsection{Necessary adjustments for other use cases}
	The MuJoCo model can easily be extended to other input devices, since 
	many different sensors like gyroscopes as well as force, torque, or touch sensors are directly available in MuJoCo. %
	Visual input (to the computer) can be implemented with cameras, that can sense RGB pictures or just depth information.
	If the input device ought to be directly manipulated by the user, one needs to adjust the user model such that it can actually use the device as intended.
	For example, to grab and use a handheld controller, the posture of the hand would have to be adjusted to fit a controller that needs to be implemented in the same physics engine.
	Furthermore, for some input devices (e.g., 
	a joystick or gamepad%
	), fine motor finger movements are necessary, which are currently not possible with our used MuJoCo model.
	
	To implement output devices and perception, one would need to add another layer after the \textit{System Dynamics} in Figure~\ref{fig:MPC_flow_basic}, which maps the ``real'' interface state to the one perceived by the user.
	For example, if the output device was a 2D screen that does not allow to directly infer depth information of the regarded scene, the output device model would need to take into account the underlying projection. %

\subsection{Interface Dynamics}
\label{sec:interfacedynamics}

Once the human input is received, the user interface needs to be updated. 
For example, a change in position of the input device should entail a movement of the controlled virtual object, e.g., the virtual cursor.
Additionally, the virtual world itself may have virtual dynamics. 
For example, throwing a virtual ball at a virtual pin may lead to that pin being knocked over. 

To formalize the whole process, we use three functions.
First, a \textit{transfer function} $f_\text{tf}$ %
transfers physical movement to virtual. 
As such, it depends on the current state of the input device, $\mathbf{x}_\text{dev}$. 
Next, the virtual dynamics come into play via the function $f_\text{vd}$%
, which takes as arguments the current state of the interface~$\mathbf{x}_\text{if}$ (e.g., cursor or button position) and the output of the transfer function.
Finally, these two components are wrapped by the wrapper function~$f_\text{if}$ into the Interface Dynamics of the considered interaction technique, which yields the updated virtual state $\mathbf{x}_\text{if}^+$, i.e.,
\begin{equation}\label{eq:xif+}
	\mathbf{x}_\text{if}^+ = f_\text{if}(\mathbf{x}_\text{if}, \mathbf{x}_\text{dev}) = f_\text{vd}\left(\mathbf{x}_\text{if},f_\text{tf}(\mathbf{x}_\text{dev})\right).
\end{equation}
Since the Interaction Dynamics is part of a \textit{nonlinear} optimal control problem, it is possible to include arbitrary complex virtual dynamics here (although continuity and smoothness of the functions are desirable).

The flip-side, i.e., if no explicit virtual dynamics are required, still fits in this framework.
In this case, $f_\text{if}$ simply is the transfer function:
\begin{equation}\label{eq:xif+novd}
	f_\text{if}(\mathbf{x}_\text{if}, \mathbf{x}_\text{dev}) = f_\text{tf}(\mathbf{x}_\text{dev}).
\end{equation}
In our case of mid-air pointing, we do not need explicit virtual dynamics and as such use~\eqref{eq:xif+novd}. 
This is due to the fact that we only simulate single aimed movements to a static target, i.e., the only change in the interface state concerns the position of the virtual cursor, which we denote by $x_\text{p}$.
This position is updated based on the transfer function that maps the (physical) end-effector position $x_{\text{ee}}$, which is perceived by the computer through the input device and thus is part of the input device state $\mathbf{x}_\text{dev}$, to the position of the virtual cursor $x_\text{p}$, which is part of $\mathbf{x}_\text{if}$.
This leads to simple transfer functions of the form:
\begin{equation}
	f_\text{tf}(x_{\text{ee}}) = x_\text{p}\text{.}%
\end{equation}

But even without virtual dynamics, solely using $f_\text{tf}$, we can encompass a variety of interaction techniques. 

First, we consider the class of virtual cursors in VR~\cite{poupyrev1996go}.
These simple interaction techniques introduce a displacement between the physical and the virtual hand of the user.
The transfer functions for these techniques can be given in an Input-Output-Space formulation.
In our case, the virtual cursor is uniquely given by an input space origin $\omega_{\text{I}} \in \mathbb{R}^3$ and an output space origin $\omega_{\text{O}} \in \mathbb{R}^3$.
The cursor position is obtained by transferring the fingertip position in input space coordinates to the output space. %
The complete transfer function for \textit{Virtual Cursor} is thus given by
\begin{equation}\label{eqn:virtual_cursor}
	f_\text{tf}(x_{\text{ee}}) = x_{\text{ee}} - \omega_{\text{I}} + \omega_O\text{.}%
\end{equation}
In particular, placing the end-effector at the input origin, i.e., $x_{\text{ee}}=\omega_{\text{I}}$, results in the cursor being at the output origin. %
The choice of the input origin~$\omega_{\text{I}}$ influences task performance.
For example, a lower input origin allows to achieve the same cursor position with a lower end-effector position, i.e., with a lowered arm, eventually resulting in more comfortable movements.
To match horizontal alignment, we define the output space such that it represents a virtual 3D space in front of the user by setting $\omega_{\text{O}} = \left( -0.1~\text{m}, 0.0~\text{m}, 0.55~\text{m} \right)$, i.e., 10~cm right and 55~cm in front of the user.

As a slightly more complex interaction technique, we select the group of \textit{Virtual Pad} techniques~\cite{andujar2007virtual}, which project the 3D fingertip position to a 2D cursor position.
The technique can be described as using a tablet placed on a table to move a cursor on a screen in front, with the differences that the tablet is indefinitely large and that there is no need to touch it.
The virtual display, i.e., the output plane on which the cursor moves, is characterized by its origin $\omega_{\text{O}} \in \mathbb{R}^3$ and normal vector $n_{\text{O}} \in \mathbb{R}^3$. 
We set this output plane to be in front of and facing the user, i.e., $\omega_{\text{O}} = \left( -0.1~\text{m}, 0.0~\text{m}, 0.55~\text{m} \right)$ and $n_{\text{O}} = \left( 0, 0, -1 \right)$.
An input plane is analogously defined by its origin  $\omega_{\text{I}} \in \mathbb{R}^3$ and normal vector $n_{\text{I}} \in \mathbb{R}^3$.
The cursor position is obtained in two steps:
First, the fingertip is projected onto the input plane by a function $\text{Proj}_\text{I}$. %
Then, this point is rotated from input to output plane orientation by a function $\text{Rot}_\text{IO}$. %
Finally, the cursor position is translated such that it lies on the output plane.
In total, the \textit{Virtual Pad} transfer function is therefore given by
\begin{equation}\label{eqn:virtual_pad}
f_\text{tf}(x_{\text{ee}}) = \text{Rot}_{\text{IO}}\left( \text{Proj}_\text{I}(x_{\text{ee}})\right) + \omega_{\text{O}}.
\end{equation}
Exact formulas for $\text{Proj}_\text{I}$ and $\text{Rot}_{\text{IO}}$ are given in Appendix~\ref{sec:appendix-virtual-pad}.

\subsubsection{Necessary adjustments for other use cases}

	In the case of pointing, the presented transfer functions can easily be modified to match different interaction techniques.
	Modeling different pointing techniques such as ray casting~\cite{lee2003evaluation} is also possible by adjusting the transfer function accordingly.
	In this case, one must add additional information of the virtual environment to the transfer function, e.g., the position and size of selectable objects.
	When creating new transfer functions, it is a good practice to implement them based on parameters, such as the input origin for the transfer functions used in this work. 
	Evaluating the interface for different parameters allows for a quick adaptation and optimization of the interaction technique.
	Since the interface dynamics can incorporate both transfer functions and virtual dynamics, one could also model techniques where the interface has its own internal state, such as driving a virtual vehicle. 
	In general, the virtual dynamics need to formalize the internal and mutual state dependencies of all interface objects that are relevant for the interaction, e.g., moving targets that need to be tracked by the virtual cursor, or interactable objects that may change their color or size depending on the context.

\subsection{Modeling Interaction as nonlinear Optimal Control Problem}
\label{sec:ocp}
Modeling human-computer interaction requires the use of dynamics that do not only change the state of the interface based on human input, but also capture biomechanics.
Due to the redundancy of the human biomechanical system, there are infinitely many body movements that can be used to execute a given interaction task.

Building on the idea of optimal human movement control~\cite{Todorov02, Umberger18}, we assume that humans aim to behave optimally with respect to an internalized cost function, subject to the dynamics of the human-computer-interaction system.
This allows us to make use of the \textit{optimal control framework} and rephrase the considered human-computer-interaction as nonlinear \textit{Optimal Control Problem (OCP)}, using an appropriate cost function as well as (discrete-time) system dynamics that describe the complete interaction loop.
Formally, this %
can be written as
\begin{equation}\label{eq:OCP}
	\begin{aligned}
		&\min_{u(\cdot)} J_\infty(\mathbf{x}_0,u(\cdot)) = \min_{u(\cdot)} \sum_{k=0}^{\infty}\ell(\mathbf{x}(k),u(k))\\
		\textnormal{such that }& \mathbf{x}(k+1) = f(\mathbf{x}(k),u(k)), \quad \mathbf{x}(0) = \mathbf{x}_0, \\
		&\mathbf{x}(k)\in\mathbb{X}, u(k) \in \mathbb{U},\quad \text{ for all }k\in\mathbb{N}.
	\end{aligned}
\end{equation}
Here, $J_\infty$ is the cost to minimize, which is defined by the \emph{stage cost} or \emph{running cost}~$\ell\colon\mathbb{X}\times\mathbb{U}\to\mathbb{R}$ that we need to design, $f\colon \mathbb{X}\times\mathbb{U}\to\mathbb{X}$ is the nonlinear, continuous state transition map that takes the current state and control and yields the subsequent state according to the system dynamics, and $\mathbf{x}(\cdot)$ denotes the overall state trajectory that results from the forward simulation of the system with initial state $\mathbf{x}_0$ and control sequence $u(\cdot)$.
State and control constraints are incorporated in the spaces~$\mathbb{X}$ (e.g., biomechanically feasible joint angles) and~$\mathbb{U}$ (e.g., maximum permissible aggregated control signal strength for each joint), respectively.

The equation $\mathbf{x}(k+1) = f(\mathbf{x}(k),u(k))$ can be written in a shorter form, analogous to the previous sections, as $\mathbf{x}^+ = f(\mathbf{x},u)$, but we kept the current time~$k$ explicitly because it occurs in~$\ell$. 
Subsequently, we show how to pour life into the abstract OCP~\eqref{eq:OCP}.

\subsubsection{System Dynamics}
The complete discrete-time system dynamics are obtained by combining the user model, input device, and interface dynamics that we have described in Sections~\ref{sec:user_models},~\ref{sec:inputoutputdevice}, and~\ref{sec:interfacedynamics}, respectively.
These dynamics map the control signal $u$ and the current state of the overall system $\mathbf{x}=(\mathbf{x}_\text{user},\mathbf{x}_\text{dev},\mathbf{x}_\text{if})$, consisting of user, device and interface states, to the next system state $\mathbf{x}^+$.
Therefore, we can formalize the system dynamics as 
\begin{equation}\label{eq:systemdynamics_complete}
	\mathbf{x}^+ = f(\mathbf{x},u) = \left(\mathbf{x}_\text{user}^+,\mathbf{x}_\text{dev}^+,\mathbf{x}_\text{if}^+\right),
\end{equation}
where the formulas for $\mathbf{x}_\text{user}^+$, $\mathbf{x}_\text{dev}^+$, and $\mathbf{x}_\text{if}^+$ are given by~\eqref{eq:usermodel}, \eqref{eq:xdev}, and~\eqref{eq:xif+}, respectively.

In particular, in our mid-air pointing use case, the state of the complete system consists of
\begin{equation}
	\label{eqn:x_complete}
	\begin{split}
		\mathbf{x}_\text{user} & =
		\begin{cases}
			x_\text{qpos},x_\text{qvel},x_\text{qacc}&\text{: joint angles, angular velocities, and accelerations}\\
			x_\sigma, x_{\dot{\sigma}}&\text{: aggregated muscle activation and their derivatives,}%
		\end{cases} \\
		\mathbf{x}_\text{dev} & =
		\begin{cases}
			x_\text{ee} & \text{: physical end-effector position, and}\\
		\end{cases}    \\
		\mathbf{x}_\text{if} & =
		\begin{cases}
			x_\text{p} & \text{: cursor position.}\\
		\end{cases}      
	\end{split}
\end{equation}

\subsubsection{Cost Function}
\label{sec:cost-function}
The cost function that is assumed to be minimized by a user during interaction needs to reflect the task requirements, goals, and intrinsically motivated objectives that can represent specific user strategies. %
Using the notation of the OCP~\eqref{eq:OCP}, the cost function is given by a stage cost function $\ell(\mathbf{x},u)$, which maps the state of the system $\mathbf{x}$ and the control signal $u$ to the respective cost.

Considering mid-air pointing as an example, the task is to reach a given target with a virtual cursor.
Reaching the target is often modeled as a terminal constraint.
For this, however, the duration of the movement must be fixed beforehand.
Since movement times vary from trial to trial and are not known in advance, they must either be estimated or calculated from experiments.
However, we want to enable the simulation of human-like movements based on a model of the interaction dynamics and the user only, without relying on experimentally observed or estimated movement duration.
Instead of using a terminal constraint, we thus follow a well-known approach and mitigate this problem by penalizing the distance between the cursor position~$x_p \in \mathbb{R}^3$ and the target position $p^\star \in \R^3$ %
at each time step~\cite{Todorov05, diedrichsen2010coordination, qian2013movement, fischer2021optimal}.
This does not only incentivize moving the cursor towards the target, but implicitly penalizes the movement duration as well, since slow movements result in higher accumulated distance costs (note the sum in~\eqref{eq:OCP}). %

Furthermore, humans are known to prefer moving with low effort~\cite{Todorov02, Li04, Guigon07}.
We implement this concept by penalizing the aggregated muscle control signal, i.e., the control vector~$u$.
As it is usually done in numerical optimization to improve the performance, we take the squared Euclidean norm, denoted by $\norm{\cdot}$ in the following. %

In addition, we introduce two different cost terms that have previously been used to model optimal human behavior. %
The first one corresponds to the well-established \textit{commanded torque change} %
\cite{kawato1993optimization, nakano1999quantitative, wada2001quantitative, wang2011optimal}, which penalizes the derivative of the commanded torques, that is, the torques that directly result from the applied motor commands.
In our case, this corresponds to the derivative\footnote{Due to the discrete-time setting, we take central/one-sided differences using \texttt{numpy.gradient}.} of the applied torque $\tau$,  %
which we denote by $\dot\tau$ in the following\footnote{We obtain $\tau$ and $\dot\tau$ from the activations via~\eqref{eq:activation2torque_DOF}.}.
The second, less frequently used cost term corresponds to the joint acceleration, which leads to smooth movements towards the target~\cite{wada2001quantitative}. 
We denote the vector of (angular) joint accelerations by $x_\text{qacc}$, which is part of~$\mathbf{x}_\text{user}$, see~\eqref{eqn:x_complete}.

With these components, we propose three different stage costs:
\begin{itemize}
	\item \textbf{DC}: Distance and Control Costs.\\
	The distance between cursor and target as well as the aggregated muscle control are penalized at each time step:
	\begin{equation}\label{eq:costs-dc}
		\ell(\mathbf{x}(k),u(k)) = \norm{x_p(k) - p^\star} +  r_1 \norm{u(k)}^2
	\end{equation}
	\item 
	\textbf{CTC}: Commanded Torque Change Cost.\\
	This cost function adds to \eqref{eq:costs-dc} a third cost term, penalizing the commanded torque change: 
	\begin{equation}\label{eq:costs-ctc}
		\ell(\mathbf{x}(k),u(k)) = \norm{x_p(k) - p^\star} +  r_1 \norm{u(k)}^2 + r_2 \norm{\dot{\tau}(k)}^2
	\end{equation}
	\item 
	\textbf{JAC}: Joint Acceleration Costs.\\
	This cost function adds to \eqref{eq:costs-dc} a third cost term, penalizing the squared joint accelerations:
	\begin{equation}\label{eq:costs-jointacc}
		\ell(\mathbf{x}(k),u(k)) = \norm{x_p(k) - p^\star} +  r_1 \norm{u(k)}^2 + r_2 \norm{x_\text{qacc}(k)}^2
	\end{equation}
\end{itemize}
The \textit{cost weights} $r_1, r_2 > 0$ define the trade-off between the different cost terms.

\subsubsection{Fitting Cost Weights}
\label{sec:fittingcostweights}

The choice of those cost weights has a significant impact on the resulting simulation trajectories (an evaluation of the effect of cost weights can be found in Section~\ref{sec:effect_costweights}).
To find the most appropriate weights for our cost functions, we need to evaluate different weight pairs $(r_1,r_2)$. 
Since we aim to generate joint movements that are as close to human movements as possible, we evaluate a cost weight pair by comparing the resulting simulation sequence of joint angles~$x_{\text{qpos}}$ to that of a sequence of joint angles~$\hat{x}_{\text{qpos}}$ obtained through a user study, considering independent joints only.
More precisely, we compute the \textit{root mean squared error (RMSE)} between simulation and experimental data, i.e., 
\begin{equation}
	\label{eq:rmse}
	\text{RMSE}\left(x_{\text{qpos}}, \hat{x}_{\text{qpos}}\right) =  \sqrt{ \frac{1}{M} \sum_{k=0}^{M-1} \norm{x_{\text{qpos}}(k)-\hat{x}_{\text{qpos}}(k)}^2},
\end{equation}
where $M$ is the number of steps of the experimental data trajectory.
To ``rate'' a weight pair, we sum the RMSE values for a given number $S$ of simulated trajectories, resulting in the following loss function used for parameter optimization:
\begin{equation}
	\label{eq:paramfitting-loss}
	\mathcal{L}_{\text{param}}(r_1, r_2) = \sum_{s=0}^{S-1} \text{RMSE}\left(x_{\text{qpos}}^{(s)}, \hat{x}_{\text{qpos}}^{(s)}\right),
\end{equation}
where $x_{\text{qpos}}^{(s)}$ denotes the simulation trajectory obtained from the cost weights $r_1$ and $r_2$, and $\hat{x}_{\text{qpos}}^{(s)}$ denotes the corresponding experimental trajectory, given a trial $s\in\{0,...,S-1\}$. %
Since each evaluation of the RMSE~\eqref{eq:rmse} requires solving a single OCP~\eqref{eq:OCP} and thus results in large computation times, we decided to use a state-of-the-art derivative-free optimization algorithm that works with a low number of function evaluations and non-convex problems, and is easy to parallelize: the \textit{Covariance Matrix Adaptation Evolution Strategy (CMA-ES)}~\cite{hansen2016cma}.

For evaluation purposes, we additionally compute the RMSE on state components other than joint angles, e.g., joint velocities or accelerations, as well as cursor positions, velocities, or accelerations.
The respective metric is defined analogously to~\eqref{eq:rmse}, with $x_{\text{qpos}}^{(s)}$ and $\hat{x}_{\text{qpos}}^{(s)}$ being replaced by the respective quantity.

\subsubsection{Necessary adjustments for other use cases}
The generalized formulation as an OCP is valid for a wide range of interactions between humans and virtual objects.
If the target user group or the interaction technique is varied, one has to modify the relevant parts of the system dynamics as described in Sections~\ref{sec:user_models},~\ref{sec:inputoutputdevice}, and~\ref{sec:interfacedynamics}.
If an interaction task different from pointing is considered, the cost function needs to be adjusted.
For example, in the case of throwing in VR, a cost penalizing the distance of a virtual ball to a target area could replace the distance cost term described above. %
The presented method to obtain user specific cost weights can be used for a variety of cost functions, but joint trajectories from user trials are necessary.
If such data is not available, one can instead use, for example, cursor trajectories instead of joint trajectories in the loss function~\eqref{eq:paramfitting-loss}.

\subsection{Simulating Movements with Model Predictive Control}\label{sec:MPC}
Since the biomechanical simulation alone has highly nonlinear dynamics, we need to solve a \textit{nonlinear} OCP.
This renders it impossible to use solvers for linear OCPs recently introduced to the HCI audience such as LQR~\citep[Ch.~7]{fischer2021optimal} or LQG~\citep[Ch.~8]{fischer2021optimal}. 
Solving nonlinear OCPs is generally quite challenging, and on longer time horizons they are often computationally intractable~\cite{Gruene08}. %
This problem can be tackled with a receding horizon approach, also known as \textit{\textit{Model Predictive Control} (MPC)}.
Due to its notable properties --- 
easy to implement, handles nonlinear constraints in contrast to the Linear Quadratic Regulator~\cite{Dorato71}, theorems guaranteeing that MPC produces sensible results~\cite{GP17,RMD17,FaGM18} ---
MPC has matured into a standard control method for linear and nonlinear dynamical systems, both from the academic and application~\cite{QIN03,VRRFN17} point of view.

\begin{figure}[h!]
	\centering
	\includegraphics[width=\linewidth]{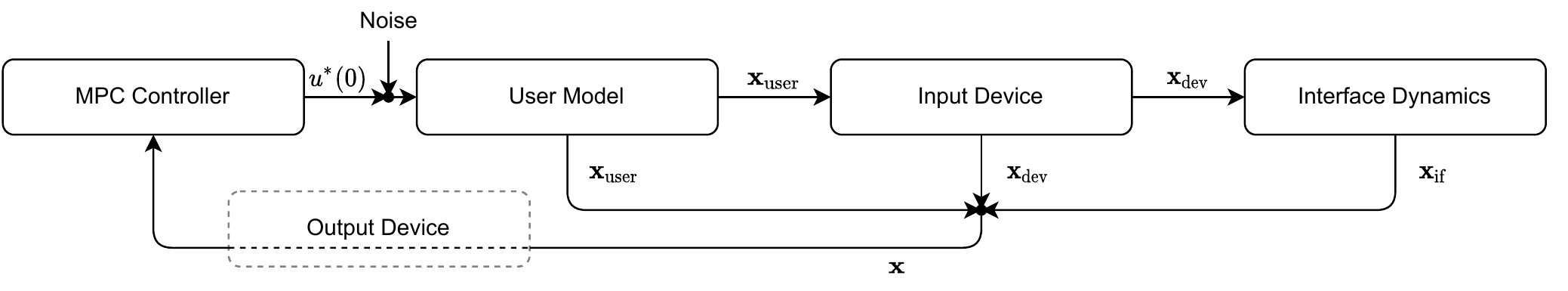}
	\caption{High-level view on the \textit{closed feedback loop}. The MPC Controller generates an optimal control signal $u^\star(0)$ which is perturbed by both constant and signal-dependent noise and then send to the User Model. The User Model updates the biomechanical simulation and yields a new user state $\mathbf{x}_\text{user}$. Based on this new user state, the Input Device then updates the device state $\mathbf{x}_\text{dev}$ and sends information to the Interface Dynamics, yielding the new interface state $\mathbf{x}_\text{if}$.
	All state components are then combined in the overall system state $\mathbf{x}$.
	The MPC Controller receives this updated state (or parts thereof) from the Output Device -- closing the feedback loop.
		}
	\label{fig:closed_loop}
\end{figure}

The main idea of MPC is \textit{complexity reduction} in time.
The solution of the OCP~\eqref{eq:OCP} is approximated by iteratively solving sub-problems of~\eqref{eq:OCP} on a much shorter time horizon. 
The first control of the resulting optimal control sequence %
is then applied to the system.
Iterating this process results in a closed-loop system, which is able to react to perturbations that may occur during execution (e.g., due to signal-dependent noise in the motor system~\cite{faisal2008noise}), without the need to handle them explicitly within each optimization step~\cite{Gruene14}.
The resulting \textit{closed feedback loop} in our framework is depicted in Figure~\ref{fig:closed_loop}.

More formally, MPC computes a \textit{feedback law} $\mu\colon\mathbb{X}\to\mathbb{U}$, which maps arbitrary states $\mathbf{x}\in\mathbb{X}$ to optimal controls $u\in\mathbb{U}$, via the following MPC algorithm:
\begin{enumerate}[topsep=0pt,itemsep=-1ex,partopsep=1ex,parsep=1ex]
	\setcounter{enumi}{-1}
	\item Given the initial state  $\mathbf{x}(0)\in\mathbb{X}$, choose the horizon length parameter $N\geq 2$ and set $n:=0$.
	\item \label{item:mpc-alg-step1} Initialize the state $\mathbf{x}_0=\mathbf{x}(n)$ and solve the following \textit{open-loop}\footnote{We call the OCP~\eqref{eq:OCP_N} open-loop to emphasize that the solution of~\eqref{eq:OCP_N} is not of feedback nature, i.e., cannot react to disturbances. 
	This ability comes from the full MPC algorithm.} 
	optimal control problem (we use the \texttt{scipy.optimize.minimize} from the Python \texttt{scipy} module\footnote{https://docs.scipy.org/doc/}, which implements the Broyden-Fletcher-Goldfarb-Shanno (L-BFGS-B) algorithm~\cite{NoceWrig06,scipybfgs}):
	\begin{equation}\label{eq:OCP_N}
		\begin{aligned}
			&\min_{u(\cdot)\in\mathbb{U}^N} J_N(\mathbf{x}_0,u(\cdot)) = \min_{u(\cdot)\in\mathbb{U}^N} \sum_{k=0}^{N-1}\ell(\mathbf{x}(k),u(k))\\
			\textnormal{such that }& \mathbf{x}(k+1) = f(\mathbf{x}(k),u(k)) \text{ for all }k\in\{0,...,N-1\}, \\ &\mathbf{x}(0) = \mathbf{x}_0, \\
			&\mathbf{x}(k)\in\mathbb{X} \text{ for all }k\in\{0,...,N\}.
		\end{aligned}
	\end{equation}
	Use the first value of the resulting optimal control sequence denoted by $u^\star(\cdot)\in\mathbb{U}^N$ for the feedback law, i.e., set $\mu(\mathbf{x}(n)):=u^\star(0)$.
	\item \label{item:mpc-alg-step2} Update the state via
	\begin{equation}\label{eq:mpc-closed-loop}
		\mathbf{x}(n+1) = f(\mathbf{x}(n),\mu(x(n))),%
	\end{equation}
	set $n:=n+1$ and go to step~\ref{item:mpc-alg-step1}.
\end{enumerate}

We specifically differentiate between~$k$ and~$n$ to distinguish open-loop dynamics ($k$) from closed-loop ones ($n$).
Design parameters include, among others, the sampling times of the state and of the control.
These parameters are hidden in the definition of~$f$, which corresponds to the system dynamics (the MuJoCo simulation in our case), and determine the resolution with which the physics are simulated, and how frequently users are assumed to be able to change their control, respectively.
In order to achieve high physical accuracy, we set the sampling time of the state %
to 2~ms.
To reflect the fact that humans are not able to adjust their behavior continuously, but only intermittently~\cite{Gawthrop2011}, we set the sampling time of the control at 40~ms (i.e., the piecewise constant control signal can be adjusted every 40~ms).

An additional design parameter introduced by the MPC algorithm is the horizon length~$N$. 
Deciding on the horizon means facing a trade-off. 
A longer horizon increases computation time, whereas a shorter horizon may lead to poor results.
For example, if $N$ is chosen too small, the cursor cannot be moved towards the target far enough to effectively reduce the total costs in the truncated horizon, i.e., the optimal control sequence $u^{\star}$ that minimizes the finite-horizon cost functional $J_{N}$ does not result in the expected behavior.
A more detailed analysis of the effect of $N$ on the resulting closed-loop trajectories is presented in Section~\ref{sec:effect_N}.
Unless stated otherwise, we set $N=8$ (i.e., 320 ms), as this value showed a good balance between performance and quality of simulation.

The L-BFGS-B algorithm was chosen as a solver for~\eqref{eq:OCP_N} due to its computation and memory efficiency and ability to include control constraints easily.
The parameters of the L-BFGS-B algorithm, which is used to solve the finite-horizon OCPs at each MPC step, are chosen as follows:
objective function tolerance \texttt{ftol}~$=10^{-6}$, gradient tolerance \texttt{gtol}~$=10^{-5}$, step size for the numerical approximation of the Jacobian \texttt{eps}~$=10^{-8}$, maximum number of objective function evaluations \texttt{maxfun}~$=10000$, and maximum number of iterations \texttt{maxiter}~$=1000$.

Previous findings suggest that human motor control signals are affected by different noise sources, e.g., sensory and motor noise~\cite{faisal2008noise, Sutton67, Schmidt79, harris1998signal, vanBeers03, Todorov05}. %
In order to create realistic human movements that also exhibit intraindividual variance similar to real users, perturbations can be included in the state-transition-map~$f$.
Note that, as the MPC is a closed-loop controller, we do not necessarily need to include the noise during optimization, i.e., the optimizer assumes that the system is deterministic.
Instead, we include noise to the applied control~$\mu(x(n))$ in step~\ref{item:mpc-alg-step1} of the MPC algorithm, i.e., before applying the second-order muscle model and proceeding with the next step.
Applying the noise in the closed loop only considerably simplifies the OCPs and allows them to be solved efficiently. %
As suggested by van Beers et al.~\cite{vanBeers03}, we add signal-dependent and constant motor noise, i.e., two Gaussians with zero mean and a standard deviation of $0.103 \cdot \mu(x(n))$ and $0.185$, respectively, to the control~$\mu(x(n))$.

\section{CFAT: A Method to Compute Maximum Voluntary Torques for Joint-Actuated Models} %
\label{sec:cfat}

Omitting real muscles in biomechanical models and replacing them with simplified muscles acting directly at the joints greatly simplifies computations, but it also creates another challenge.
It is unclear how strong these simplified muscles need to be. 
Since the relative strength of each actuator has a large impact on how it needs to be actuated~\cite{jiang2019synthesis, Yu18}, an appropriate choice of the \textit{maximum voluntary torques} is crucial to generate biomechanically plausible movements.
We therefore need to define the torque ranges of all actuators, i.e., the maximal positive and negative torques that can be applied at each DOF.

The natural approach to identify the torques humans apply during interaction would be to use existing Inverse Dynamics tools, as implemented in OpenSim.
However, such tools obtain the complete \textit{inter-segmental} torques acting on both independent and dependent joints, including passive forces, e.g., due to spring-dampers.
In addition, the dependent joints cannot be actively actuated, but their torques emerge implicitly from the torques applied to the independent joints, i.e., %
the results from Inverse Dynamics cannot be used to determine the maximum voluntary torques at the independent joints\footnote{A comparison of applied torques obtained from Inverse Dynamics and CFAT can be found in the Appendix~\ref{app:cfat}.}.
Instead of relying on Inverse Dynamics, we thus apply a method similar to \textit{Computed Muscle Control (CMC)}~\cite{Thelen06}, which yields the sequence of muscle excitations that accounts for experimentally observed movements, given a fully muscle-actuated biomechanical model.

Starting with an initial posture from experimental data, the goal of our method to \textit{compute feasible applied torques (CFAT)} %
is to find the sequence of applied torques that best explains the sequence of joint postures observed during an experiment.
Due to the curse of dimensionality, we solve a sequence of optimization problems, one for each time step, as opposed to an optimization problem covering the entire motion, minimizing the following loss function: %
\begin{equation}\label{eq:CFC_loss}
	\mathcal{L}_{\text{CFAT}}(\tau) = \alpha e_{\text{qpos}}(\tau) + \beta e_{\text{qvel}}(\tau) + \gamma e_{\text{qacc}}(\tau).
\end{equation}
Here, the error terms $e_{\text{qpos}}(\tau)$, $e_{\text{qvel}}(\tau)$, and $e_{\text{qacc}}(\tau)$ denote the Euclidean distance between the one-step MuJoCo forward simulation with applied torques $\tau$ and the corresponding user data at this time step, in terms of joint angles, velocities, and accelerations, respectively (only incorporating the independent joints).
According to our experience, penalizing an appropriate combination of joint angles, velocities, and accelerations turned out to be necessary to guarantee stability -- choosing the weights $\alpha = 1000$, $\beta = 50$, and $\gamma = 0.01$ showed good results in our case.
After each optimization, one forward step is taken in the MuJoCo environment using the computed optimal torque. %
The resulting joint angles and velocities in the next time step are then used as initial values for the subsequent optimization, which returns the next optimal torques, and so on.
Using this CFAT tool, we thus obtain a sequence of \textit{applied torques} that result in the original user trajectory when sequentially applied at the DOFs of the biomechanical model.
Additionally, for each trial, CFAT yields the initial activations $\sigma_0$ and their derivatives $\dot{\sigma}_0$ used in our muscle model described in Section~\ref{sec:musclemodel}. 

We clean the obtained torques from outliers by removing those that deviate more than three standard deviations from the respective mean. 
The vectors of maximum positive and negative torques, $\tau^{+}$ and $\tau^{-}$, are then determined as the component-wise maximum and minimum of the computed torques $\tau$ of all considered movements.

For technical reasons, the maximum and minimum torques are normalized such that the larger of both equals one for each DOF, and the resulting values are used as boundaries for the control $u$.\footnote{Note that, because we use the muscle model described in~\ref{sec:musclemodel}, the applied controls $u$ might differ from the activation $x_{\sigma}$. However, given that $\Delta t < \sqrt{t_e t_a}$ holds for the second-order muscle dynamics~\eqref{eq:activationmodel}, the activation $x_{\sigma}$ cannot exceed the applied controls $u$ in absolute terms. It is thus reasonable to impose the normalized torque boundaries on $u$ instead of $x_{\sigma}$.} %
The positive \textit{scaling ratio} vector $g = \max(\left|\tau^{-}\right|, \left|\tau^{+}\right|)$, with maximum taken component-wise, is then used as a gain vector, mapping the normalized activations $x_{\sigma}$ to the applied torques $\tau$ as in~\eqref{eq:activation2torque_DOF}. %

It should be noted that the CFAT tool requires reference user data to measure how ``human-like'' a simulated joint trajectory is. 
In this work, we used the data from our user study, which was explicitly recorded for the considered interaction task.
However, the obtained torque ranges should be appropriate for related interaction techniques and tasks as well, as was recently shown for the case of mid-air keyboard typing~\cite{Hetzel21}.
If major changes to the user model are made, such as modifying the physiology, running CFAT on new reference data is recommended.

\section{Use Case: ISO Pointing in VR}\label{sec:use-case}
As a use case, we demonstrate the applicability of our simulation framework to mid-air pointing in VR.
In Section~\ref{sec:usecase-definitions}, we describe the considered task and techniques. 
In Sections~\ref{sec:userstudy} and~\ref{sec:derived-user-models}, we proceed with a description of the user study that we conducted to collect data for the user model generation and the evaluation of our simulation.
Finally, in Section~\ref{sec:study-based-simulation}, we explain how our approach can be used to replicate individual trials of the user study.

\subsection{Target User Group, Interaction Techniques, and Interaction Task}
\label{sec:usecase-definitions}
Our \textit{target user group} includes healthy adults of average size and body shape.
Therefore, we do not need to make special adjustments to the biomechanical user model.
Nonetheless, since we did not have user models beforehand and aim to compare our simulation trajectories to those obtained from our user study, we derive user models that match the biomechanical properties of the participants in the user study, as described in Section~\ref{sec:derived-user-models}.
Technically, our target user group thus corresponds to those six participants (see Section~\ref{sec:user-study-participants}). %

We are interested in how well our model can synthesize human movement given different \textit{interaction techniques}.
As input device we use a motion capturing system, which tracks the position of an LED marker that is placed on the tip of the right index finger, modeled in MuJoCo as described in Section~\ref{sec:inputoutputdevice}.
Using the notation introduced in Section~\ref{sec:interfacedynamics}, we investigate transfer functions without any additional virtual dynamics. %

For each of the two interaction technique classes \textit{Virtual Cursor}~\eqref{eqn:virtual_cursor} and \textit{Virtual Pad}~\eqref{eqn:virtual_pad}, we define a basic variant in which the input space is at the same position as the output space, i.e., $\omega_I = \omega_O$. %
For the virtual cursor, this means that the cursor always matches the position of the fingertip (i.e., the transfer function is the identity function), and for the virtual pad, the cursor is the orthogonal projection of the fingertip onto the input/output plane.
Therefore, we refer to these techniques as \textit{Virtual Cursor Identity/ID} and \textit{Virtual Pad Identity/ID}, respectively.
For both classes, we also consider an ``ergonomic'' condition, where the input space is at a lower, more comfortable height\footnote{In a small preliminary study, we tried different input options for both techniques, and the variants we consider here proved suitable to reach all targets comfortably.}, denoted as \textit{Virtual Cursor Ergonomic} and \textit{Virtual Pad Ergonomic} in the following.
The input and output normal vectors $n_{\text{I}}$ and $n_{\text{O}}$, respectively, are selected in such a way that the planes face the user and coincide for both interaction techniques.
Details on the input and output spaces of all considered techniques are given in Table~\ref{tbl:input-origins}.

\begin{table}[htb]
	\centering
	\begin{tabular}{|l|c|c|}
		\hline
		Technique & Input Origin (relative to shoulder) & Input Normal Vector\\
		\hline \hline
		Virtual Cursor Identity & $\left(-0.1~\text{m}, 0.0~\text{m}, 0.55~\text{m}\right)$ & --	\\
		Virtual Cursor Ergonomic & $\left(-0.1~\text{m}, -0.4~\text{m}, 0.45~\text{m}\right)$ & -- \\
		Virtual Pad Identity &  $\left(-0.1~\text{m}, 0.0~\text{m}, 0.55~\text{m}\right)$ & $\left(0, 0, -1\right)$ \\
		Virtual Pad Ergonomic & $\left(-0.1~\text{m}, -0.3~\text{m}, 0.55~\text{m}\right)$ & $\left(0, 0, -1\right)$\\ \hline
		\end{tabular}
		\caption{Interaction techniques used in the user study and simulations. All parameters are given in coordinates with respect to the right shoulder. The output origin is fixed at $\left( -0.1~\text{m}, 0.0~\text{m}, 0.55~\text{m} \right)$, and the output normal vector is given as $n_{\text{O}}=(0, 0, -1)$.}
	\label{tbl:input-origins}
\end{table}

\begin{figure}
	\begin{center}
		\subfloat{\includegraphics[width=0.5\linewidth, clip]{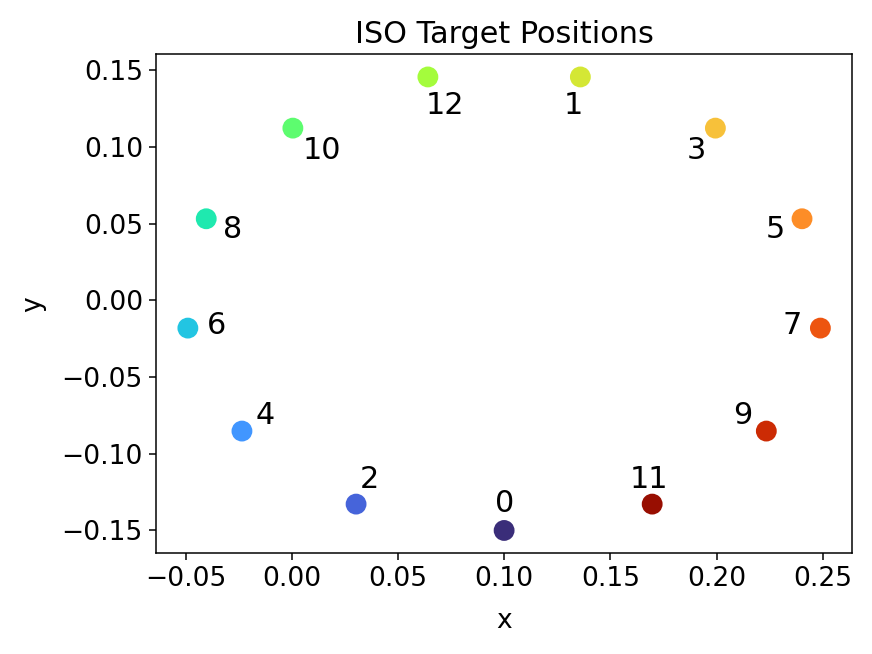}}%
		\caption{The thirteen targets of the ISO 9241-9 standard pointing task, which are displayed separately in ascending order. The task is to move the cursor towards the active target and keep it inside until the next target is shown.}
		\label{fig:isotargets}
	\end{center}
\end{figure}

\begin{figure}
	\centering
	\begin{tabular}{cc}
		\adjustbox{valign=b}{\begin{tabular}{@{}c@{}}
				\subfloat[User study\label{fig:user_study}]{%
					\includegraphics[width=.344\linewidth]{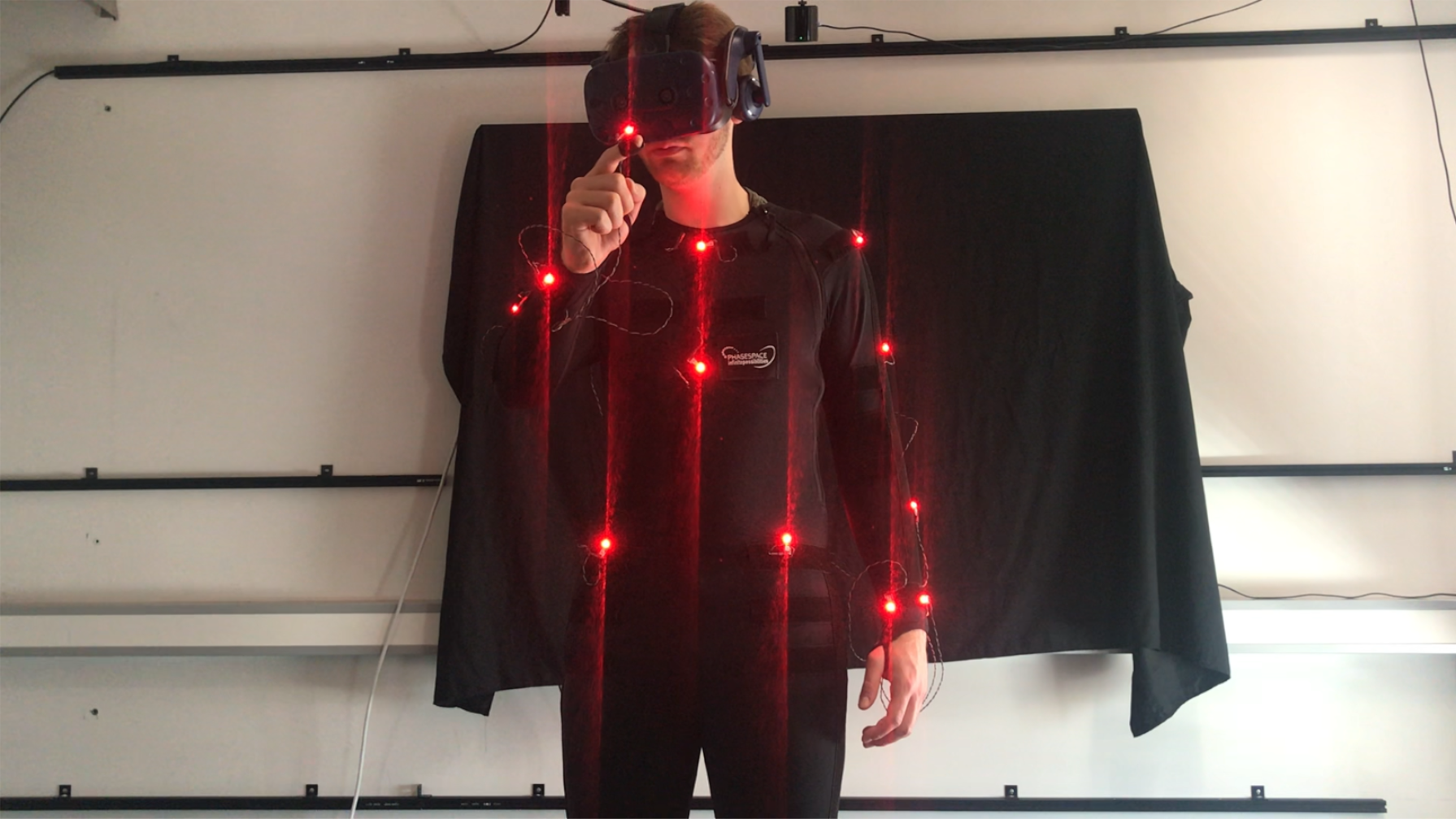}}  \\ 
				\subfloat[VR scene\label{fig:vr_scene}]{%
					\includegraphics[width=.344\linewidth]{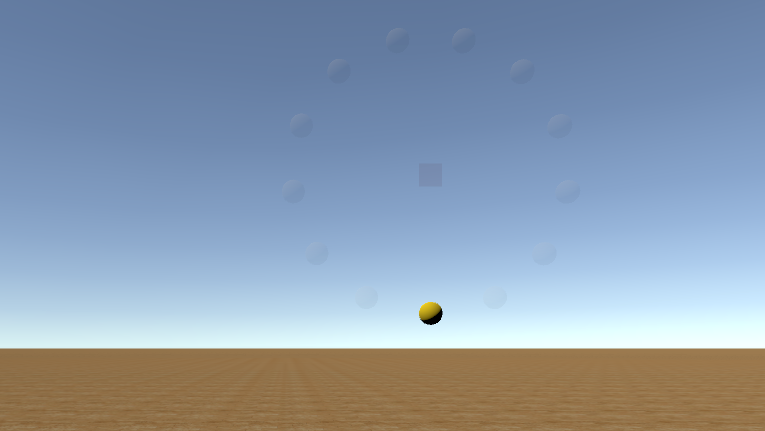}}
		\end{tabular}}
		&   
		\adjustbox{valign=b}{\subfloat[MuJoCo Simulation\label{fig:simulation}]{%
				\includegraphics[width=.615\linewidth]{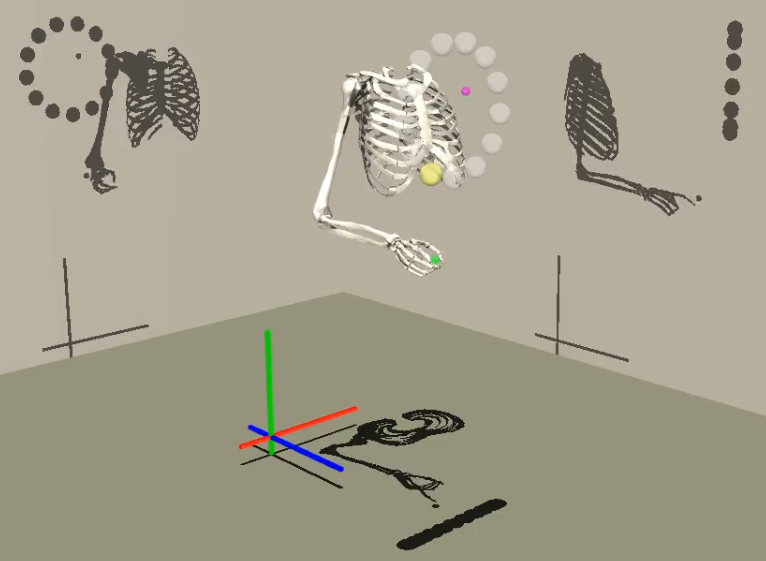}}}   
		
	\end{tabular}
	\caption{\textbf{(a)} Image showing the user study. Motion is captured by tracking the movements of the LEDs. \textbf{(b)} The VR scene, which is perceived via a head-mounted display, shows the targets spheres (active: yellow; inactive: gray). \textbf{(c)} The MuJoCo simulation of the Virtual Cursor Ergonomic technique. The xyz-axes are colored in red, green, and blue, respectively. Important objects are displayed as follows. Green sphere: The virtual marker placed at the (physical) end-effector, i.e., the tip of the right index finger. Purple sphere: The virtual cursor (after applying the considered transfer function). Yellow sphere: active target. Gray spheres: inactive targets. The shadows show the orthogonal projections of the model for each dimension.}
	\label{fig:simulation_model_complete}
\end{figure}

Our \textit{interaction task} is based on the discrete Fitts' Law paradigm, following the ISO 9241-9 standard.
13 targets with a diameter of 5~cm were placed on a circle of 30~cm diameter, resulting in an index of difficulty of 2.8 bits (cf.~Figure~\ref{fig:isotargets}).
The center of the circle is placed 55 cm in front and 10 cm to the right of the right shoulder.
We chose this placement, since most interactions with the right hand take place on the right side of the body.
In each trial, the task is to move the virtual cursor as quickly and accurately as possible towards the active target, which is represented as a yellow sphere with a diameter of 5~cm (cf.~Figure~\ref{fig:vr_scene}), and then hold within the target. 
As soon as the cursor reaches the target (with a velocity lower than 0.5 m/s to avoid early termination in case of overshoot), the next target according to the ISO 9241-9 standard is displayed after 500 ms. %

\subsection{User Study}
\label{sec:userstudy}

We ran a user study for several reasons.
First, the obtained experimental data can be used to create user-specific variants of the default biomechanical model introduced in Section~\ref{sec:user_models}.
For example, in Section~\ref{sec:cfat}, we introduce CFAT as a tool to identify the maximum voluntary torques at each DOF, given experimentally observed user trajectories.
Second, having user data allows to evaluate the quality and realism of simulated movements against observed human motion. 
In particular, it can be used as reference data to compare simulations for different cost functions and weights, which allows to identify the cost function parameters that best replicate observed behavior (see Section~\ref{sec:fittingcostweights}). %

We therefore asked participants to perform the task described above, using the presented interaction techniques.

\subsubsection{Participants}
\label{sec:user-study-participants}

We recruited 6 participants (Mean Age=28.8, SD=6.6, 4 Male, all right-handed) from our local university campus for the study.
Half of the participants had previous experience of interaction in VR, and no participants suffered from perceptual or neuromotor impairments.
In the following, we refer to the different users as U1,..., U6.
All four interaction techniques are varied within subjects.

\subsubsection{Apparatus and Procedure}
We used a Phasespace X2E\footnote{https://www.phasespace.com/x2e-motion-capture/} motion capture system with a full-body suit to track the participants' movements at 240Hz.
The movements of the upper extremity and torso were continuously tracked by 14 optical markers placed at anatomical landmarks.
Participants were immersed in Virtual Reality using a HTC Vive Pro VR headset\footnote{https://www.vive.com/de/product/vive-pro/}. %
The setup is shown in Figure~\ref{fig:user_study}.
The VR scene and experimental setup were implemented in Unity3D\footnote{https://unity.com} using the SteamVR plugin\footnote{https://valvesoftware.github.io/steamvr\_unity\_plugin/} (cf.~Figure~\ref{fig:vr_scene}). 
We aligned the coordinate systems of Phasespace and Unity as follows.
We placed a Phasespace marker at the origin of a HTC Vive Pro VR controller.
We then performed wanding of the interaction space using this controller, creating a set of 3D point pairs in both coordinate systems.
We calculated a rigid transform between both coordinate systems using translation between the centroids to compute the translation component of the transformation, and the singular value decomposition to compute the rotation between the Phasespace and SteamVR coordinates~\cite{sorkine2017least}.

Participants interacted with the VR scene using an end-effector marker placed at the tip of their right index finger.
The movements are tracked in the Phasespace coordinate system.
The cursor and target positions were only converted to the VR coordinate system right before the visualization.
During the experiment, we logged the motion capture data and the experimental meta-data, as well as the timestamps at which the targets were hit.

Participants were informed about the ISO pointing task described in Section~\ref{sec:usecase-definitions}.
Since we were interested in arm-only movements, participants were also instructed to only move their arm, while keeping the rest of the body as still as possible.
This is important because the torso in our biomechanical model cannot move.
Substantial torso movements would therefore distort the comparison between user and simulation.

Since the main objective of the user study was to collect movement data for different interaction techniques, we only tested a single index of difficulty to limit the impact of fatigue. %
After recording a T-pose for model scaling, participants put on the HMD and performed several movements for each interaction technique. 
During a warm-up phase, each interaction technique was trained for at least 30 movements.
Afterwards, all participants performed the complete ISO task consisting of 13 subsequently shown targets 5 times per interaction technique, resulting in 65 movements per interaction technique and user, or 1560 movements in total.
In order to reduce fatigue, participants were asked to take a break of one minute between interaction techniques.

\subsubsection{Processing Data and Inverse Kinematics}
\label{sec:preprocessing-inversekinematics}
The raw motion capture data is preprocessed %
according to the common conventions for biomechanical analyses~\cite{bachynskyi2014}.
The marker data is first cleaned from artifacts caused by marker occlusions and reflections based on the condition values delivered by the motion capture system, and then by filtering out outliers (i.e., segments with a difference of more than four standard deviations from the mean). %
The resulting gaps in the data are linearly interpolated, while keeping track of the gaps.
Afterwards, the data is smoothed using a Kalman filter~\cite{Kalman}, and divided into individual aimed movements using the target switch times from the experiment.

We then run the OpenSim Inverse Kinematics (IK) tool for each movement of any considered participant and interaction technique individually. 
This tool computes the joint angles for each frame of motion capture data through solving an optimization problem. 
To this end, it applies the kinematic constraints and freely modifies the independent joint coordinates of the model to minimize the IK loss function, which is the weighted sum of squared distances between all virtual and the corresponding experimental markers. 
We use a larger weight for the end-effector marker than for the other markers, as it is critical to the considered pointing task to track the end-effector as accurately as possible.

In the experimental data, the time spans between target switch and movement onset differ substantially between trials. %
Since we are not interested in modeling reaction times, we decided to remove these frames from user data. %
To this end, we determine movement onset as the time at which the acceleration of the cursor reaches 1~$\text{m}/\text{s}^2$ for the first time.
We also removed trials that started too early (i.e., the cursor left the previous target before the new target appeared), and movements of exceptional length (i.e., the movement duration deviated more than three standard deviations from the average duration for the considered participant and interaction technique) from the dataset.
In total, 158 out of 1560 recorded trials were removed, which is equivalent to $10.1\%$.\footnote{114 of these trials are due to participants 2 and 5 occasionally starting their movements before the target switch.}

Note that we have different time scales in simulation ($2$~ms) and data ($1/240$~s $\approx 4.17$~ms).
To be able to compare user and simulation trajectories on a moment-by-moment basis, we therefore align the two time series by applying linear interpolation on the user data.

\subsection{Customized Models}		
\label{sec:derived-user-models}
In the following, we explain how the generic user model described in Section~\ref{sec:user_models} is adjusted, both in terms of its biomechanical properties and in terms of the cost weights, which determine the trade-off between the constituents of the cost functions introduced in Section~\ref{sec:cost-function}.

First, we scale the models to match the kinematic and inertial properties of each participant of our user study using the OpenSim scaling tool. 
This tool computes ratios between pairs of markers recorded for a static posture in the experiment and the corresponding virtual markers attached to the model, only using the markers attached at the anatomical landmarks. 
These ratios are then used to scale the respective body segments.
We ensure good quality of model scaling and marker adjustment by visually inspecting the resulting models with respect to experimental data.
The scaling is then transferred from the OpenSim model to the MuJoCo model. 

In addition, we adjust the joint limits to include all joint angles corresponding to the movement data of the respective participant.
This is necessary because joint ranges are enforced in MuJoCo only via ``soft'' constraints, that is, high opponent forces are applied to postures outside the permissible region, which would reduce the reliability of the CFAT tool described in Section~\ref{sec:cfat}.
However, it is important to note that the joint angles obtained from Inverse Kinematics (see Section~\ref{sec:preprocessing-inversekinematics}) are inherently dependent on the joint boundaries from the original OpenSim model (which can be found in the Appendix~\ref{tab:joint-limits}).
This makes large deviations very unlikely.\footnote{Indeed, all user-specific joint limits were within a range of $\pm5$ degrees around the default model values.}%

After scaling, we obtain the maximum voluntary torques for each user by running CFAT for all available movements.
An overview of the computed maximum and minimum torques is given in Table~\ref{tab:max-torques}.
We use the initial activations $\sigma_0$ and their derivatives $\dot{\sigma}_0$ also obtained by CFAT as valid initial values for the muscle dynamics used in our simulations as described in Section~\ref{sec:musclemodel}.

\begin{table}[!ht]
	\centering
	\resizebox{.99\textwidth}{!}{%
		\begin{tabular}{|c|c|c|c|c|c|c|c|c|c|c|c|c|} 
			\hline
			\rule{0pt}{10pt}\noindent
			\multirow{3}{*}{Joint} & \multicolumn{12}{c|}{Torque Ranges (Nm)} \\
			\cline{2-13}
			\rule{0pt}{10pt}\noindent
			& \multicolumn{2}{c|}{U1} & \multicolumn{2}{c|}{U2} & \multicolumn{2}{c|}{U3} & \multicolumn{2}{c|}{U4} & \multicolumn{2}{c|}{U5} & \multicolumn{2}{c|}{U6} \\
			\cline{2-13}
			\rule{0pt}{10pt}\noindent
			& $\tau^-$ & $\tau^+$ & $\tau^-$ & $\tau^+$ & $\tau^-$ & $\tau^+$ & $\tau^-$ & $\tau^+$ & $\tau^-$ & $\tau^+$ & $\tau^-$ & $\tau^+$ \\
			\hline \hline
			\rule{0pt}{10pt}\noindent
			EA & $-12.74$ & $\mathbf{16.12}$ & $-22.08$ & $\mathbf{26.12}$ & $-14.92$ & $\mathbf{19.20}$ & $-14.33$ & $\mathbf{18.38}$ & $-10.99$ & $\mathbf{15.16}$ & $-21.64$ & $\mathbf{26.73}$ \\
			\rule{0pt}{10pt}\noindent
			SE & $-8.61$ & $\mathbf{20.43}$ & $-6.91$ & $\mathbf{18.36}$ & $-9.19$ & $\mathbf{20.92}$ & $-7.07$ & $\mathbf{15.49}$ & $-4.66$ & $\mathbf{17.08}$ & $-10.05$ & $\mathbf{17.82}$ \\
			\rule{0pt}{10pt}\noindent
			SR & $-\mathbf{3.35}$ & $0.70$ & $-\mathbf{4.28}$ & $1.00$ & $-\mathbf{3.88}$ & $0.71$ & $-\mathbf{4.03}$ & $0.98$ & $-\mathbf{3.54}$ & $1.37$ & $-\mathbf{5.11}$ & $2.41$ \\
			\rule{0pt}{10pt}\noindent
			EF & $0.25$ & $\mathbf{5.08}$ & $-0.17$ & $\mathbf{5.36}$ & $0.21$ & $\mathbf{5.88}$ & $0.48$ & $\mathbf{5.54}$ & $0.42$ & $\mathbf{4.81}$ & $-0.92$ & $\mathbf{6.42}$ \\
			\rule{0pt}{10pt}\noindent
			PS & $-\mathbf{1.82}$ & $1.71$ & $-\mathbf{1.36}$ & $1.15$ & $-\mathbf{3.06}$ & $2.73$ & $-\mathbf{0.81}$ & $0.58$ & $-\mathbf{4.01}$ & $3.68$ & $-\mathbf{1.42}$ & $1.14$ \\
			\rule{0pt}{10pt}\noindent
			WD & $-\mathbf{2.11}$ & $2.00$ & $-\mathbf{1.60}$ & $1.35$ & $-\mathbf{1.98}$ & $1.72$ & $-\mathbf{0.95}$ & $0.57$ & $-\mathbf{1.87}$ & $1.64$ & $-\mathbf{1.36}$ & $1.07$ \\
			\rule{0pt}{10pt}\noindent
			WF & $-\mathbf{1.86}$ & $0.78$ & $-\mathbf{1.52}$ & $0.72$ & $-\mathbf{1.76}$ & $0.71$ & $-\mathbf{1.24}$ & $0.41$ & $-\mathbf{1.77}$ & $1.02$ & $-\mathbf{1.36}$ & $0.43$ \\
			\hline
		\end{tabular}%
	}
	\caption{Joint %
		torque ranges obtained through CFAT (that is, components of $\tau^{-}$ and $\tau^{+}$, respectively) of the joints that are directly actuated.
		The unsigned bold values are used as respective scaling ratios $g$.
		(EA: Shoulder elevation angle; SE: Shoulder elevation; SR: Shoulder rotation; EF: Elbow flexion; PS: Pronation/Supination; WD: Wrist deviation; WF: Wrist flexion.)
	}
	\label{tab:max-torques} 
\end{table}

To obtain reasonable cost weights, we perform parameter fitting as described in Section~\ref{sec:fittingcostweights} for each user and interaction technique.
That is, we identify cost weights i.e., user strategies, that best explain observed user behavior, both in terms of general behavior and intraindividual variance.
This is in contrast to previous approaches, where parameters were fitted to replicate a single user trajectory~\cite{muller2017control,fischer2021optimal}. 
To ensure computational efficiency, we create simulation trajectories for five different movement directions from the ISO task, and compute the RMSE in terms of joint angles (cf.~Equation~\eqref{eq:rmse}) between each simulation trajectory and the respective reference user trajectory.
The loss function used for the cost weight fitting is thus given by Equation~\eqref{eq:paramfitting-loss} with $S=5$.
As described in Section~\ref{sec:fittingcostweights}, we use CMA-ES as a derivative-free solver.
We omit motor noise during the parameter fitting, since the resulting stochastic outcome for a given set of parameters would considerably complicate the parameter search.

In cases where the optimization did not converge, we ran CMA-ES for 24 hours for each setup and took the parameter set with the lowest RMSE.
The resulting cost weights for each user and interaction technique are listed in Table~\ref{tab:cost_parameters} in the Appendix.

\subsection{Simulation}\label{sec:study-based-simulation}
Our method cannot only be used to replicate existing movements, but also to predict movements in arbitrary conditions (i.e., for different interaction techniques, tasks, and user models). %
To evaluate the performance of our approach, however, we need to simulate movements with the same ``prerequisites'' as the users in the study we are comparing to.
This includes the kinematic and inertial properties of the body as well as its initial joint configuration, which should coincide between simulation and user study.

Each aimed movement that was carried out in the user study is simulated separately.
That is, for a given \textit{reference user trajectory} (also referred to as \textit{Baseline U1/.../U6}) we generate a corresponding simulation trajectory using the corresponding user model as described in Section~\ref{sec:derived-user-models} and the same interaction technique that was used in the study, i.e., we use the user-specifically scaled MuJoCo model and relevant cost weights.
The simulation is shown in Figure~\ref{fig:simulation}, where our model performs a task with the Virtual Cursor Ergonomic interaction technique.

To ensure a fair comparison, we then set the torso position and orientation to that of the participant at movement onset.
As mentioned above, the torso is fixed during simulation.
Next, we set the initial state (including joint angles and velocities, aggregated muscle activations and their derivatives, and the virtual state of the interface, i.e., the cursor position), %
to the initial values of the reference user trajectory.
We then synthesize the aimed movement using our MPC method.
We want to emphasize that our simulation does \textit{not} explicitly depend on the duration of the corresponding user movement.
Instead, the receding time horizon approach allows to simulate arbitrarily long movements.
Since comparing the resulting trajectories to that of the user study requires them to have equal length, we need to adjust the simulation trajectory to match the movement time of the participant in the particular trial. %
Therefore, the simulation stops when the movement time of the respective trial is reached.

This simulation is performed for all trials that passed the preprocessing, resulting in a total of $1402$ simulation trajectories.

\section{Results}\label{sec:results}
In the following, we compare the ISO task trajectories resulting from our simulation to those observed during the user study described in Section~\ref{sec:userstudy}.

In Section~\ref{sec:comparison-cost-functions}, we first compare the three proposed cost functions regarding their ability to replicate and predict human movement trajectories.
Using the Joint Acceleration Costs (JAC), which turn out to be most suitable for simulating human pointing movements, we show in Section~\ref{sec:simulation-vs-user} that our simulation predicts user trajectories with an accuracy that is comparable to or even better than between-user comparisons, while making use of biomechanically plausible joint postures. %
In Section~\ref{sec:effect_costweights}, we show that the predicted trajectories continuously depend on the choice of the cost weights $r_{1}$ and $r_{2}$, paving the road to creating new simulated users, ``tailored'' to some desired movement characteristics such as speed.
Finally, in Section~\ref{sec:effect_N} we discuss the effect of the MPC horizon~$N$ and provide some general thumb rule on how to choose this hyperparameter.

For qualitative evaluation, we mainly focus on the following six quantities: cursor position and velocity time series, which are orthogonally projected onto the direct path between initial and target position, %
as well as joint angles and velocities for both shoulder rotation and elbow flexion, as these are two of the most impactful joints for the considered mid-air movements.
The angle and velocity plots of the five remaining joints are shown in the Appendix~\ref{sec:supplementary-material}. 

\subsection{Comparison of Cost Functions: Joint Acceleration Costs best predict Human Motion}\label{sec:comparison-cost-functions}

As described in Section~\ref{sec:cost-function}, we use the following stage costs to simulate human movement in the ISO pointing task:
\begin{itemize}
	\item \textbf{DC}: Distance and Control Costs~\eqref{eq:costs-dc},
	\item \textbf{CTC}: Distance, Control, and Commanded Torque Change Cost~\eqref{eq:costs-ctc},
	\item \textbf{JAC}: Distance, Control, and Joint Acceleration Costs~\eqref{eq:costs-jointacc}.
\end{itemize}

For each cost function, participant, and interaction technique, the respective cost weights $r_1$ (weight for control costs) and $r_2$ (weight for commanded torque change or joint acceleration costs) are optimized to match joint angles between simulation and user data, as described in Section~\ref{sec:derived-user-models}.
The resulting parameter values are shown in Table~\ref{tab:cost_parameters} in the Appendix. %
We evaluate the accuracy of our simulations in terms of predicted cursor and joint trajectories, both qualitatively and quantitatively.

\begin{figure}[h!]
	\centering
	\subfloat{\includegraphics[width=0.33\linewidth, clip]{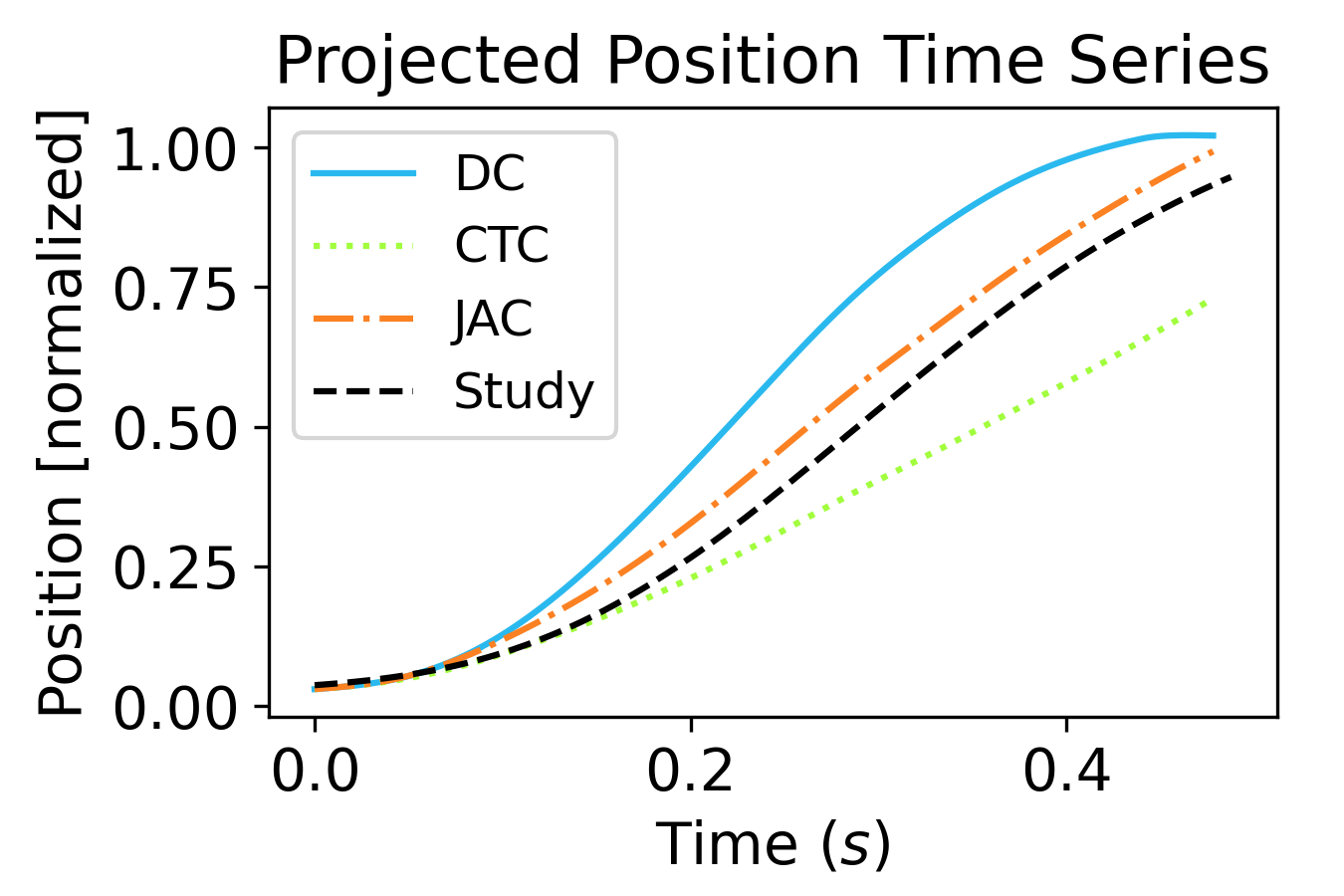}}
	\subfloat{\includegraphics[width=0.33\linewidth, clip]{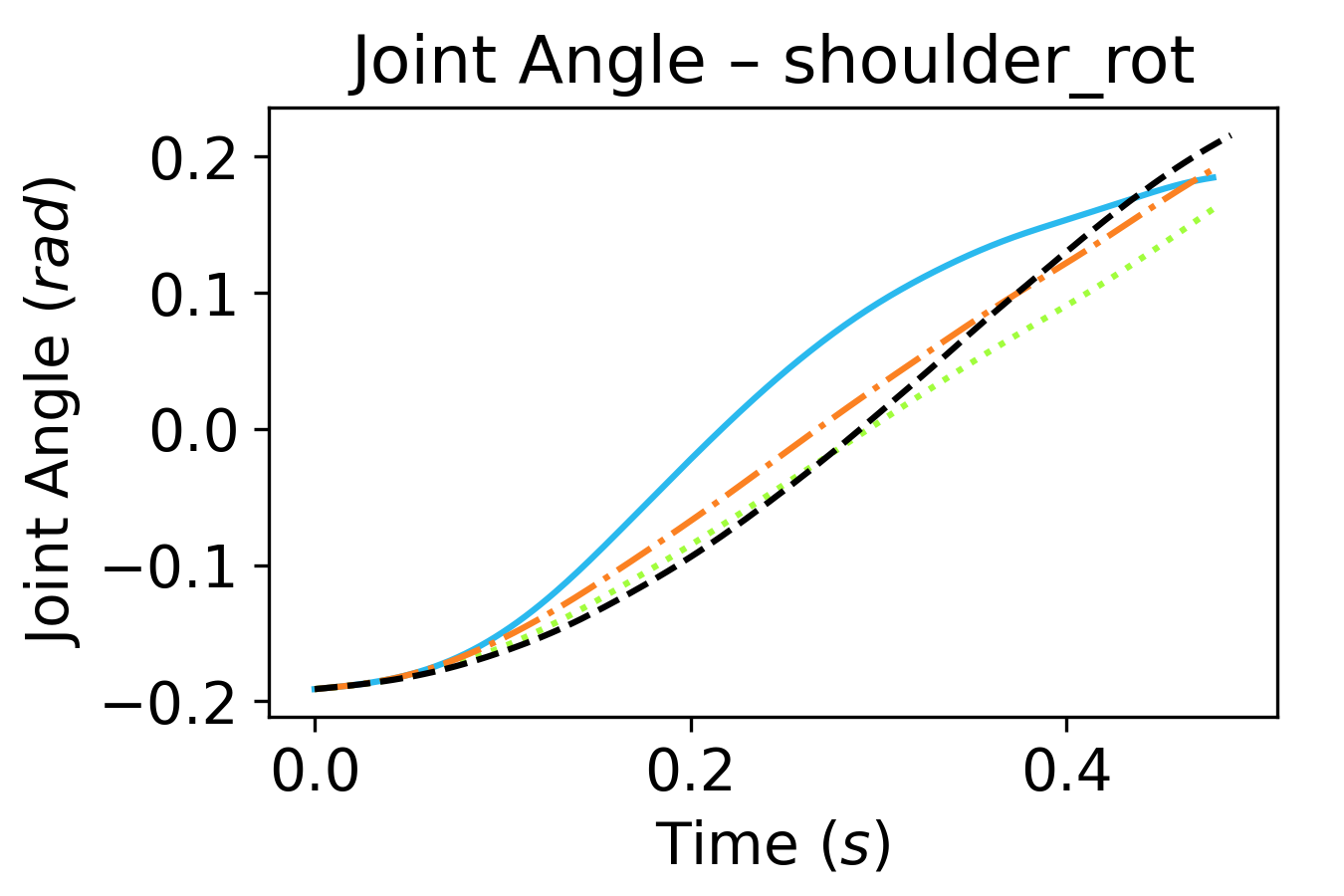}}
	\subfloat{\includegraphics[width=0.33\linewidth, clip]{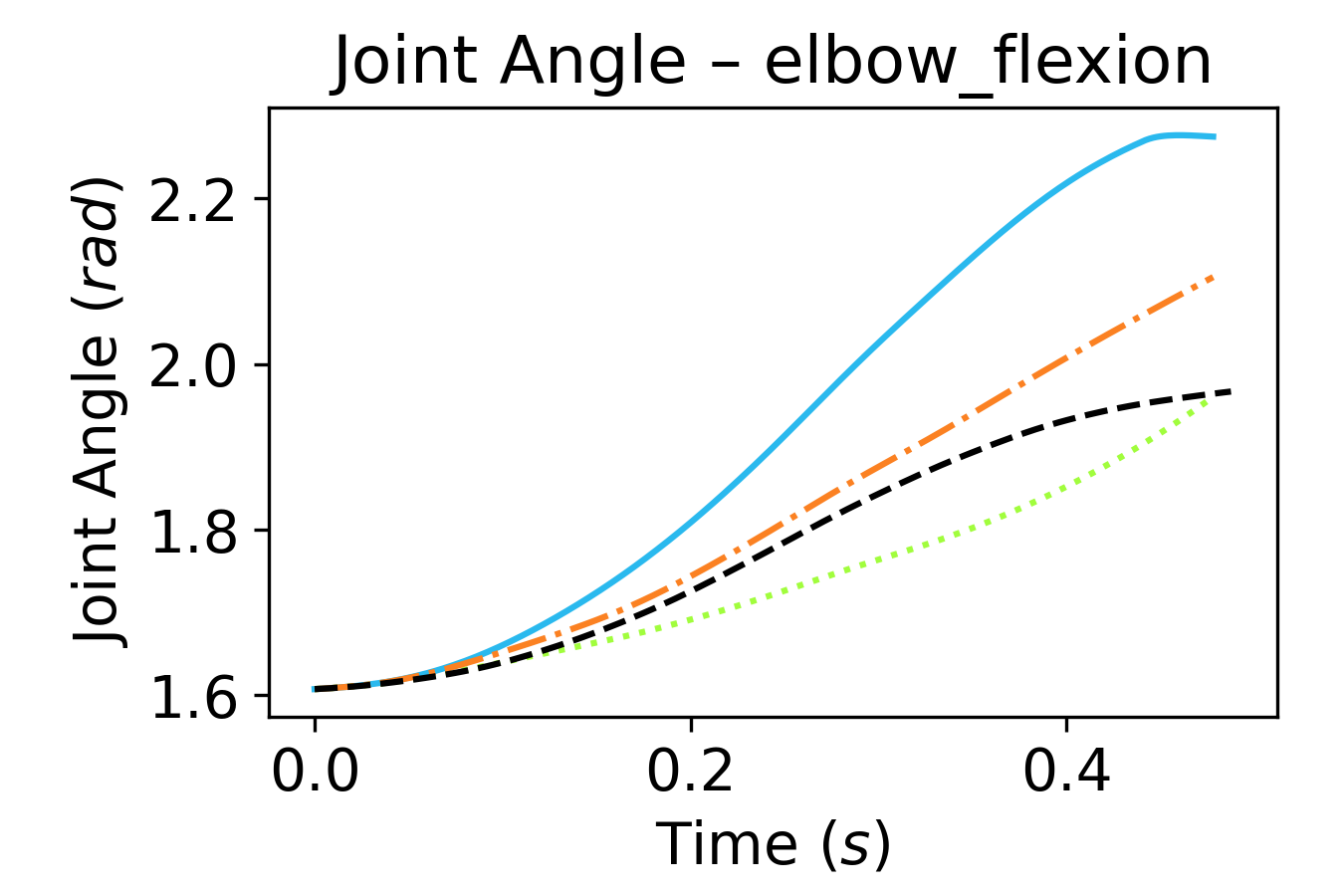}}\\
	\subfloat{\includegraphics[width=0.33\linewidth, clip]{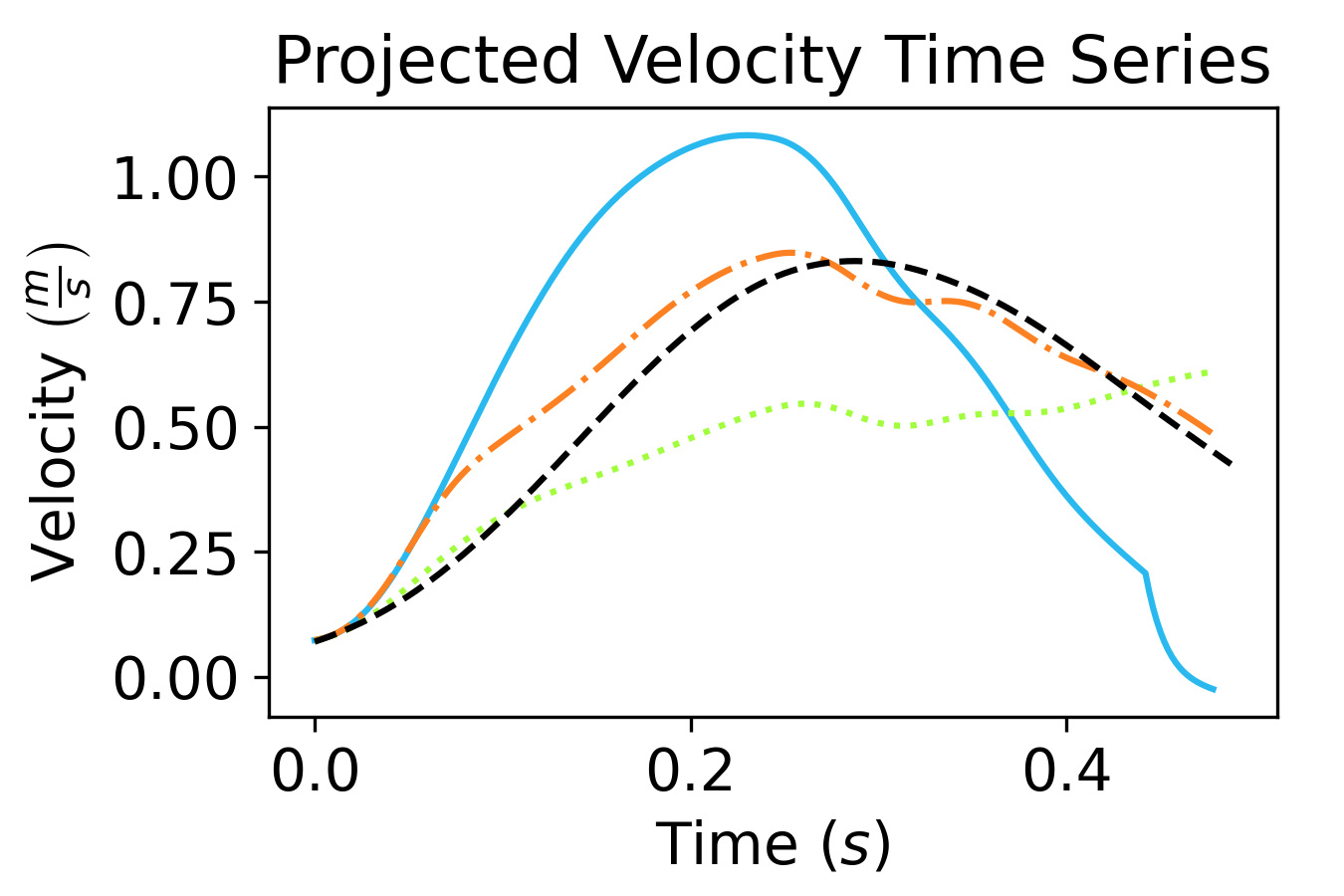}}
	\subfloat{\includegraphics[width=0.33\linewidth, clip]{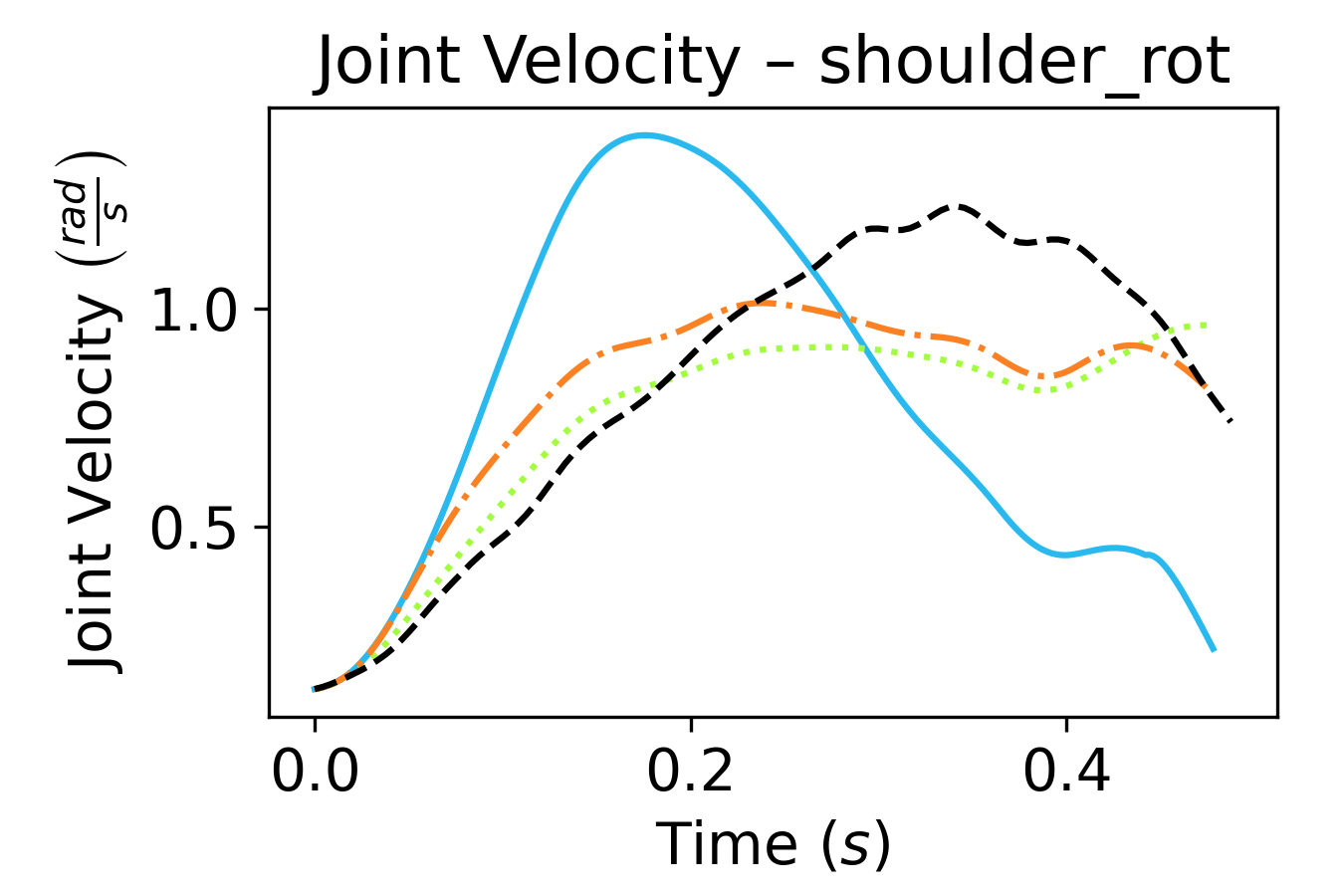}}
	\subfloat{\includegraphics[width=0.33\linewidth, clip]{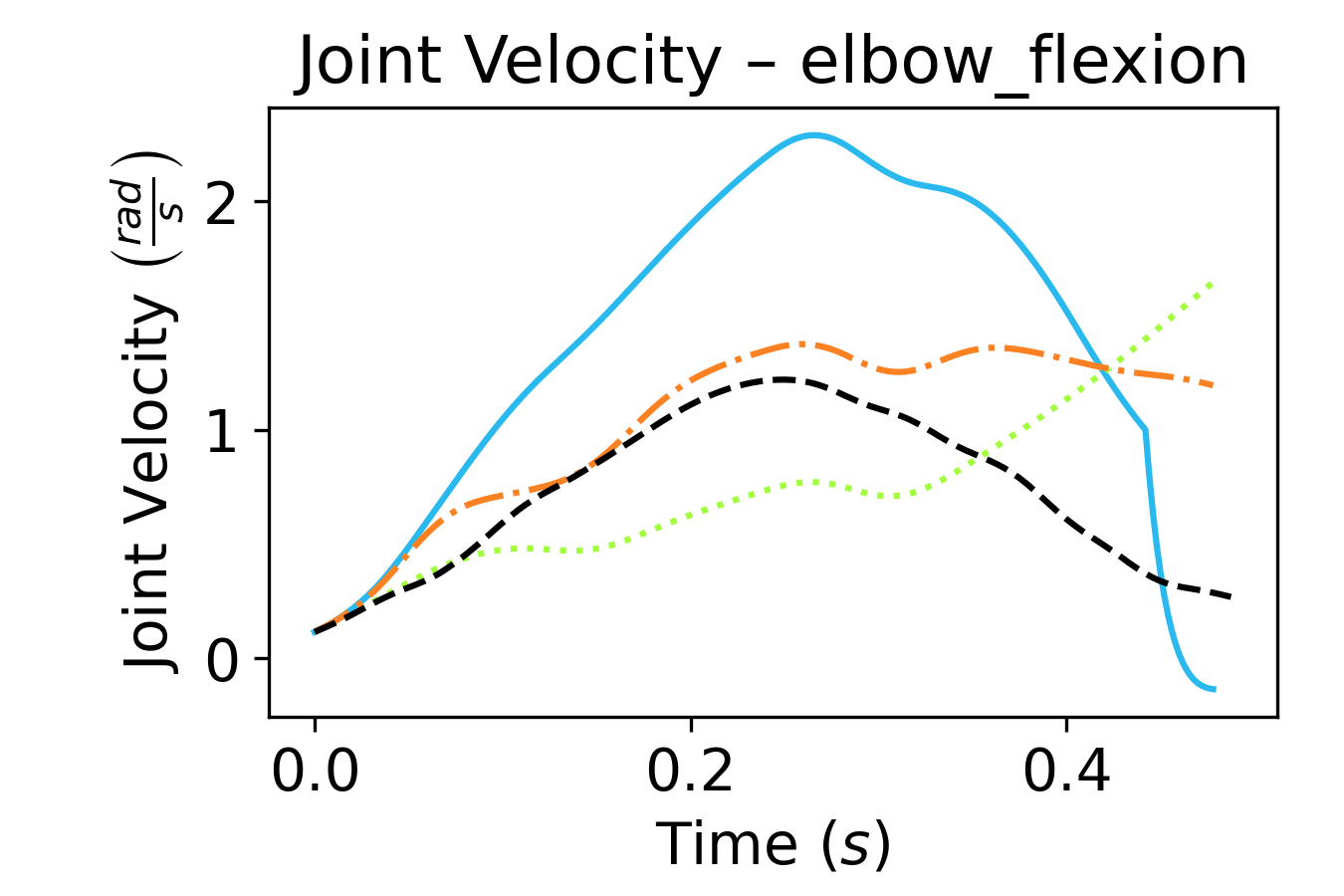}}
	\caption{Projected cursor and joint trajectories for one trial of U4 for the Virtual Pad Identity technique. The Joint Acceleration Costs (JAC; orange dashdotted lines) qualitatively explain observed user behavior best.}
	\label{fig:vgl_qual_proj}
\end{figure}

There are clear qualitative differences between the three cost functions, as shown in Figure~\ref{fig:vgl_qual_proj} for an exemplary user study trial (black dashed lines; U4, Virtual Pad Identity, first movement from target 7 to target 8).

DC (blue solid lines) exhibits the highest velocities both in joint and cursor space, resulting in movements that are slightly faster than humans. 
The peak velocity tends to be too large, and for some trials, corrective submovements are required towards the end of the movement.

With CTC (green dashed lines), there is a considerable undershoot of the aimed target, with the cursor often not reaching the target at all within simulation time. 
As can be seen in the bottom left and right plots of Figure~\ref{fig:vgl_qual_proj}, penalization of commanded torque change seems to impose too restrictive constraints on the underlying joint dynamics, resulting in velocity time series of both elbow flexion and cursor that differ considerably from the typical bell-shaped velocity profiles observed in the user data.

In contrast, the simulation trajectories obtained from JAC (orange dash-dotted lines) match the human trajectories better: there are only slight differences between simulation and study in the projected cursor position and velocity profiles; it outperforms the other variants in terms of elbow flexion, and outperforms DC in shoulder rotation.
Similar results can be obtained for the elevation angle, while pronation/supination as well as wrist deviation and flexion are predicted well by any of the considered cost functions (see Figure~\ref{fig:vgl_qual_proj_otherjoints} in the Appendix).

\begin{figure}[h!]
	\centering
	\subfloat{\includegraphics[width=0.33\linewidth, clip]{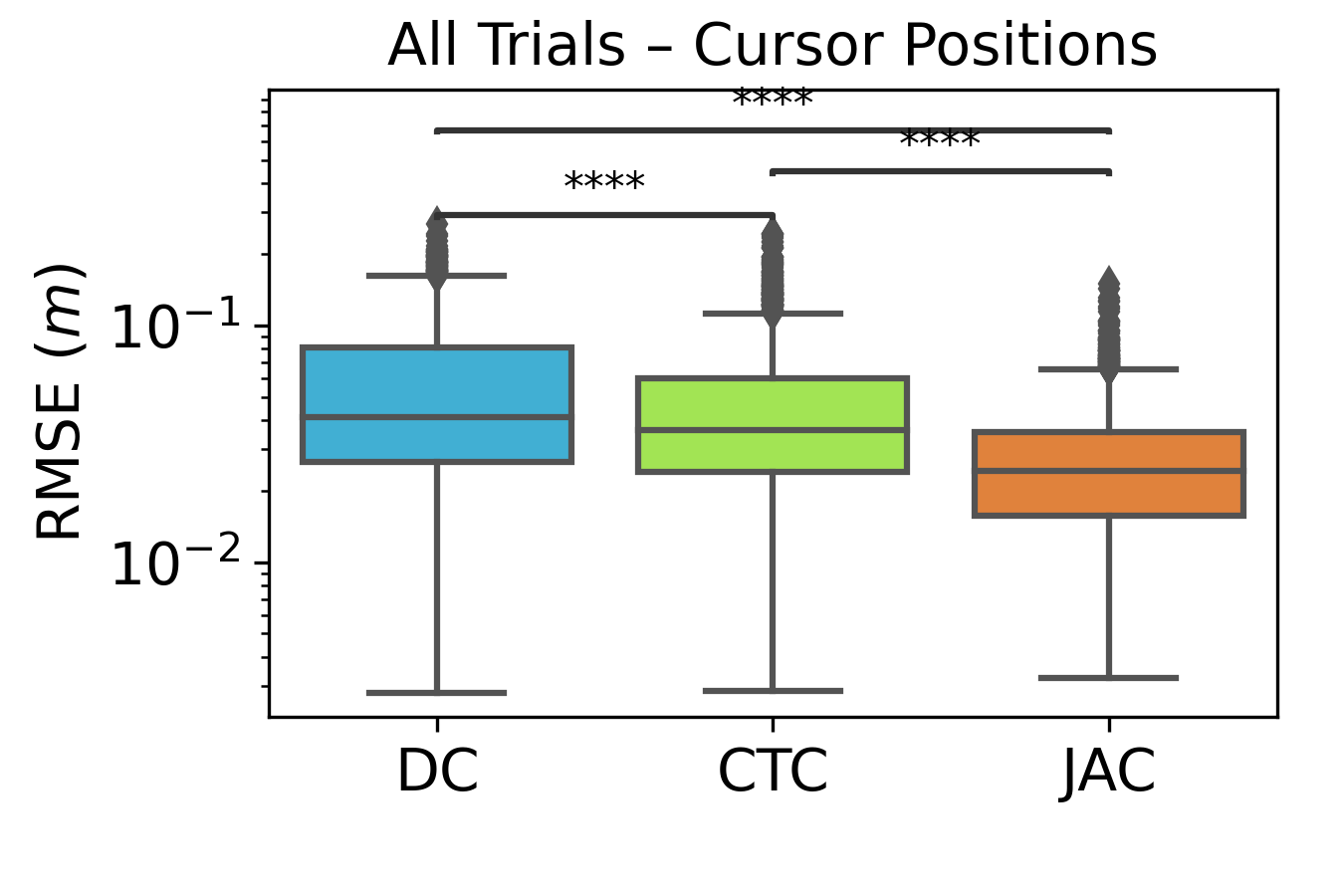}}	
	\subfloat{\includegraphics[width=0.33\linewidth, clip]{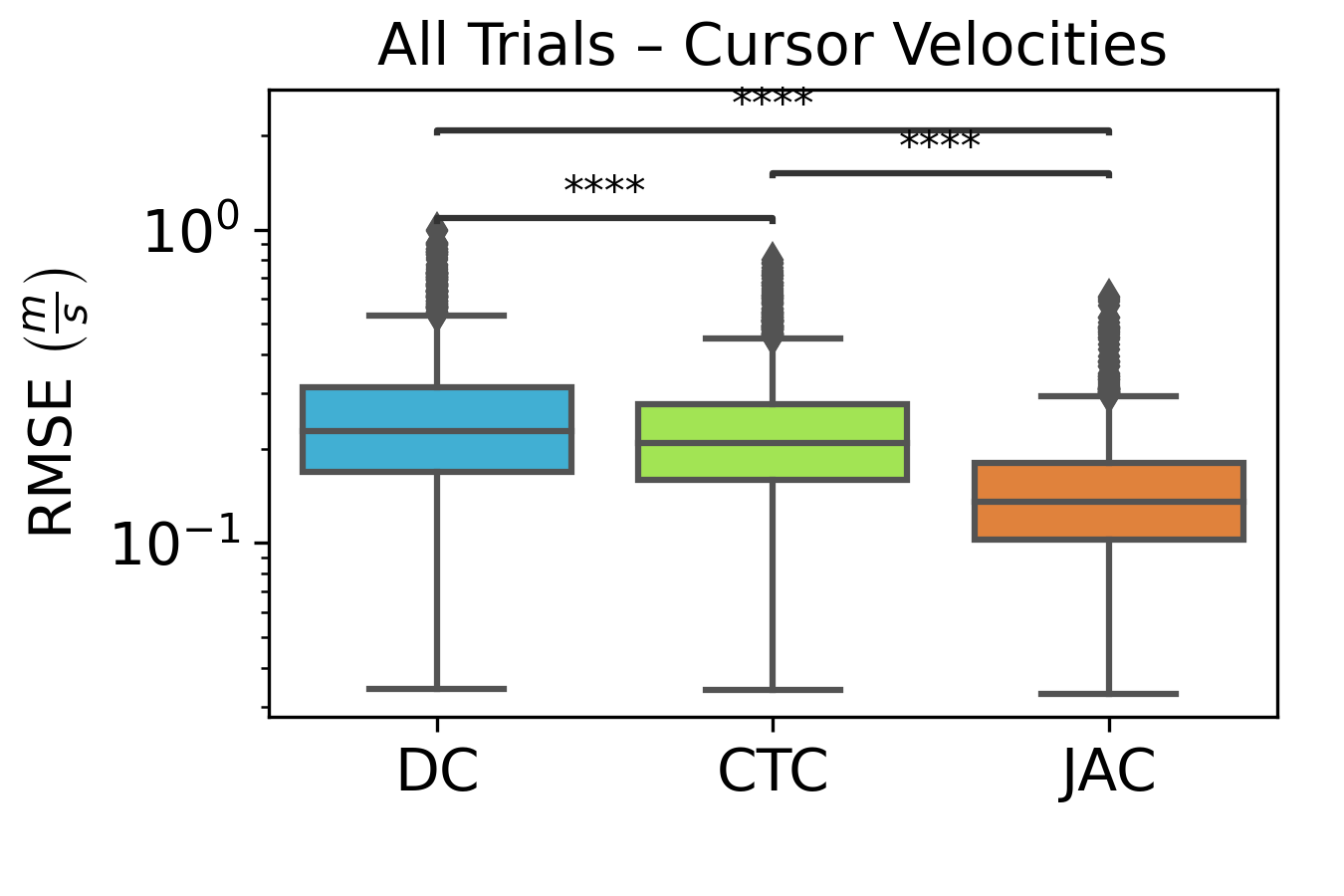}}	
	\subfloat{\includegraphics[width=0.33\linewidth, clip]{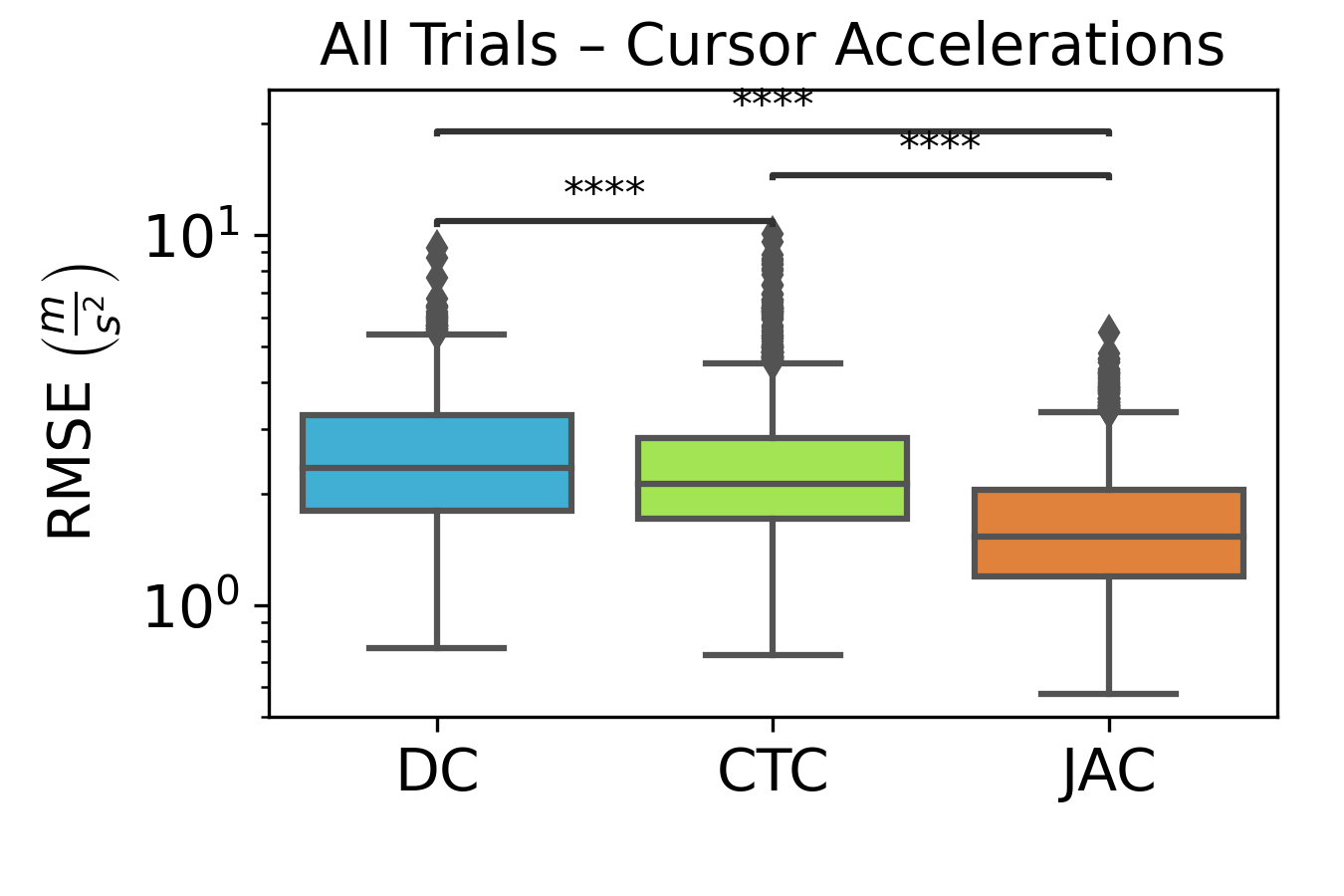}}\\	\subfloat{\includegraphics[width=0.33\linewidth, clip]{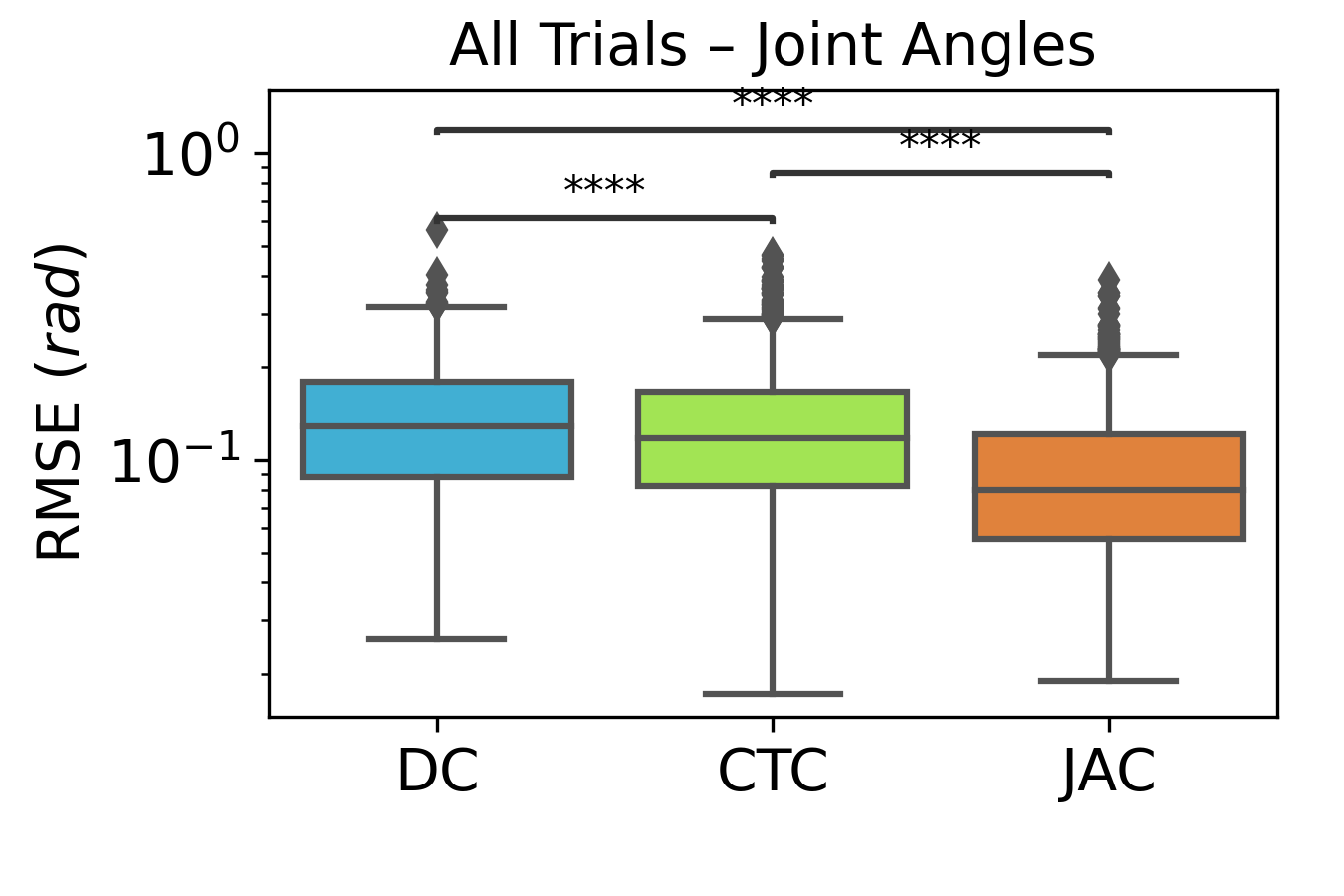}}	
	\subfloat{\includegraphics[width=0.33\linewidth, clip]{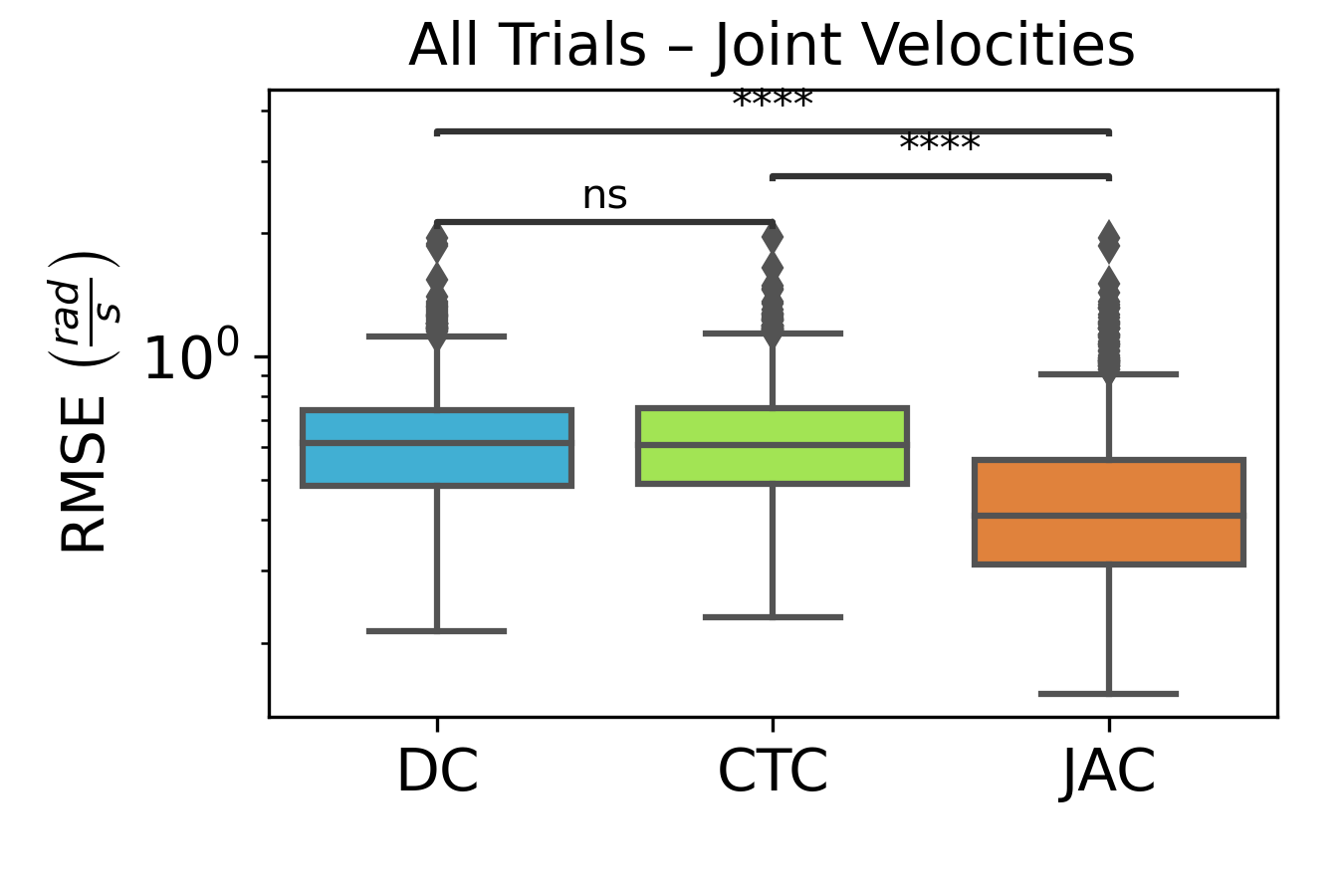}}	
	\subfloat{\includegraphics[width=0.33\linewidth, clip]{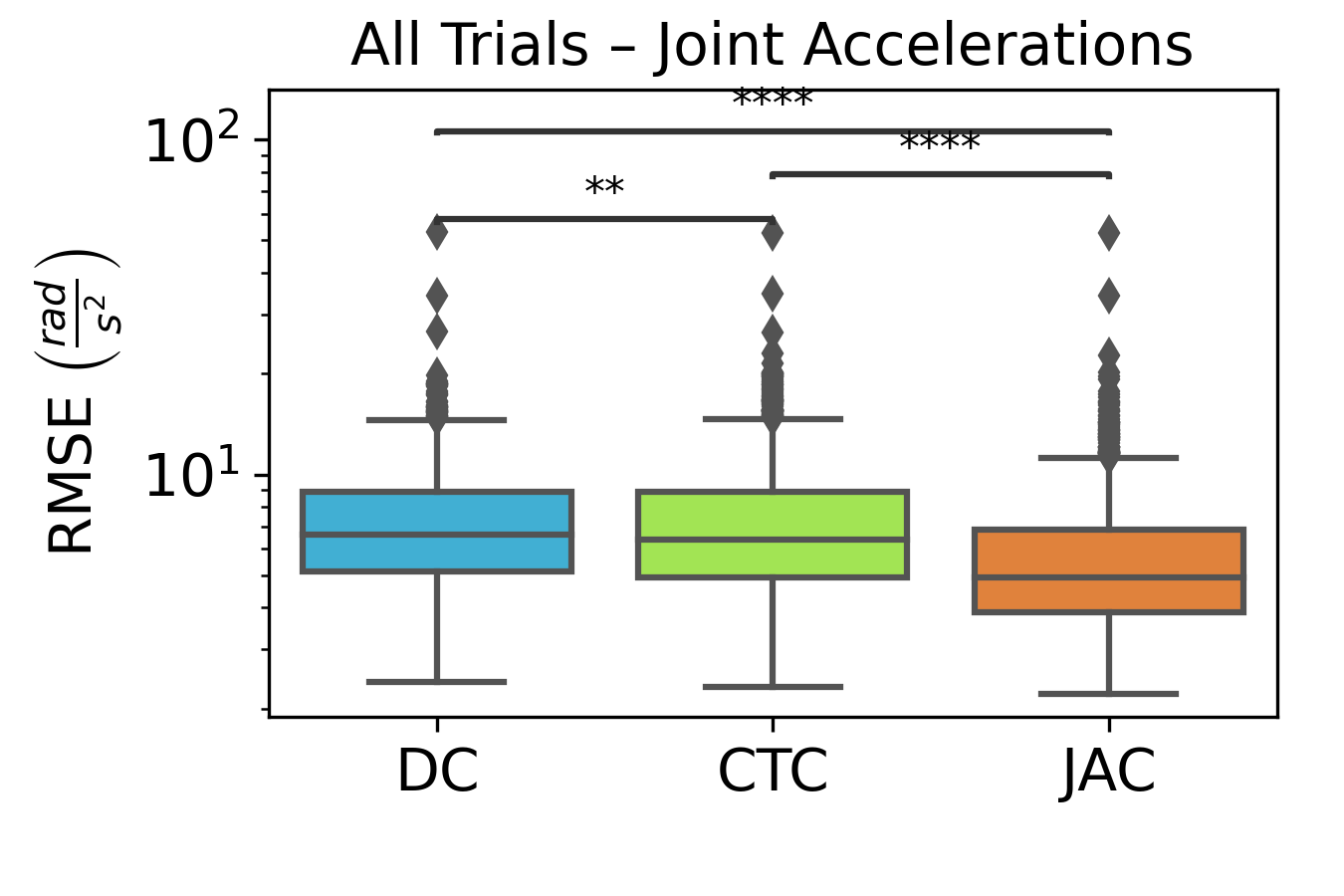}}
	\caption{Comparison of the different cost functions. The boxplots show the RMSE of all trials.}
	\label{fig:cost-comparison}
\end{figure}

For quantitative comparison, boxplots containing the RMSEs of all ISO pointing movements for each considered cost function are shown in Figure~\ref{fig:cost-comparison}, considering both cursor (top row) and joint space (bottom row).
A breakdown of the cursor position and joint angle boxplots by invididual users can be found in Figure~\ref{fig:cost-comparison_peruser} in the Appendix.
Kolmogorov-Smirnov tests showed that for each of the three cost functions, none of the considered RMSE distributions fits the assumption of normality (all values $p<0.0001$). 
Thus, we carried out the non-parametric Wilcoxon Signed Rank tests with Bonferroni corrections.

For the following statements, details on the results of the statistical tests are provided in Table~\ref{tab:cost-comparison}.
The simulation trajectories generated with JAC~\eqref{eq:costs-jointacc} replicate the respective user study trajectories significantly better than those generated with DC. %
The JAC trajectories also significantly outperform the CTC trajectories in terms of RMSE.  %
Comparing CTC to DC, some RMSE quantities yield significant differences in favor of CTC, %
while for others, the cost function has no or only a small significant effect. %
\begin{table}[!ht]
	\centering
	\resizebox{.99\textwidth}{!}{%
		\begin{tabular}{|c|c|c|c|c|c|c|} 
			\hline
			\multirow{2}{*}{} & \multicolumn{3}{c|}{Cursor $Z$-scores} & \multicolumn{3}{c|}{Joint $Z$-scores} \\
			\cline{2-7}
			\rule{0pt}{10pt}\noindent
			& position & velocity & acceleration & angle & velocity & acceleration \\
			\hline \hline
			\rule{0pt}{10pt}\noindent
			JAC vs.\ DC ($p<0.0001$) & $-23.5$ & $-25.4$ & $-25.8$ & $-24.8$ & $-27.2$ & $-28.9$ \\
			\hline
			\rule{0pt}{10pt}\noindent
			JAC vs.\ CTC ($p<0.0001$) & $-20.2$ & $-22.3$ & $-21.5$ & $-21.8$ & $-25.8$ & $-26.4$ \\
			\hline
			\rule{0pt}{10pt}\noindent
			\multirow{2}{*}{CTC vs.\ DC} & $-11.5$ & $-9.2$ & $-8.5$ & $-7.1$ & $-0.7$ & $-3.5$ \\
			& $p<0.0001$ & $p<0.0001$ & $p<0.0001$ & $p<0.0001$ & $p=0.52$ & $0.001<p<0.01$ \\
			\hline
		\end{tabular}%
	}
	\caption{$Z$-scores and $p$-values of the comparisons between the three considered cost functions, using Wilcoxon Signed Rank tests with Bonferroni corrections.
	}
	\label{tab:cost-comparison} 
\end{table}

We thus conclude that, although both CTC~\eqref{eq:costs-ctc} and JAC~\eqref{eq:costs-jointacc} open the door to a better fit through an additional weight parameter $r_{2}$, by far the best results in terms of replicating observed human trajectories is obtained by JAC~\eqref{eq:costs-jointacc}, which we focus on in the following.

\subsection{Simulation vs. Users: MPC is able to simulate User Movement in Mid-Air Pointing}%
\label{sec:simulation-vs-user}

We compare the movements generated by our simulation with JAC to those from the user study in terms of both projected cursor trajectories and joint postures. %
In particular, we show that
\begin{enumerate}
	\hyperrefitem[item:res-bio-plausible]{} our simulated movements exhibit biomechanically plausible joint movements,
	\hyperrefitem[item:res-between-user]{} the produced cursor and joint trajectories predict human movements within between-user variability, and
	\hyperrefitem[item:res-individual]{} the method can predict motion of individual users.
\end{enumerate}

\begin{figure}%
	\centering
	\begin{tikzpicture}
		\node[anchor=north west] (img1) at (0, 0) {
			\subfloat{\includegraphics[width=0.5\linewidth, clip]{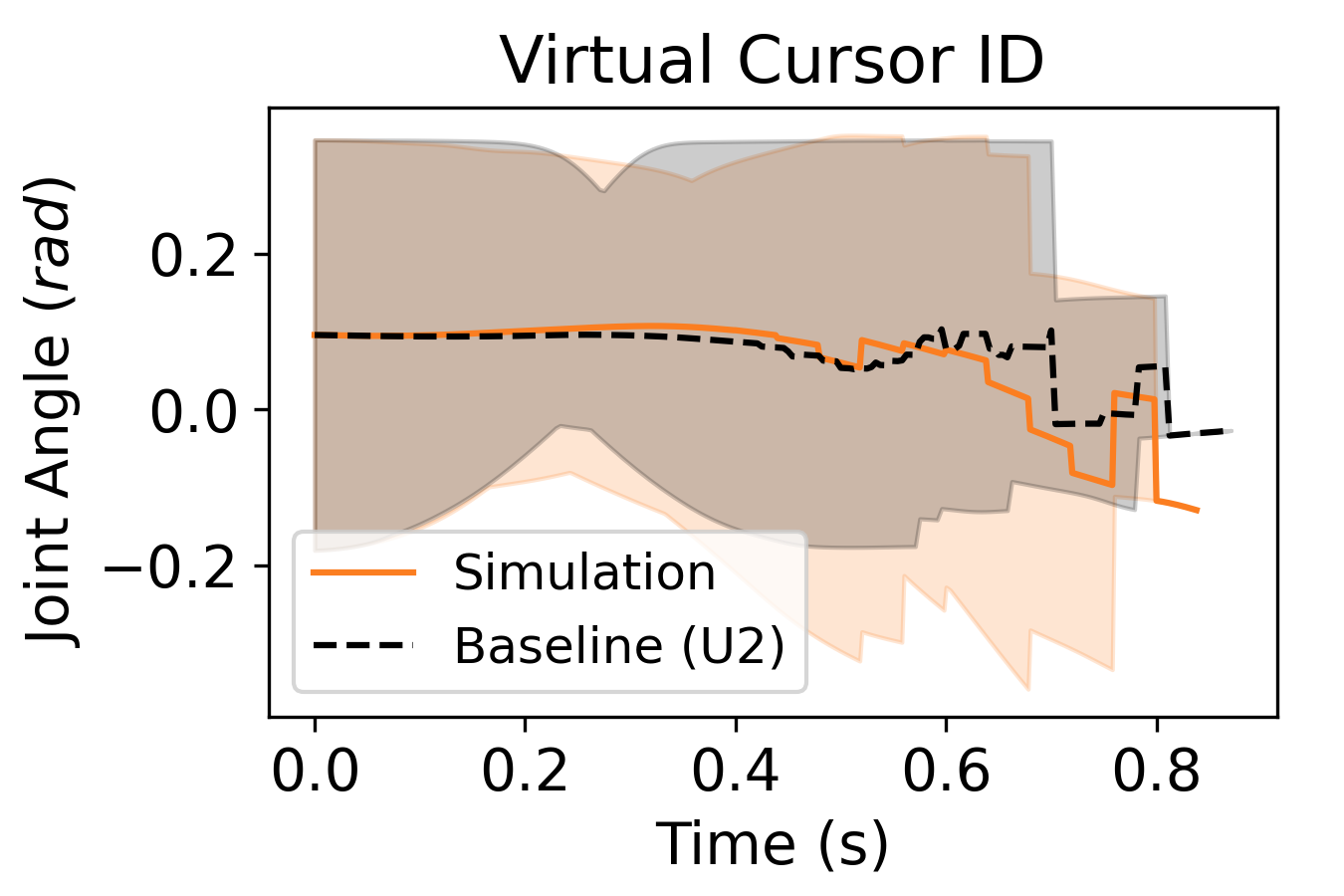}}
		};
		\node[anchor=north west] (img2) at (0.5\linewidth, 0) {
			\subfloat{\includegraphics[width=0.5\linewidth, clip]{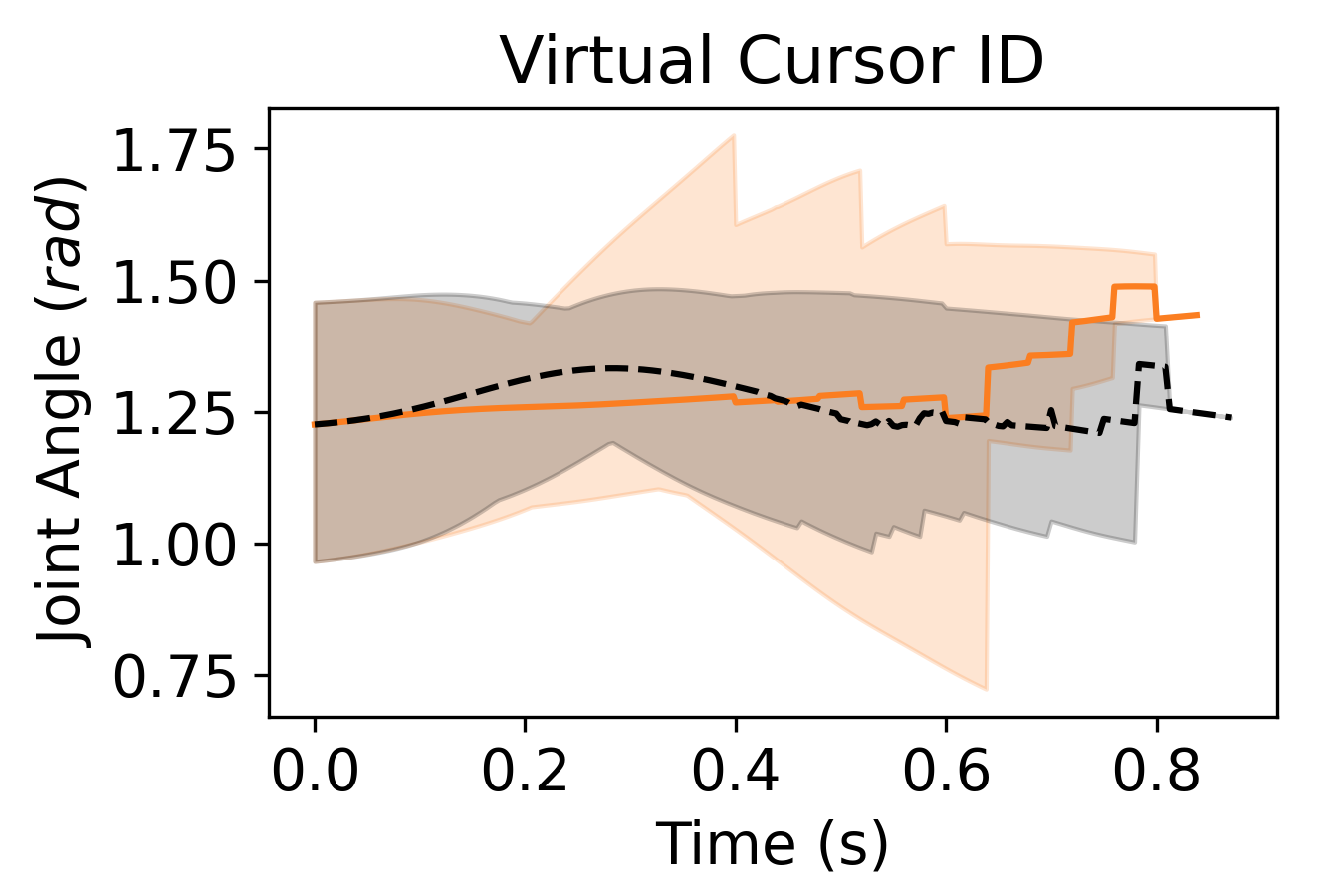}}
		};
		\node[anchor=north west] (img3) at (0, -5.25) {
			\subfloat{\includegraphics[width=0.5\linewidth, clip]{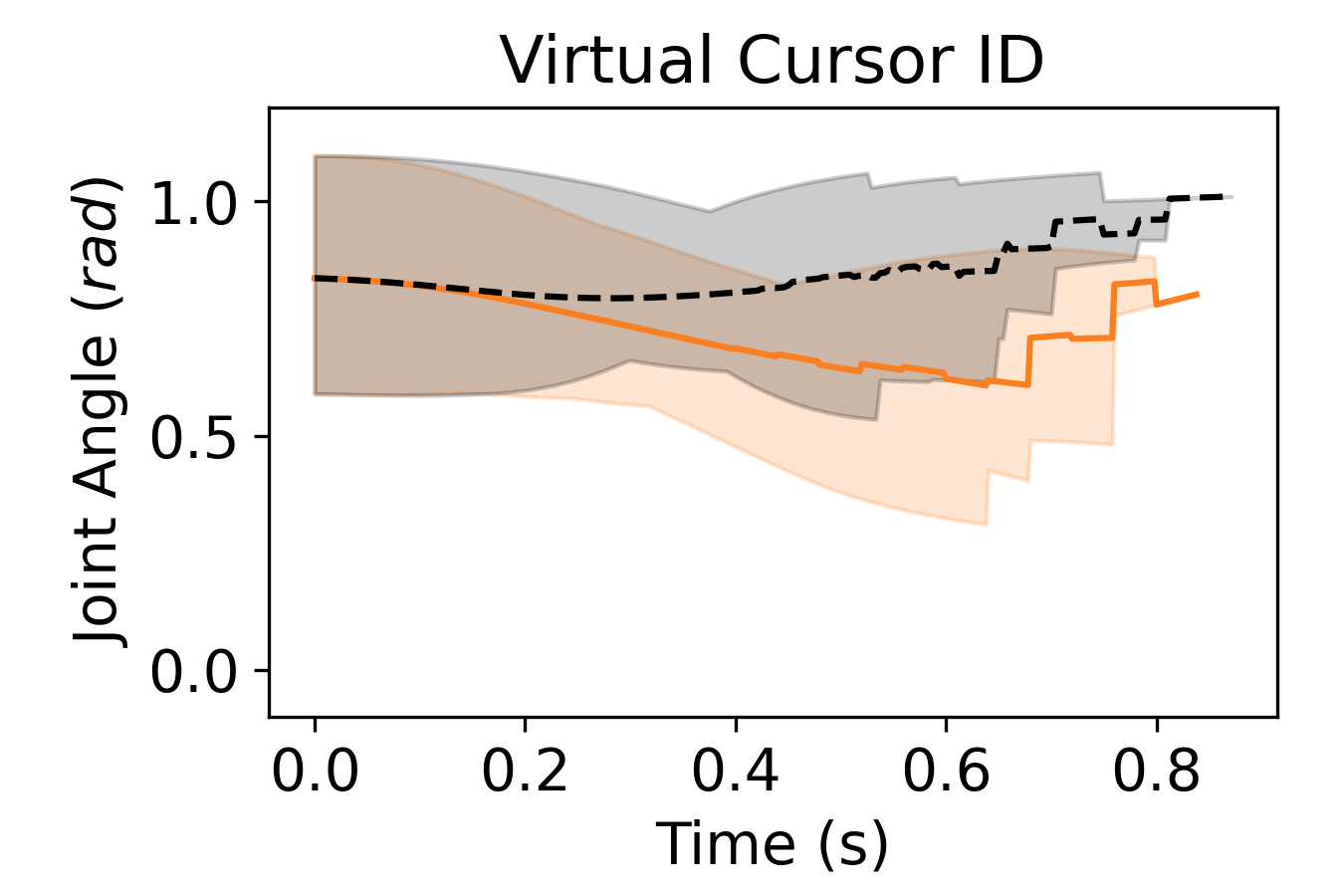}}
		};
		\node[anchor=north west] (img4) at (0.5\linewidth, -5.25) {
			\subfloat{\includegraphics[width=0.5\linewidth, clip]{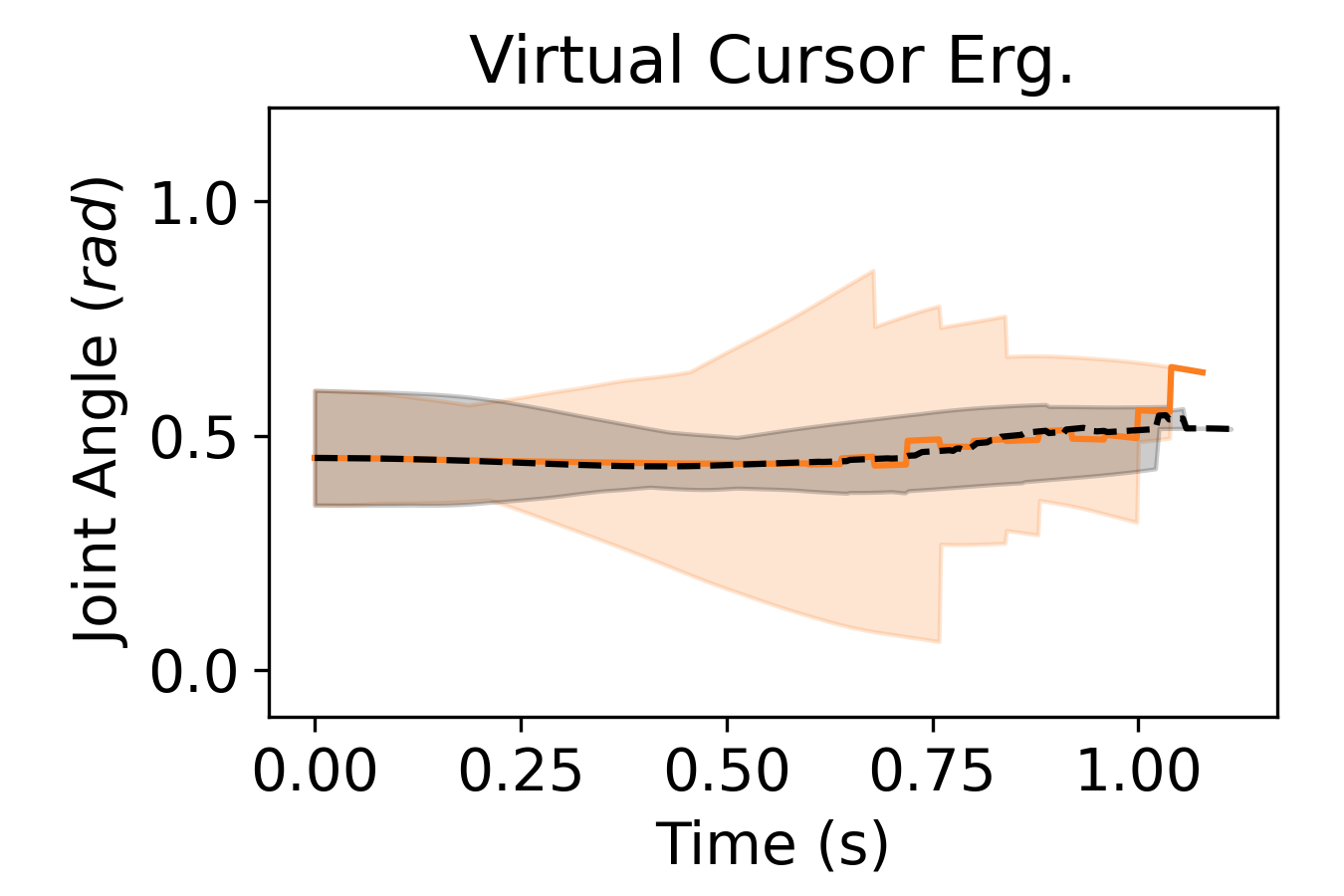}}
		};
	
		\node[anchor=center, align=center, xshift=0.2cm, yshift=-0.2cm] at (img1.north west) {\Large{\textbf{a.}}};
		\node[anchor=center, align=center, yshift=-0.2cm] at (img1.north) {\large{\textbf{Joint angles -- Shoulder rotation}}};
		\node[anchor=center, align=center, yshift=-0.2cm] at (img2.north) {\large{\textbf{Joint angles -- Elbow flexion}}};
		
		\node[anchor=center, align=center, xshift=0.2cm, yshift=-0.2cm] at (img3.north west) {\Large{\textbf{b.}}};
		\node[anchor=center, align=center, yshift=-0.2cm] at (img3.north) {\large{\textbf{Joint angles -- Shoulder elevation}}};
		\node[anchor=center, align=center, yshift=-0.2cm] at (img4.north) {\large{\textbf{Joint angles -- Shoulder elevation}}};
	\end{tikzpicture}
	\caption{\textbf{a.} The joint angle ranges predicted by our simulation for different movements in the ISO task (orange solid lines) match those observed in our user study (black dashed lines) fairly well.
	The mean of all movements of a single participant/user model (U2) is shown together with the entire value ranges.
	\textbf{b.}	In the ISO task, the Virtual Cursor Identity technique (black dashed lines in left plot) requires considerably higher shoulder elevation angles than the Virtual Cursor Ergonomic technique (black dashed lines in right plot). This characteristic difference is captured well by our simulation (orange solid lines). 
	}
	\label{fig:JAC_jointranges}
\end{figure}
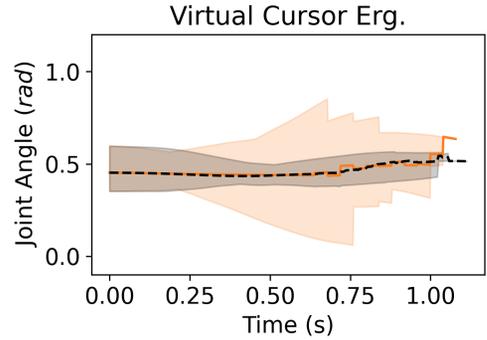

\newcounter{resultssubsections}
\subsubsection*{(\stepcounter{resultssubsections}\theresultssubsections\label{item:res-bio-plausible}) Our simulated movements exhibit biomechanically plausible joint movements}
Figure~\ref{fig:JAC_jointranges} shows the shoulder rotation, shoulder elevation, and elbow flexion angles for one exemplary user along with the corresponding simulation data.
The user's mean angles over time (black dashed lines) are captured well by our simulation (orange solid lines) for each joint.
In addition, the range of joint angles applied during any of the considered movements (black area) exhibits the same structure as in our simulation (orange area). 
It should be noted that these ranges only make up a relatively small portion of the admissible model joint ranges (see Table~\ref{tab:joint-limits}).
The plots for the remaining four joints are shown in Figure~\ref{fig:CursorID_jointranges_otherjoints} in the Appendix.

There are also characteristic differences in the joint ranges when simulating different interaction techniques. 
Due to its shifted input space, the Virtual Cursor Ergonomic technique allows the participant to perform movements to arbitrary directions using considerably lower shoulder elevation angles than needed for the Virtual Cursor Identity technique, as can be inferred from the bottom plots in Figure~\ref{fig:JAC_jointranges} (black dashed lines, black areas). %
These technique-dependent movement characteristics are captured by our simulation, which predicts comparable joint ranges for both techniques (orange solid lines, orange areas).

In summary, our proposed MPC simulation is capable of generating motions that are plausible from a biomechanical perspective.

\begin{figure}%
	\centering
	\subfloat{\includegraphics[width=0.33\linewidth, clip]{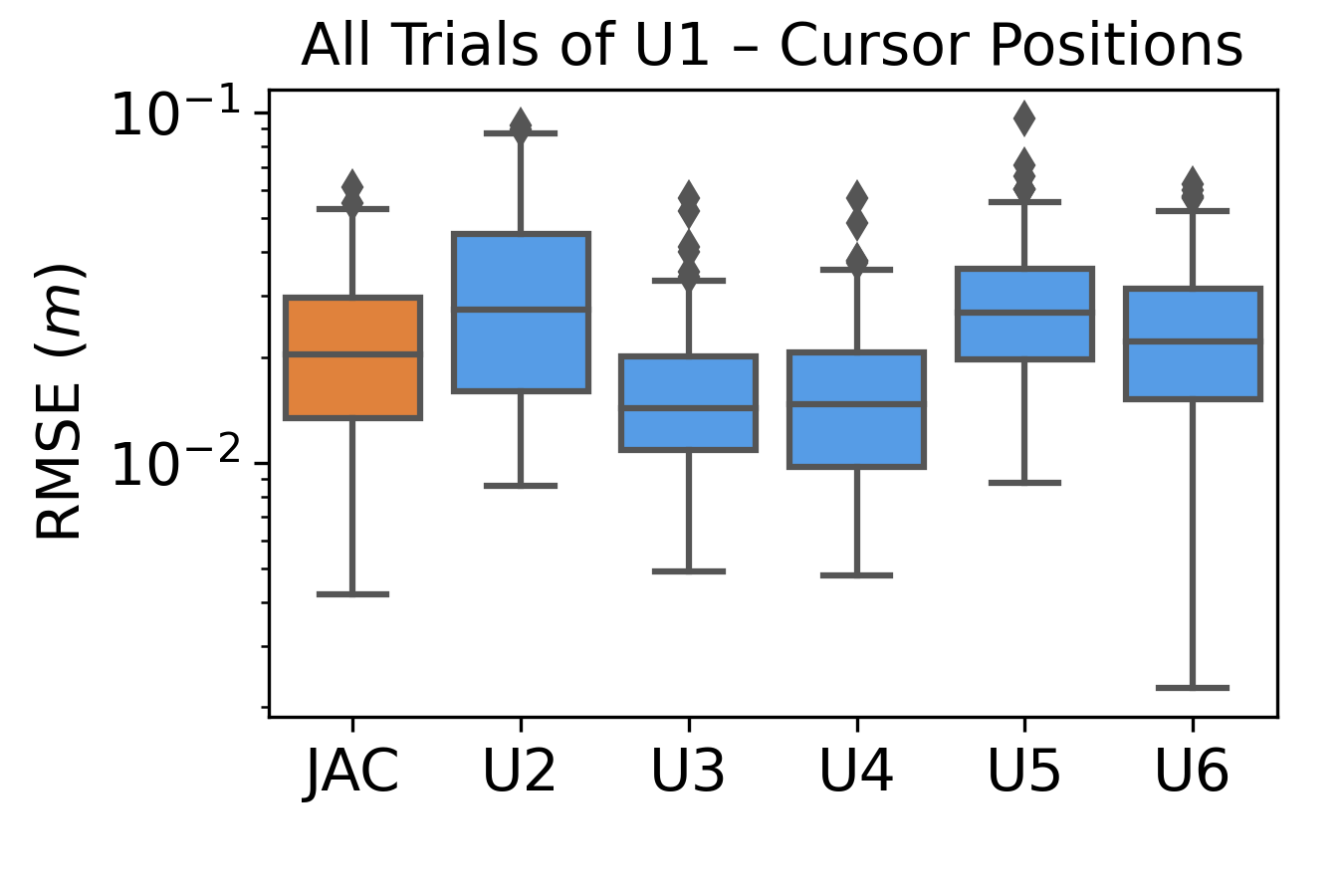}}
	\subfloat{\includegraphics[width=0.33\linewidth, clip]{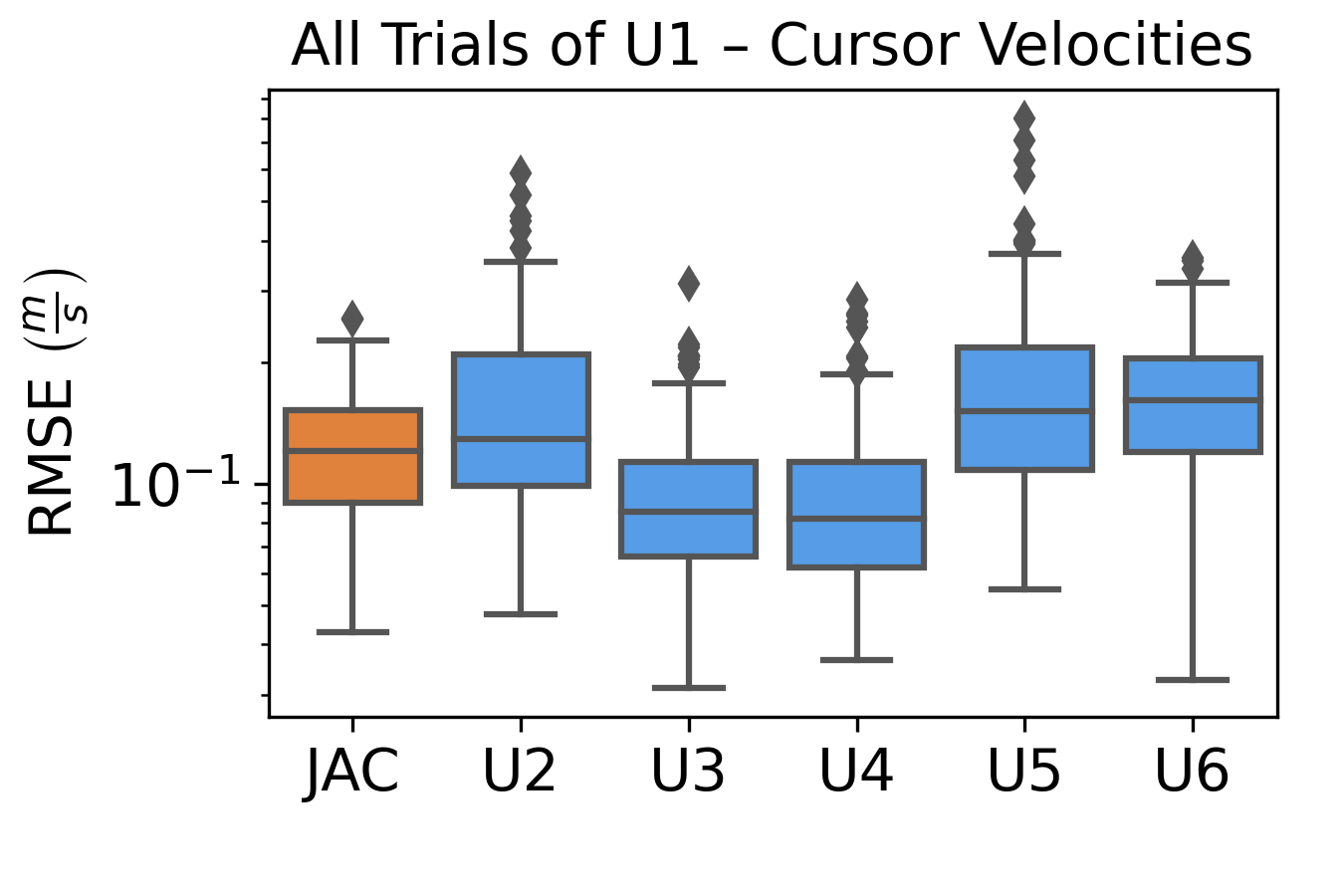}}
	\subfloat{\includegraphics[width=0.33\linewidth, clip]{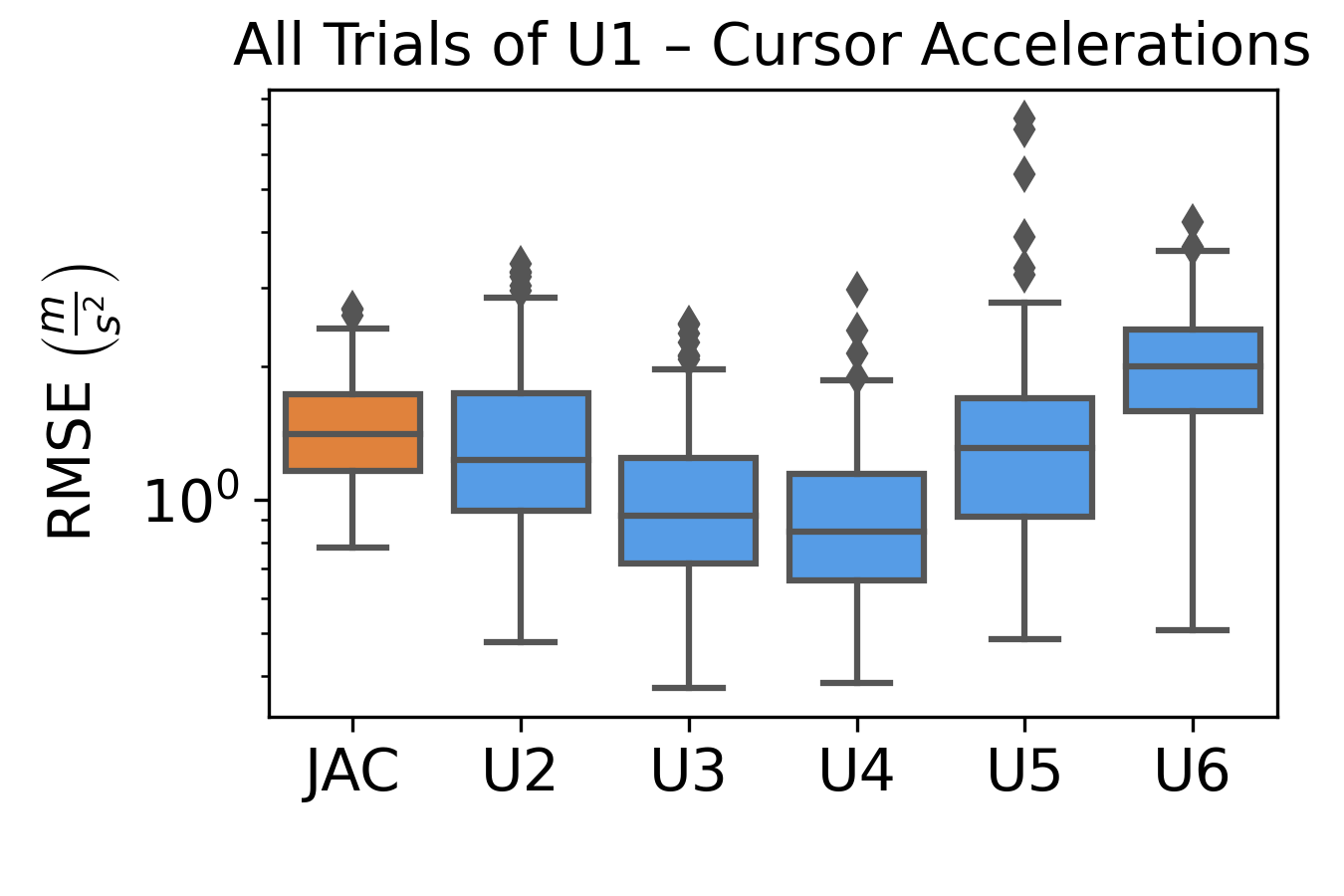}}\\
	\subfloat{\includegraphics[width=0.33\linewidth, clip]{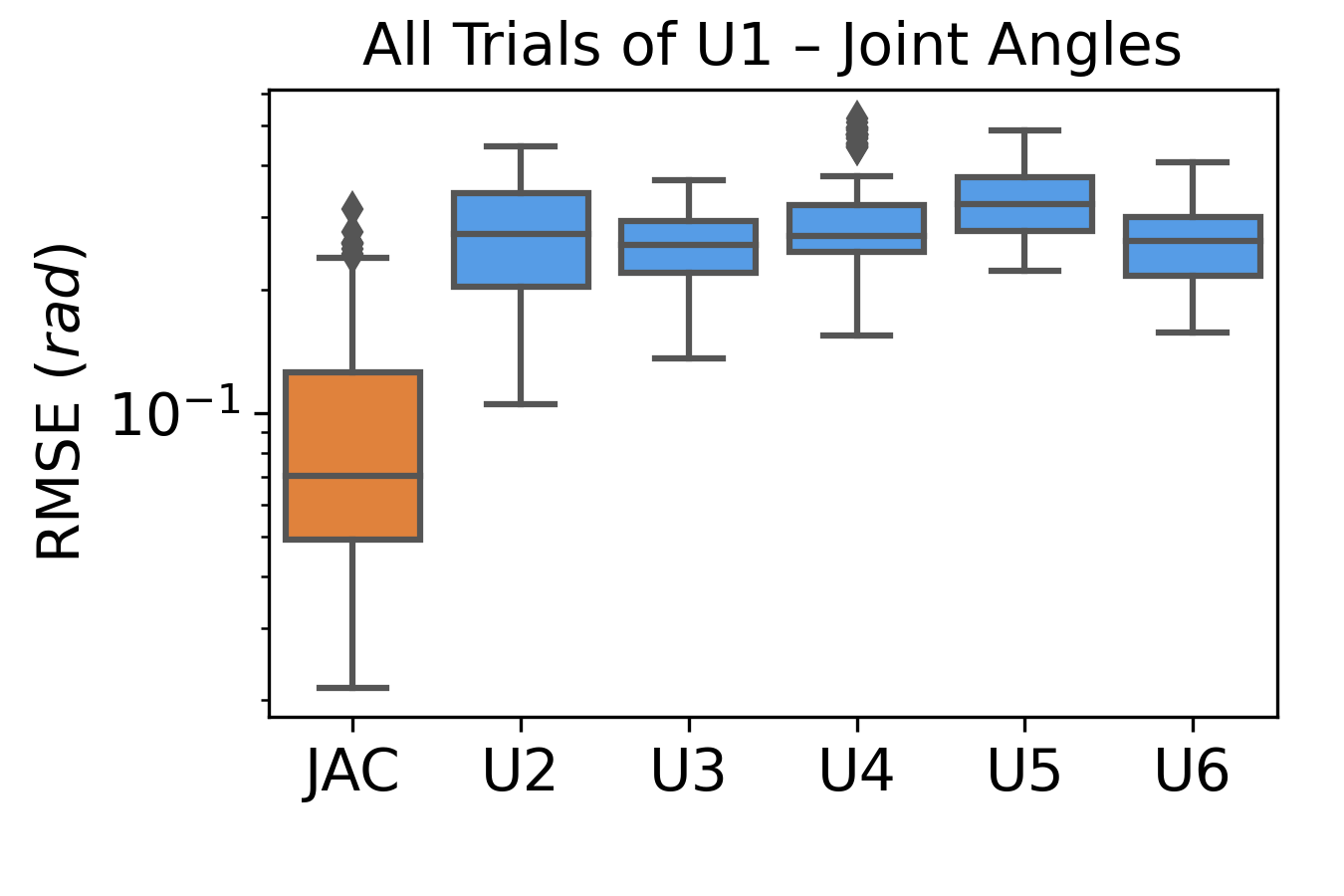}}
	\subfloat{\includegraphics[width=0.33\linewidth, clip]{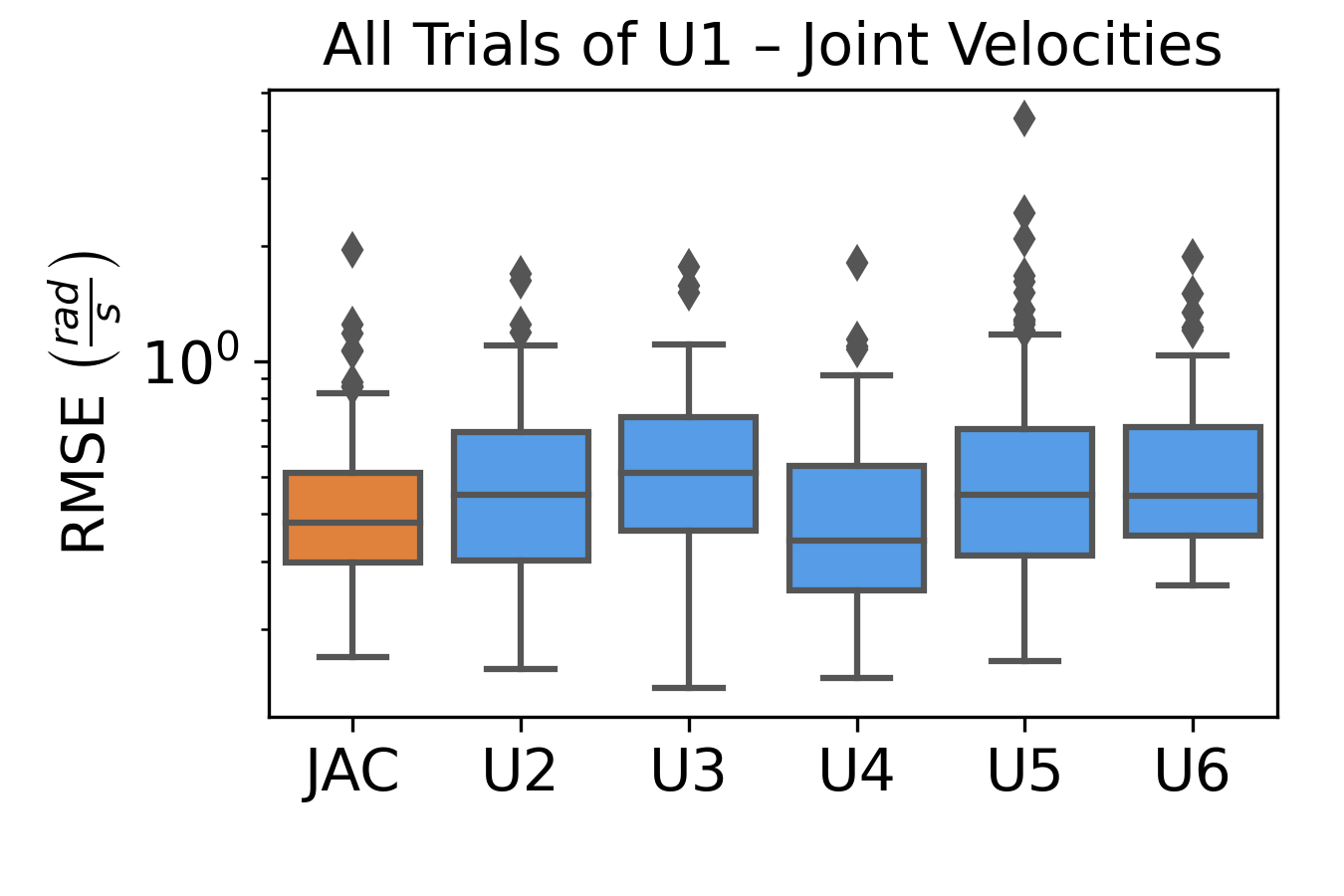}}
	\subfloat{\includegraphics[width=0.33\linewidth, clip]{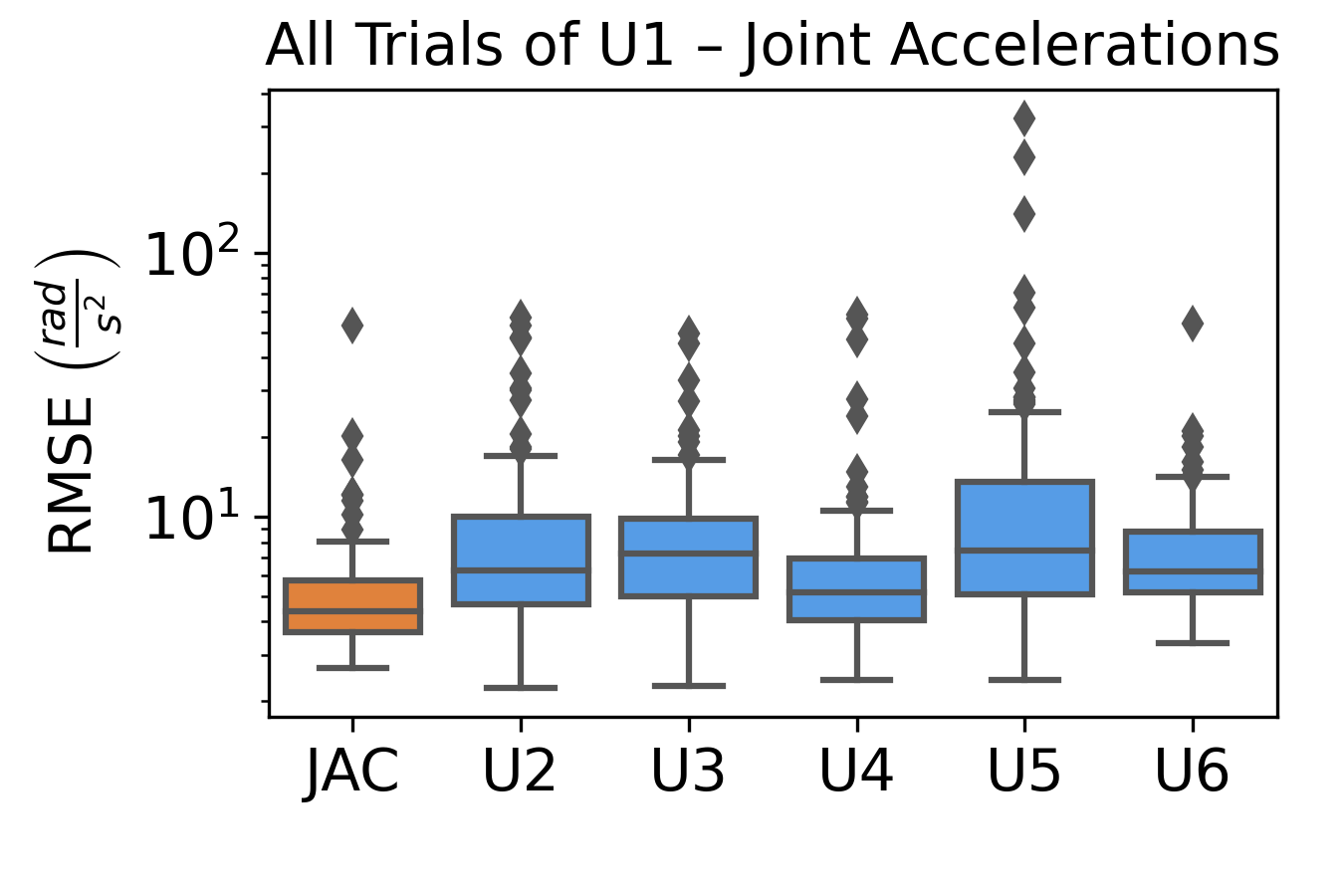}}
	
	\caption{RMSE between our simulation and U1 (orange whiskers), as well as between the remaining participants U2-U6 and U1 (blue whiskers), for cursor positions, velocities, and accelerations, as well as (aggregated) joint angles, velocities, and accelerations. Each whisker includes RMSE values for all interaction techniques and trials.}
	\label{fig:U1FIXED_quant}
\end{figure}

\begin{figure}%
	\centering
	\subfloat{\includegraphics[width=0.33\linewidth, clip]{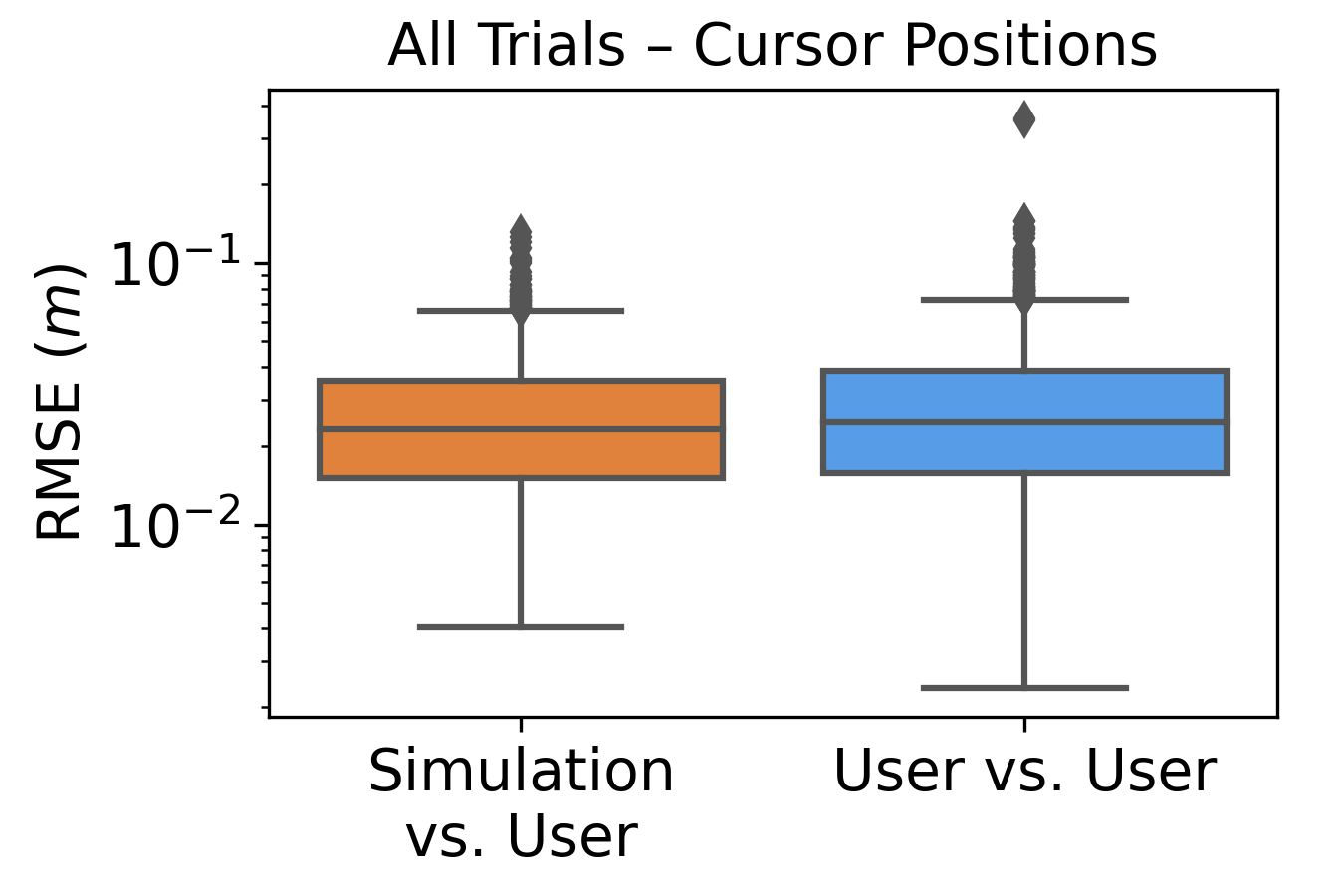}}
	\subfloat{\includegraphics[width=0.33\linewidth, clip]{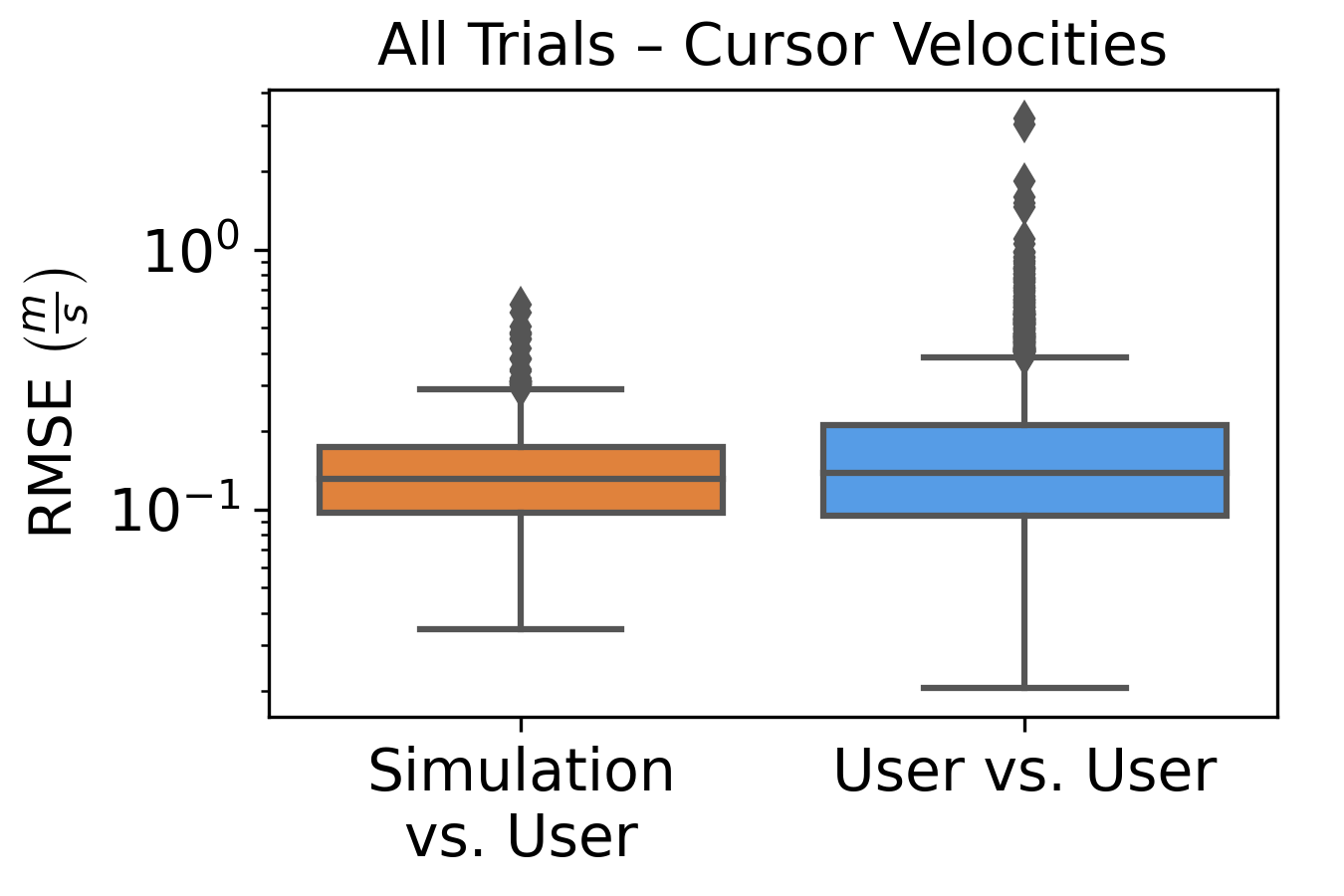}}
	\subfloat{\includegraphics[width=0.33\linewidth, clip]{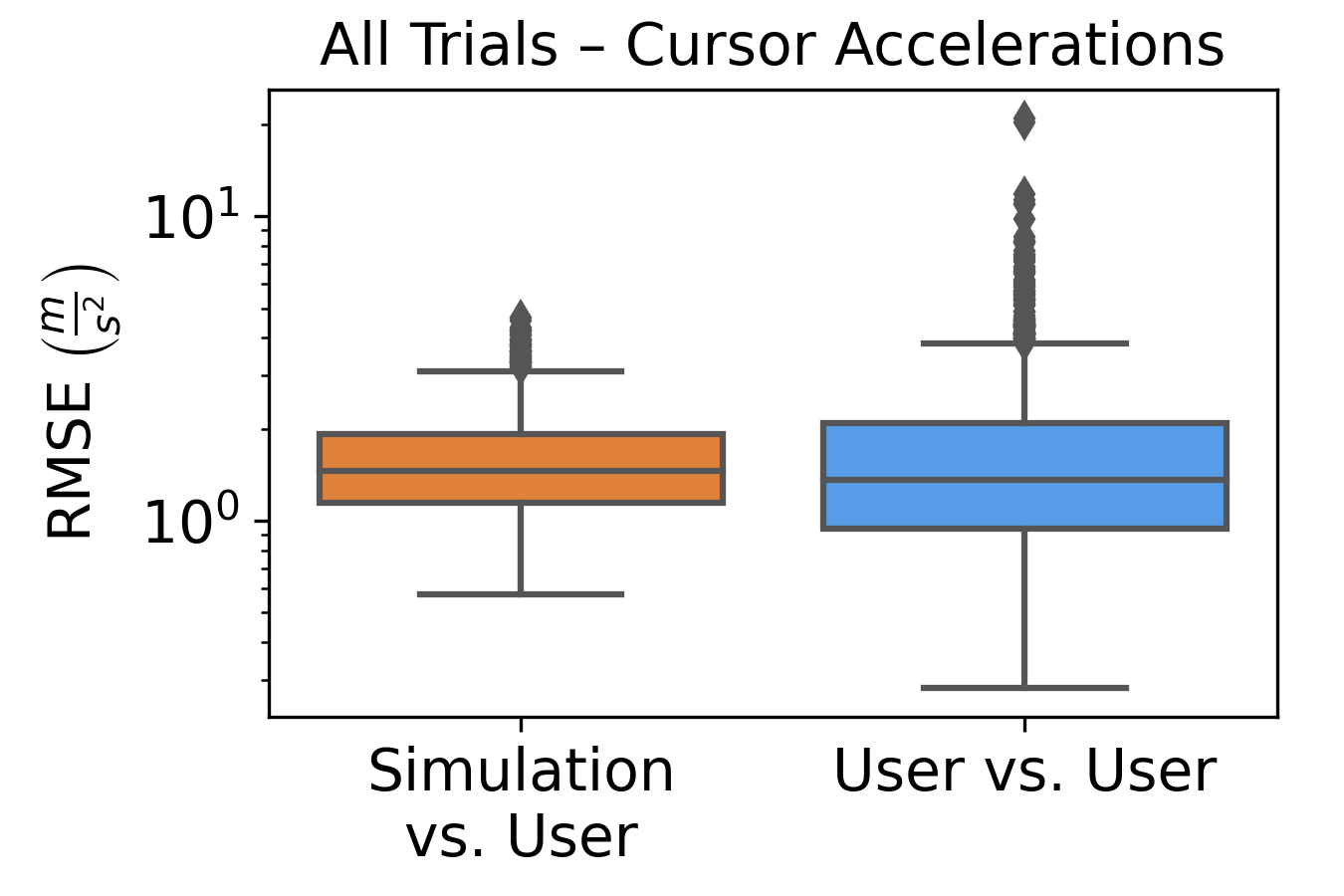}}\\
	\subfloat{\includegraphics[width=0.33\linewidth, clip]{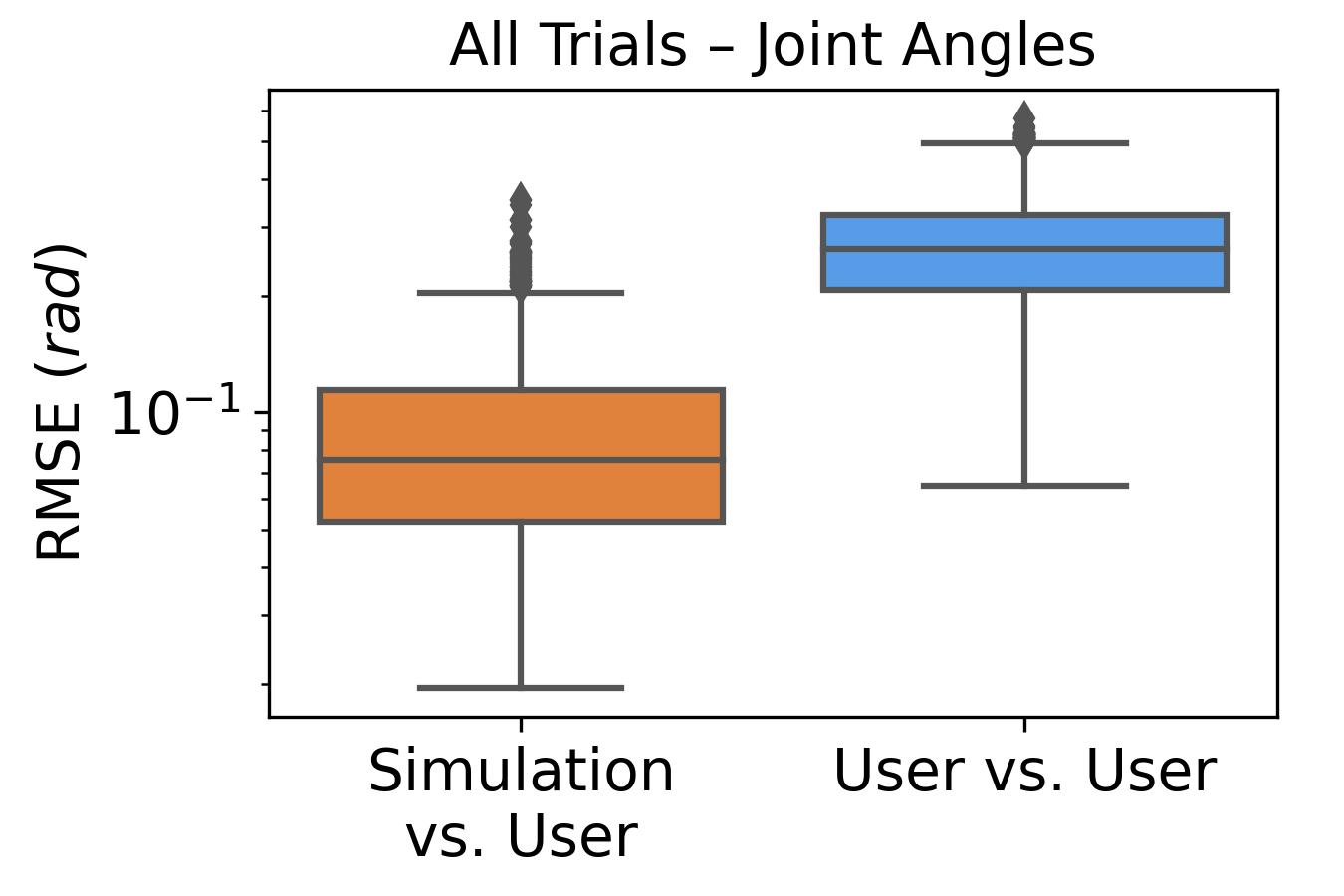}}
	\subfloat{\includegraphics[width=0.33\linewidth, clip]{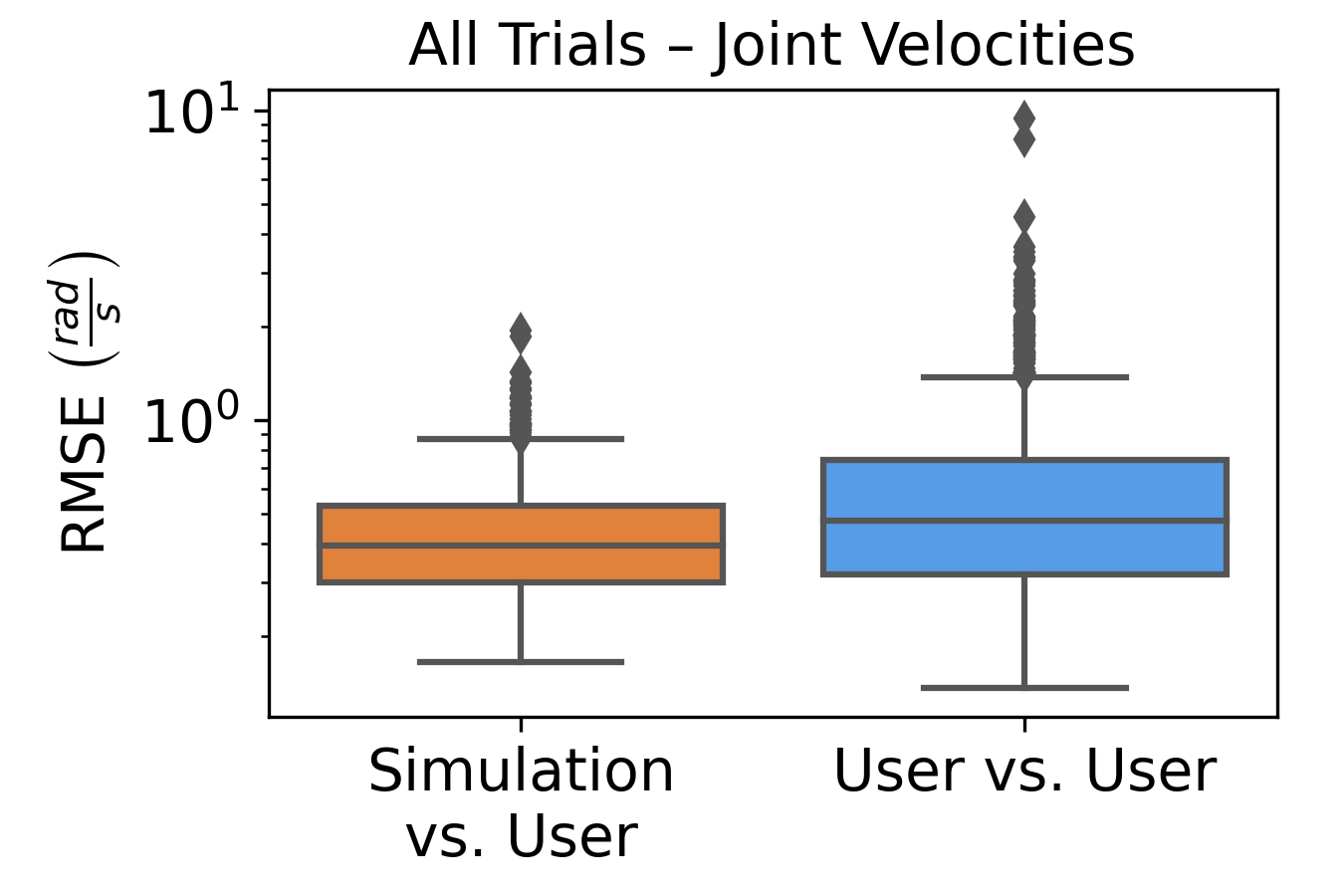}}
	\subfloat{\includegraphics[width=0.33\linewidth, clip]{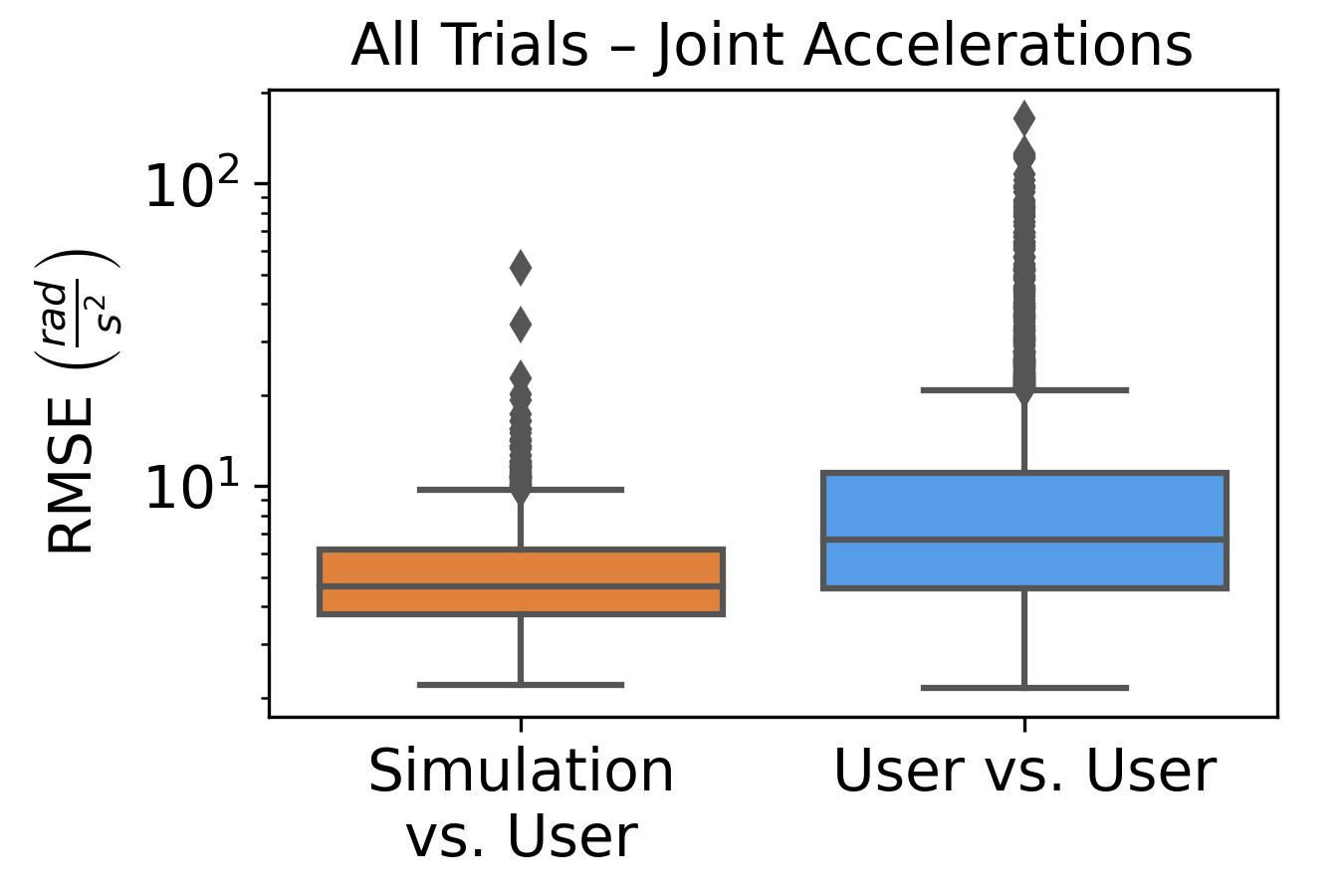}}
	
	\caption{RMSE comparisons, where all user study trials of a single participant (e.g., U1) are compared to either movements predicted by our simulation (\textit{Simulation vs. User}, orange whiskers) or by the remaining five participants (e.g., U2-U6) (\textit{User vs. User}; blue whiskers). %
	In contrast to Figure~\ref{fig:U1FIXED_quant}, the trials of all 6 participants are once used as baseline. 
	}
	\label{fig:ALL_quant}
\end{figure}

\subsubsection*{(\stepcounter{resultssubsections}\theresultssubsections\label{item:res-between-user}) The produced cursor and joint trajectories predict human movements within between-user variability}
We argue that the movements JAC generates are within between-user variability. %
To this end, we first predict the movements of U1 with JAC, for all trials and interaction techniques, and compare the similarity in terms of RMSE with how well the trajectories from the remaining participants U2-U6 match those of U1. 
The results are displayed in Figure~\ref{fig:U1FIXED_quant}.
Comparing the orange ``JAC'' whisker to the blue U2-U6 whiskers, Figure~\ref{fig:U1FIXED_quant} shows that the movement trajectories generated by JAC are well within the RMSE ranges of other users, i.e., within the between-user variability. 
What stands out are the low RMSE values of JAC in the joint angles. 
It should be noted that this comparison is slightly biased because our simulation is necessarily initialized with the same joint angles as U1, while the other users might have started in slightly different postures.
We extend this procedure, i.e., to infer the movements of one participant from JAC and from the respective remaining participants, to all participants, and combine the participants' RMSE values in the blue ``User vs.\ User'' whiskers in Figure~\ref{fig:ALL_quant}.\footnote{In the pre-processing of experimental data we had to remove all movements to two targets (T4, T10) of user U5 using the Virtual Pad Ergonomic technique. For technical reasons, we also needed to drop the eight trials starting at one of these targets from our quantitative analysis, resulting in a total of 1394 trials used for these RMSE computations.}
The plots in Figure~\ref{fig:ALL_quant} show that the results for the special case of inferring U1's movement can be extended to all participants.

We therefore conclude that our simulation predicts movements of a given user not worse than other users on average, %
i.e., our simulation trajectories are within the between-user variability.

\begin{figure}%
	\centering
	\subfloat{\includegraphics[width=0.33\linewidth, clip]{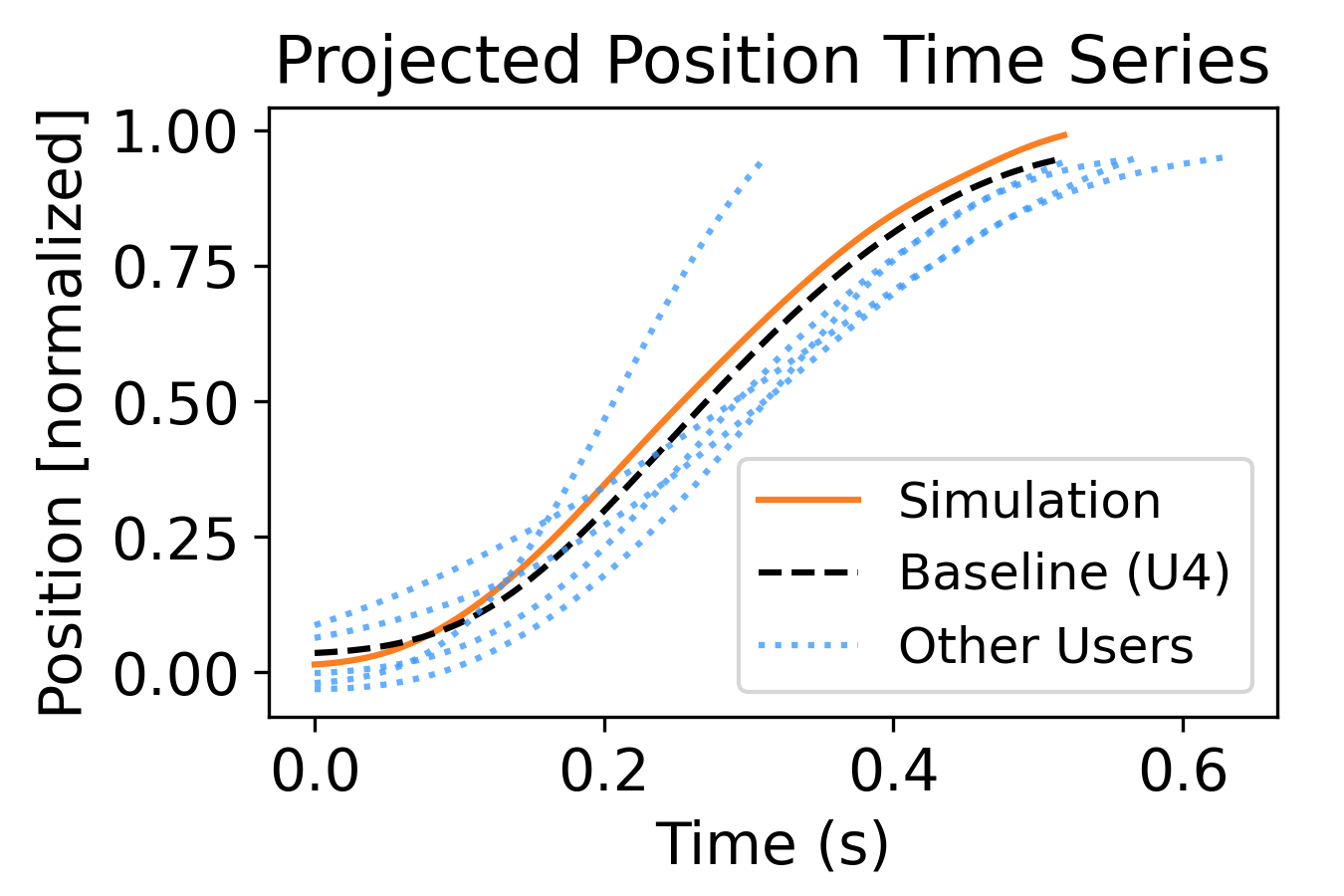}}
	\subfloat{\includegraphics[width=0.33\linewidth, clip]{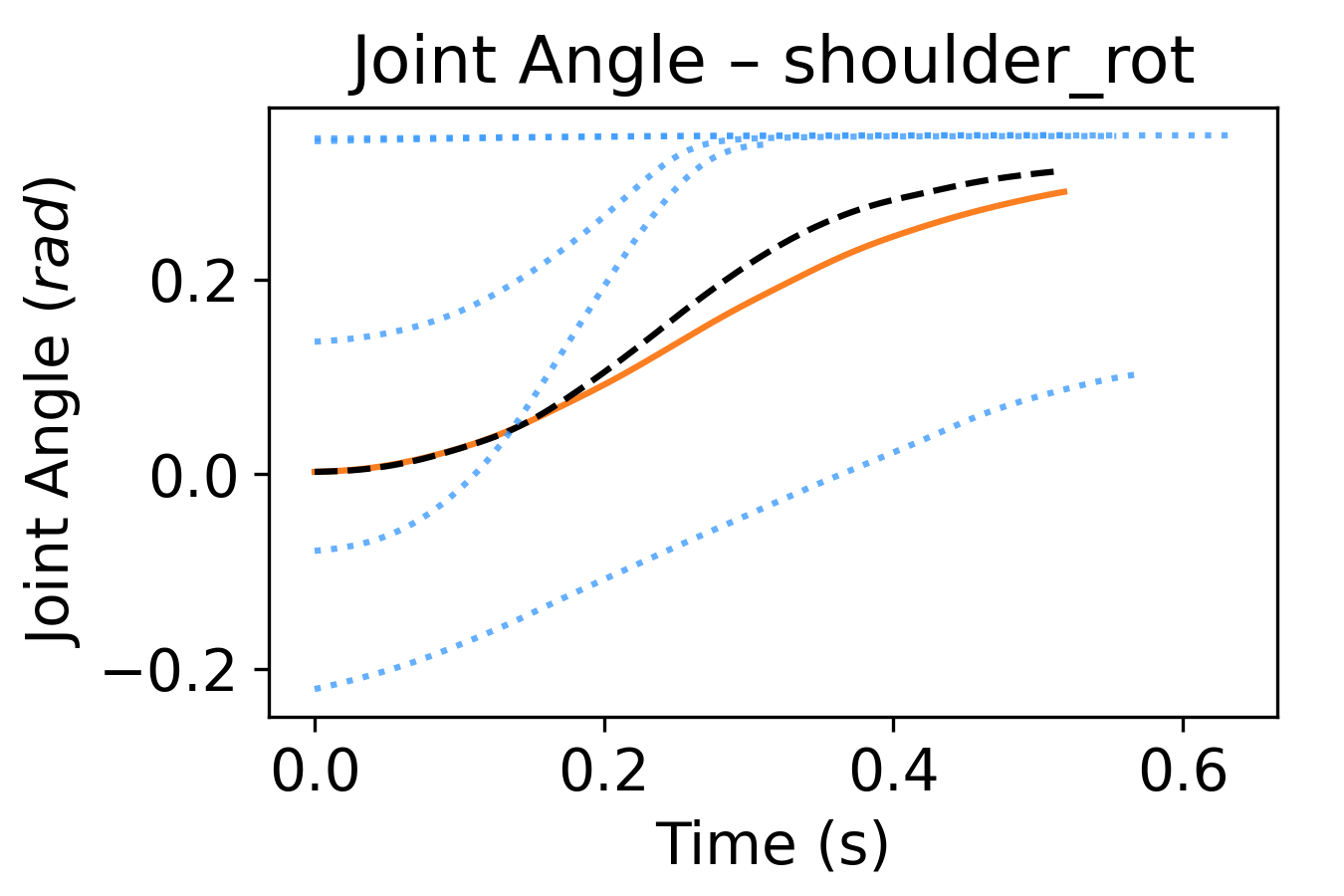}}
	\subfloat{\includegraphics[width=0.33\linewidth, clip]{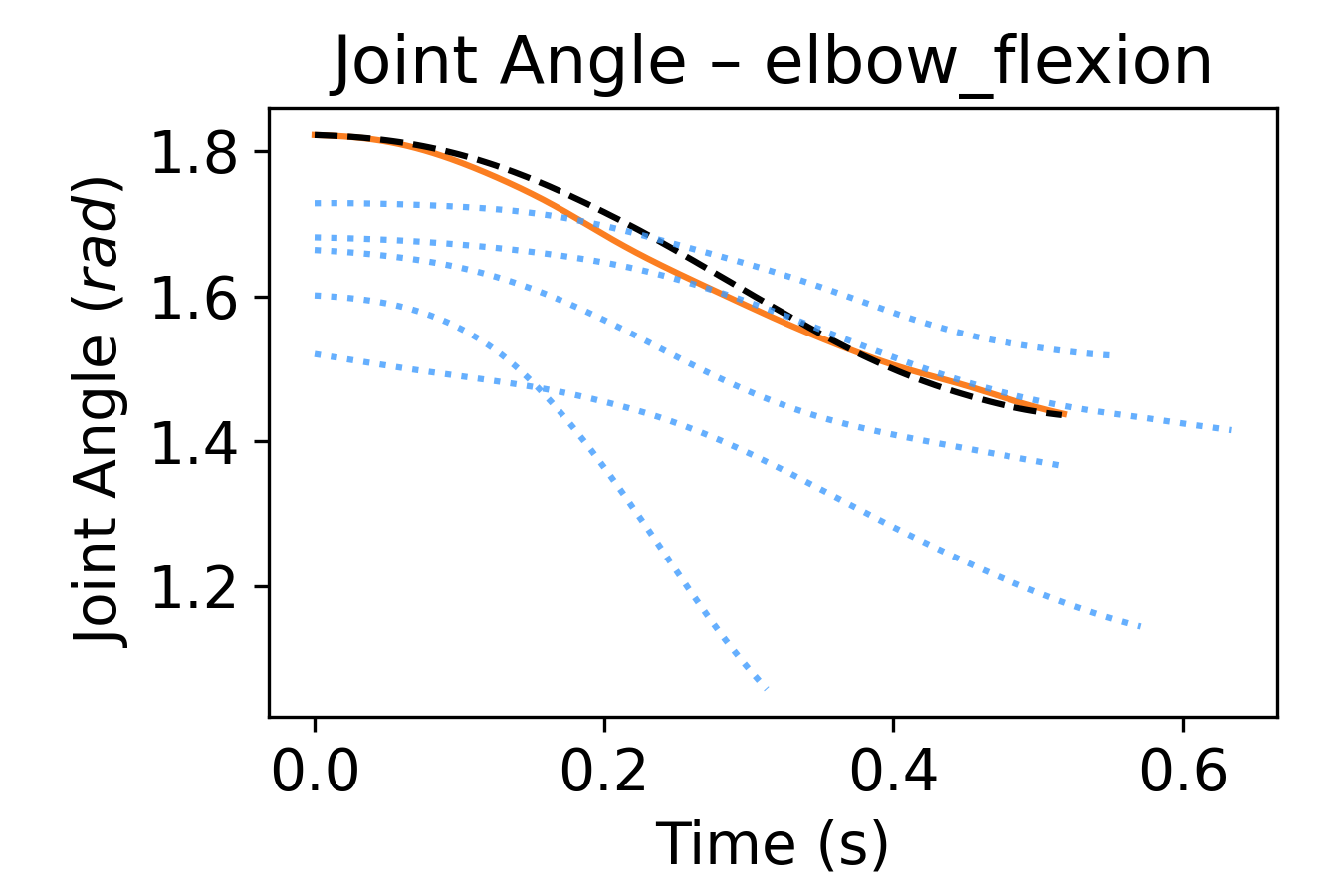}}\\
	
	\subfloat{\includegraphics[width=0.33\linewidth, clip]{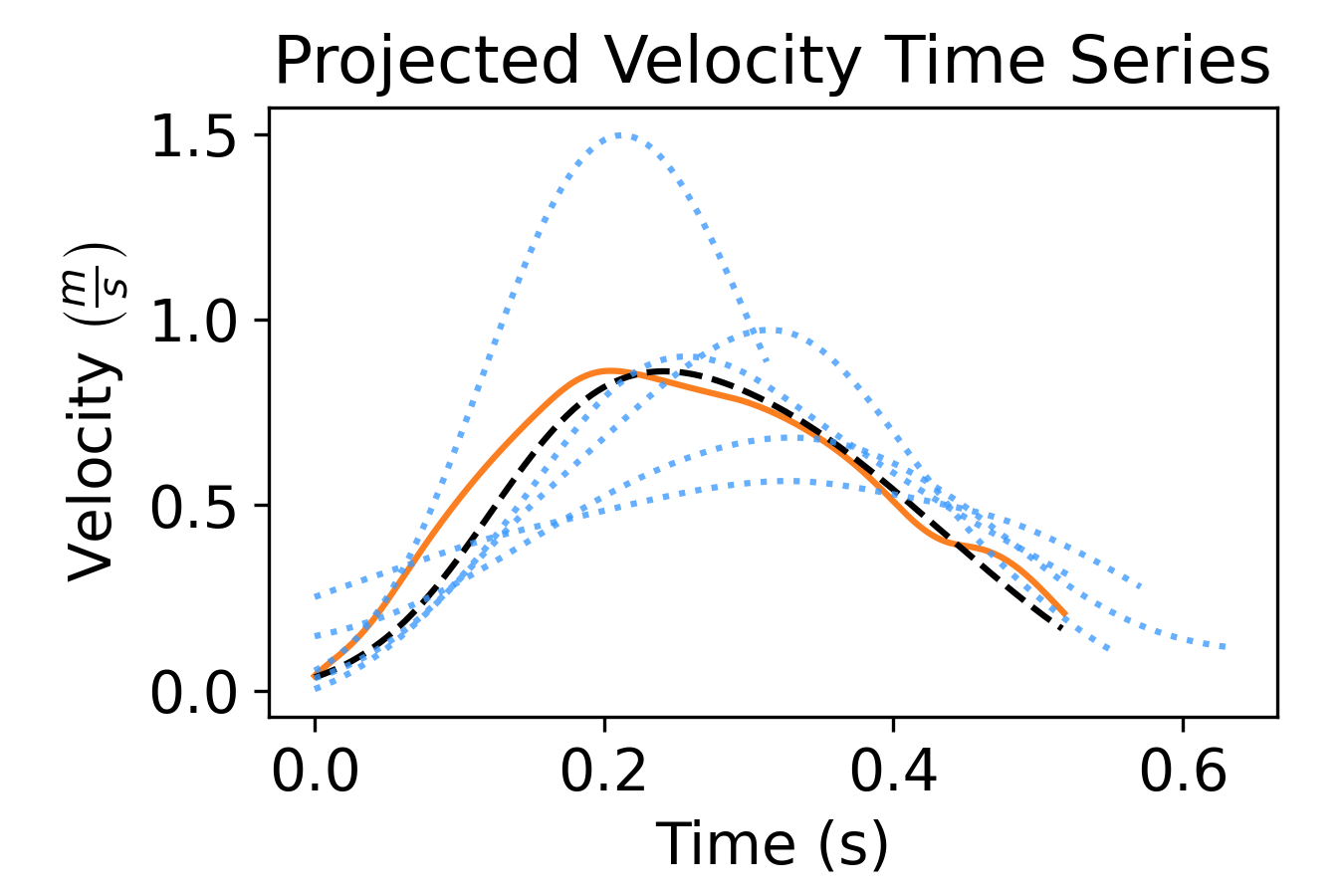}}
	\subfloat{\includegraphics[width=0.33\linewidth, clip]{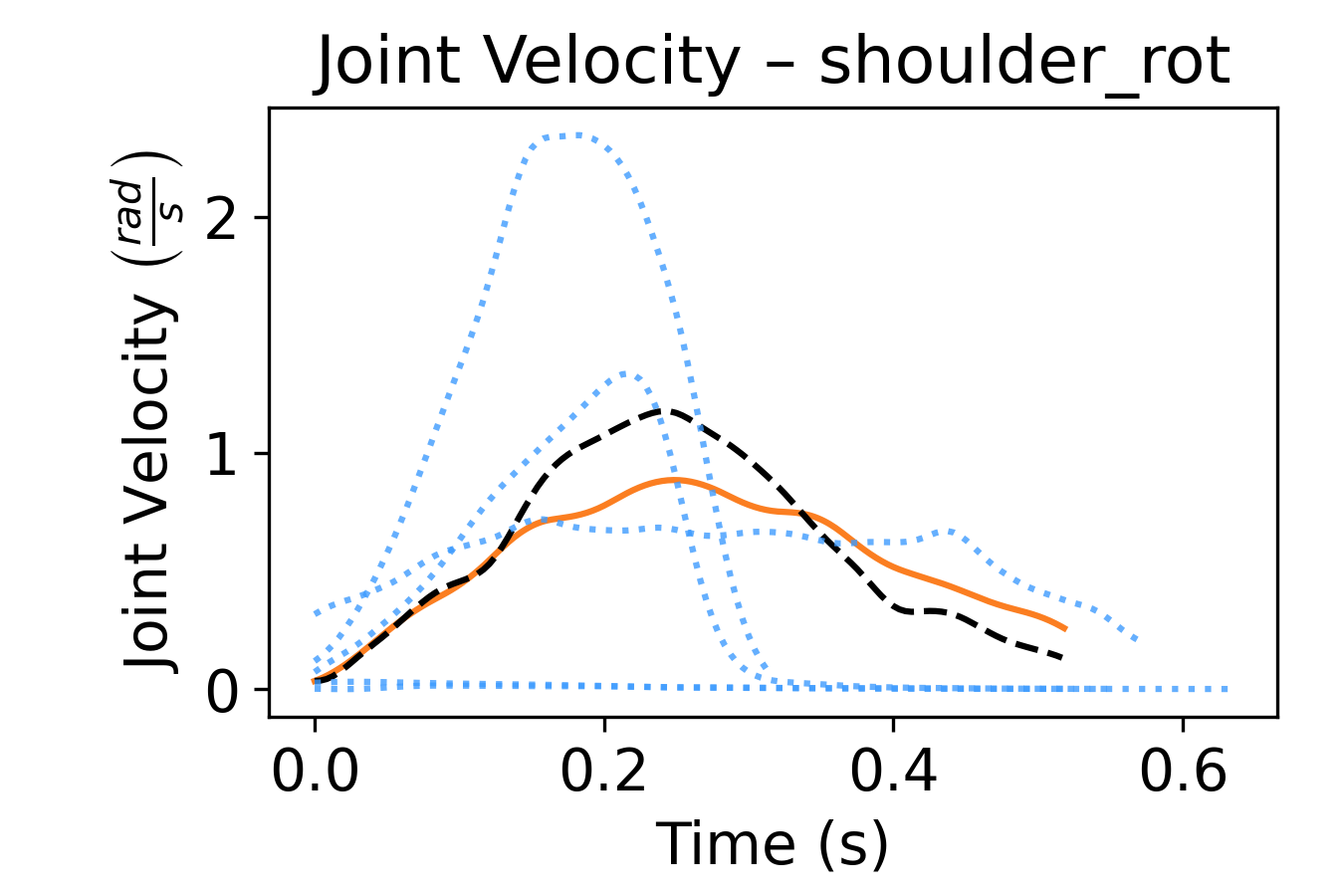}}
	\subfloat{\includegraphics[width=0.33\linewidth, clip]{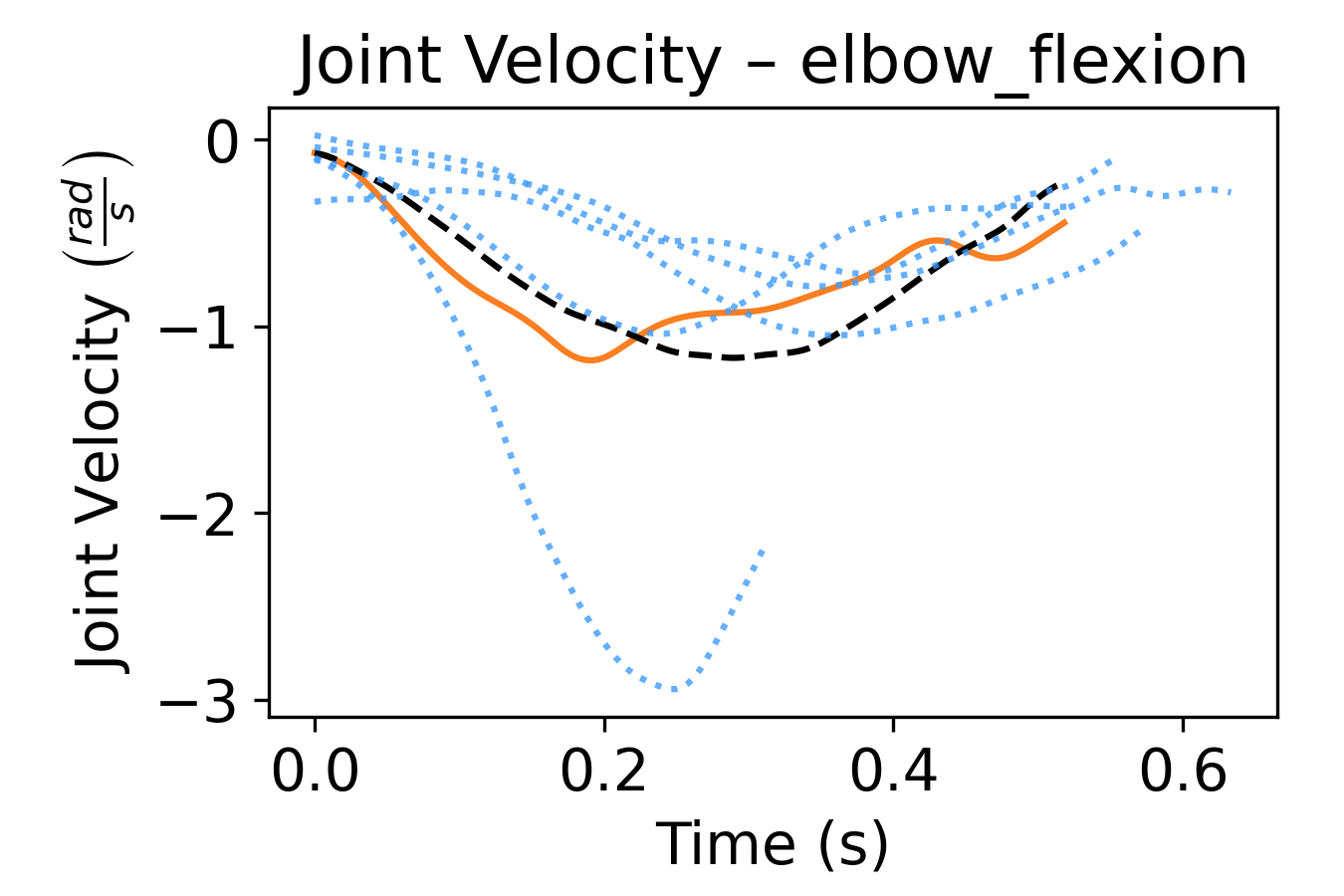}}
	
	\caption{Given an interaction technique (here: Virtual Pad Ergonomic) and a movement direction (here: movements from target 1 to target 2), the characteristic cursor and joint trajectories of an individual user (here: U4, black dashed lines; trajectories of the remaining users are shown as blue dotted lines for comparison) can be predicted by our simulation (orange solid lines).}
	\label{fig:PadErgonomic_qual}
\end{figure}

\subsubsection*{(\stepcounter{resultssubsections}\theresultssubsections\label{item:res-individual}) We can predict movement of individual users}
Besides these quantitative comparisons, we analyze how well we can predict movement characteristics of individual users.
In Figure~\ref{fig:PadErgonomic_qual}, both the simulation (orange solid lines) and the corresponding reference user trajectories (black dashed lines; for details see Section~\ref{sec:study-based-simulation}) are shown for an exemplary trial from the user study (U4, Virtual Pad Ergonomic, third movement from target 1 to target 2).
We also show the respective trajectories of the remaining users for this trial (blue dotted lines).
Figure~\ref{fig:PadErgonomic_qual} (left column) shows that we can match the characteristic projected position and velocity time series of a specific user.
Moreover, the target is reached within a single ballistic movement, and the velocity time series exhibit the bell-shaped velocity profile typically observed in aimed movements~\cite{Morasso81}.
Figure~\ref{fig:PadErgonomic_qual} (middle and right columns) illustrate how the simulation is able to distinguish the baseline user from the remaining participants in terms of shoulder rotation and elbow flexion angles and velocities.
Similar results can be observed for the remaining joints and interaction techniques, as shown in Appendix~\ref{sec:simulation-vs-user-appendix}. %

In summary, these findings show that our simulation generates biomechanically plausible joint postures, it is capable of predicting human trajectories that are within the between-user variability of the respective interaction technique, and it allows to replicate characteristic movement patterns of individual users.

\subsection{Effects of the Cost Weights and User Model Generation}\label{sec:effect_costweights}
Being able to replicate characteristic movement patterns of individual users, one natural follow-up question is how good the proposed method is in generating \textit{new} users. 
Of course, we can adjust joint ranges, but even if they are kept the same, the cost weights $r_1$ and $r_2$ of JAC (cf.~\ref{eq:costs-jointacc}) allow for significant customization, which we demonstrate in this section.

The most important insight is that the movement trajectories exhibit a \textit{continuous dependence} on $r_1$ and $r_2$, i.e., if $r_1$ and/or $r_2$ change only slightly, the movement trajectories (of cursor and joints) also change only slightly. 
With this, we provide indicators how the cost weights should be changed in order to generate, e.g., slower movements. 
To this end, we start by analyzing the effects of $r_1$ and $r_2$ on the movement trajectories. 

To avoid distorting the effects, we omit the motor noise from all simulations in this section.

\begin{figure}[!h]
	\subfloat{\includegraphics[width=0.33\linewidth, clip]{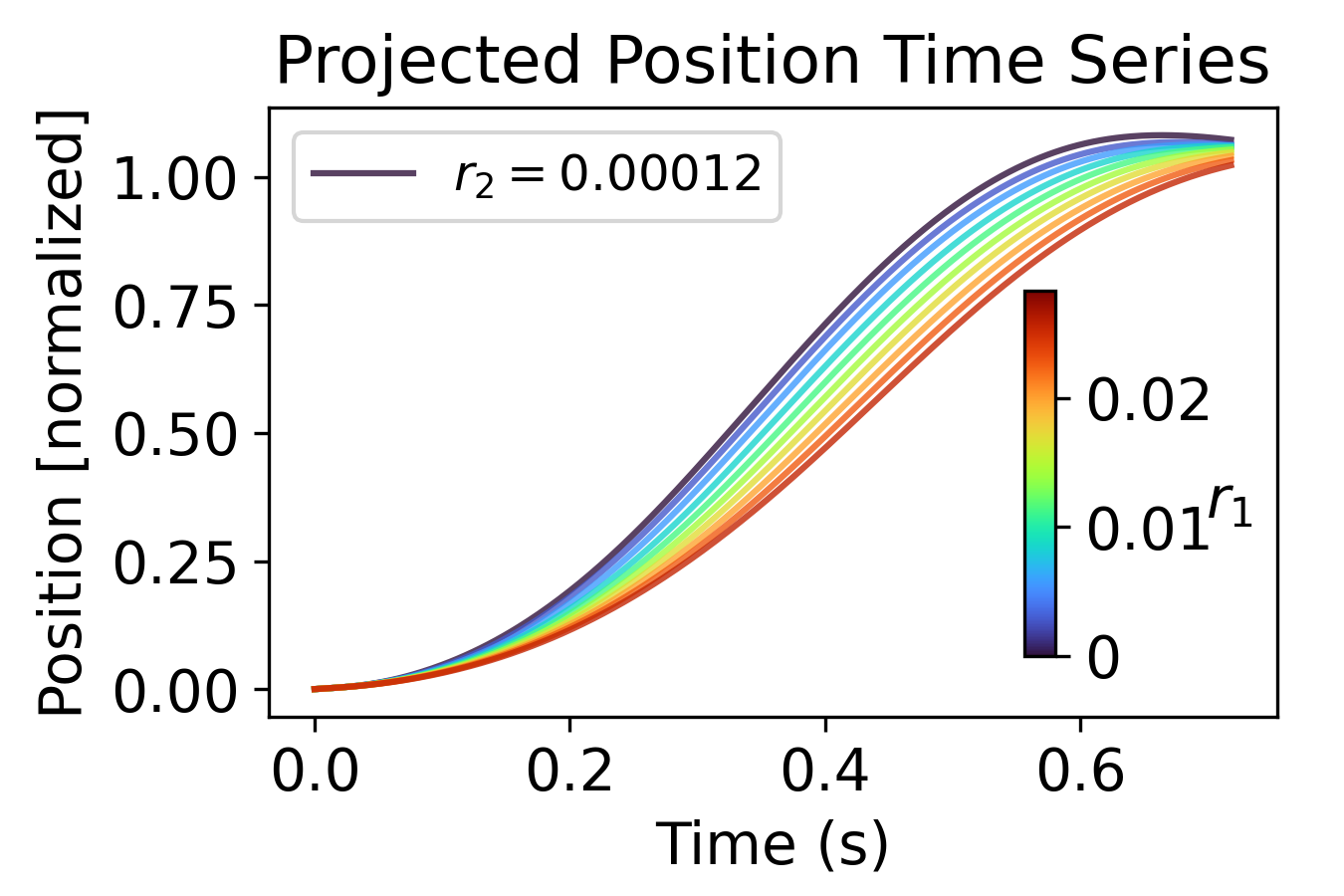}}
	\subfloat{\includegraphics[width=0.33\linewidth, clip]{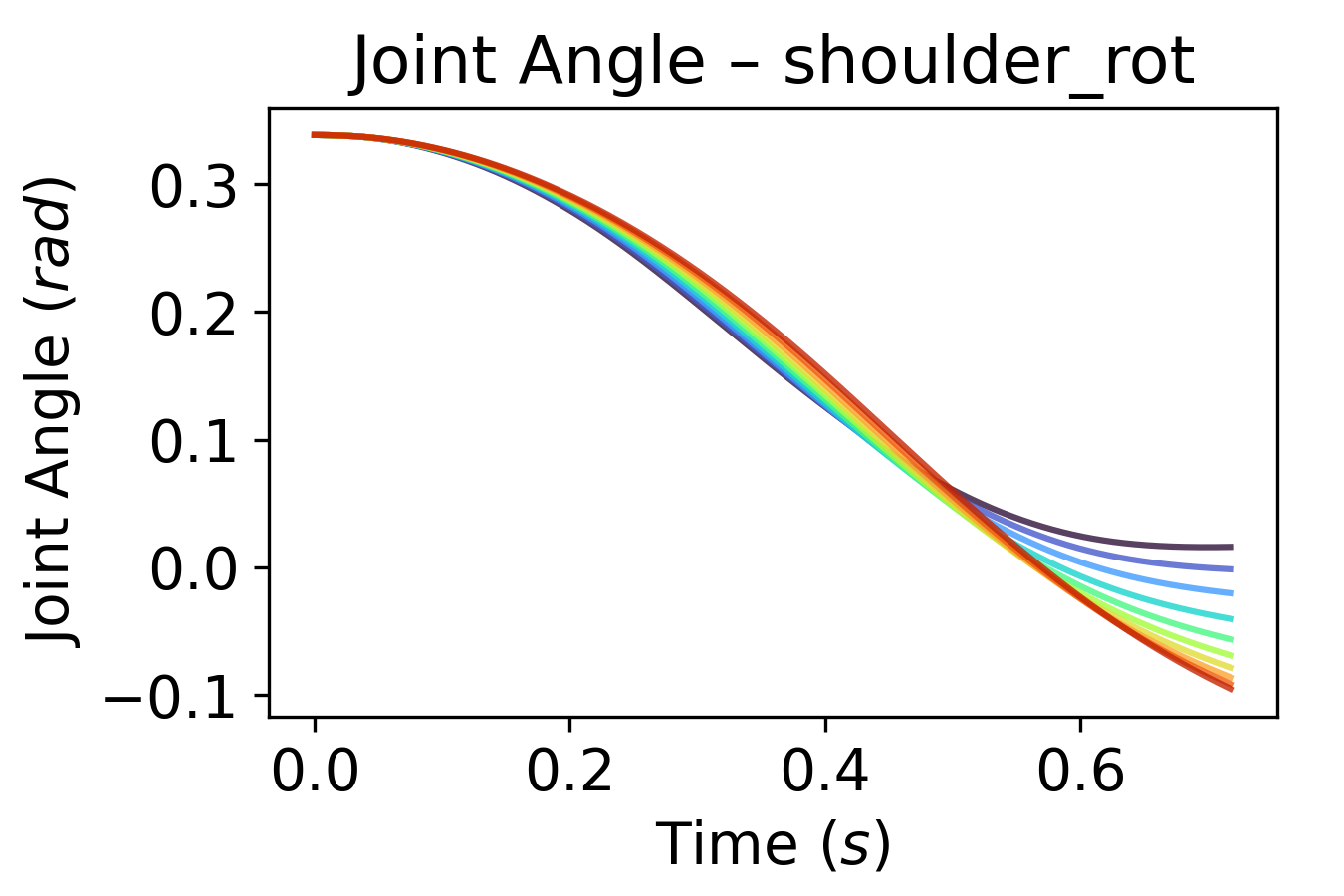}}
	\subfloat{\includegraphics[width=0.33\linewidth, clip]{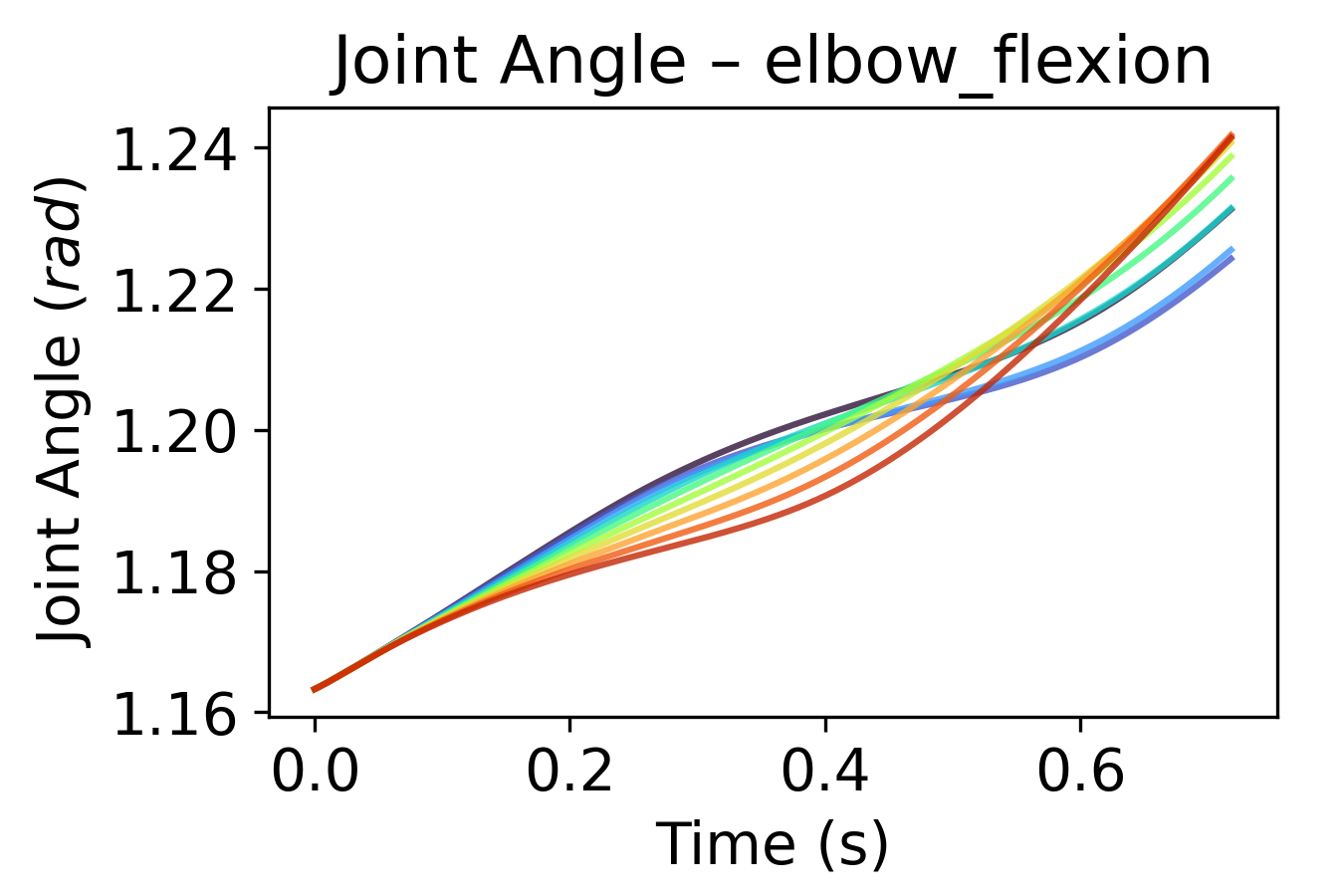}}\\
	\subfloat{\includegraphics[width=0.33\linewidth, clip]{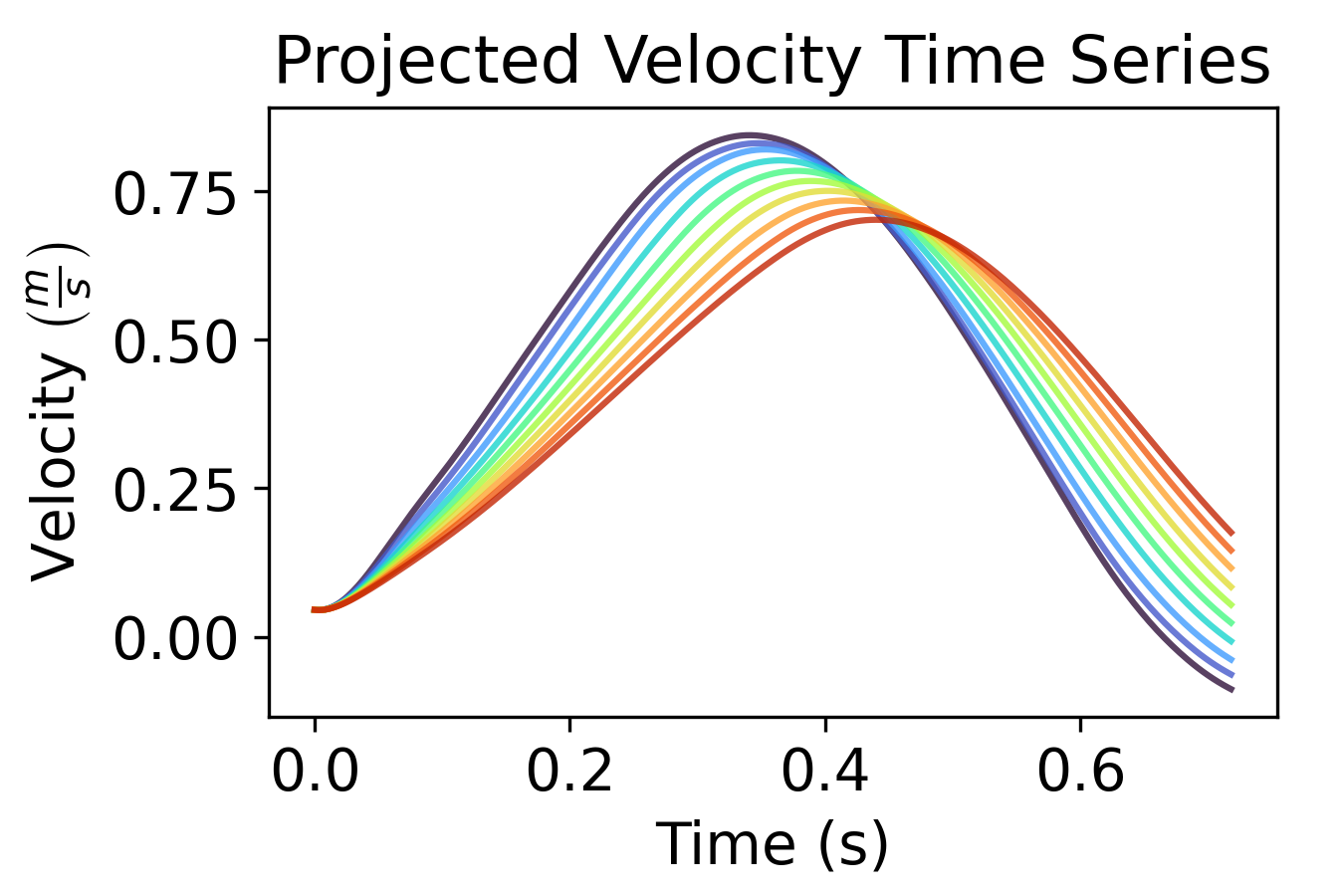}}
	\subfloat{\includegraphics[width=0.33\linewidth, clip]{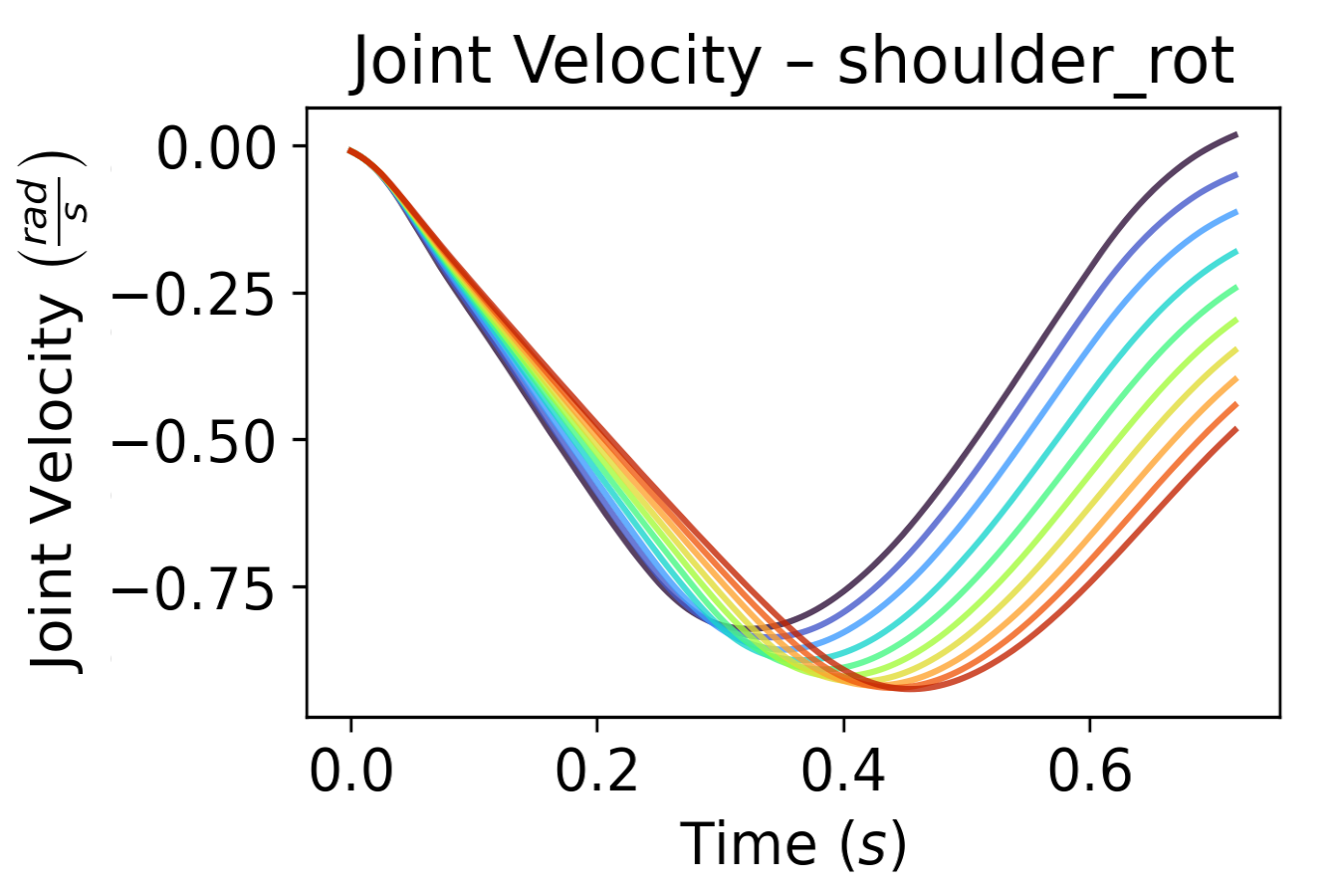}}	\subfloat{\includegraphics[width=0.33\linewidth, clip]{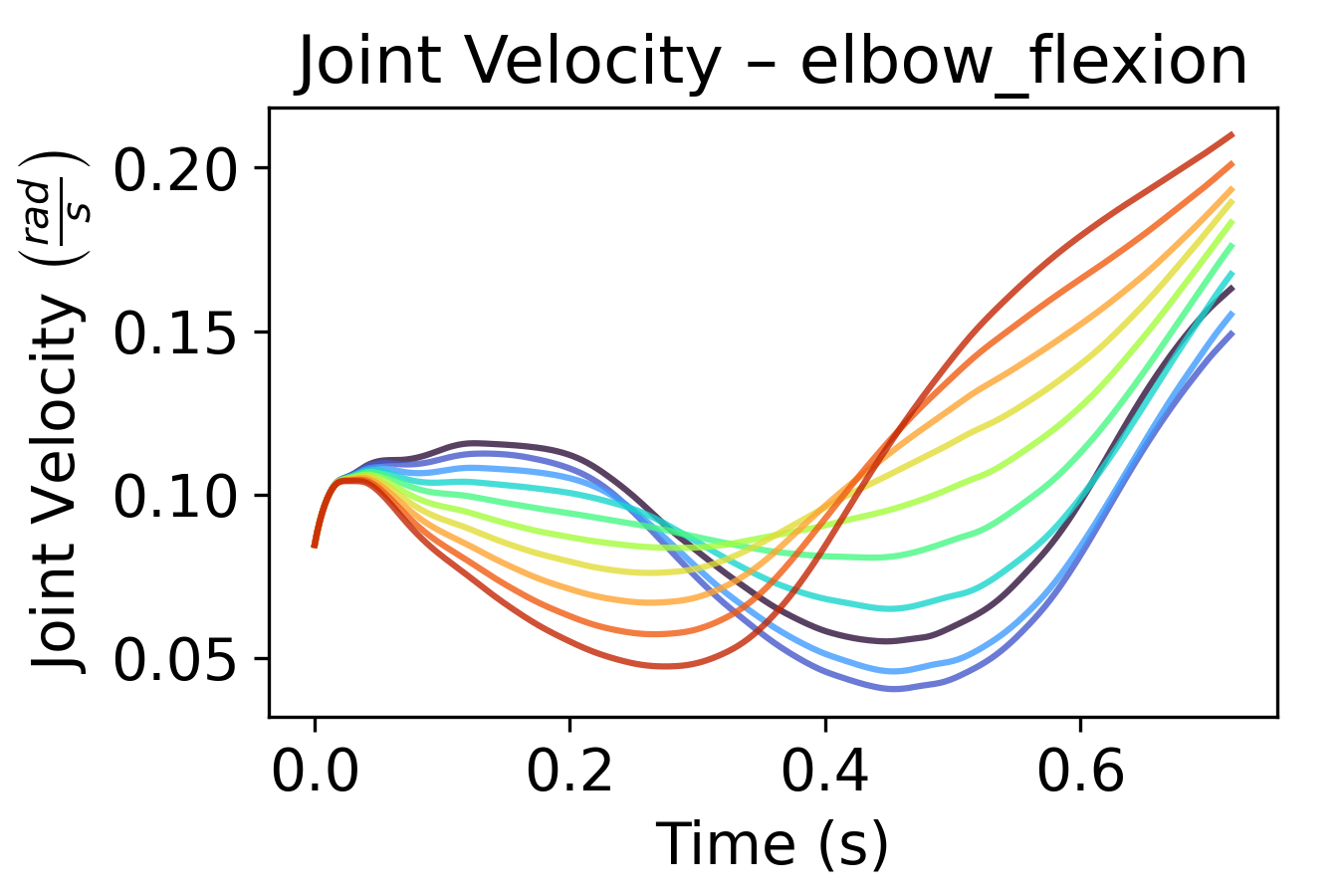}}
	
	\caption{Trajectories of the simulation for different cost weights $r_1$. It is clearly visible that an increase of $r_1$ results in slower, delayed movements towards the targets. Joint patterns are also (slightly) affected by the choice of $r_{1}$.}
	\label{fig:accjoint_r1_qual}
\end{figure}
	
\begin{figure}[!h]
	\subfloat{\includegraphics[width=0.33\linewidth, clip]{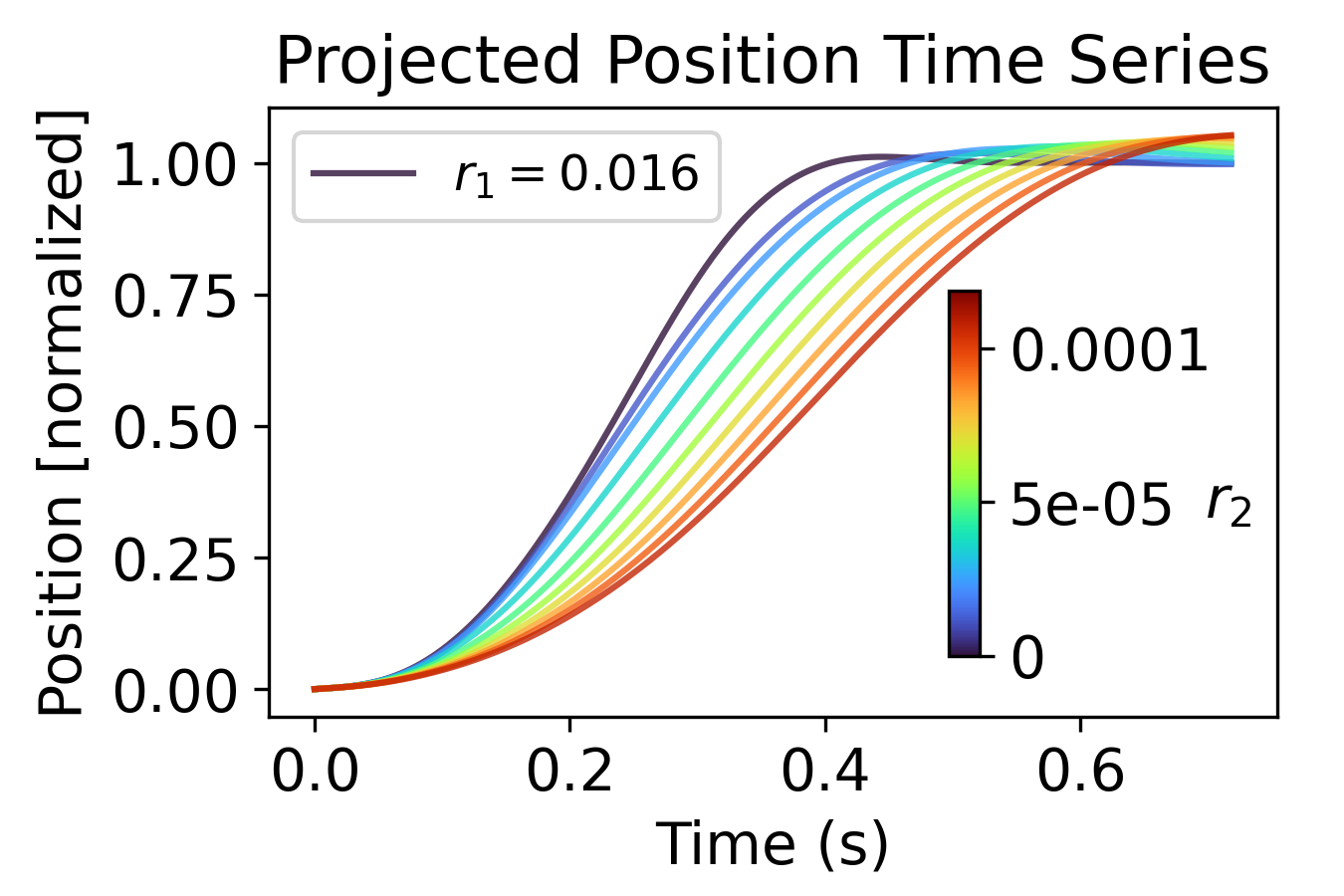}}
	\subfloat{\includegraphics[width=0.33\linewidth, clip]{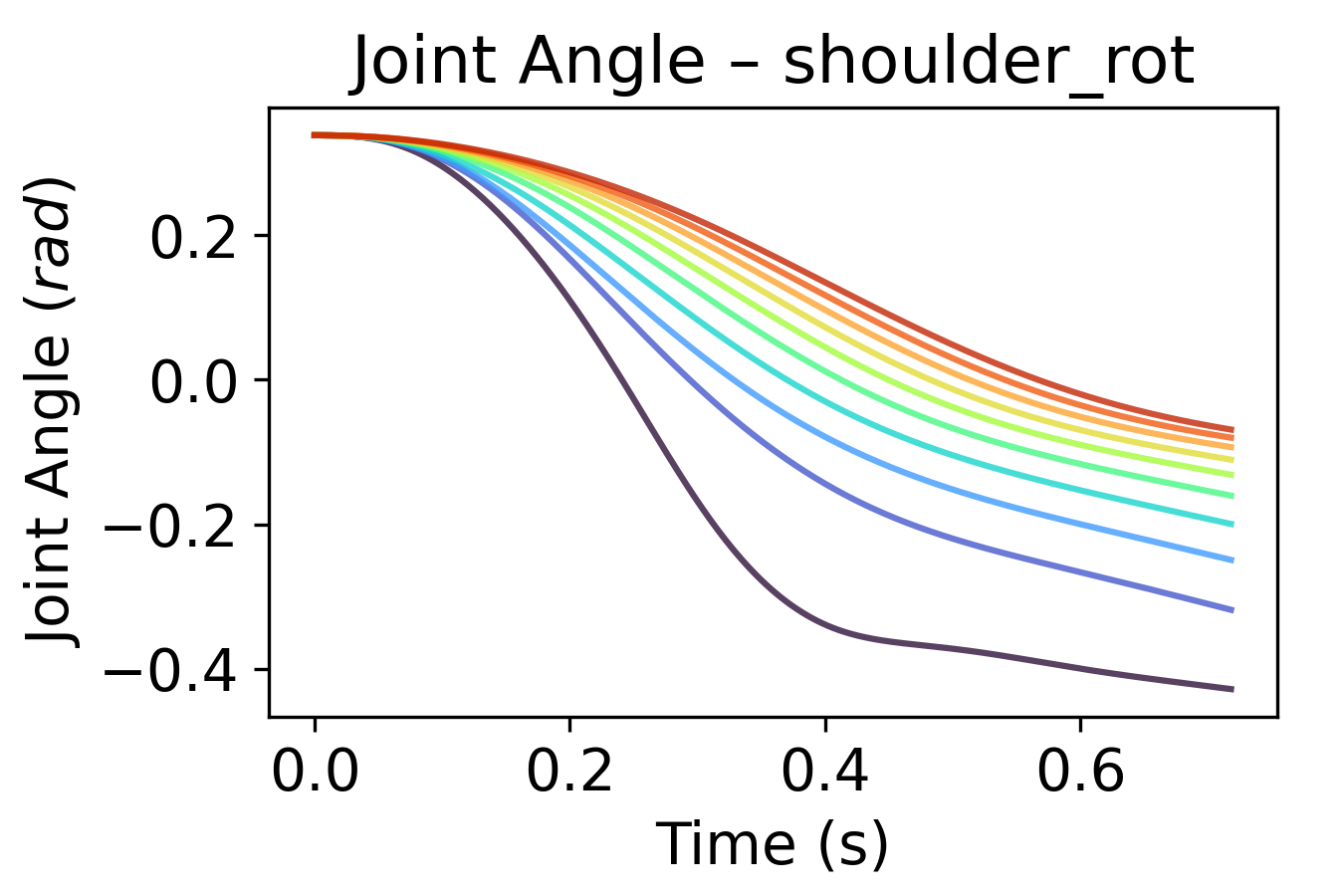}}
	\subfloat{\includegraphics[width=0.33\linewidth, clip]{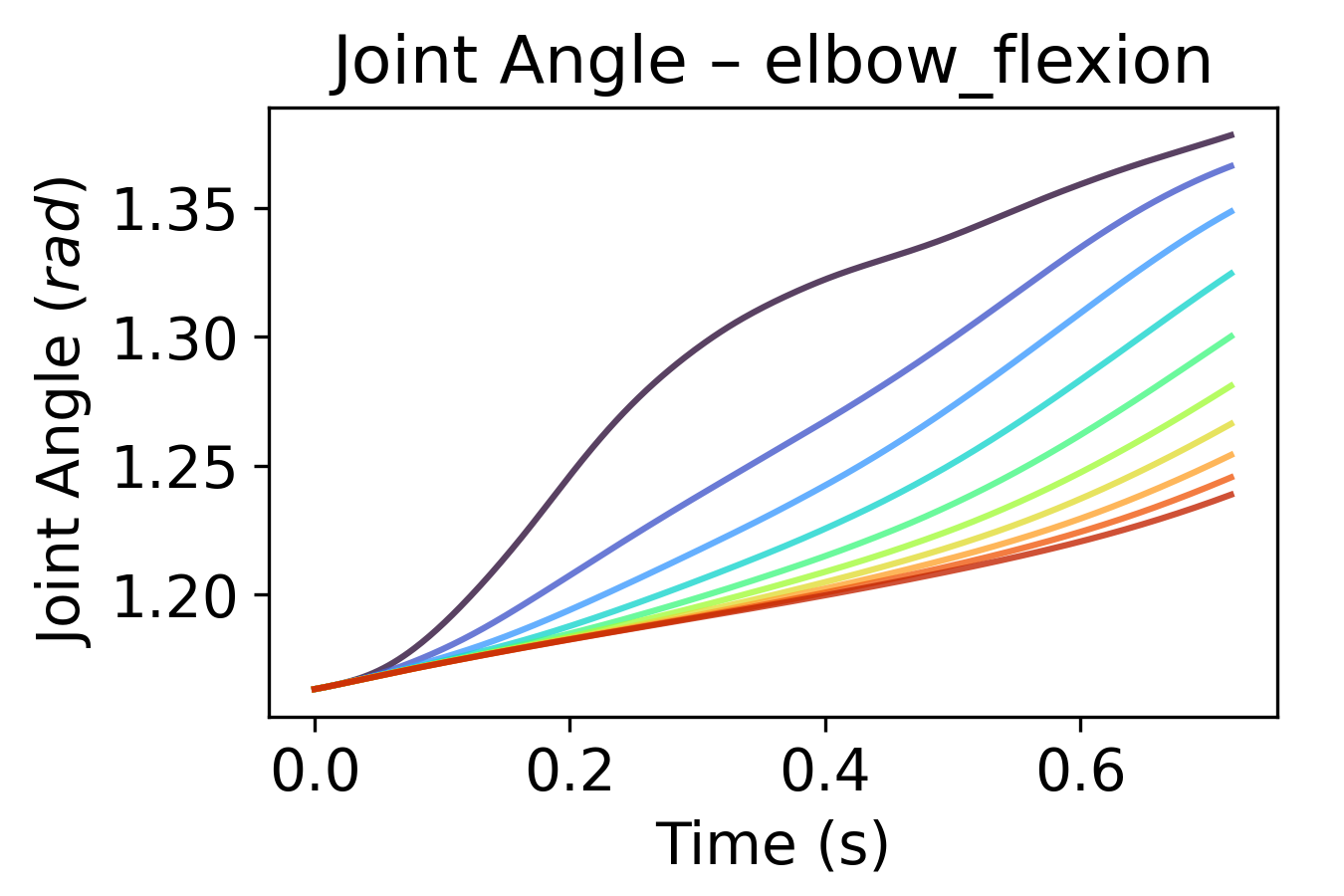}}\\
	\subfloat{\includegraphics[width=0.33\linewidth, clip]{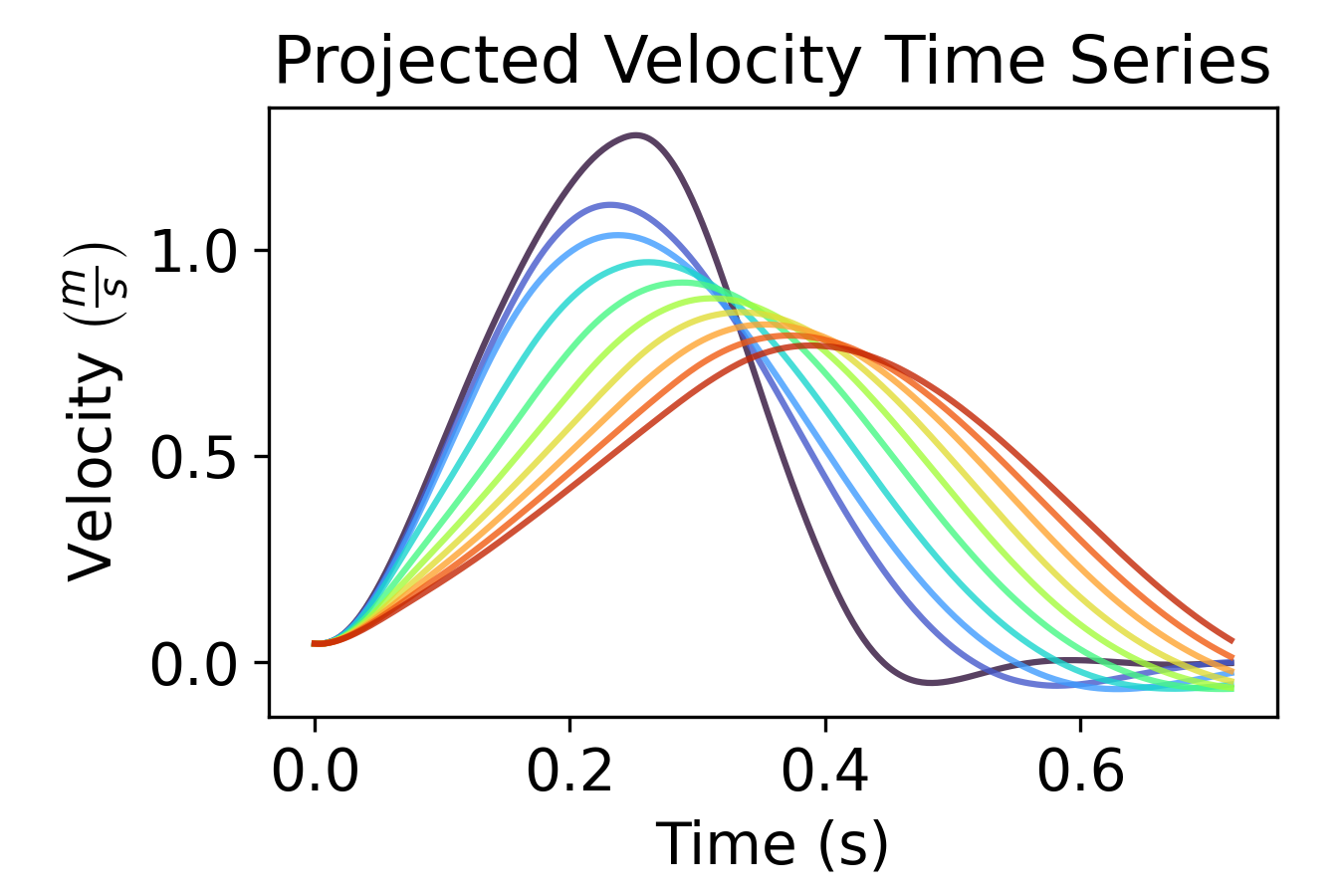}}
	\subfloat{\includegraphics[width=0.33\linewidth, clip]{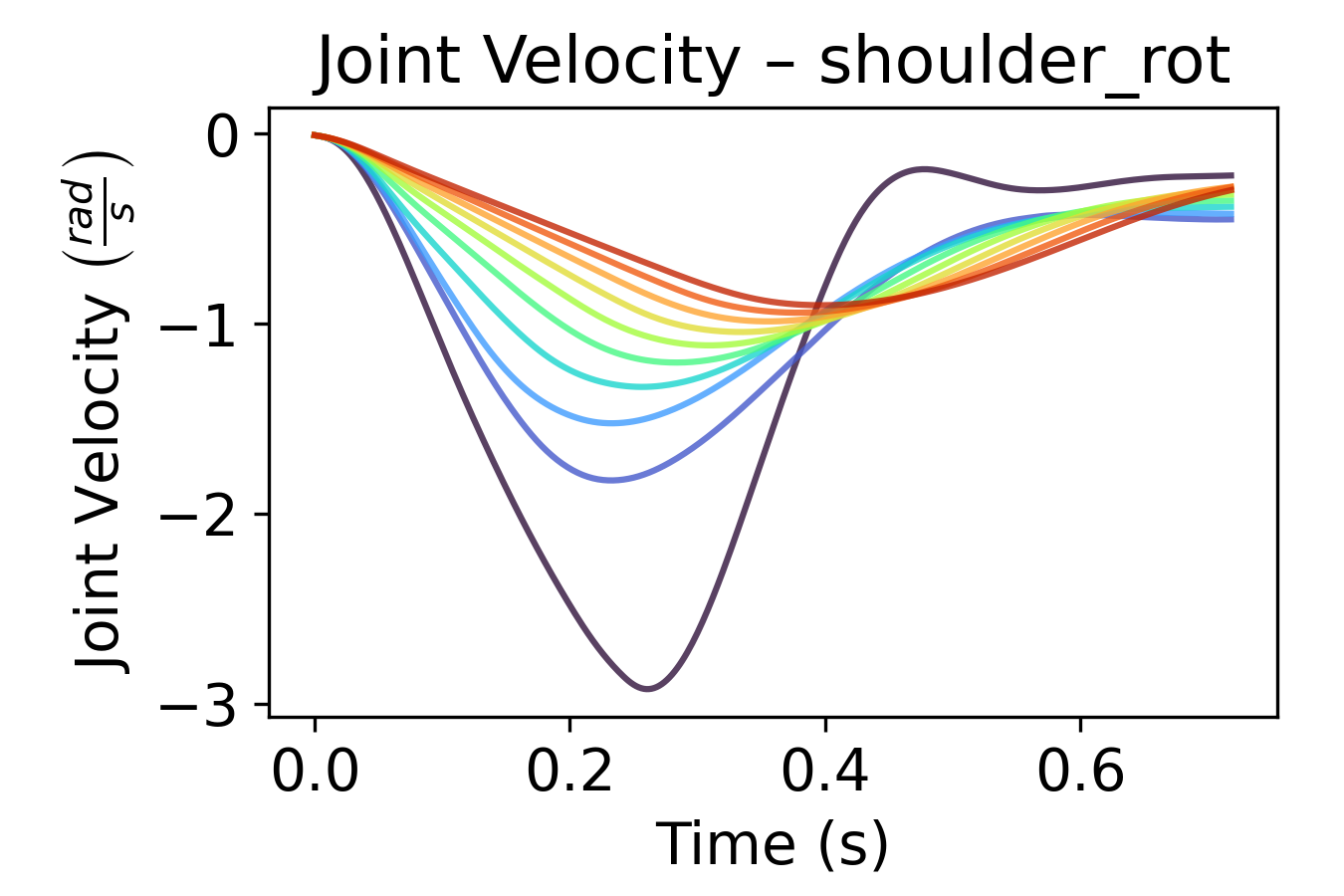}}	\subfloat{\includegraphics[width=0.33\linewidth, clip]{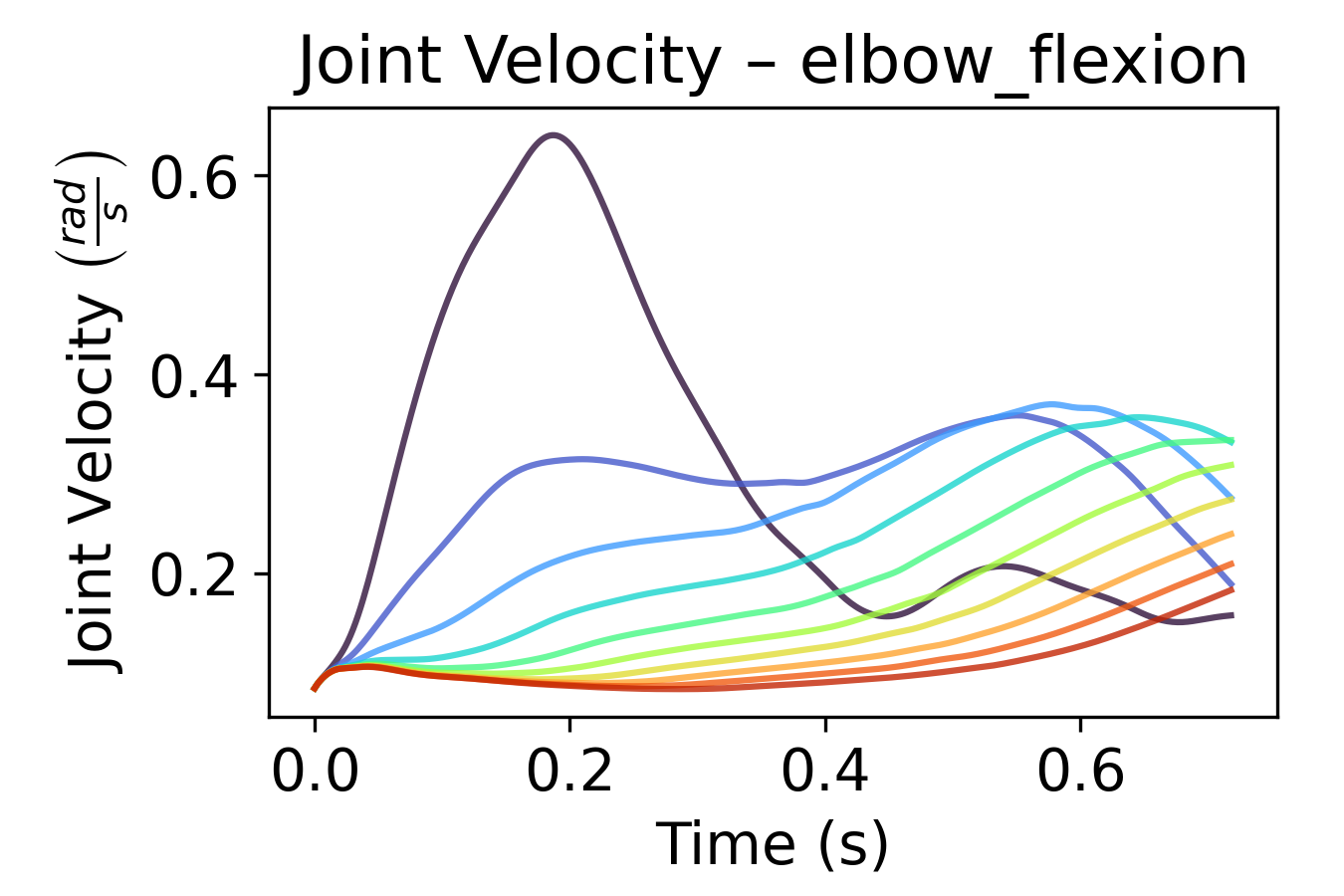}}
	
	\caption{Trajectories of the simulation for different cost weights $r_2$. The clear cursor and joint profiles that can be observed for $r_{2}=0$ (black lines) become more and more smooth as $r_{2}$ increases, resulting in a similar effect as observed for $r_{1}$.}
	\label{fig:accjoint_r2_qual}
\end{figure}

In Figure~\ref{fig:accjoint_r1_qual}, projected cursor and joint trajectories are shown for 10 different values of the control costs weight $r_{1}$, with constant joint acceleration costs weight $r_{2}=0.00012$ (U4, Virtual Cursor Ergonomic, first movement between targets 4 and 5).
There is a clear decrease in peak velocity, as $r_{1}$ increases, resulting in considerably slower movements with target reached later. 
Moreover, the projected cursor velocity profile becomes more right-skewed, which can be explained by the increased incentive to apply lower torques per time step (note that the penalization of squared control signals incentives the use of multiple small control signals instead of one large control signal). 
This suppressive effect of increased control costs can also be observed in the shoulder rotation plots, where larger $r_{1}$ values result in considerably less negative joint velocities at the beginning of the movement, which need to be compensated later in order to reach the target.
The elbow flexion is also affected by changes in the control costs weight, however, with comparably small impact (note the considerably smaller joint angle range).
Note that for very large $r_{1}$ values, the only relevant objective is to reduce the control cost, i.e., it is optimal to apply no controls during the entire movement, which causes the arm to fall.

In Figure~\ref{fig:accjoint_r2_qual}, the effects of the joint acceleration costs weight $r_{2}$ are shown for the same participant, interaction technique, and trial, using a fixed intermediate control costs weight $r_{1}=0.016$.
The effect on projected cursor trajectories is qualitatively comparable to that of $r_{1}$, albeit considerably more pronounced. 
Without joint acceleration costs, i.e., $r_{2}=0$ (black lines), the projected velocity time series exhibits a very high peak velocity that is compensated by a corrective submovement starting after $\sim450$ milliseconds, resulting in a very fast movement towards the target. %
For shoulder rotation and elbow flexion, the changes in both joint angles and velocities strongly decrease as $r_{2}$ increases. In particular, the characteristic joint patterns that can be observed for $r_{2}=0$ (black lines) become more and more flattened as joint accelerations are more penalized.

The effects of the two cost weights on the remaining joints are depicted in Figures~\ref{fig:accjoint_r1_qual_otherjoints} and~\ref{fig:accjoint_r2_qual_otherjoints} in the Appendix, respectively. %

\begin{figure}[h!]
	\centering
	\subfloat{\includegraphics[width=0.66\linewidth]{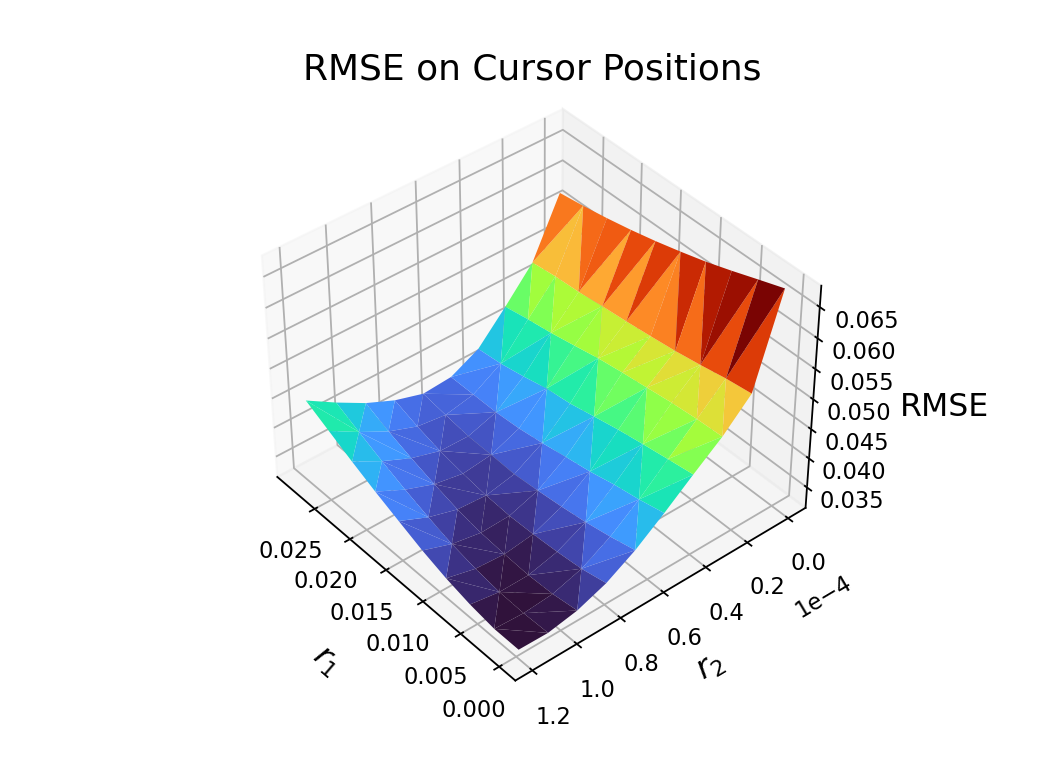}}
	
	\caption{Surface plot showing the performance of the simulation with different cost weights $r_{1}$ and $r_{2}$. The z-axis corresponds to the mean RMSE on cursor positions across all trials of U4 using the Virtual Cursor Ergonomic technique.}
	\label{fig:cost_weight_rmse}
\end{figure}

To examine, how well a certain cost weight pair fits one specific user, the surface plot in Figure~\ref{fig:cost_weight_rmse} shows a simulation vs. user comparison (U4, Virtual Cursor Ergonomic) for different combinations of cost weights.
Here, the z-axis shows the mean RMSE used to measure the similarity between simulation and user data trajectories (see Section~\ref{sec:fittingcostweights}) with respect to cursor position across all trials of U4 using the Virtual Cursor Ergonomic technique.
As can be seen, the surface is clearly convex.
While convexity of the RMSE with respect to the cost weights can neither be guaranteed nor expected in general, it ensures that there exists a unique global optimum, towards which most OCP solvers (including the one we use) are guided.
In the considered case, the global optimum is located at $r_{1}=0$ and $r_{2}\approx1.1e-4$ (black colored valley).\footnote{Note, that this does not exactly coincide with the cost weight pair found by our fitting process. This is due to the facts, that during the fitting, we use RMSE based on joint angles, and that out of computational reasons, we only considered five movements to evaluate a cost weight pair.} %

In summary, our analysis of cost weight effects shows that both the cursor and joint trajectories predicted by the closed MPC loop continuously depend on the cost weight parameters. %
This has two major advantages:
First, it allows to generate new and reasonable user models by tweaking the cost weight parameters.
Second, it supports the parameter fitting process in finding cost weights that best reflect specific user behavior.

\subsection{Effects of the MPC Horizon}\label{sec:effect_N}

To better understand the impact of the MPC horizon $N$, which may need manual adjustments depending on the task under consideration, in the following we analyze how the simulated movements change with increasing horizon.

We also omit the motor noise from all simulations done in this section to avoid distorting the effect of $N$.

\begin{figure}[!h]
	\subfloat{\includegraphics[width=0.33\linewidth, clip]{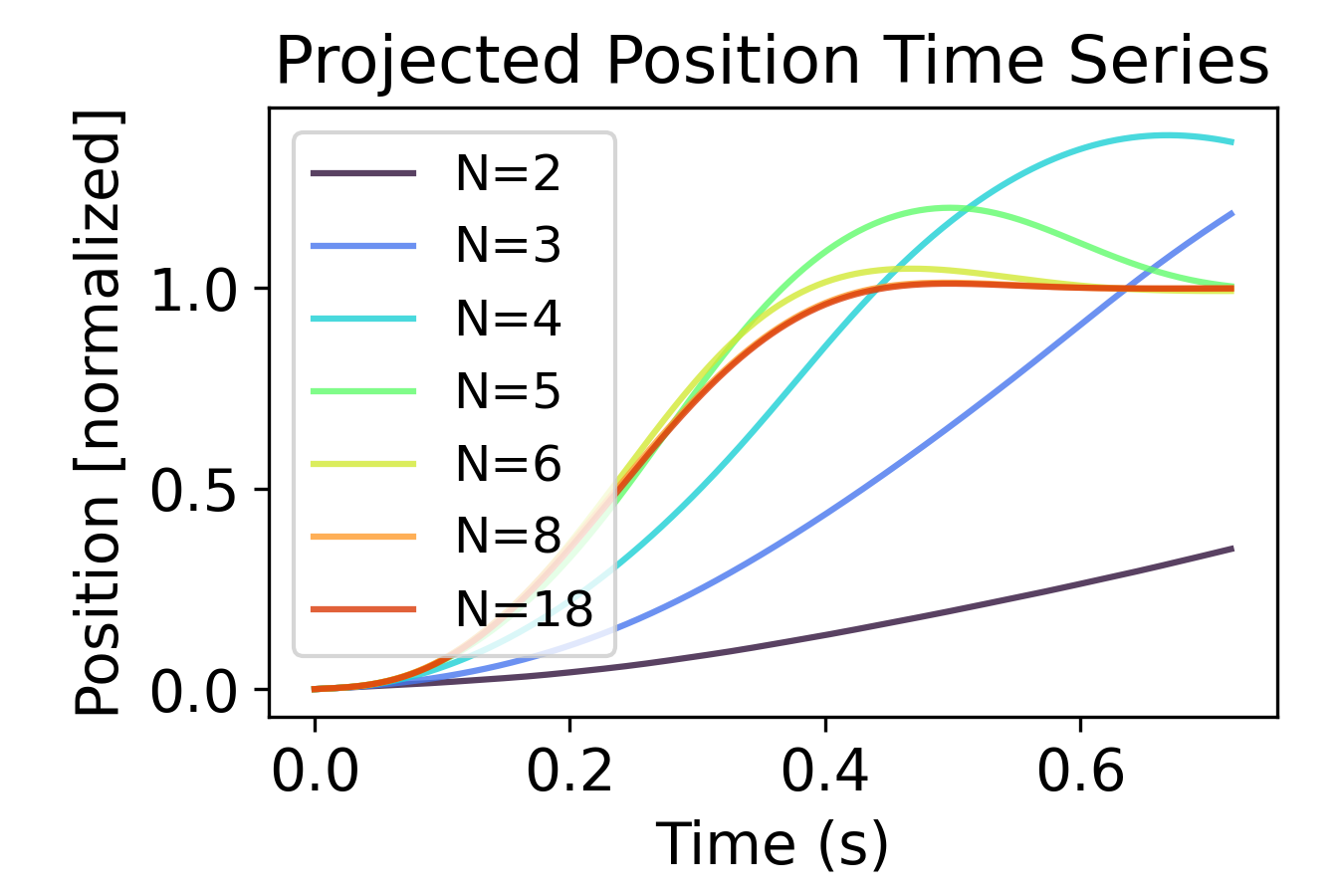}}
	\subfloat{\includegraphics[width=0.33\linewidth, clip]{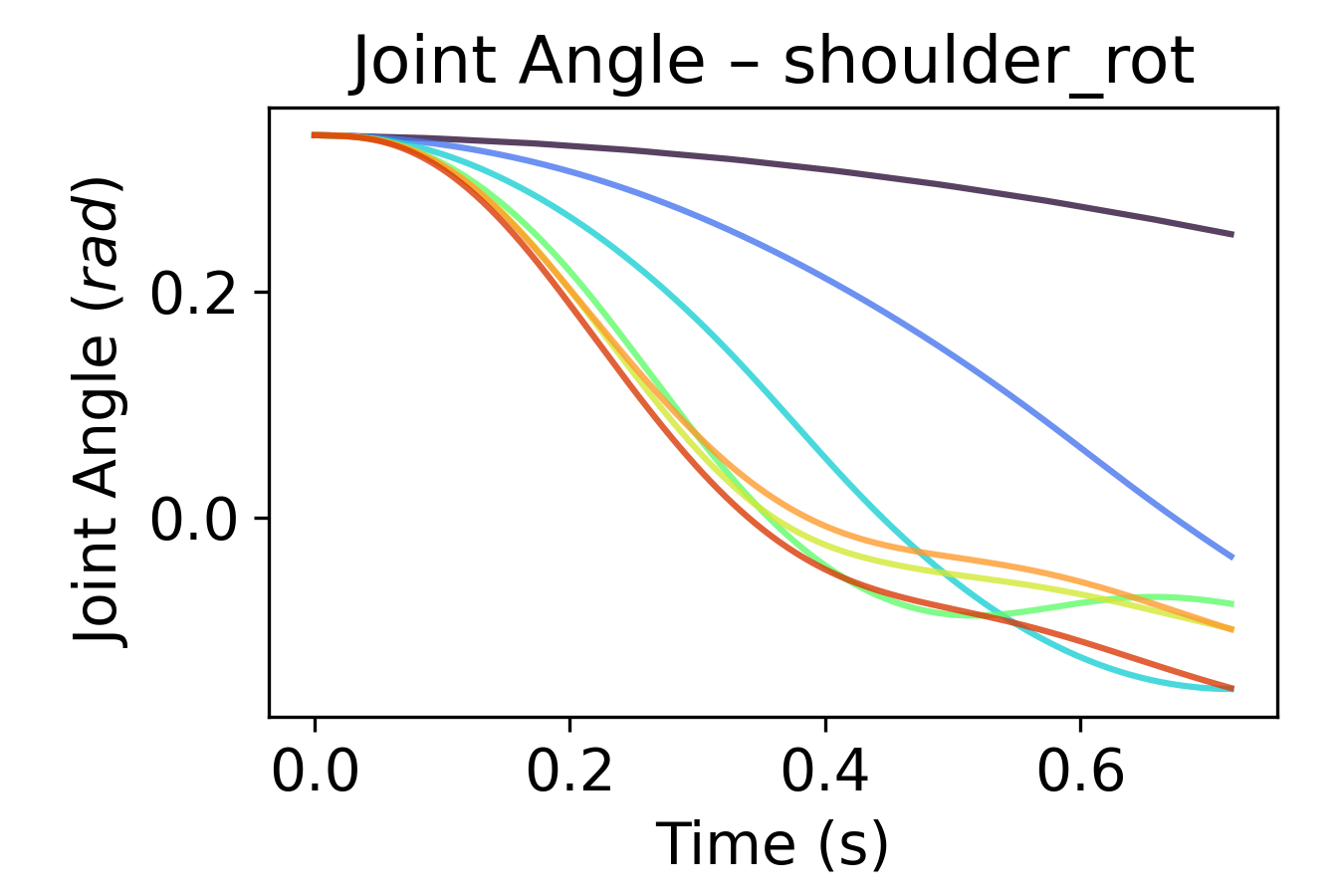}}
	\subfloat{\includegraphics[width=0.33\linewidth, clip]{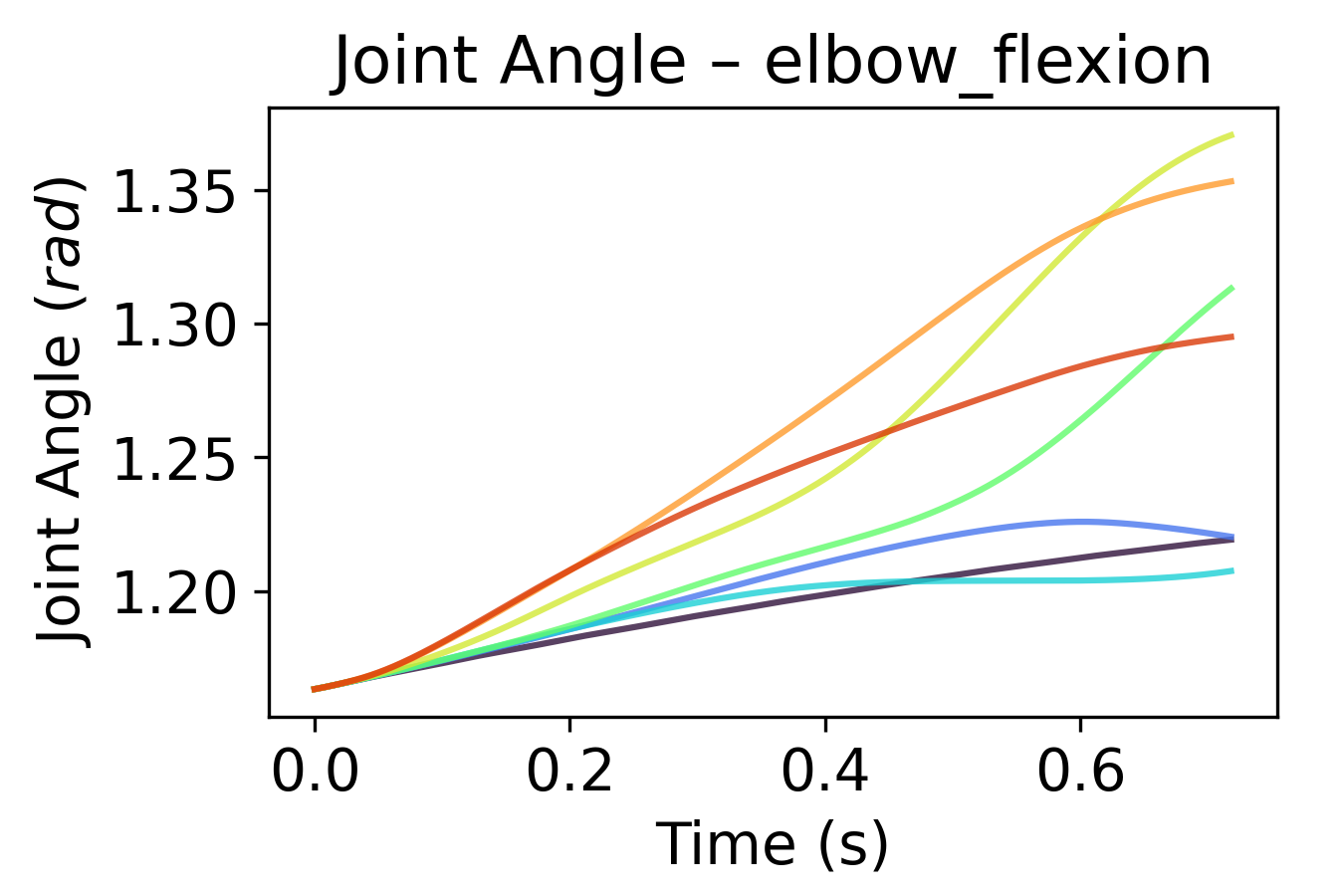}}\\
	\subfloat{\includegraphics[width=0.33\linewidth, clip]{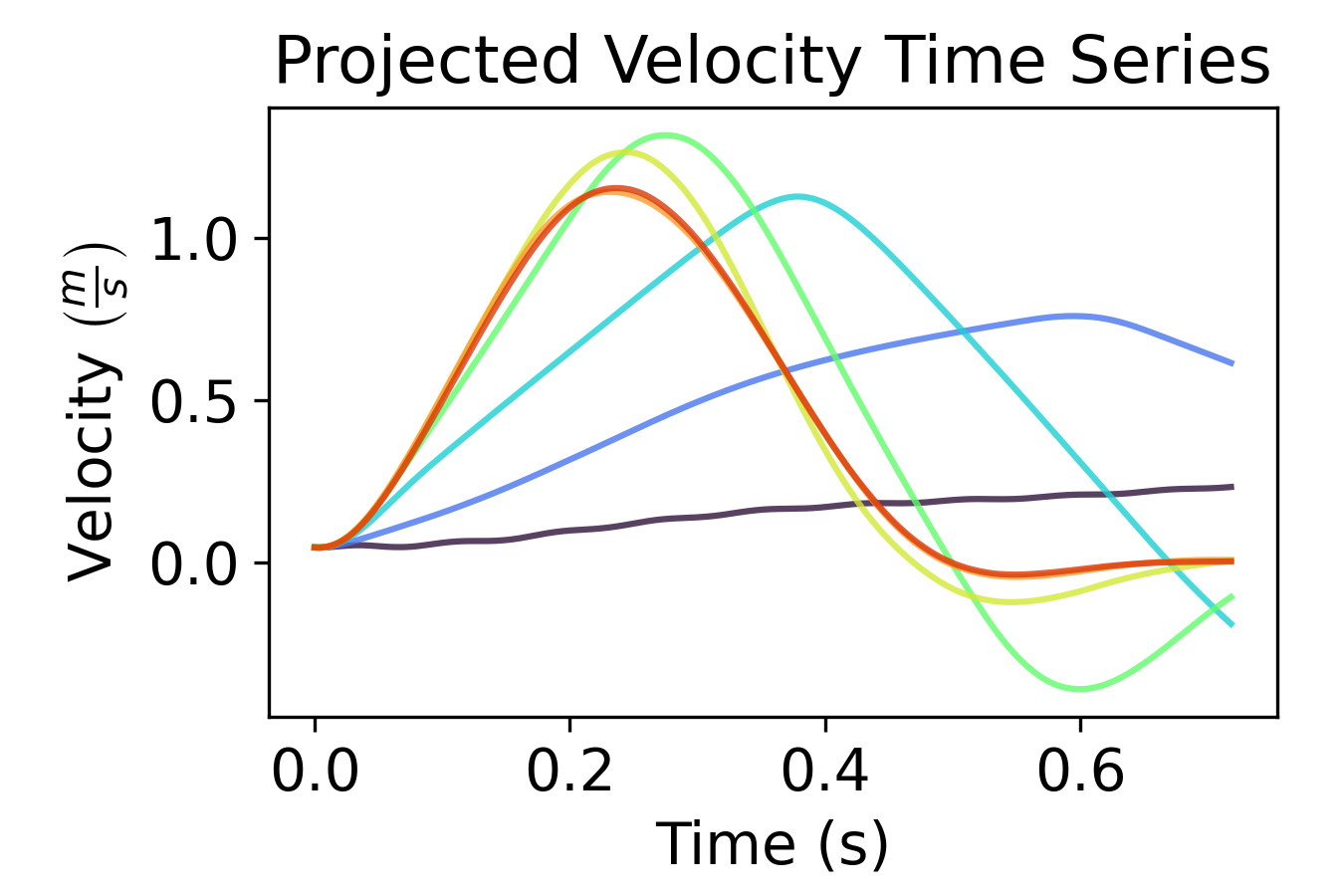}}
	\subfloat{\includegraphics[width=0.33\linewidth, clip]{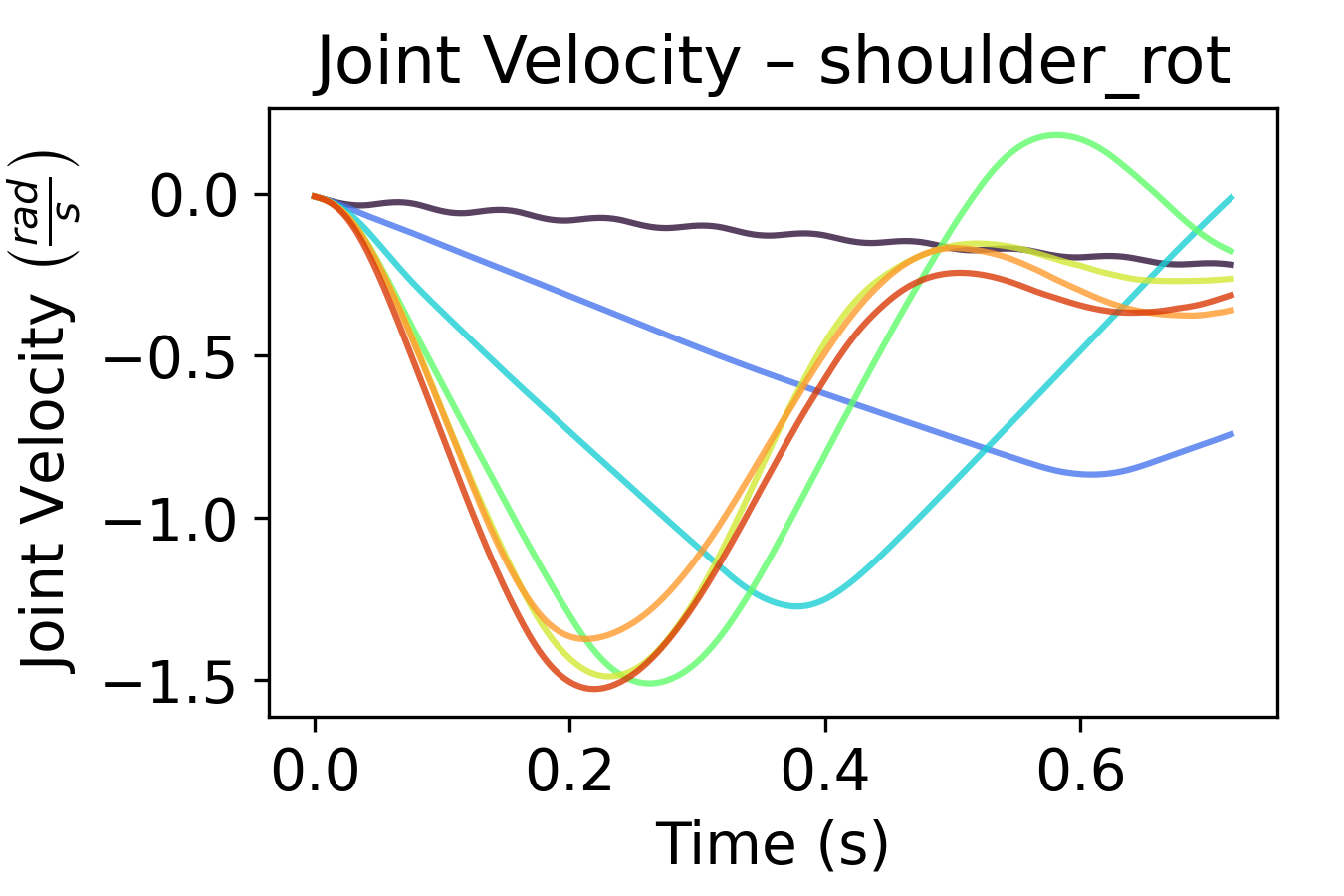}}	\subfloat{\includegraphics[width=0.33\linewidth, clip]{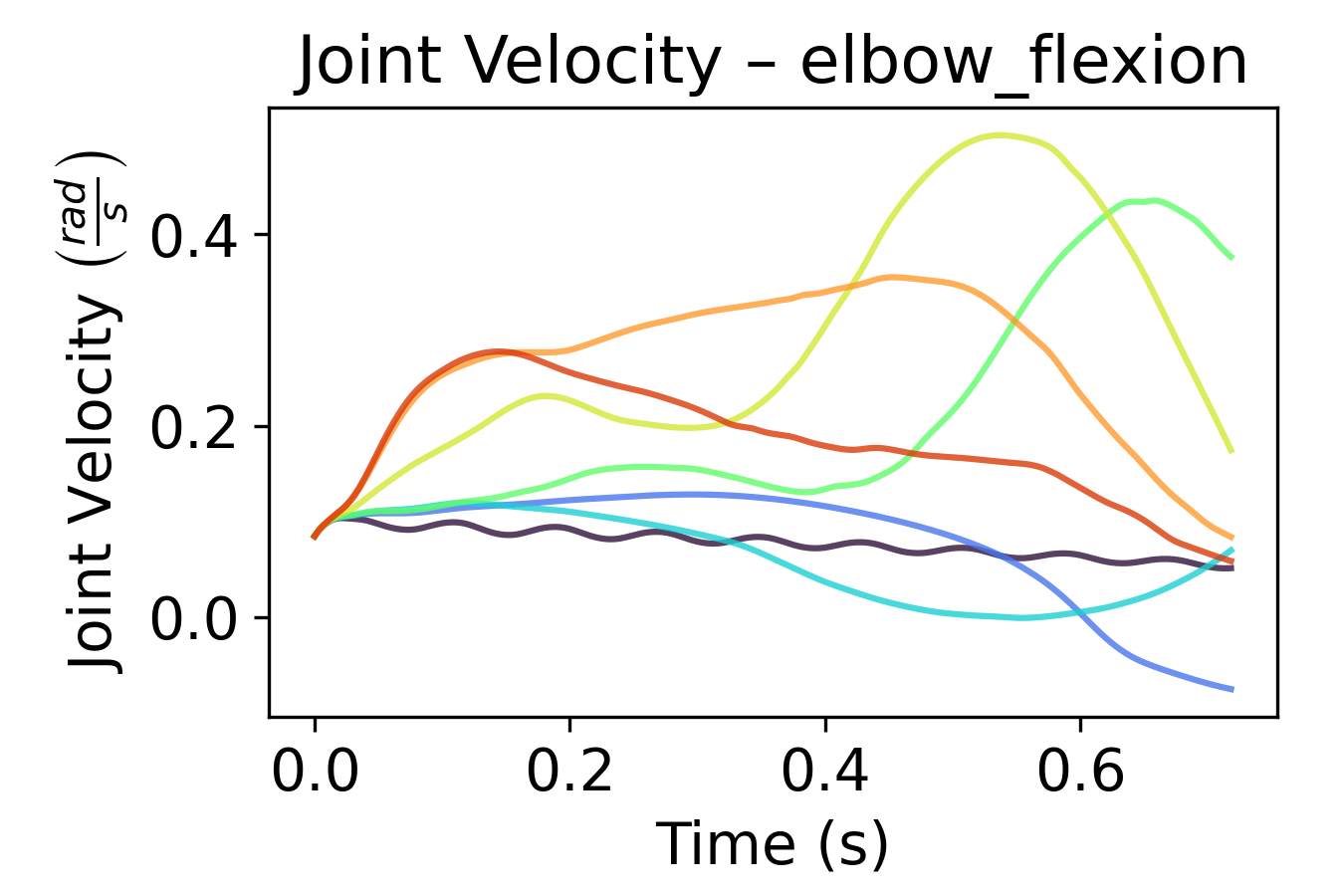}}
	
	\caption{Projected cursor and joint trajectories of one trial for varying MPC horizon $N$, using the Joint Acceleration Costs (JAC), without control noise.}
	\label{fig:accjoint_N_qual}
\end{figure}

As depicted in the top left plot of Figure~\ref{fig:accjoint_N_qual} for the same trial as used in Section~\ref{sec:effect_costweights}, the MPC horizon N, which determines how many future steps are taken into account to select the control at a certain time step, has a considerable effect on the resulting closed-loop trajectories. Choosing $N$ too small results in movements that are either too slow to reach the target at all ($N=2$, black line), cross the target but do not return within reasonable time ($N=3,4$, blue lines), or exhibit some considerable overshoot ($N=5$, grass green line). Starting from $N=8$, the differences in cursor and joint trajectories are quite small and hardly visible anymore. This suggests that a prediction horizon of $8\cdot40$ ms $=320$ ms is sufficient to adequately solve our optimal control problem via MPC.

\begin{figure}[h!]
	\centering
	\subfloat{\includegraphics[width=0.5\linewidth, clip]{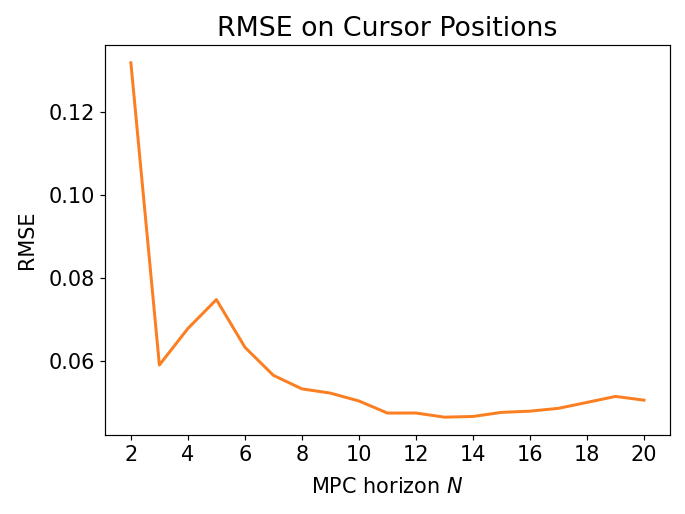}}
	
	\caption{Effect of the MPC horizon $N$ on the mean RMSE between JAC and user data in terms of cursor positions, considering all trials of U4 with the Virtual Cursor Ergonomic technique. The performance clearly deteriorates when the MPC horizon $N$ is set too low (i.e., below $N=8$). 
	}
	\label{fig:n_rmse}
\end{figure}

This is also confirmed by the quantitative comparison of MPC horizons shown in Figure~\ref{fig:n_rmse}, where we again computed the mean RMSE on cursor positions, %
considering all trials of U4 using the Virtual Cursor Ergonomic technique, for different MPC horizons.
The performance of our simulation clearly deteriorates when the MPC horizon $N$ is set too low, i.e., $N\leq6$. Interestingly, user trajectories are best explained by an intermediate MPC horizon of $N=13$, while a too large horizon ($N\geq16$) results in slightly worse mean RMSE for this interaction technique. 
Since the computational time exponentially increases with $N$, we decided to use the lowest MPC horizon replicating human behavior sufficiently well for our simulations, that is, $N=8$.

\section{Discussion and Limitations}\label{sec:discussion}
\begin{figure}[h!]
	\centering
	\subfloat{\includegraphics[width=0.5\linewidth]{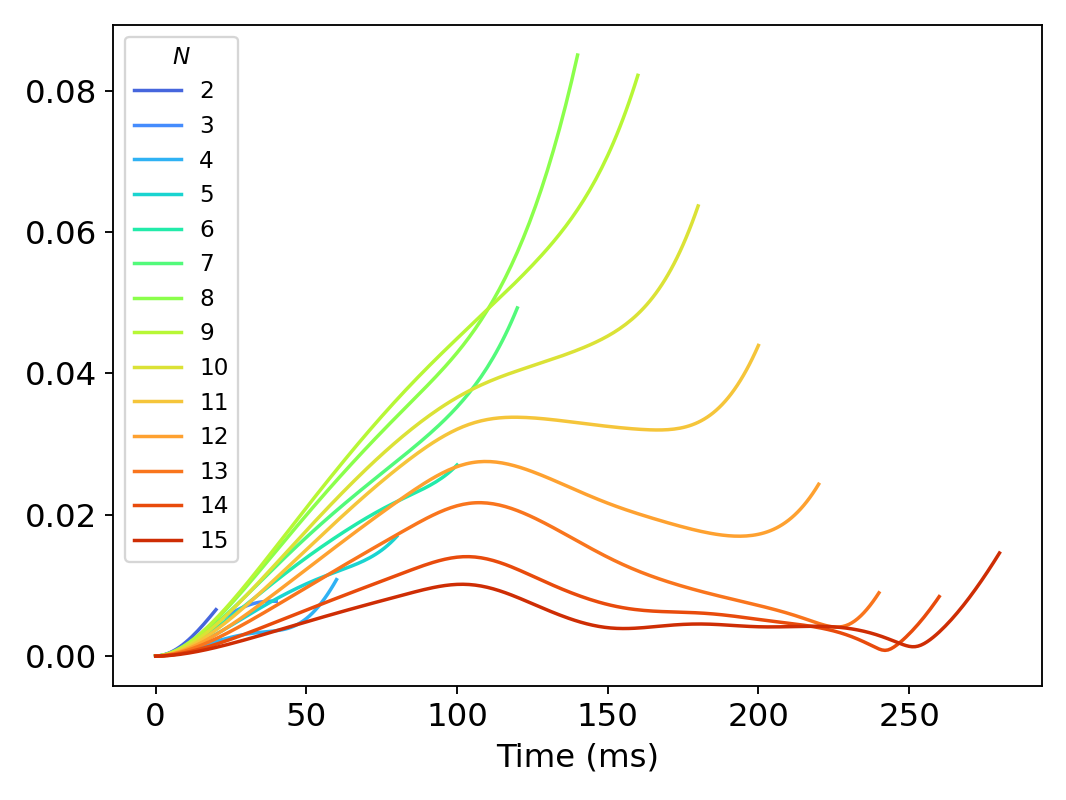}}
	\subfloat{\includegraphics[width=0.5\linewidth]{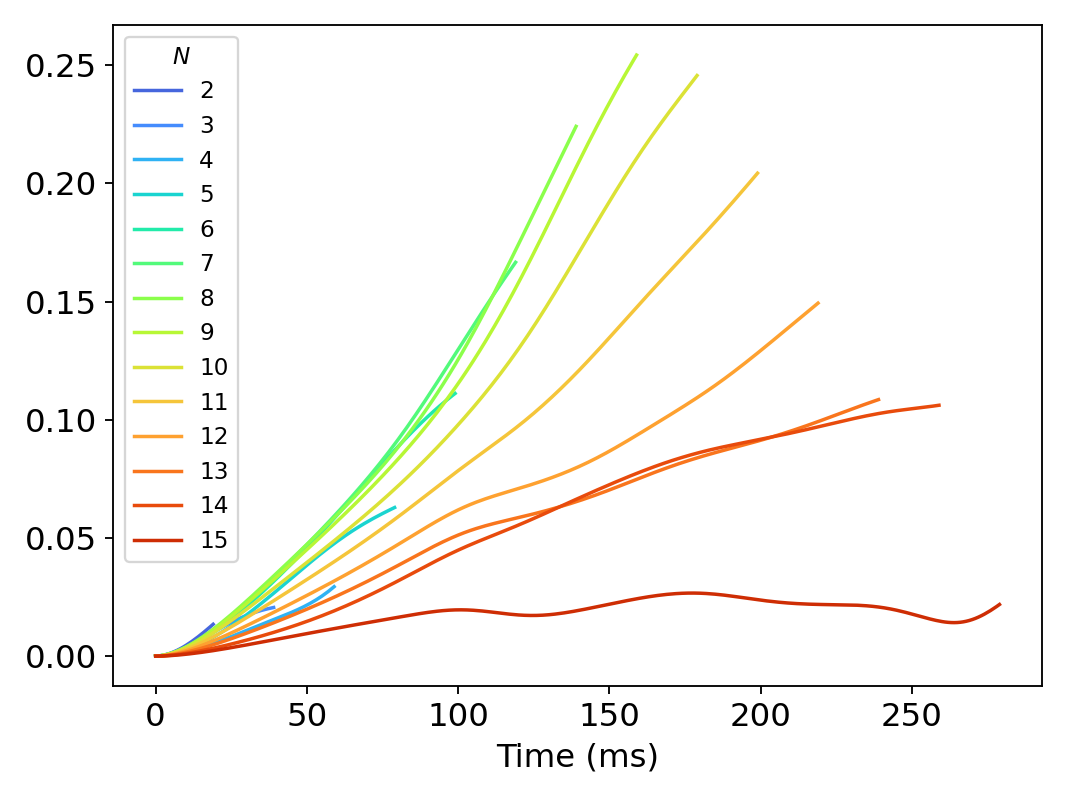}}
	
	\caption{Euklidean norm between the cursor position (left) and the aggregated joint angles (right) of open-loop trajectories with different MPC horizons $N$ and a reference open-loop trajectory for $N=16$. 
	The \textit{Turnpike} property is reflected in the fact that the trajectories stay longer in the vicinity of the reference as the horizon $N$ increases. All simulations use the same trial of U4 and the Virtual Cursor Identity technique (defined in Section~\ref{sec:usecase-definitions}).}
	\label{fig:turnpike}
\end{figure}

We have seen in the previous chapter that MPC can be successfully used to simulate biomechanically plausible movements with an accuracy at least equal to that between different users (Section~\ref{sec:simulation-vs-user}).
In addition, it was shown that our method can predict the movements of individual users, and furthermore, simulations of ``new'' users or with different interaction techniques can be easily created.

The fact that our simulation better reproduces baseline joint angle trajectories than other users (Figure~\ref{fig:ALL_quant}) is influenced by our choice of initial posture (see Section~\ref{sec:study-based-simulation}).
To allow a fair comparison, our simulation, which depends on the user model under consideration, requires an initial posture that yields a valid initial cursor position.
Therefore, the simulations performed for the evaluation depend on the user data.

We have also seen in Section~\ref{sec:effect_N} that, once agreed on the cost function (in our case, JAC), the choice of the MPC horizon~$N$ is \textit{the} key governing factor. 
In practice, some trial and error is needed, but also possible here, to control the tradeoff between computation time and quality of the movement trajectory.
To this end, we have provided practical guidance in Section~\ref{sec:effect_N}. 
Once we have such an~$N$, then MPC ``works'' for this application.

MPC has been used successfully for other applications; it is no coincidence that it is performing well in this particular setting. 
In this paper we have relied on user data and performed an extensive analysis on the performance of MPC. 
However, if one wants to very quickly check whether the effort of using MPC will be worthwhile in their application, a good indicator is the so-called \textit{turnpike property}.
It states that the computed optimal trajectories remain close to the so-called \textit{optimal steady-state}, most of the time. 
As the name suggests, it is the optimal state to remain in 
with respect to stage costs~$\ell$.
In our JAC case, it is determined by the state costs (e.g., keeping the cursor near the target) and control or effort costs (e.g., using a joint posture in the process that is not strenuous or tiring).
Usually this optimal steady-state is not known analytically, so the first step is to approximate it numerically.
This can be done by solving the optimal control problem~\eqref{eq:OCP_N} over the time frame of interest. 
Then, instead of only applying the first value of the resulting optimal control sequence as in the MPC algorithm (Section~\ref{sec:MPC}), one computes the so-called \textit{open-loop optimal trajectory} by simply applying the full optimal control sequence. 
For this purpose, the parameter~$N$ should be chosen as large as possible -- in our following example, we have chosen $N=16$. %
Then compare this result to open-loop optimal trajectories for smaller $N$, by plotting their difference to the reference trajectory (large $N$).
This is visualized in Figure~\ref{fig:turnpike}.

Figure~\ref{fig:turnpike} shows the typical turnpike behavior: 
First, for all $N$ the corresponding open-loop trajectories at some point start deviating more and more from the reference trajectory. 
Second, this happens later and less pronounced the higher~$N$ is. 
The turnpike property is an essential part in theorems that guarantee that MPC produces ``sensible'' results, i.e., MPC ``works''~\cite{faulwasser2014}.
This assurance provides trust and reliability, is one of the great strengths of MPC, and is the reason why it is used, e.g., in industry applications~\cite{QIN03}.

Similar guarantees exist for other controllers, such as LQR or LQG from~\cite{fischer2021optimal}. 
However, they cannot handle nonlinear system dynamics (like we have here).
Possibly even more important for HCI, they cannot directly handle state and control constraints, but have to resort to soft constraints via penalty terms in the cost function.
These constraints include joint angle and/or torque limits imposed by the human biomechanics, body position constraints when interfering with physical objects and devices (e.g., desk, gamepad, or monitor), and boundaries on the location of the virtual objects (e.g., due to fixed bounding boxes of application windows or limited screen size).

If, on the other hand, guarantees take a back seat, RL most certainly enters the picture. 
RL is very powerful, can handle nonlinearities and even state and control constraints, and thus is used in many applications. 
In particular, RL can be used to control a biomechanical model as well~\cite{fischer2021reinforcement}. 
However, computationally it can be very demanding to compute a single optimal movement. 
This is because RL learns a \textit{policy} to then sample movements.\footnote{We note that getting the policy can be advantageous depending on the use case.}
However, this policy may have to be completely retrained if the interaction technique is changed, e.g., a different input origin for the Virtual Pad is used. 
In contrast, MPC acts as a complexity reducer in time: instead of directly solving an optimization problem for the full movement time, in MPC we consider subproblems on smaller time horizons, which are much easier to solve due to exponentially increasing complexity the longer the movement lasts. 

From our point of view, it is therefore promising to combine the respective strengths of MPC and RL (or, more generally, Machine Learning techniques) in future applications. 
Following the current trend, MPC is more and more augmented by data-driven techniques (especially from the control theory community). 
To put it in the words of Benjamin Recht (UC Berkeley, ``Reflections on the Learning-to-Control Renaissance'' Keynote at the 2020 IFAC World Congress): 
\textit{``I still remain baffled by how Model Predictive Control is consistently underappreciated. [...] there are really very few people, especially in the Machine Learning community, who are trying to analyze why MPC is so successful, especially when it's coupled with Machine Learning.''}

\section{Conclusion}\label{sec:conclusion}

We have presented a method to simulate movements during interaction with computers, combining biomechanical modeling with Model Predictive Control.
This methods allows to predict not only summary statistics such as movement time or success rate, but kinematic and dynamic quantities of both physical and virtual objects, including cursor positions, joint angles and velocities, and aggregated muscle recruitments, on a moment-by-moment basis. %
Our framework allows easy combination of biomechanical models of the human body with arbitrary interaction techniques and tasks, which was not possible with existing approaches that relied on linear optimal control methods.
As a use case, we have applied our approach to a joint-actuated state-of-the art model of the upper extremity, and considered four mid-air interaction techniques for the task of ISO pointing in VR.
We have shown that MPC is able to simulate user movement in mid-air pointing both in terms of cursor and joint trajectories, with an accuracy within between-user variability.
Comparing three different cost functions, the combination of distance, control, and joint acceleration costs (JAC) was shown to best explain the movements observed in our user study.

In a practical sense, our method allows to predict movement of a given individual user, and can be used to generate customized models of new users. 
Moreover, we provide advice on how to leverage our approach to simulate interactions for different target user groups, interaction techniques, or interaction tasks.
We have also introduced \textit{CFAT}, a novel tool to compute the applied torques underlying a given joint angle sequence, which we have used to infer the maximum voluntary torques that were applied in the considered ISO pointing task. %

Our presented framework assumes complete observation, i.e., complete knowledge of the system state (e.g., joint angle and cursor position).
In reality, humans observe their environment, e.g. through proprioception or visual perception, and have to deal with limited information, e.g., they have to \textit{estimate} the target position from an image on a screen.
To control such a system with imperfect observation, it would be an interesting future task to add an \textit{observer}.

The combination of biomechanical simulations with MPC could open the door for online optimization and customization of user interfaces and interaction techniques, based on the predictions of a ``digital twin'' simulation running in the background.
The ability to evaluate the entire interaction loop between humans and computers in terms of efficiency or ergonomics with the help of realistic simulations could allow for partial replacement of costly and time-consuming user studies in the future.
We therefore hope that the optimal control perspective on interaction, together with a biomechanical simulation of interaction via MPC, will become an important tool for the HCI community in the future.

\bibliographystyle{ACM-Reference-Format}
\bibliography{bibliography}

\clearpage

\appendix
\counterwithin{figure}{section}
\counterwithin{table}{section}

\include{appendix}

\end{document}

%% file: appendix.tex
\section{Virtual Pad Transfer Function}\label{sec:appendix-virtual-pad}
To obtain the cursor position of the Virtual Pad, we have to project the end-effector position $x_{\text{ee}}$ onto the input plane and transfer it onto the output plane.
The projection onto the input plane is given by %
\begin{equation}
	\text{Proj}_\text{I}(x_{\text{ee}}) =  x_{\text{ee}} - \left(\left(x_{\text{ee}}-\omega_I\right) \cdot n_I\right)n_I\text{,}
\end{equation}
where the operator $\cdot$ denotes the inner product, $\omega_I$ is the input space origin, and $n_I$ is the input space normal.\footnote{While the projected position could be denoted in 2D coordinates of the input plane, the projection maps onto the position in global 3D coordinates to be able to transfer this point correctly onto the output plane.}
To rotate this new point correctly, we compute the rotation matrix $R$ that rotates the input normal vector $n_I$ such that it equals the output normal vector $n_O$:
		\begin{equation}
			\begin{aligned}
				R = (1-C) aa^\top+
				\left( \begin{array}{c c c}
					C& -a_3\overline{C} & a_2\overline{C}\\
					a_3\overline{C}&  C&  -a_1\overline{C}\\
					-a_2\overline{C}& a_1\overline{C}& C  
				\end{array}\right)\text{,}
			\end{aligned}
		\end{equation}
		where $a$ is the rotation axis, i.e.,
		\begin{equation}
			a = (a_1,a_2,a_3)^\top = \frac{n_{\text{I}} \times n_O}{\norm{n_{\text{I}} \times n_O}},
		\end{equation}
		$aa^\top$ is the \textit{dyadic product} of $a$, and $C$ and $\overline{C}$ are the cosine and sine of the angle between the normals, i.e.,
		\begin{equation}
			C = \frac{n_{\text{I}} \cdot n_O}{\norm{n_I}\norm{n_O}},\quad \overline{C} = \sqrt{1-C^2}.
		\end{equation}
	Since rotation is defined around the global origin, we subtract $\omega_I$ before rotating, i.e., for some point $y\in\R^3$:
	\begin{equation}
		\text{Rot}_{\text{IO}}(y) = R (y-\omega_I).
	\end{equation}
		As a last step, we need to translate back onto the output plane by adding the output origin $\omega_O$. 
		The overall transfer function is thus given by
		\begin{equation}
			f_\text{tf}(x_{\text{ee}}) = \text{Rot}_{\text{IO}}(	\text{Proj}_\text{I}(x_{\text{ee}})) + \omega_{\text{O}} = R (x_{\text{ee}} - \left(\left(x_{\text{ee}}-\omega_I\right) \cdot n_I\right)n_{\text{I}} - \omega_I) + \omega_O\text{.}%
		\end{equation}

\clearpage

\section{Supplementary Figures and Tables}\label{sec:supplementary-material}
\subsection{MuJoCo Model}
\begin{table}[!ht]
	\centering
	
	\begin{tabular}{|c|c|c|c|c|} 
		\hline
		\rule{0pt}{10pt}\noindent
		\multirow{2}{*}{Joint} & \multicolumn{4}{c|}{Angle Ranges (deg/rad)} \\
		\cline{2-5}
		\rule{0pt}{10pt}\noindent
		& \multicolumn{2}{c|}{Min.} & \multicolumn{2}{c|}{Max.} \\
		\hline \hline
		\rule{0pt}{10pt}\noindent
		\textbf{E}levation \textbf{A}ngle & $-90$ & $-\frac{1}{2}\pi$ & $130$ &$\frac{13}{18}\pi$   \\
		\rule{0pt}{10pt}\noindent
		\textbf{S}houlder \textbf{E}levation & $0$ &$0$ & $180$ &$\pi$  \\
		\rule{0pt}{10pt}\noindent
		\textbf{S}houlder \textbf{R}otation & $-90$ &$-\frac{1}{2}\pi$ & $20$ &$\frac{1}{9}\pi$ \\
		\rule{0pt}{10pt}\noindent
		\textbf{E}lbow \textbf{F}lexion & $0$ &$0$ & $130$ &$\frac{13}{18}\pi$ \\
		\rule{0pt}{10pt}\noindent
		\textbf{P}ronation/\textbf{S}upination & $-90$ &$-\frac{1}{2}\pi$ & $90$ &$\frac{1}{2}\pi$\\
		\rule{0pt}{10pt}\noindent
		\textbf{W}rist \textbf{D}eviation & $-10$ &$-\frac{1}{18}\pi$ & $25$ &$\frac{5}{36}\pi$\\
		\rule{0pt}{10pt}\noindent
		\textbf{W}rist \textbf{F}lexion & $-70$ &$-\frac{7}{18}\pi$ & $70$ &$\frac{7}{18}\pi$\\
		\hline
	\end{tabular}
	
	\caption{Default joint angle ranges of the used upper extremity model~\cite{fischer2021reinforcement}.}
	\label{tab:joint-limits} 
\end{table}

\subsection{CFAT}
\label{app:cfat}
\begin{figure}[!h]
	\includegraphics[width=\linewidth]{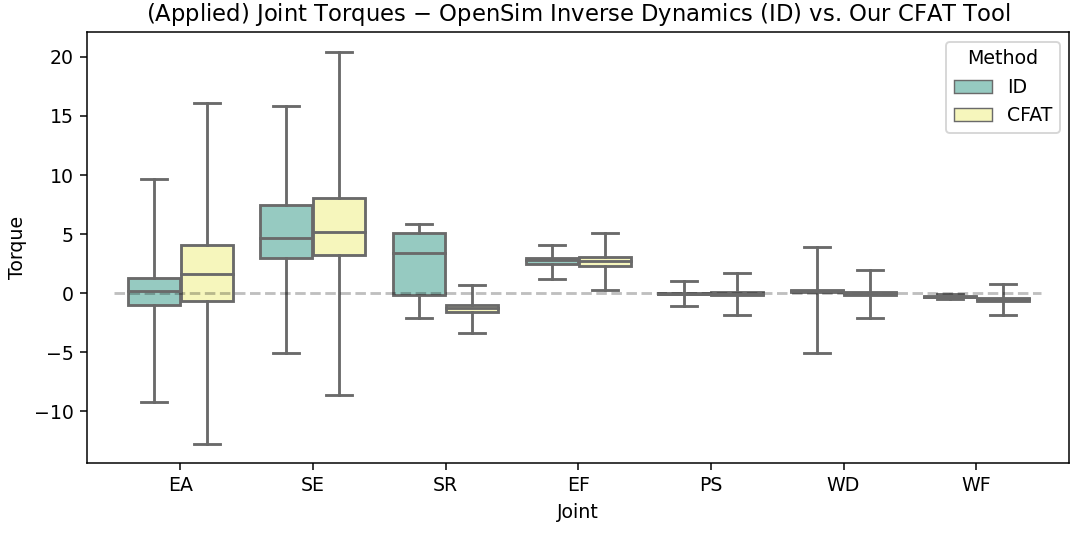}
	\caption{Joint torques of all participants and interaction techniques, both computed using OpenSim Inverse Dynamics (ID, green) and our proposed CFAT tool (yellow). For each DOF, the colored boxes show the respective interquartile ranges ($25\%$ to $75\%$ quantiles) and the whiskers correspond to the minimum and maximum torques after removing some outliers (see Section~\ref{sec:cfat}).
	}
	\label{fig:ID_vs_CFAT_torqueranges}
\end{figure}

\clearpage

\subsection{Comparison of Cost Functions}

\begin{table}[h!]
	\centering
	\includegraphics[clip,trim=1.5cm 11cm 0.3cm 2.5cm,width=1.00\textwidth]{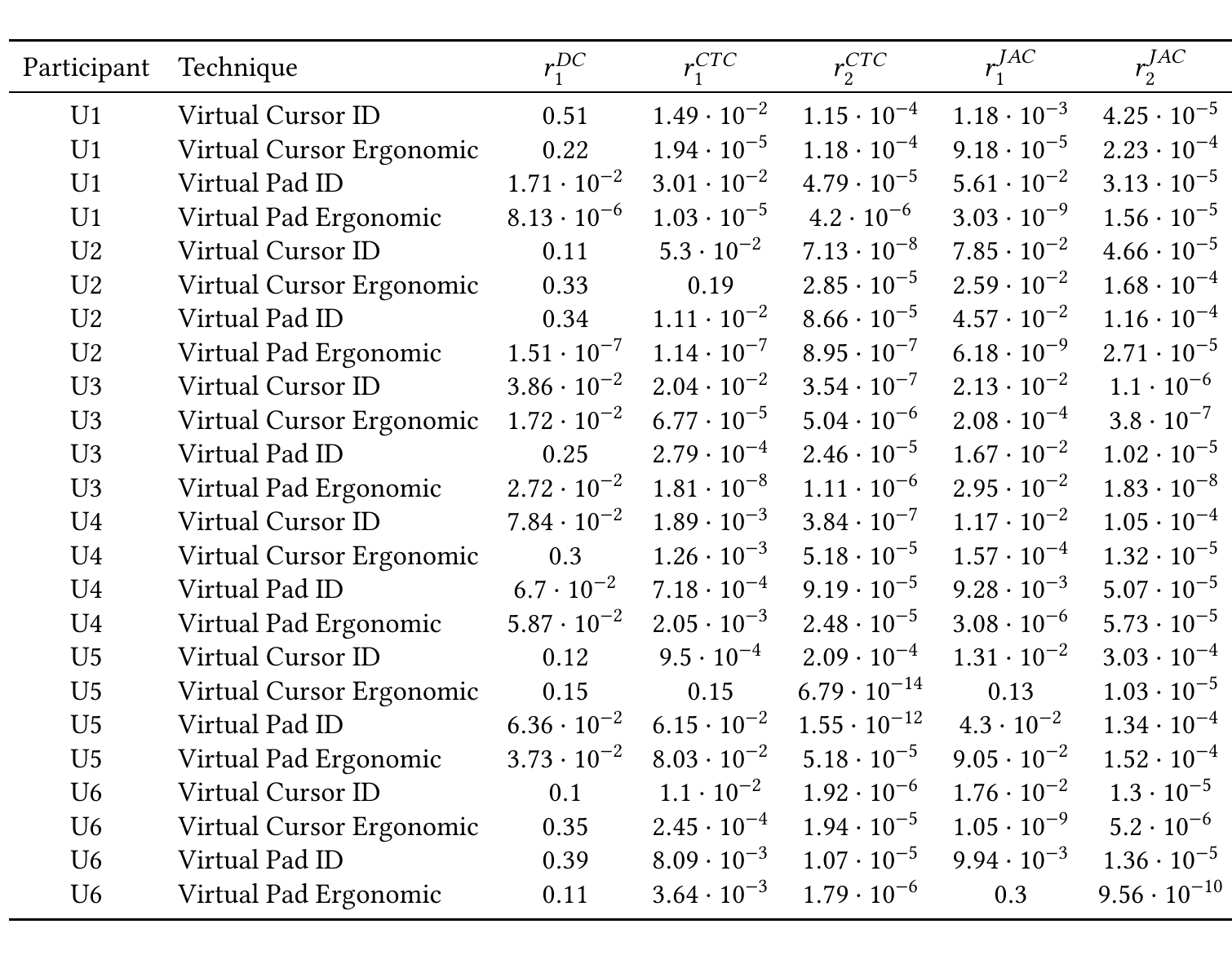}
	\caption{The cost weights obtained by the CMA-ES parameter optimization for each participant, condition, and cost function. %
	}
	\label{tab:cost_parameters}
\end{table}

\begin{figure}[h!]
	\centering
	\subfloat{\includegraphics[width=0.2\linewidth, clip]{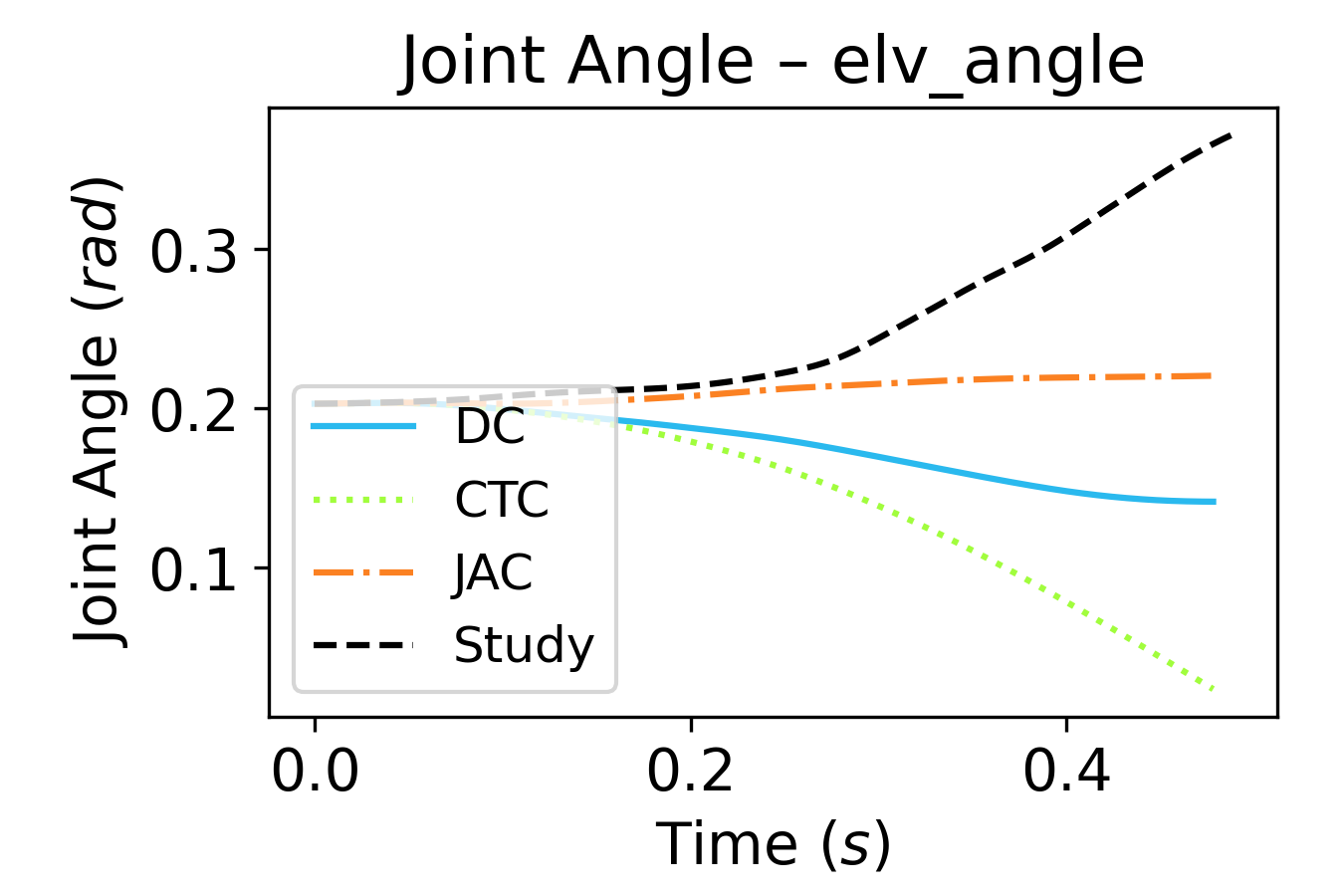}}
	\subfloat{\includegraphics[width=0.2\linewidth, clip]{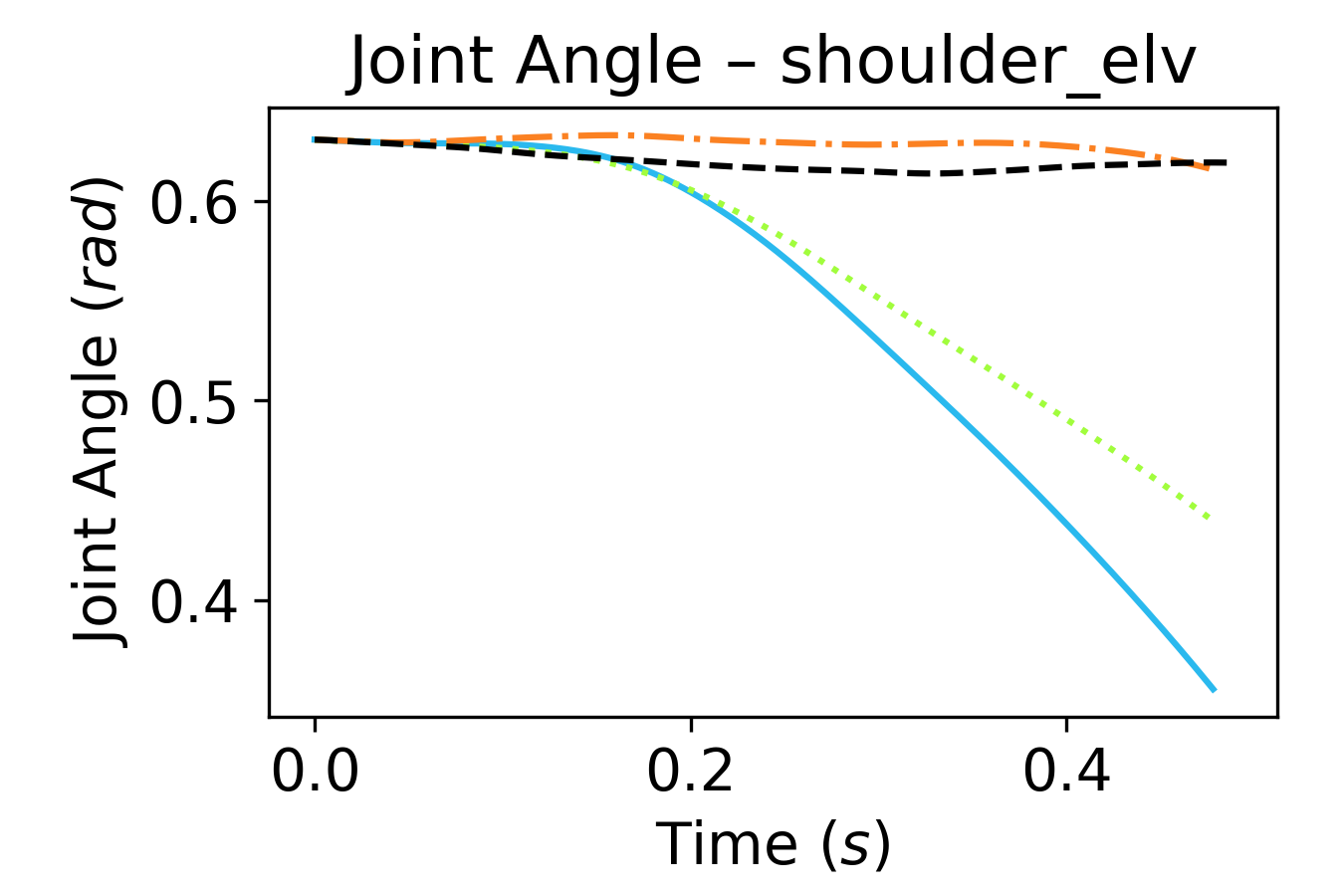}}
	\subfloat{\includegraphics[width=0.2\linewidth, clip]{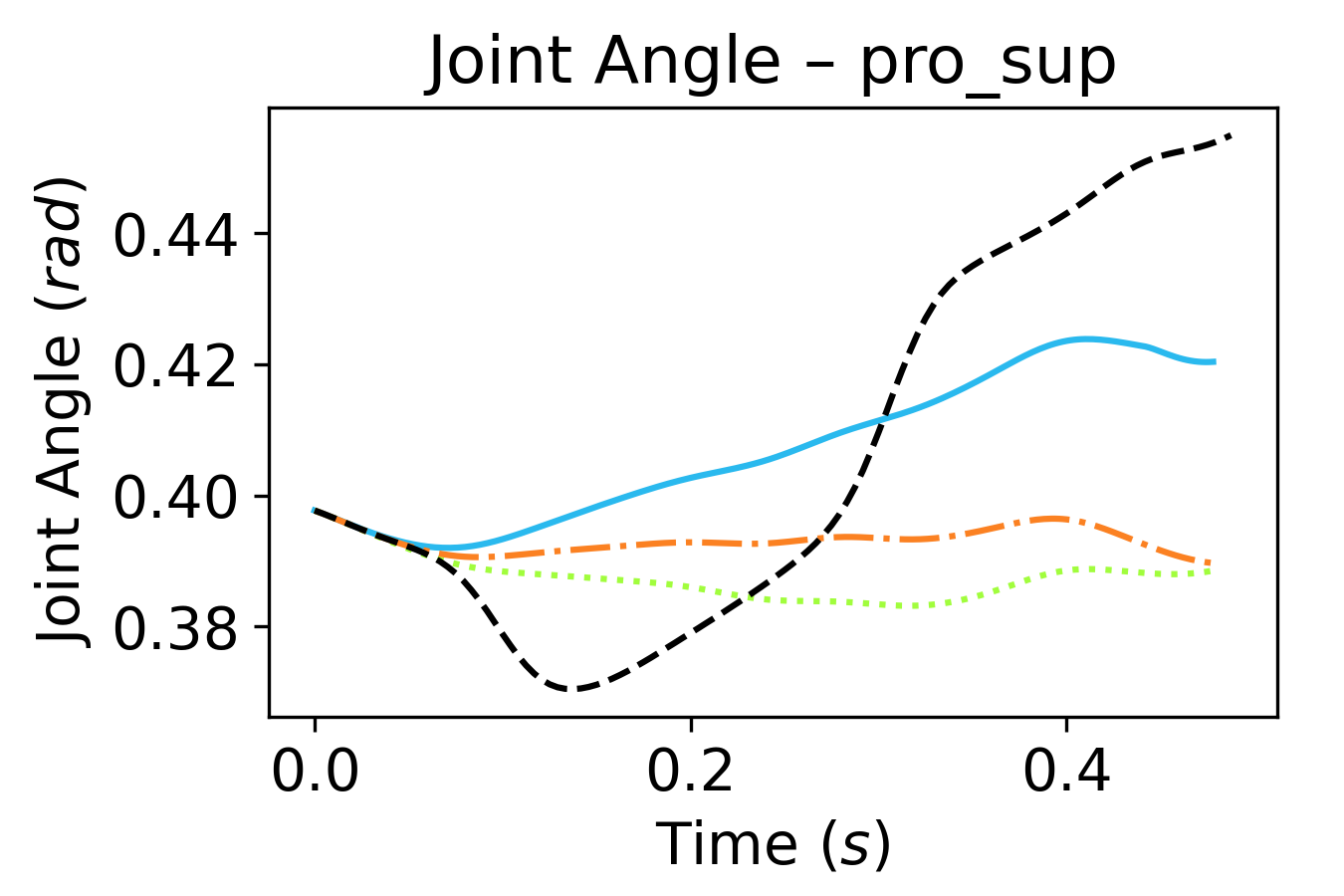}}
	\subfloat{\includegraphics[width=0.2\linewidth, clip]{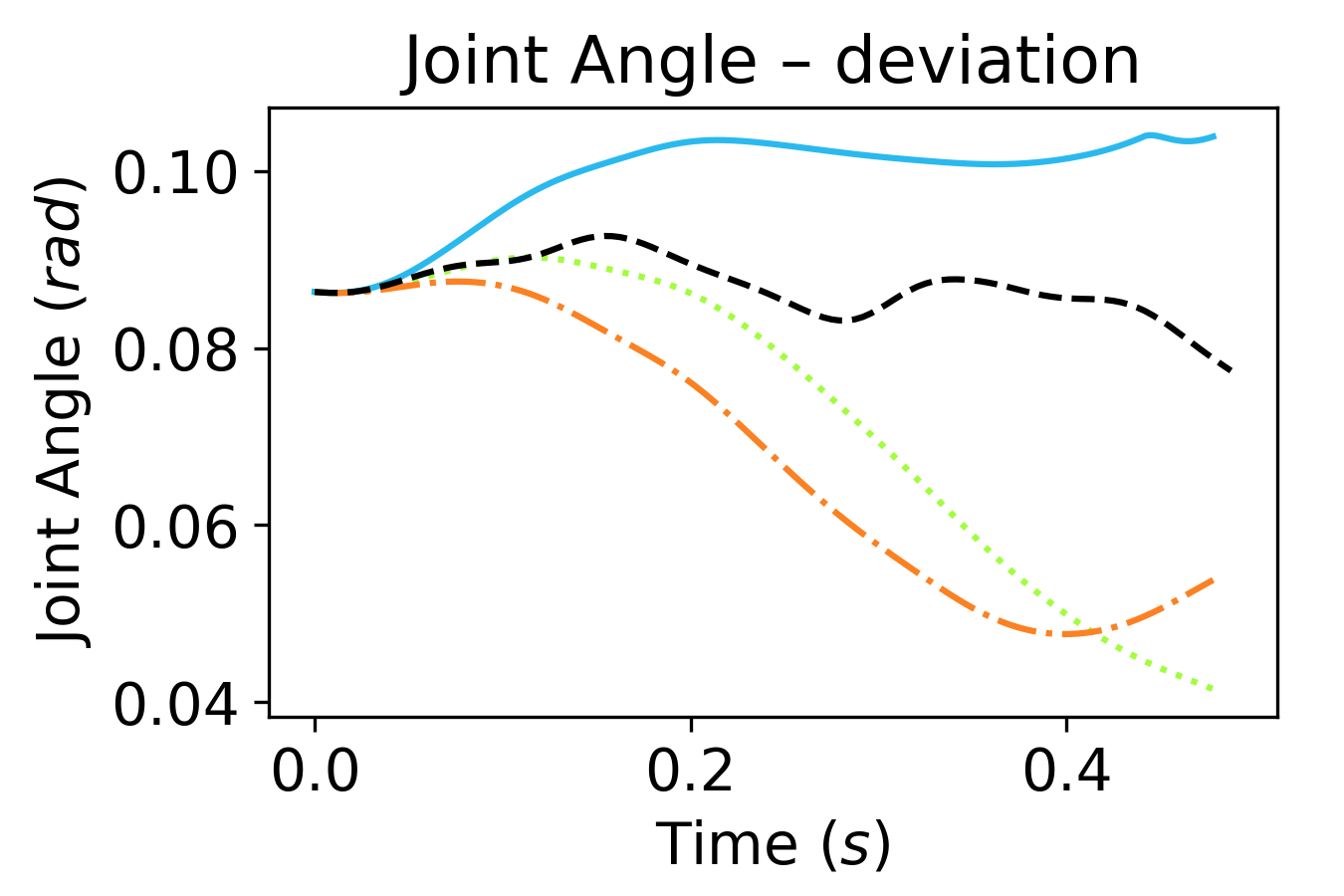}}
	\subfloat{\includegraphics[width=0.2\linewidth, clip]{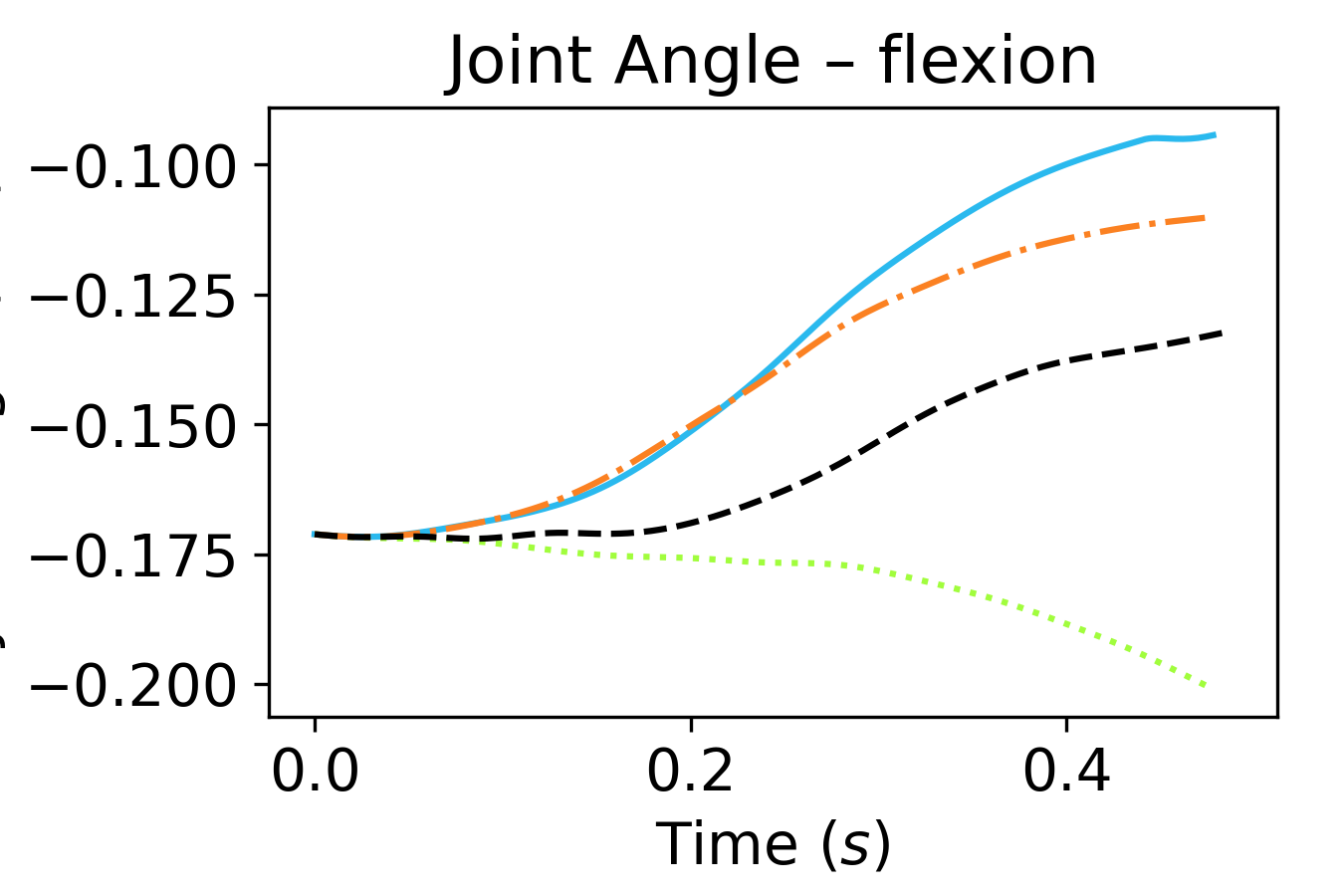}}\\
	\subfloat{\includegraphics[width=0.2\linewidth, clip]{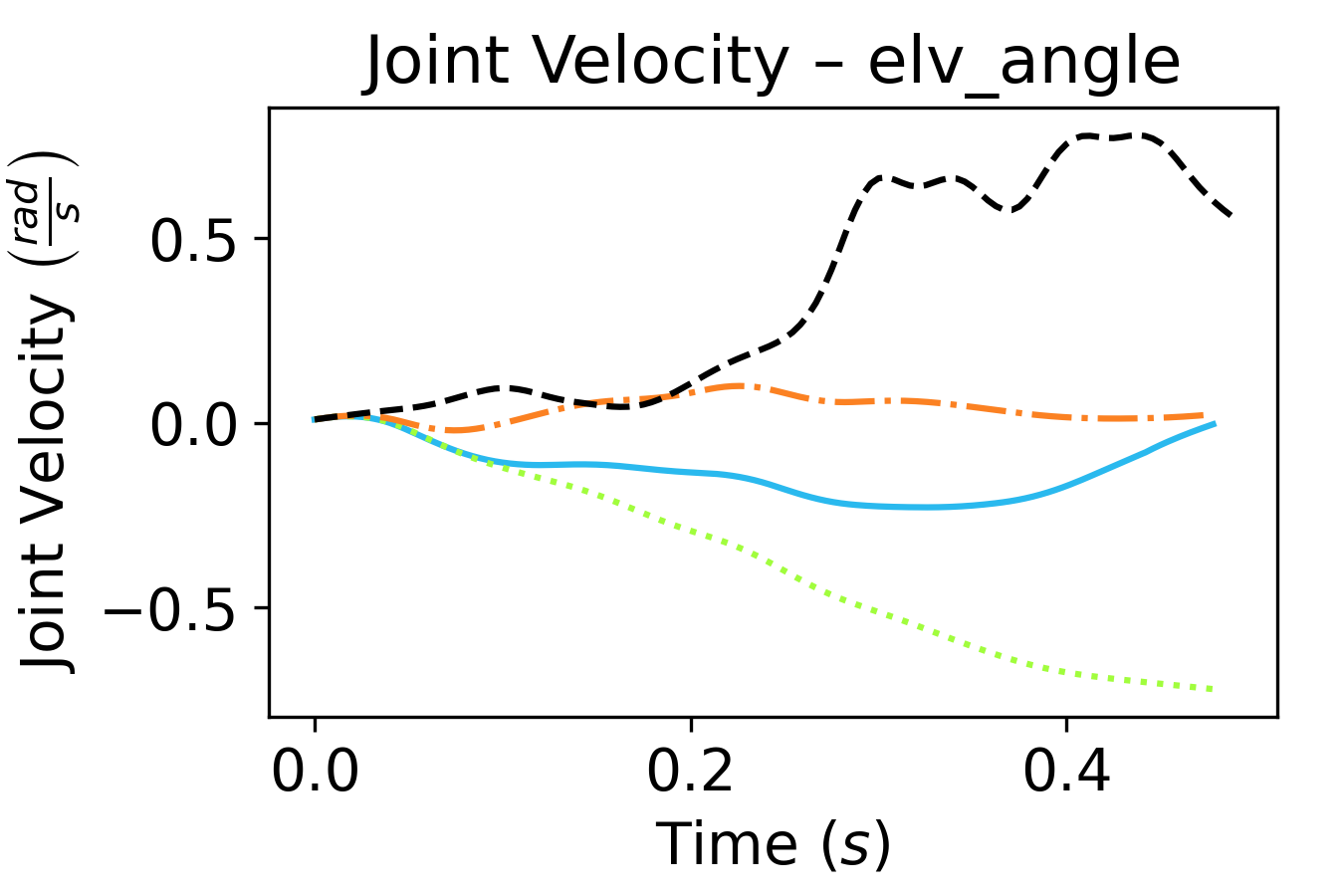}}
	\subfloat{\includegraphics[width=0.2\linewidth, clip]{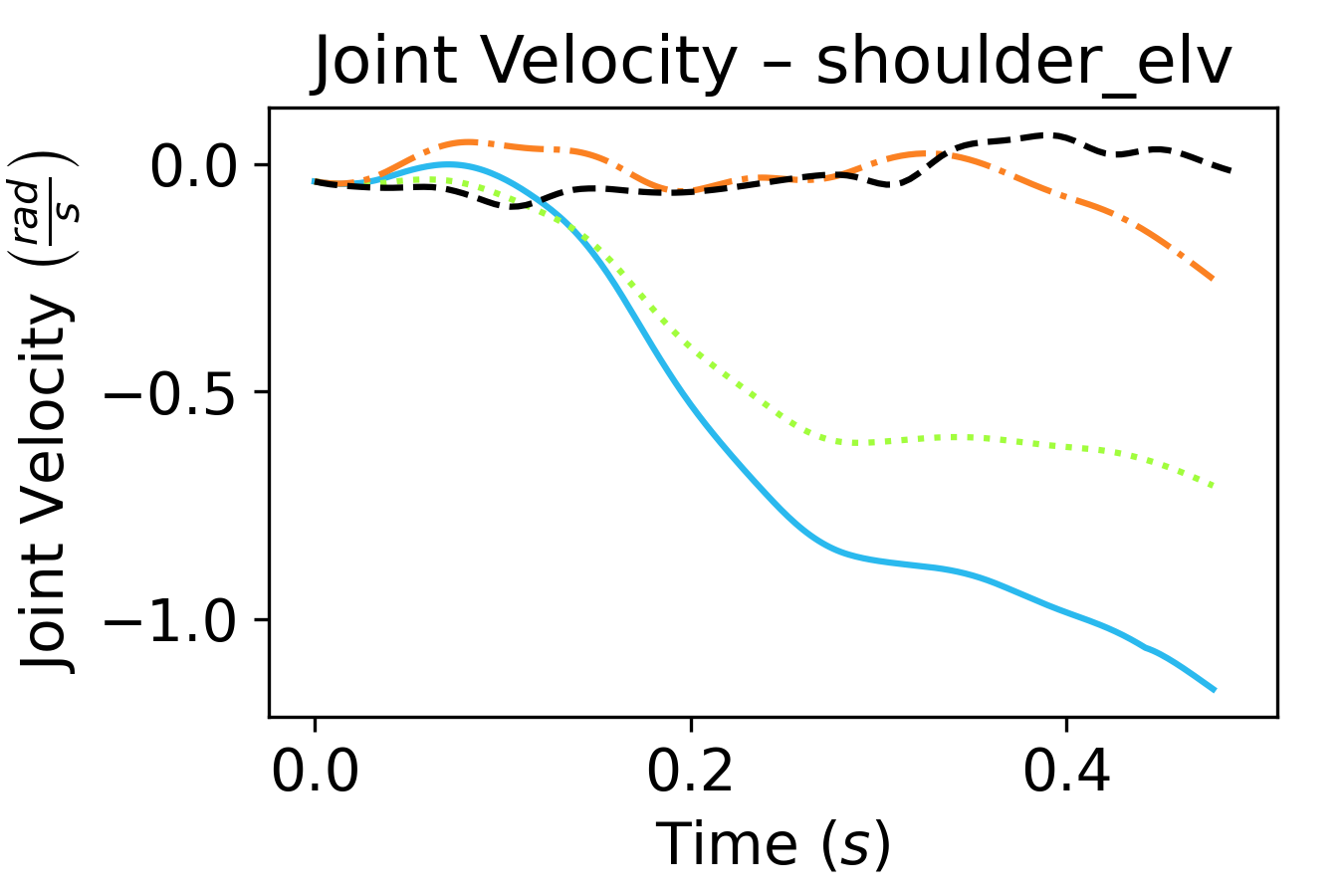}}
	\subfloat{\includegraphics[width=0.2\linewidth, clip]{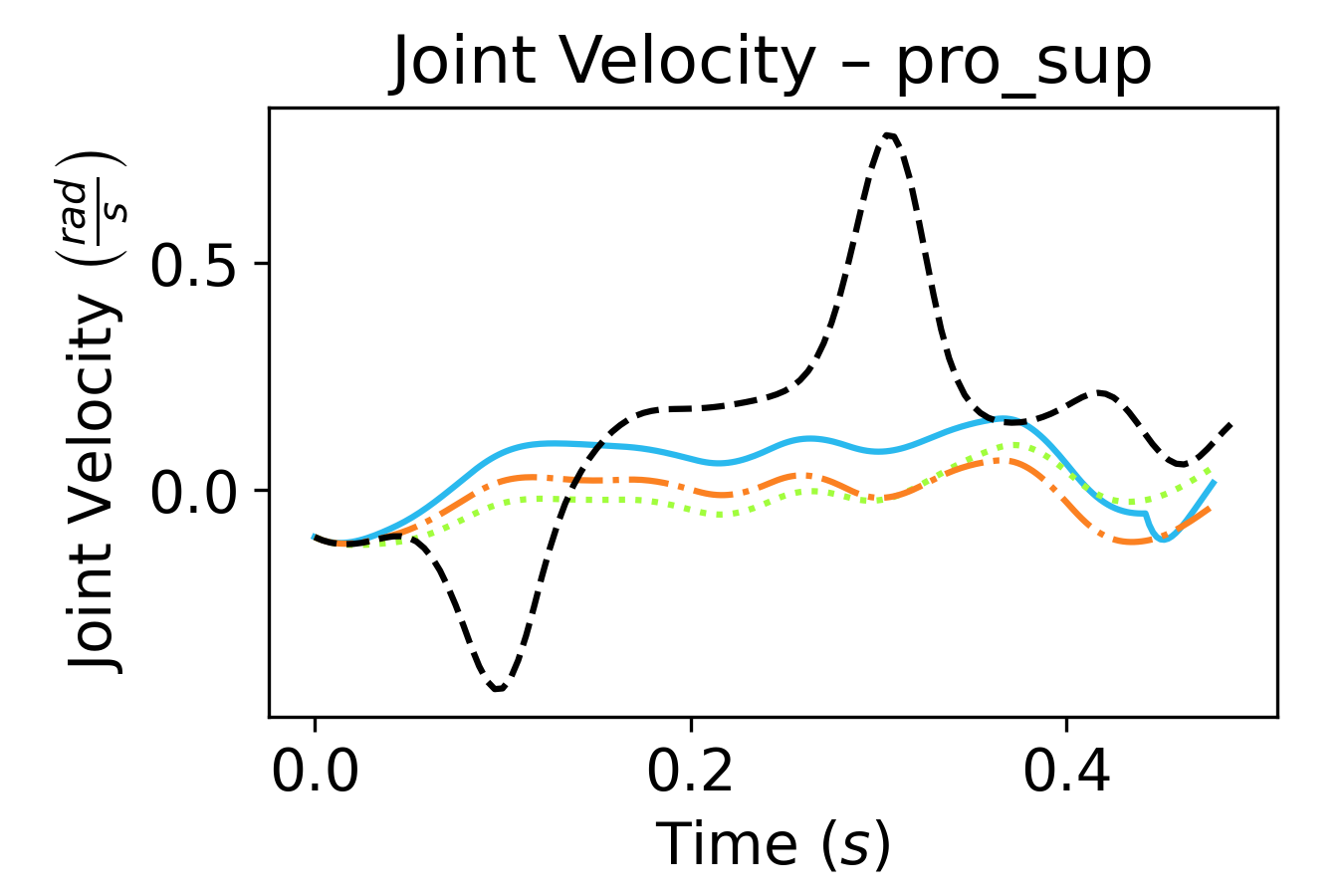}}
	\subfloat{\includegraphics[width=0.2\linewidth, clip]{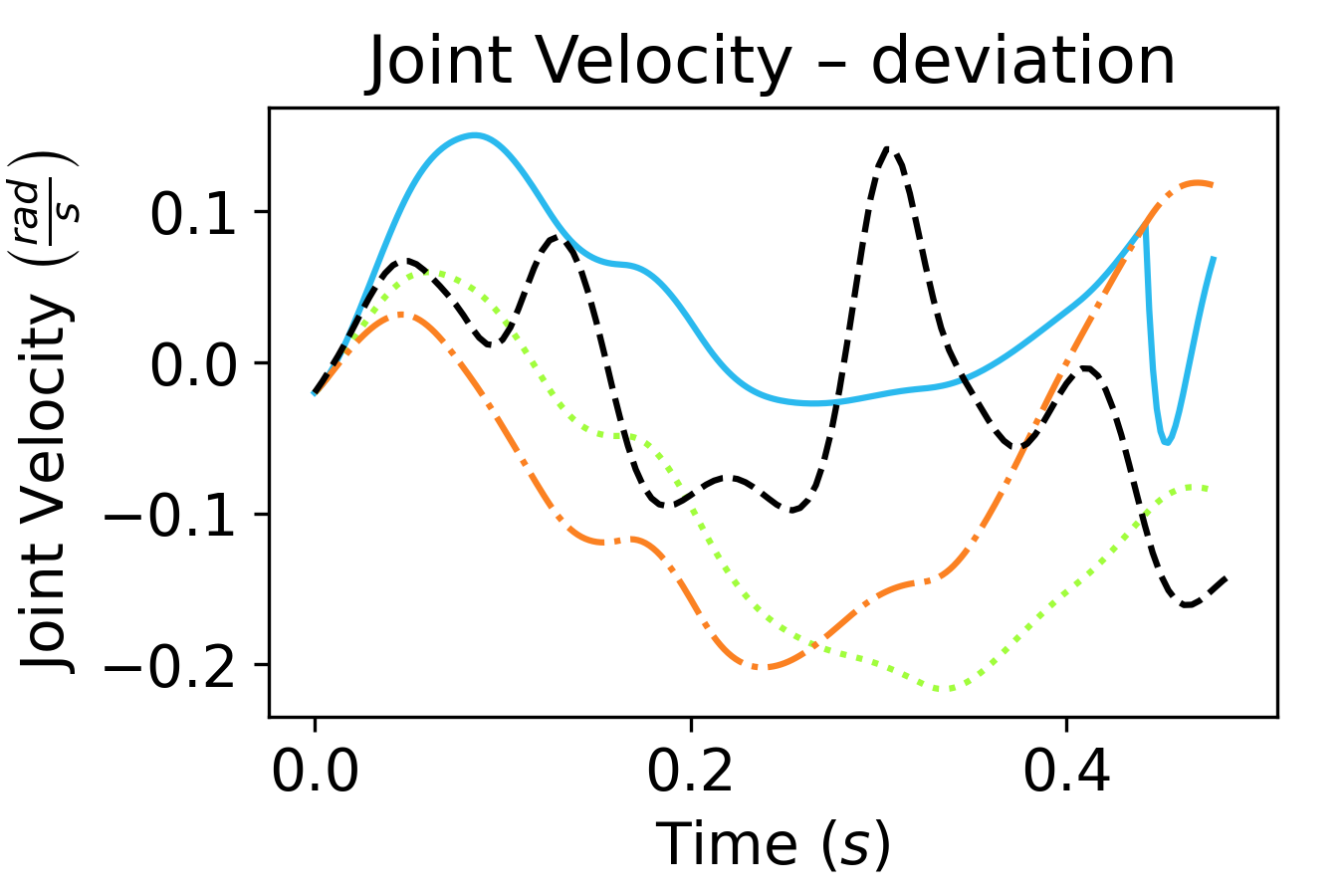}}
	\subfloat{\includegraphics[width=0.2\linewidth, clip]{figures/plots/ALLCOSTS/U4/Standing_Pad_ID_ISO_15_plane/flexion_Tr7_pos}}
	\caption{Projected joint trajectories for one trial of U4 for the Virtual Pad Identity technique. The Joint Acceleration Costs (JAC; orange dashdotted lines) qualitatively explain observed user behavior best. For pro\_sup, deviation, and flexion, all considered cost function show a good fit to observed user behavior (note the small angle ranges for these joints).
	Remaining joints are shown in Figure~\ref{fig:vgl_qual_proj}.
	}
	\label{fig:vgl_qual_proj_otherjoints}
\end{figure}

\begin{figure}[h!]
	\centering
	\subfloat{\includegraphics[width=\linewidth, clip]{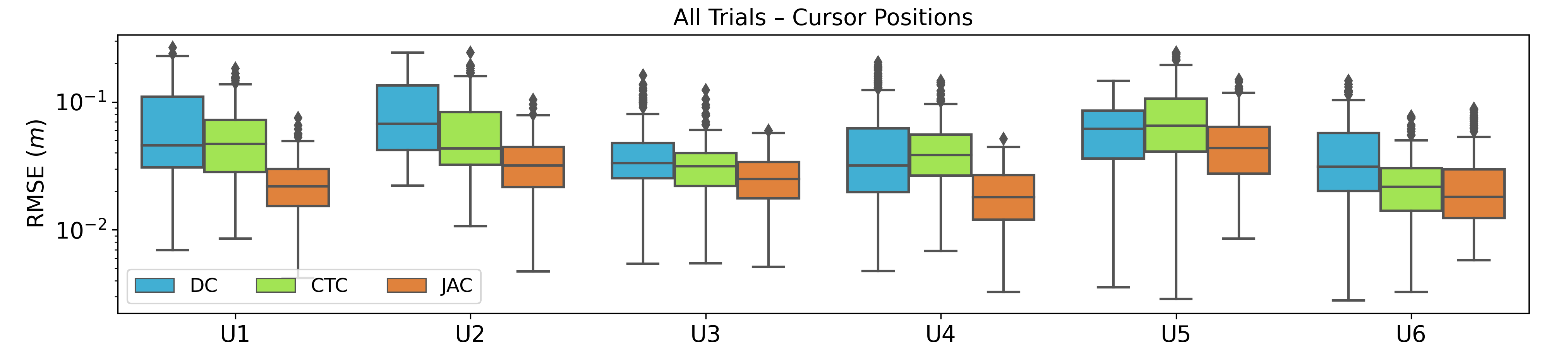}}\\
	\subfloat{\includegraphics[width=\linewidth, clip]{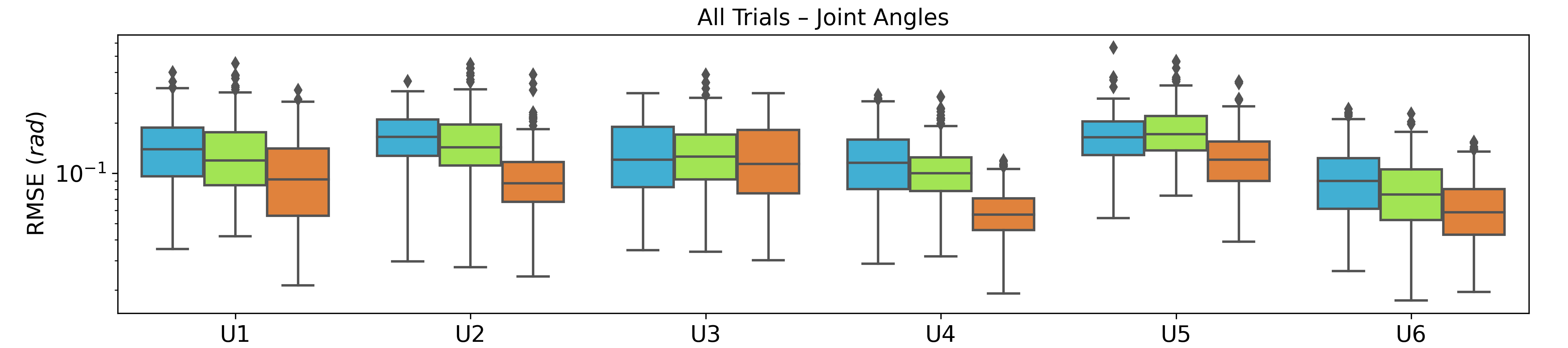}}\\
	\caption{Comparison of the different cost functions in terms of cursor positions and joint angles, separated by user. The boxplots show the RMSE of all trials.}
	\label{fig:cost-comparison_peruser}
\end{figure}

\clearpage
\subsection{MPC can Simulate User Movements}\label{sec:simulation-vs-user-appendix}

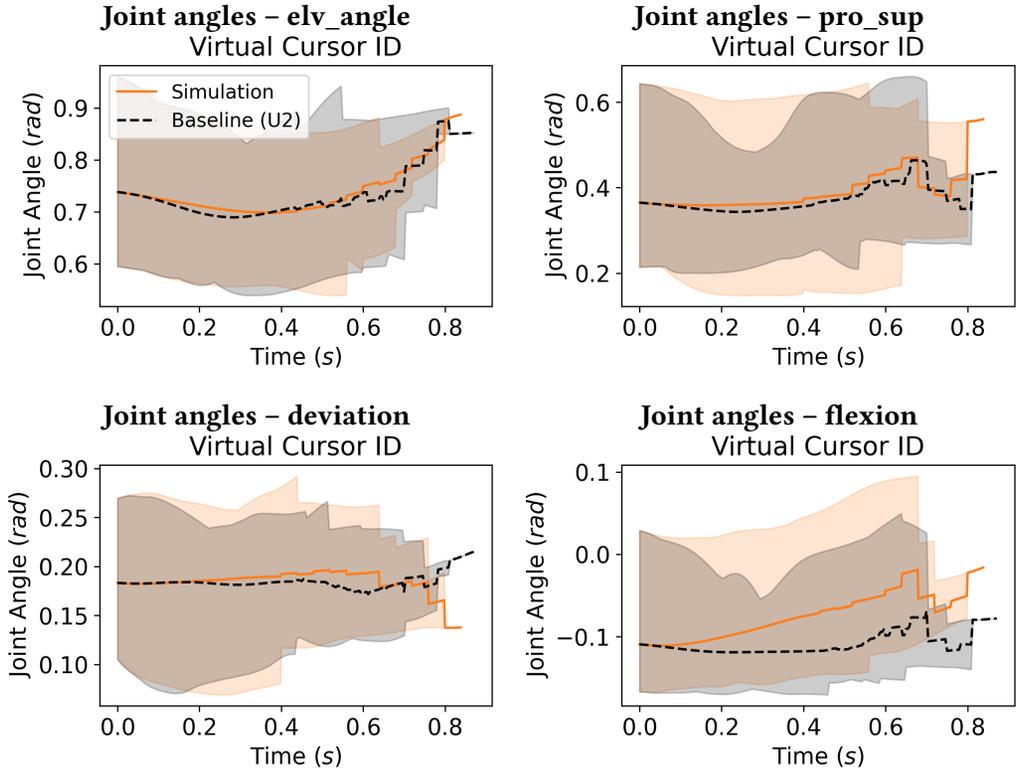
\begin{figure}[!h]
		\begin{tikzpicture}
		\node[anchor=north west] (img1) at (0, 0) {
			\includegraphics[width=0.5\linewidth, 	clip]{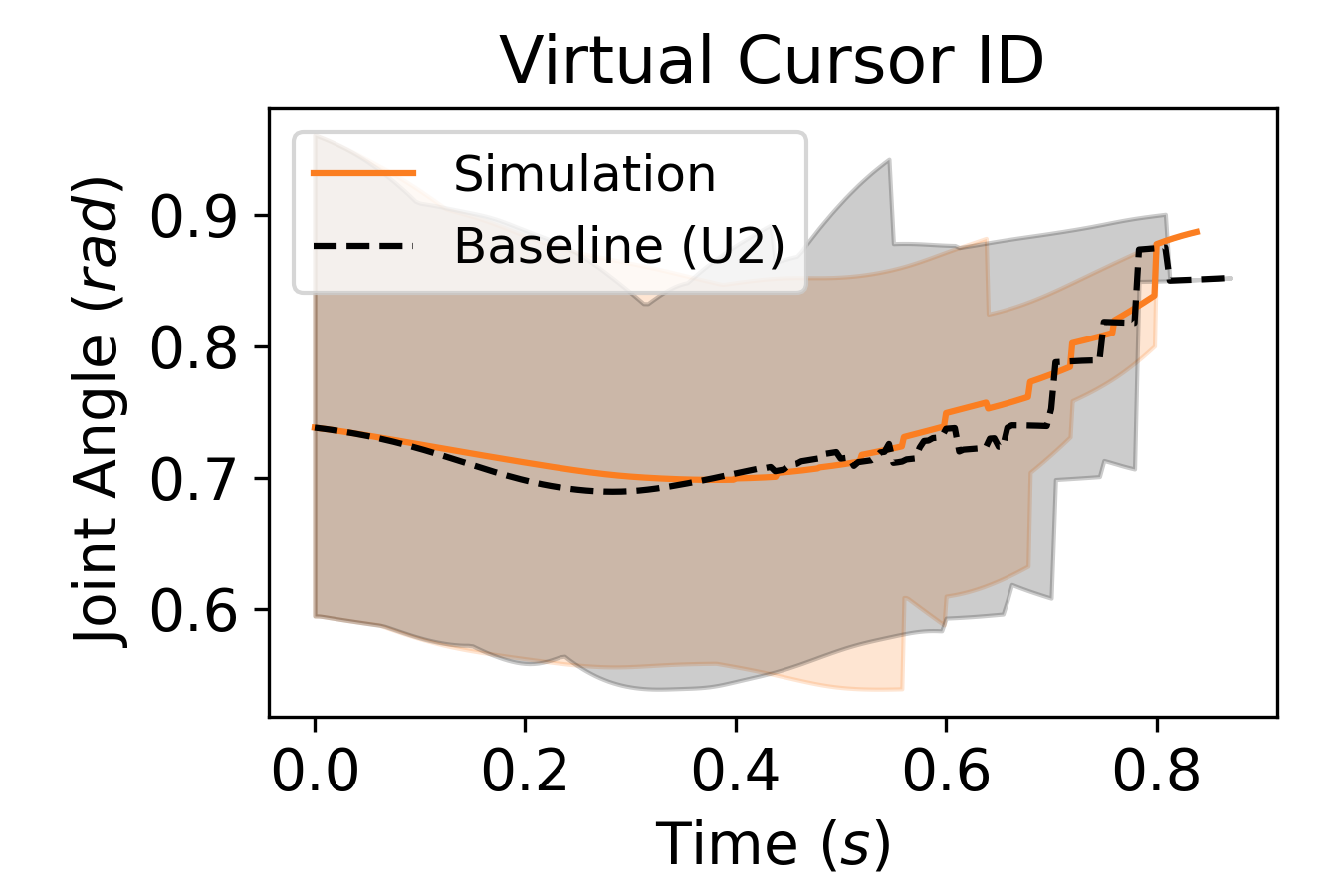}
		};
		\node[anchor=north west] (img2) at (0.5\linewidth, 0) {
			\includegraphics[width=0.5\linewidth, 	clip]{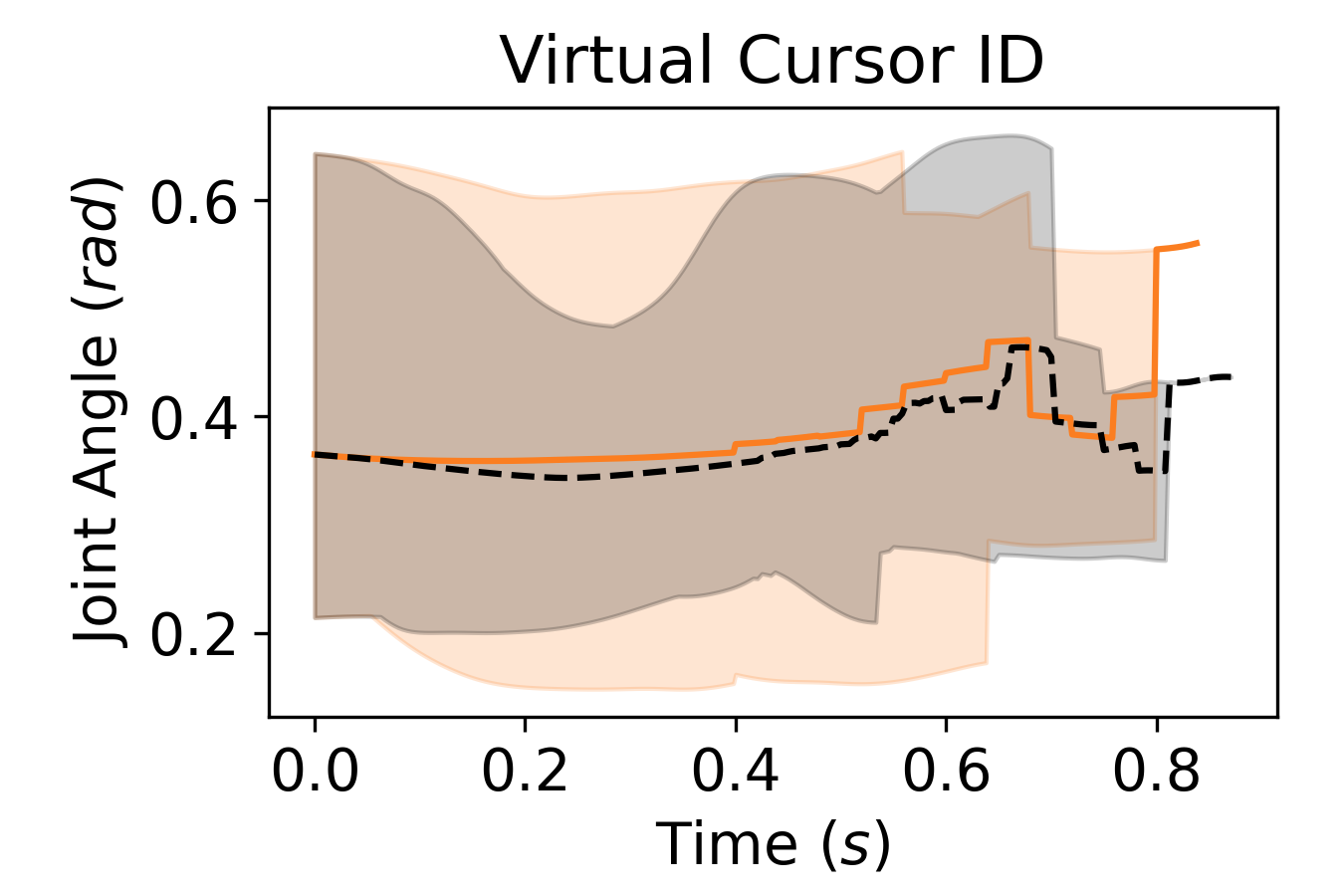}
		};
		\node[anchor=north west] (img3) at (0, -5.25) {
			\includegraphics[width=0.5\linewidth, 	clip]{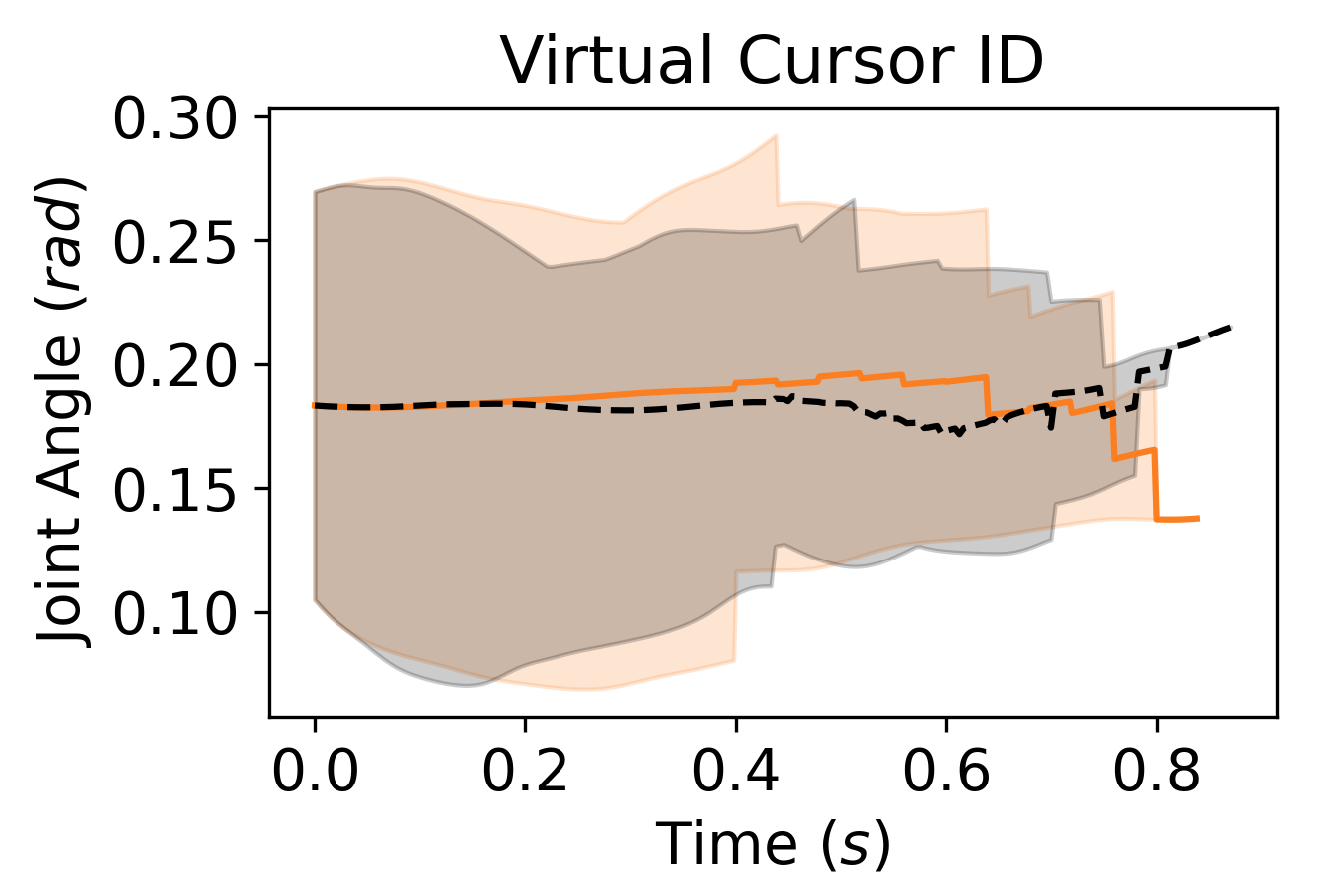}
		};
		\node[anchor=north west] (img4) at (0.5\linewidth, -5.25) {
			\includegraphics[width=0.5\linewidth, 	clip]{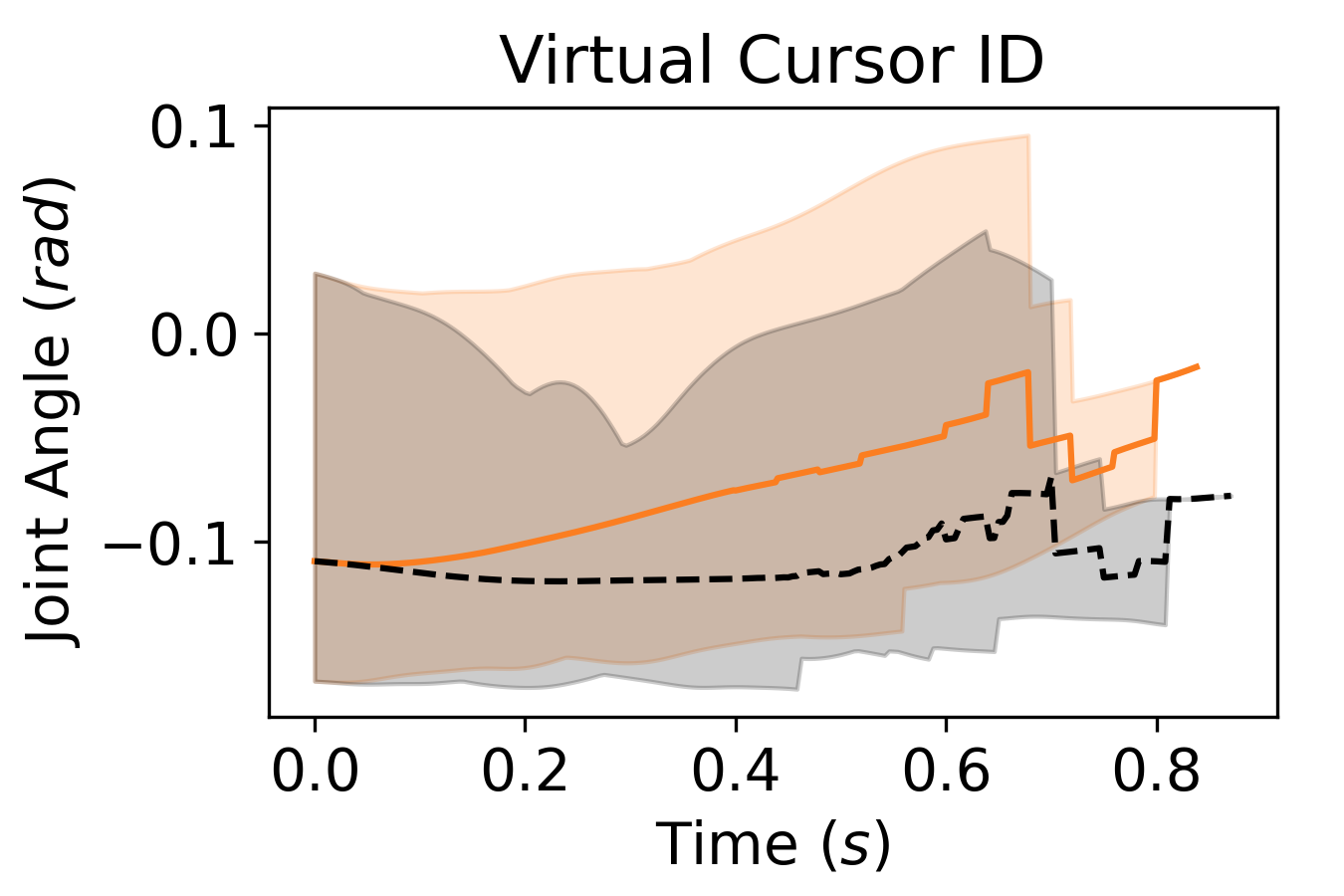}
		};
		
		\node[anchor=center, align=center, yshift=-0.05cm] at (img1.north) {\large{\textbf{Joint angles -- elv\_angle}}};
		\node[anchor=center, align=center, yshift=-0.05cm] at (img2.north) {\large{\textbf{Joint angles -- pro\_sup}}};
		
		\node[anchor=center, align=center, yshift=-0.05cm] at (img3.north) {\large{\textbf{Joint angles -- deviation}}};
		\node[anchor=center, align=center, yshift=-0.05cm] at (img4.north) {\large{\textbf{Joint angles -- flexion}}};
	\end{tikzpicture}
	
		\caption{The joint angle ranges predicted by our simulation for different movements in the ISO task (orange solid lines) match those observed in our user study (black dashed lines) fairly well.
		The mean of all movements of a single participant/user model (U2) is shown together with the entire value ranges.
		Remaining joints are shown in Figure~\ref{fig:JAC_jointranges}.
		}
	\label{fig:CursorID_jointranges_otherjoints}
\end{figure}

\begin{figure}[h!]
	\centering

	\subfloat{\includegraphics[width=0.2\linewidth, clip]{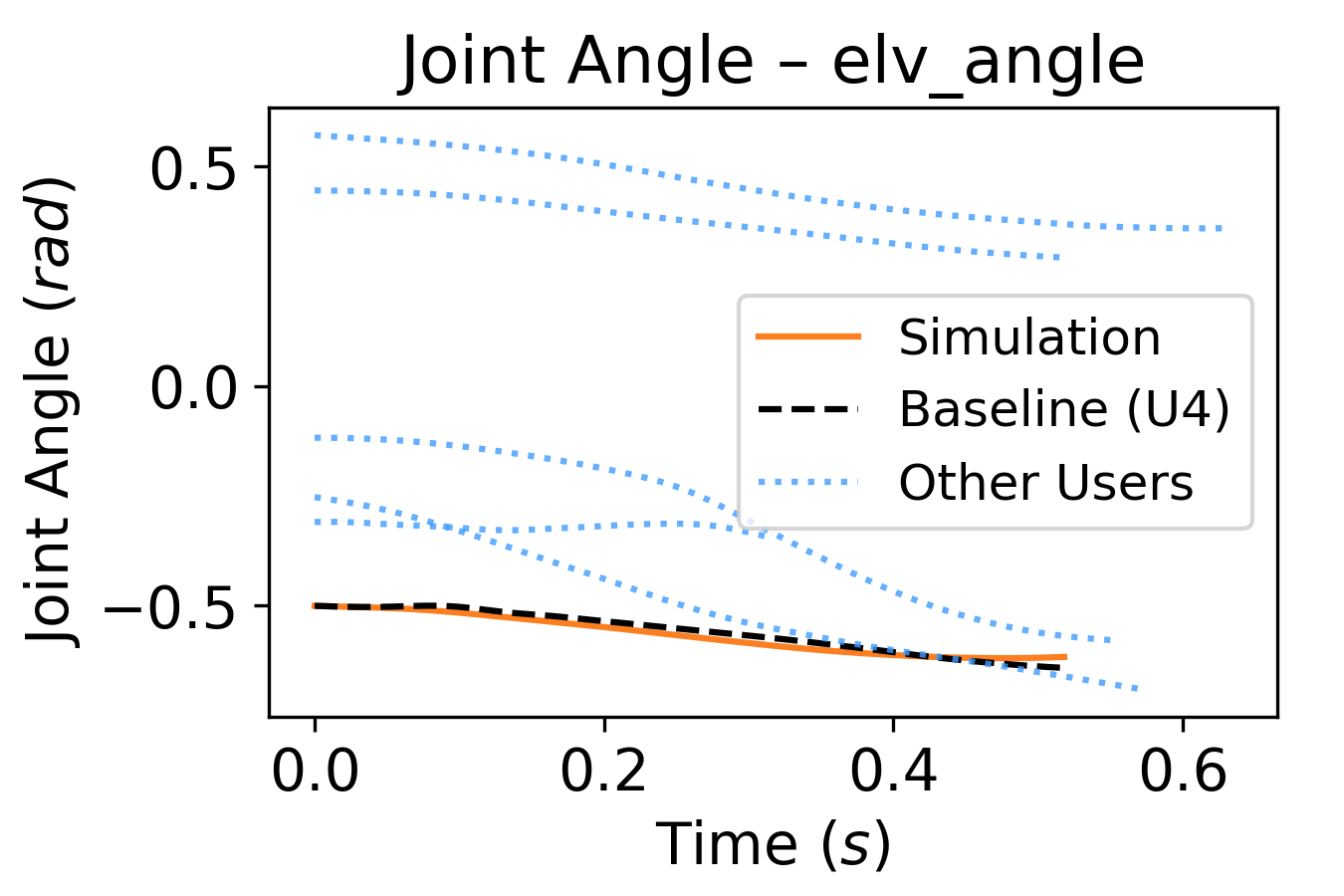}}
	\subfloat{\includegraphics[width=0.2\linewidth, clip]{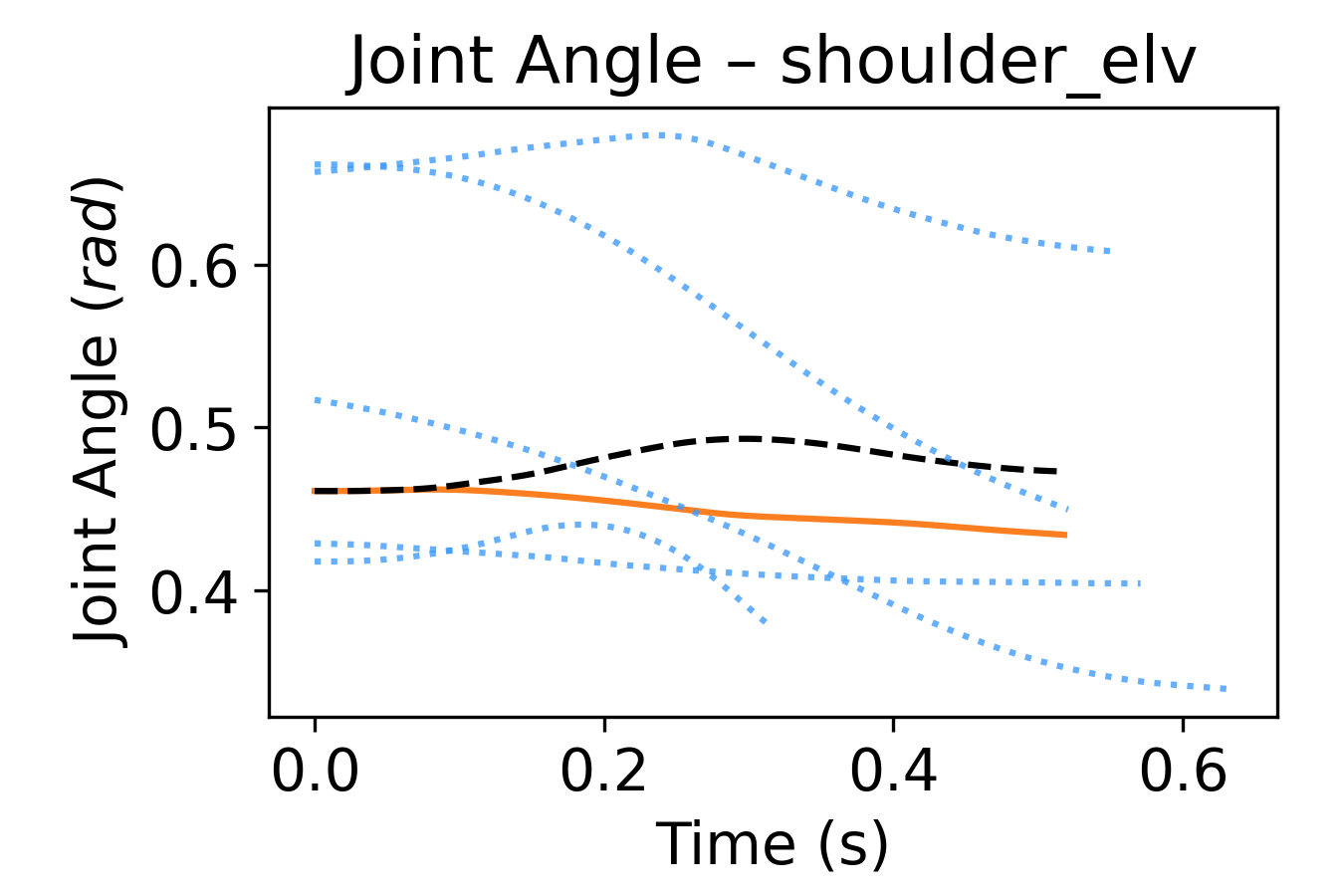}}
	\subfloat{\includegraphics[width=0.2\linewidth, clip]{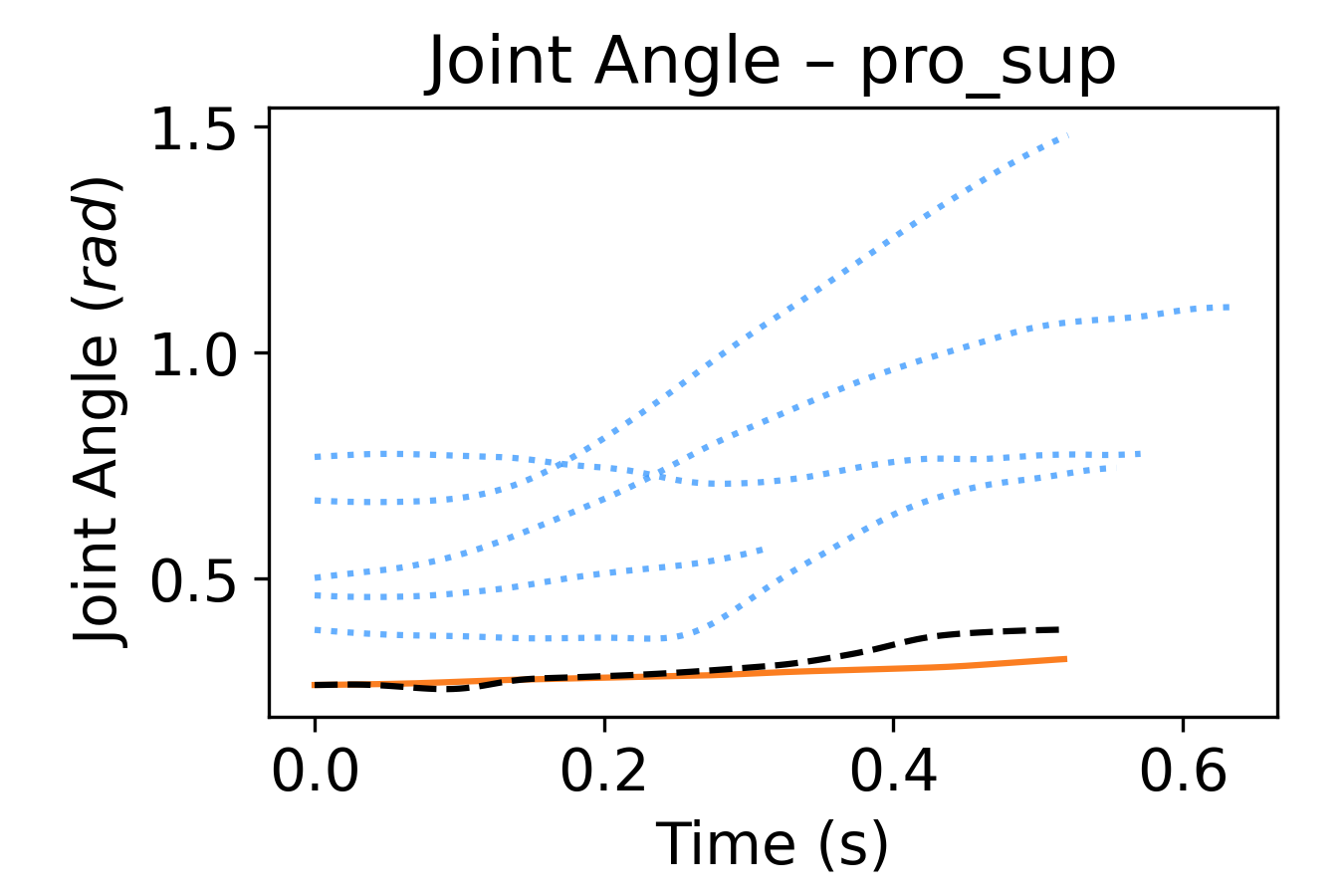}}
	\subfloat{\includegraphics[width=0.2\linewidth, clip]{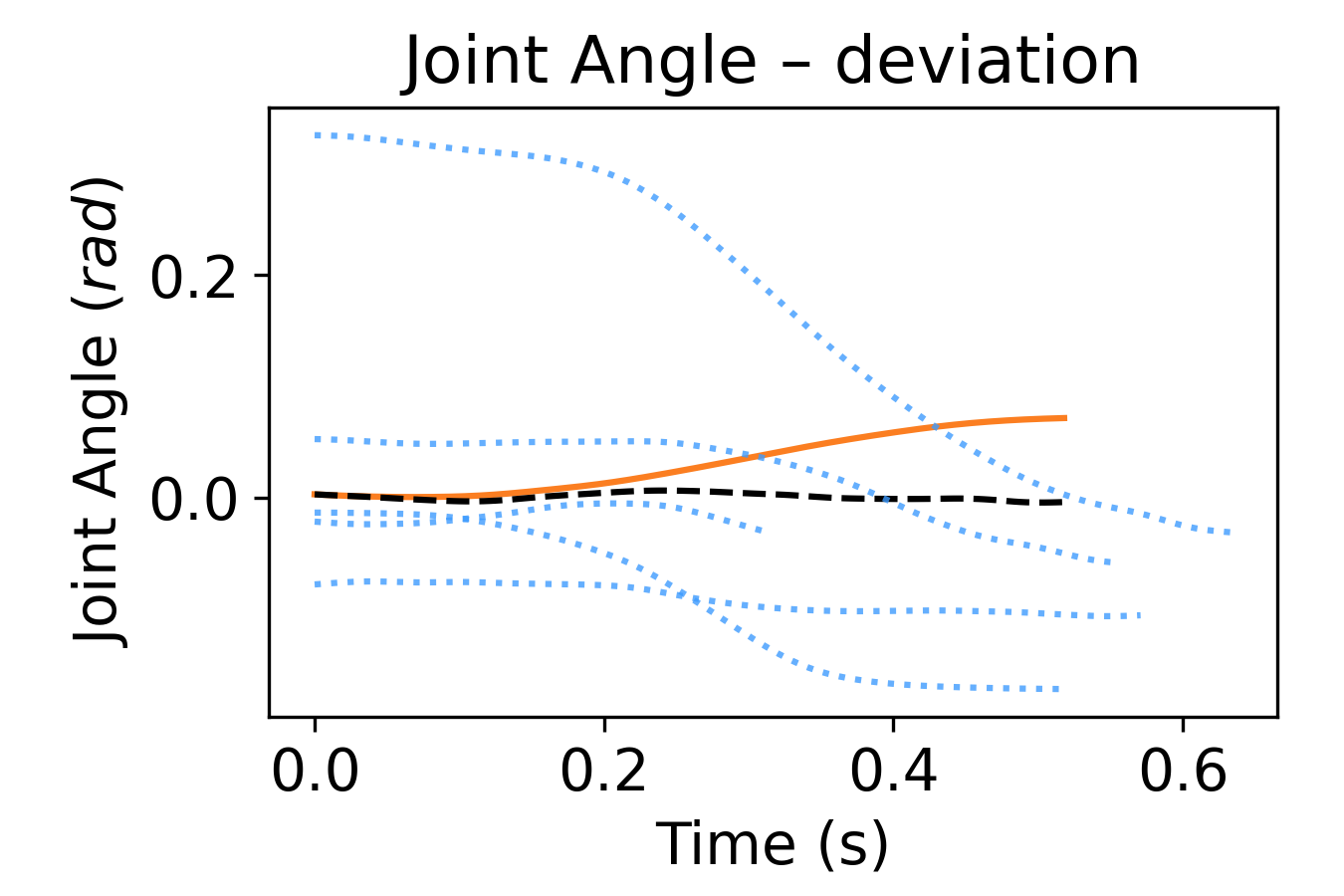}}
	\subfloat{\includegraphics[width=0.2\linewidth, clip]{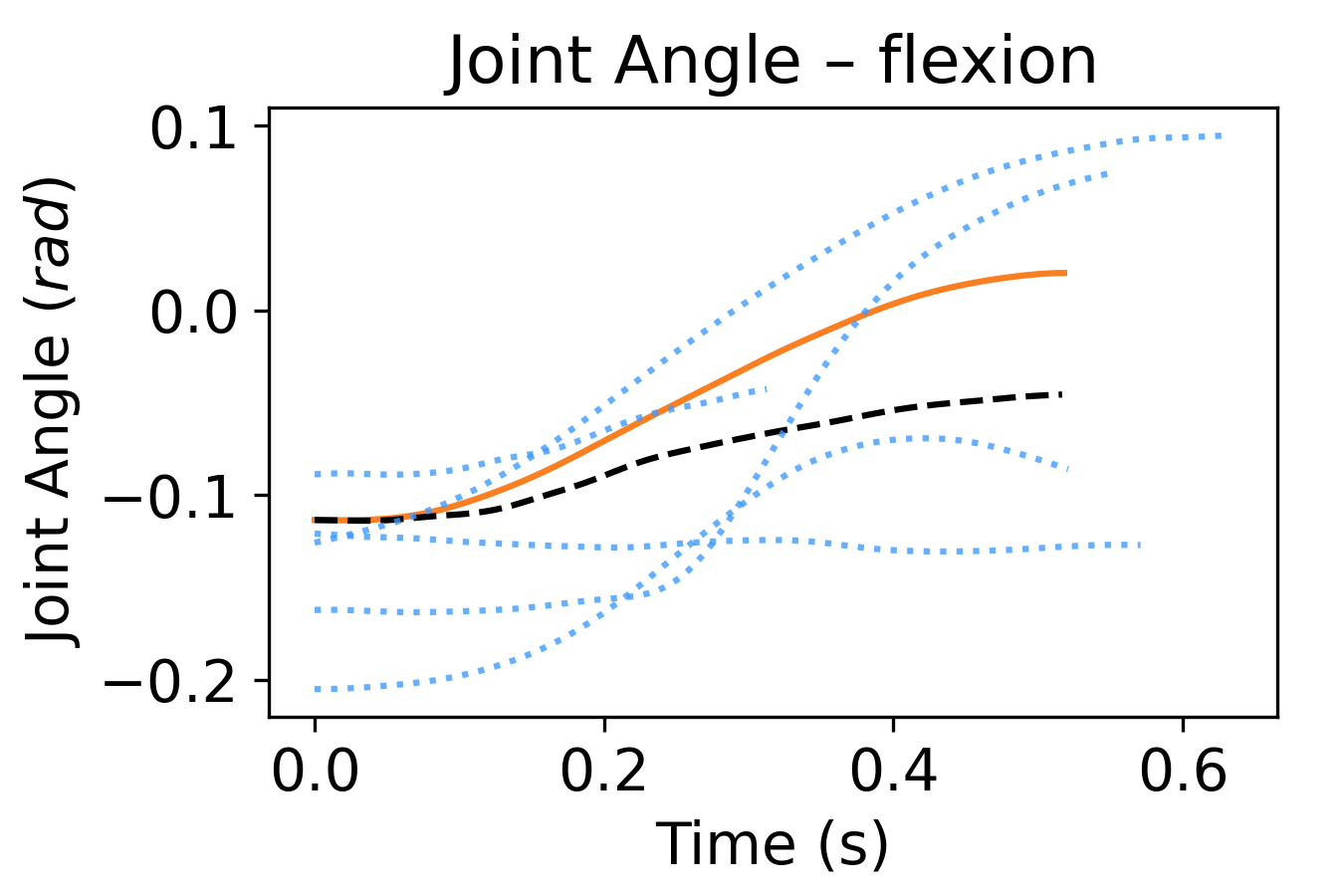}}\\
	
	\subfloat{\includegraphics[width=0.2\linewidth, clip]{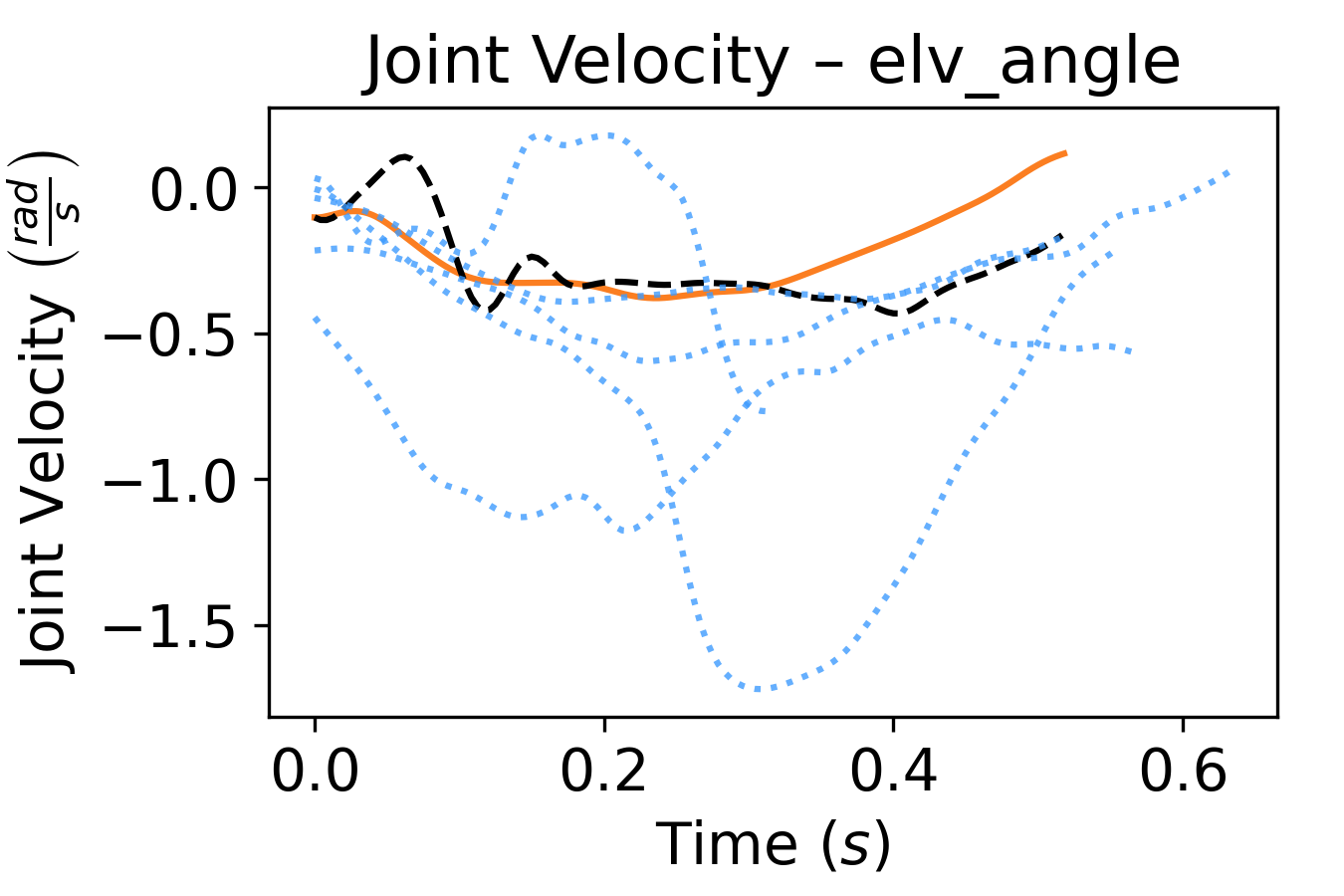}}
	\subfloat{\includegraphics[width=0.2\linewidth, clip]{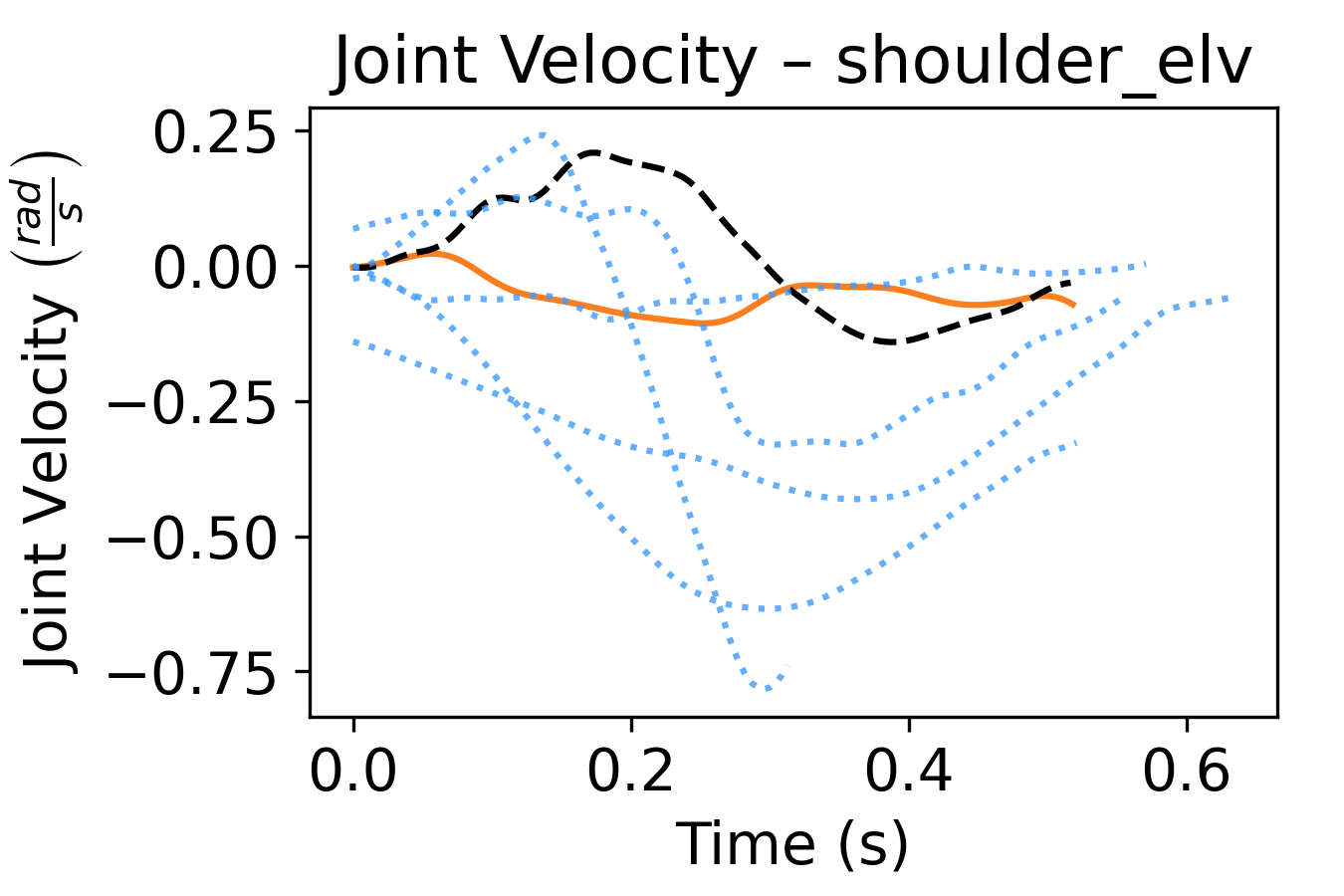}}
	\subfloat{\includegraphics[width=0.2\linewidth, clip]{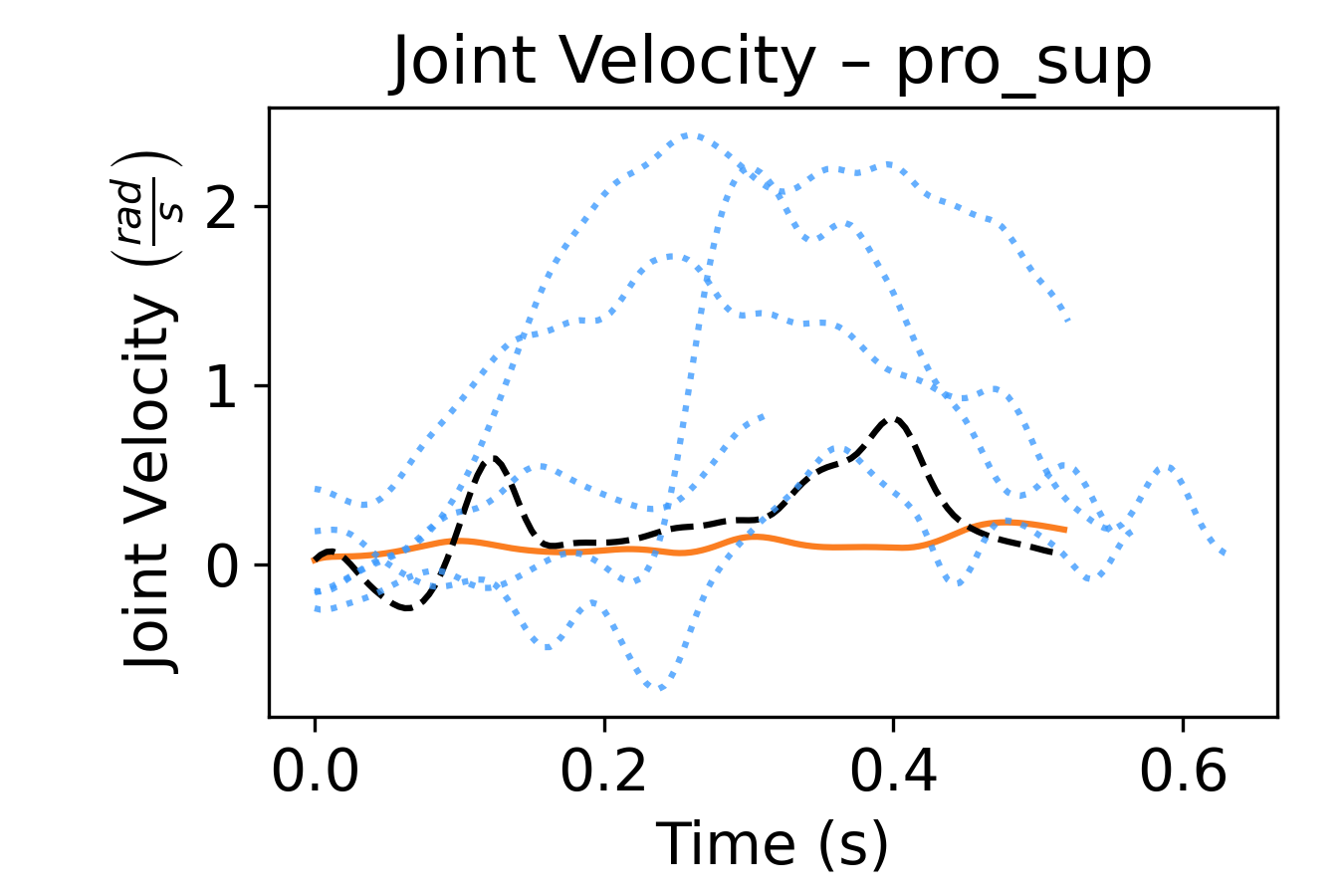}}
	\subfloat{\includegraphics[width=0.2\linewidth, clip]{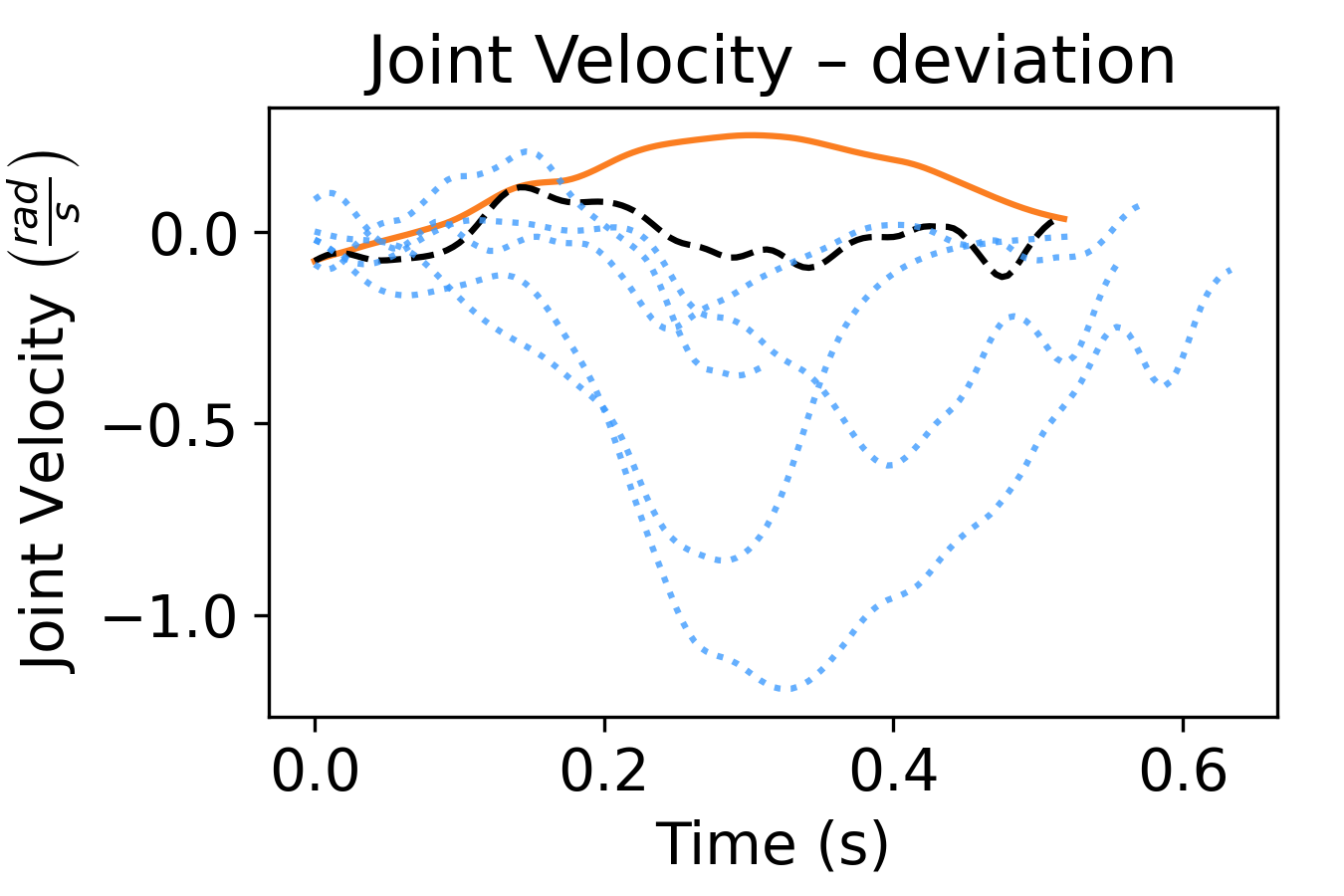}}
	\subfloat{\includegraphics[width=0.2\linewidth, clip]{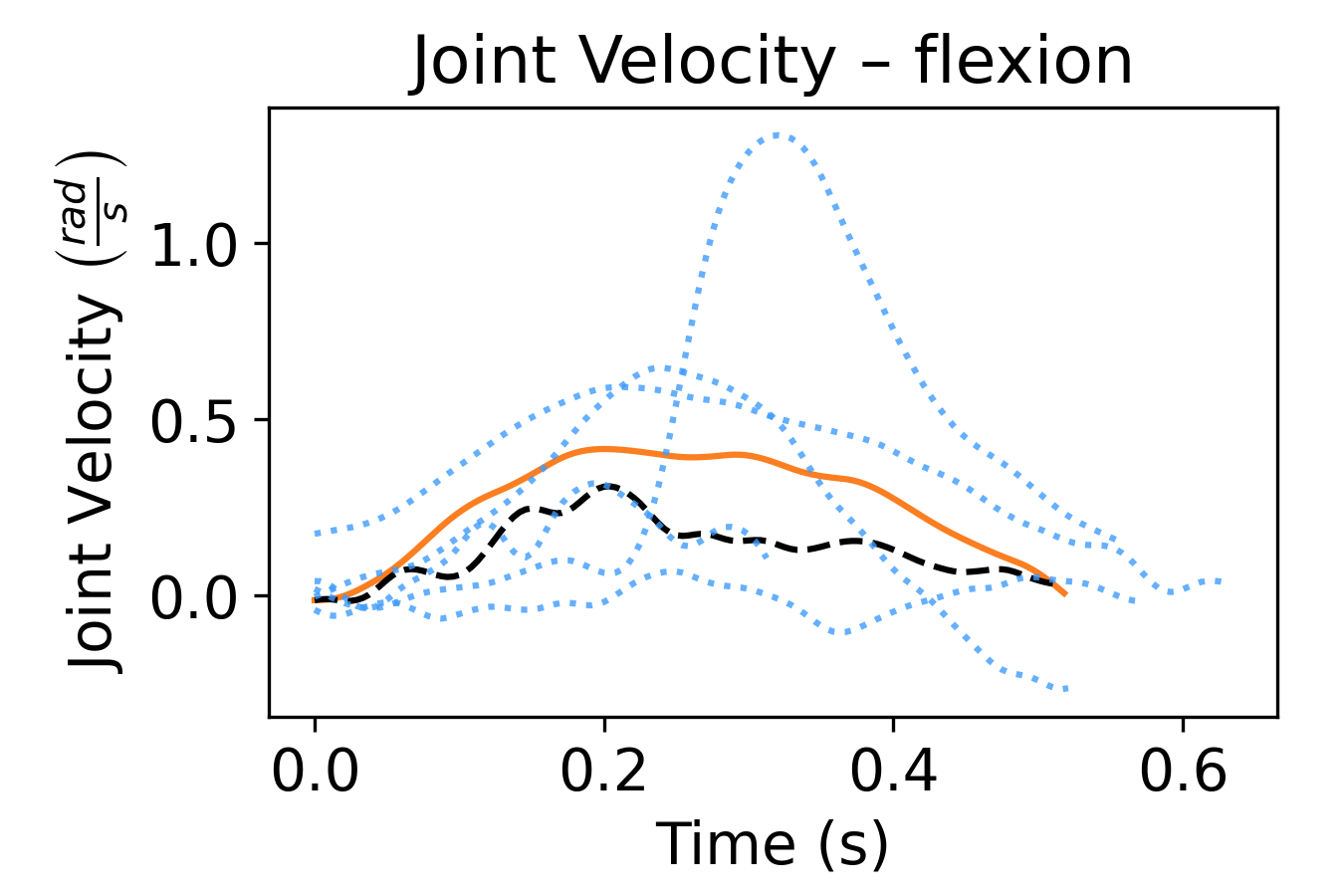}}
	
	\caption{Given an interaction technique (here: Virtual Pad Ergonomic) and a movement direction (here: movements from target 1 to target 2), the characteristic cursor and joint trajectories of an individual user (here: U4, black dashed lines; trajectories of the remaining users are shown as blue dotted lines for comparison) can be predicted by our simulation (orange solid lines). Remaining joints are shown in Figure~\ref{fig:PadErgonomic_qual}.}
	\label{fig:PadErgonomic_qual_all}
\end{figure}

\begin{figure}[h!]
	\centering
	\subfloat{\includegraphics[width=0.25\linewidth, clip]{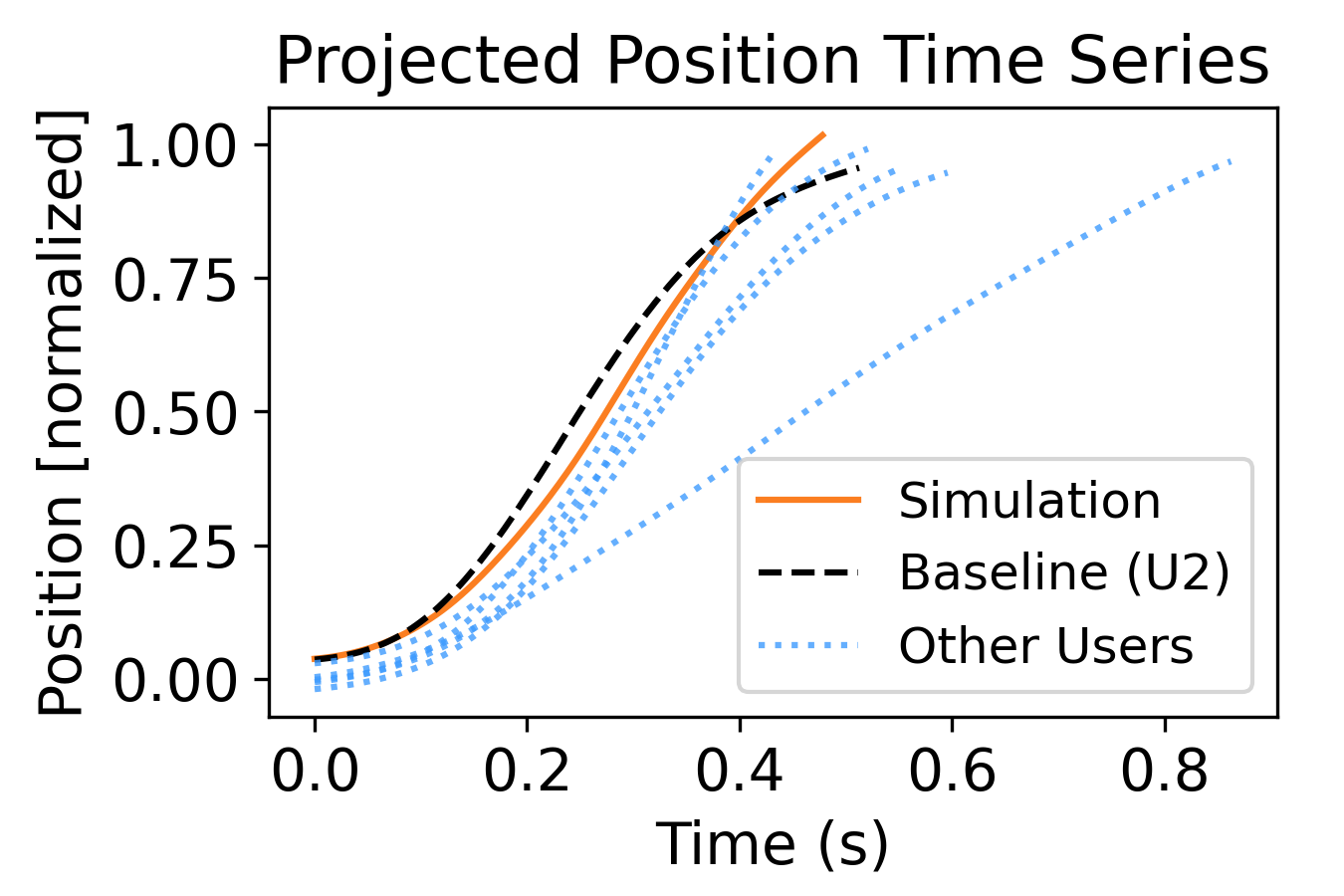}}
	\subfloat{\includegraphics[width=0.25\linewidth, clip]{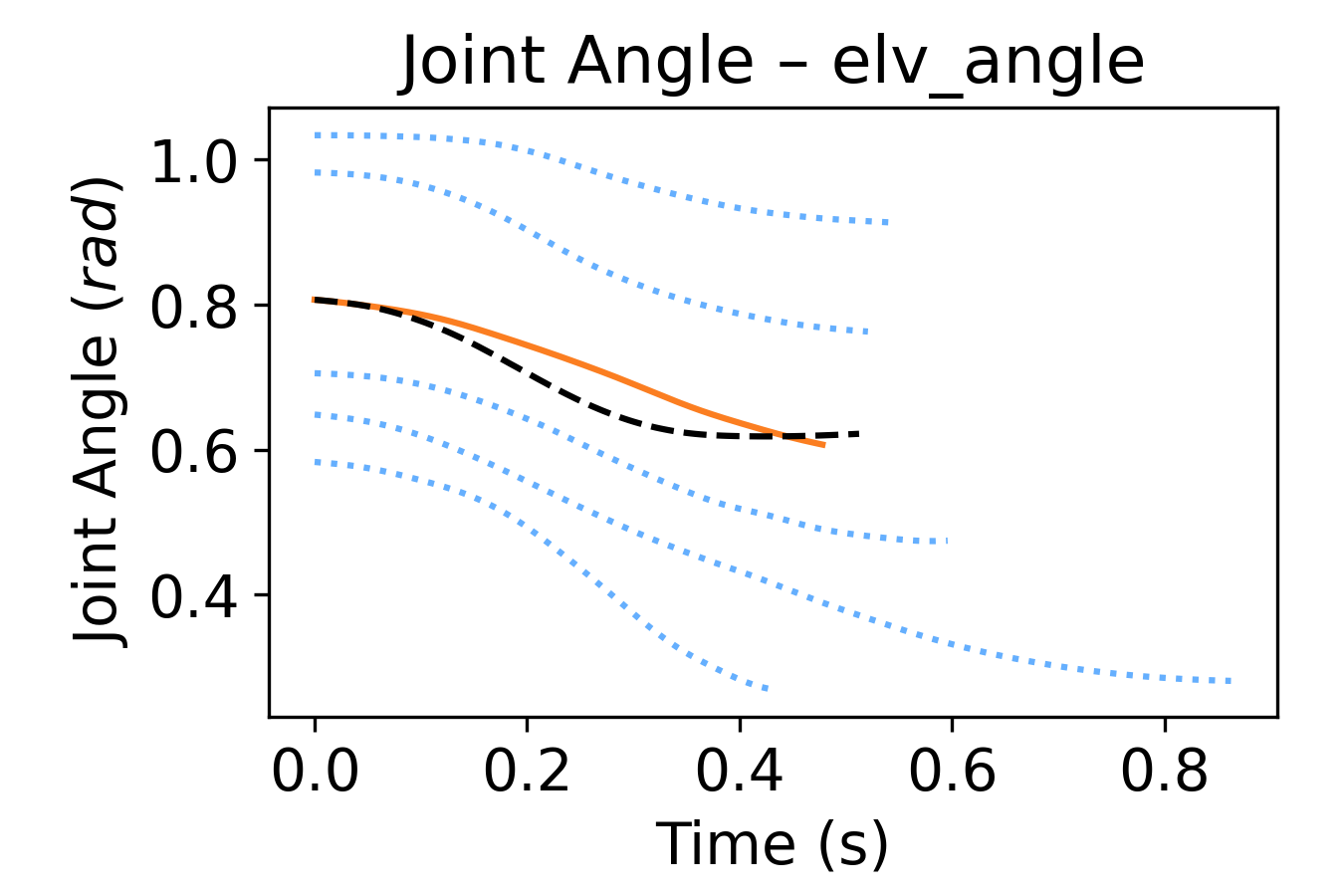}}
	\subfloat{\includegraphics[width=0.25\linewidth, clip]{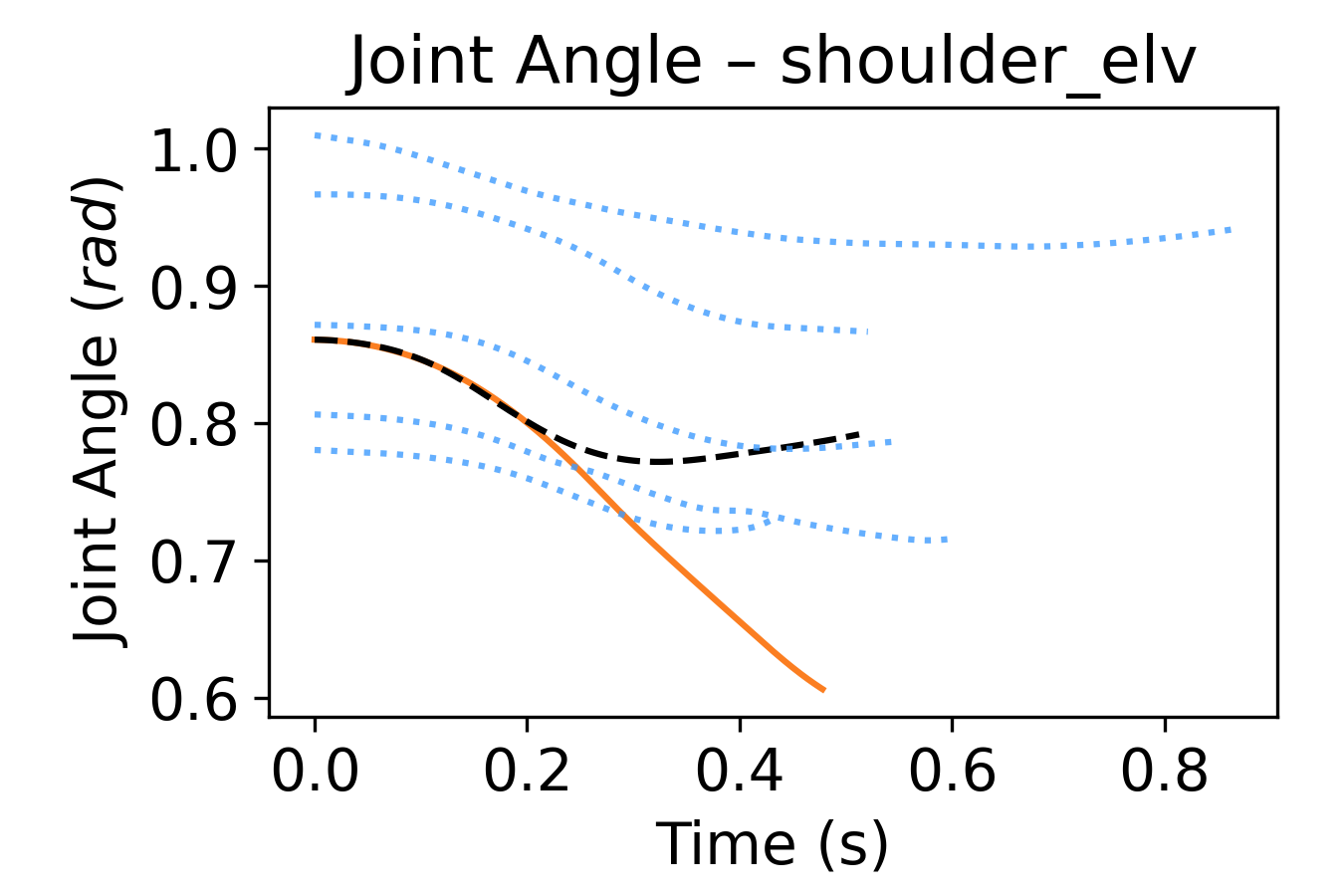}}
	\subfloat{\includegraphics[width=0.25\linewidth, clip]{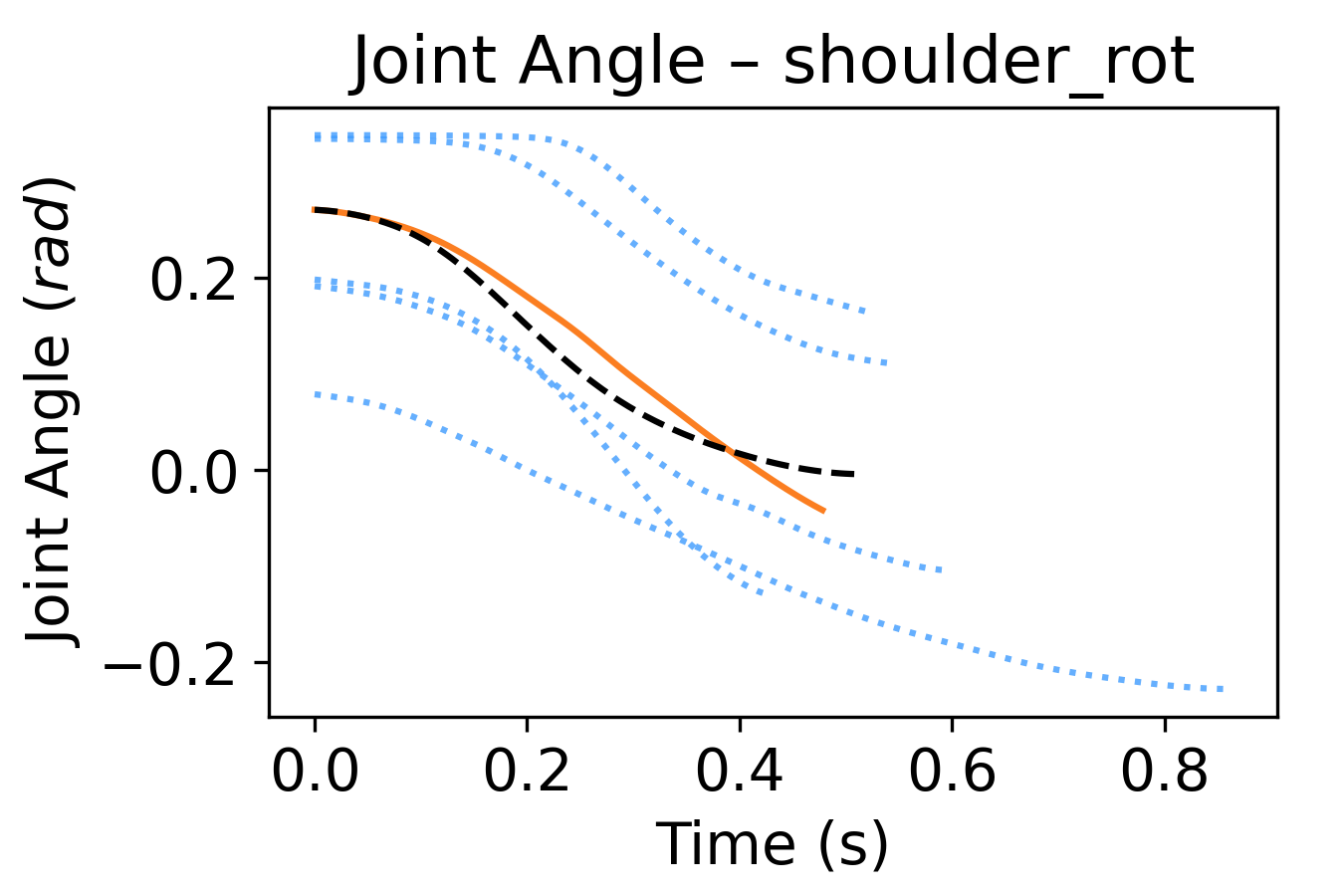}}\\
	
	\subfloat{\includegraphics[width=0.25\linewidth, clip]{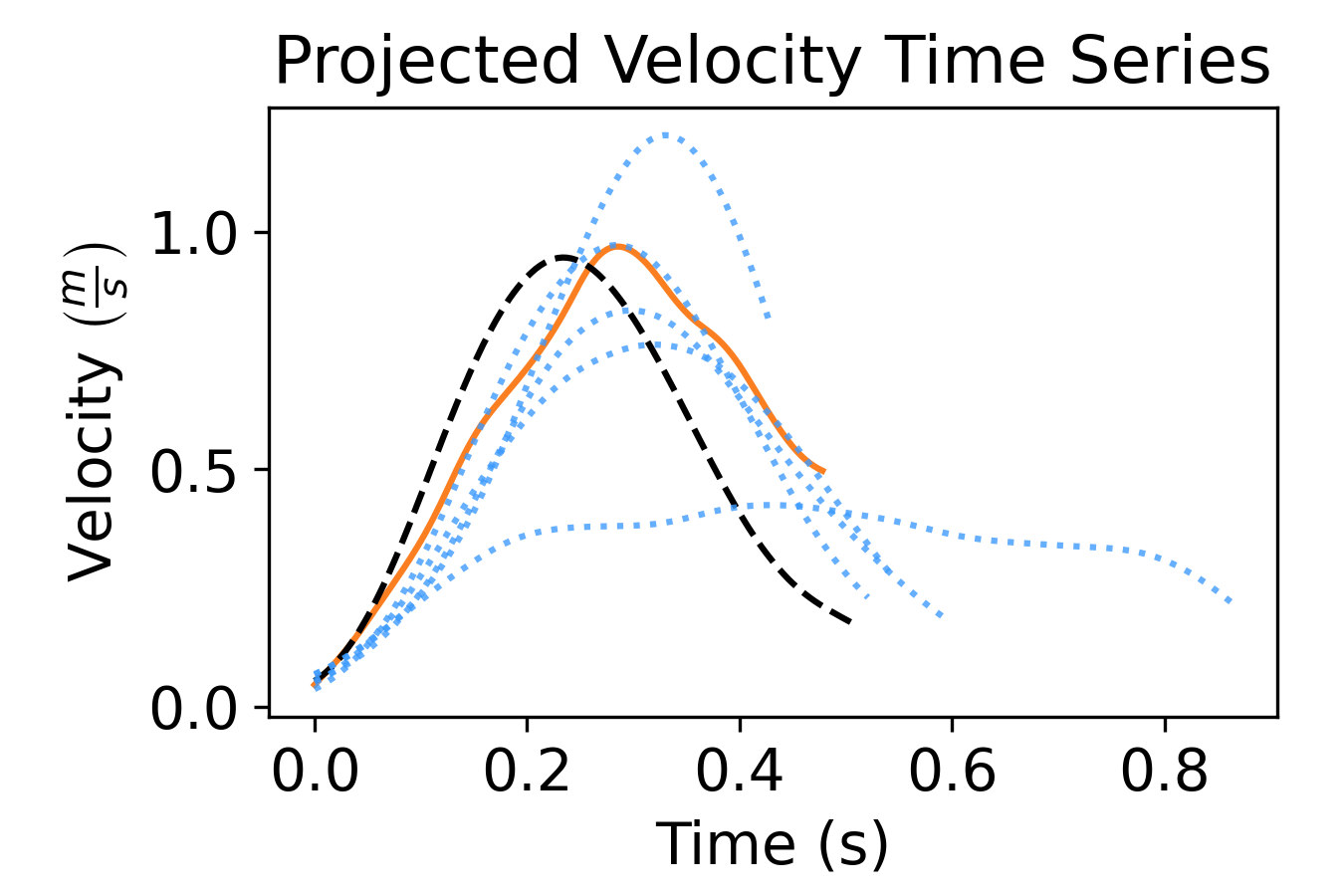}}
	\subfloat{\includegraphics[width=0.25\linewidth, clip]{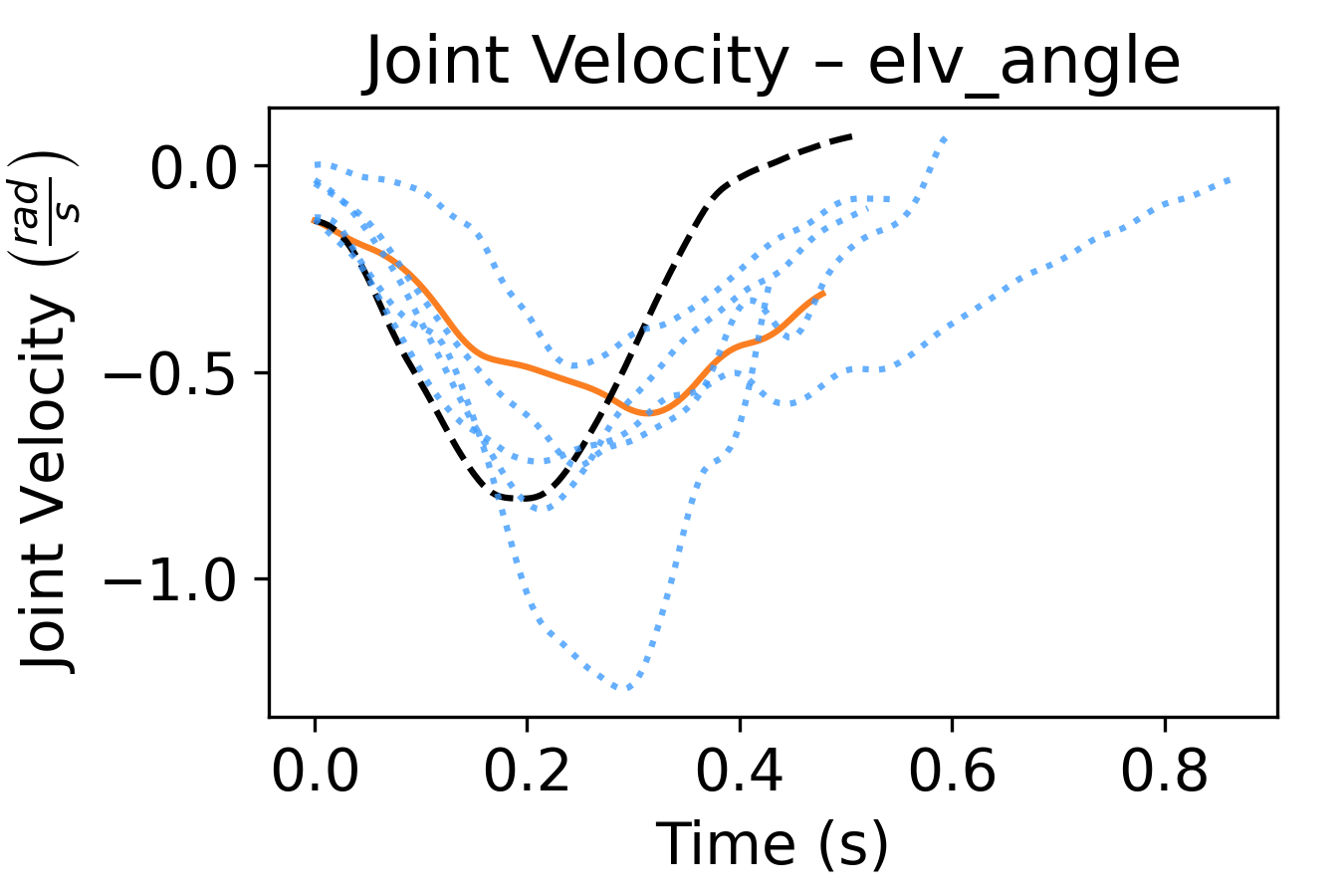}}
	\subfloat{\includegraphics[width=0.25\linewidth, clip]{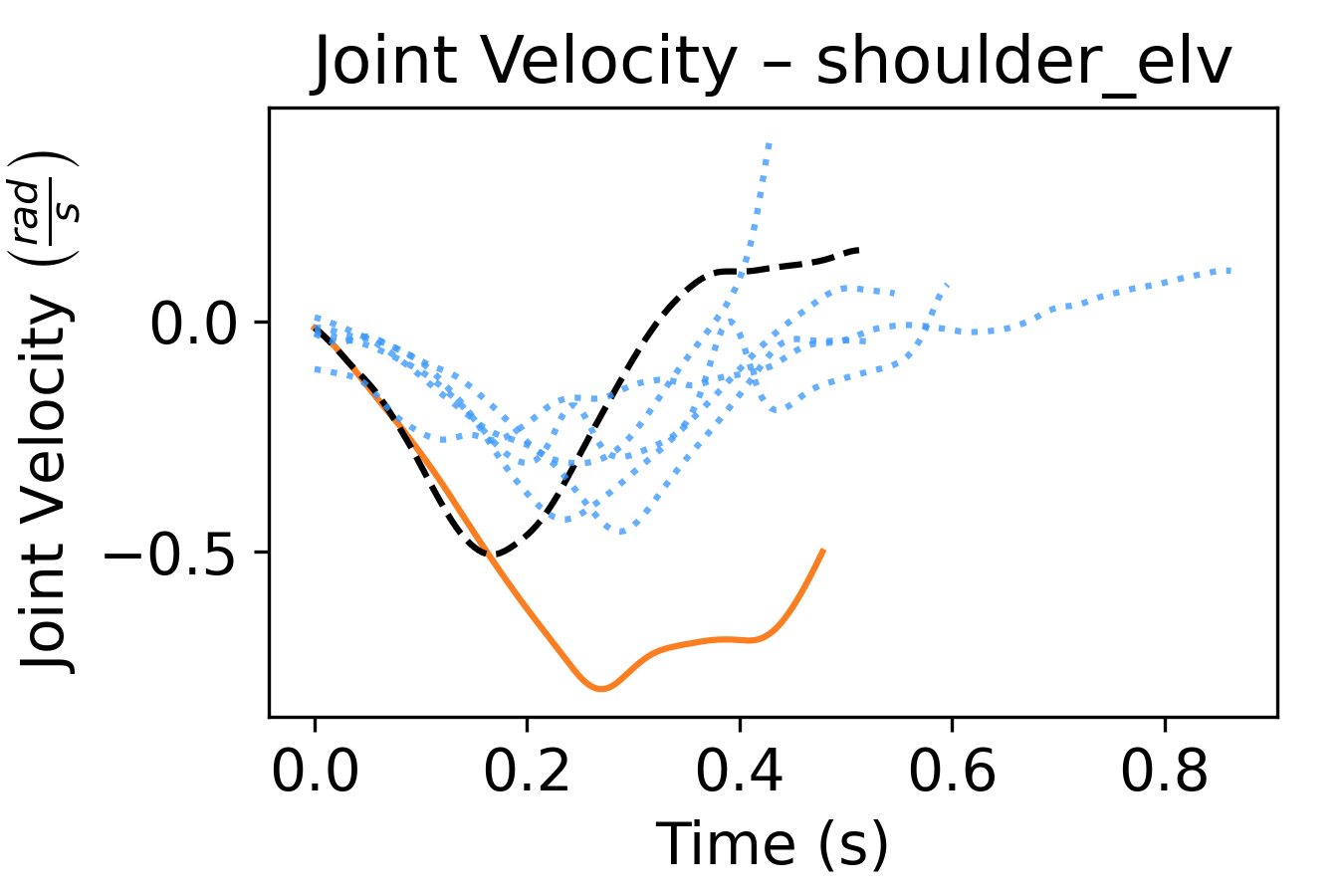}}
	\subfloat{\includegraphics[width=0.25\linewidth, clip]{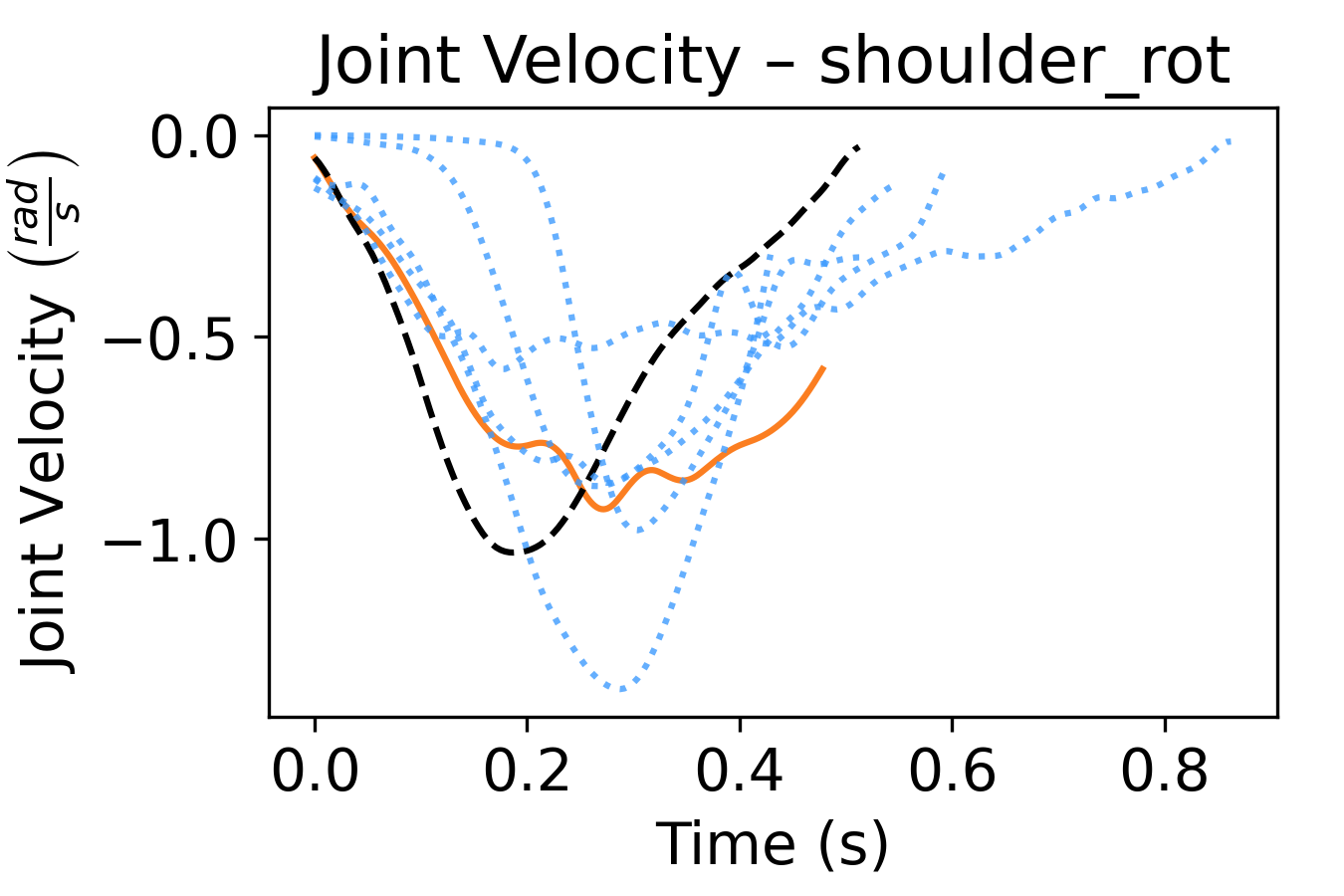}}\\
	
	\subfloat{\includegraphics[width=0.25\linewidth, clip]{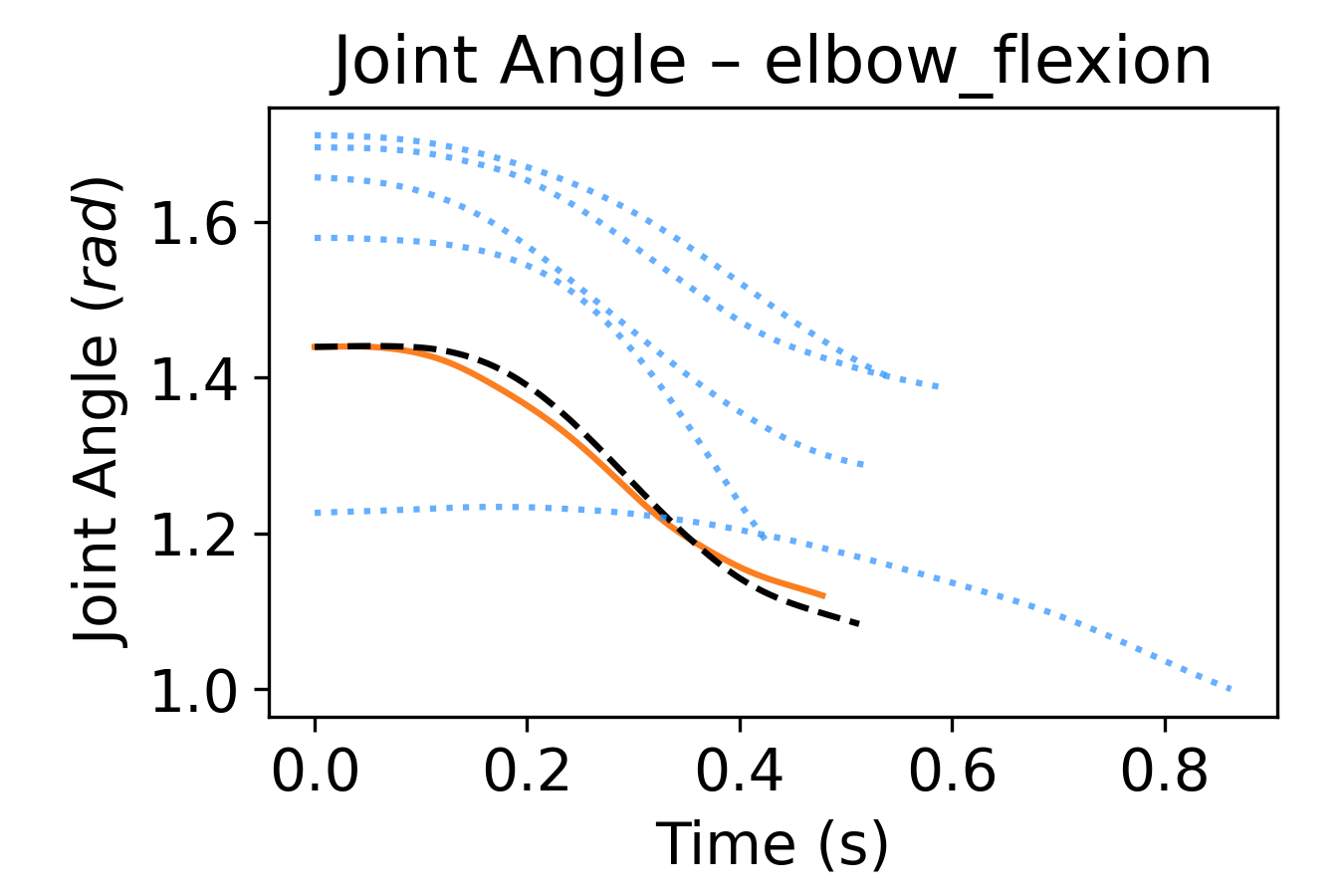}}
	\subfloat{\includegraphics[width=0.25\linewidth, clip]{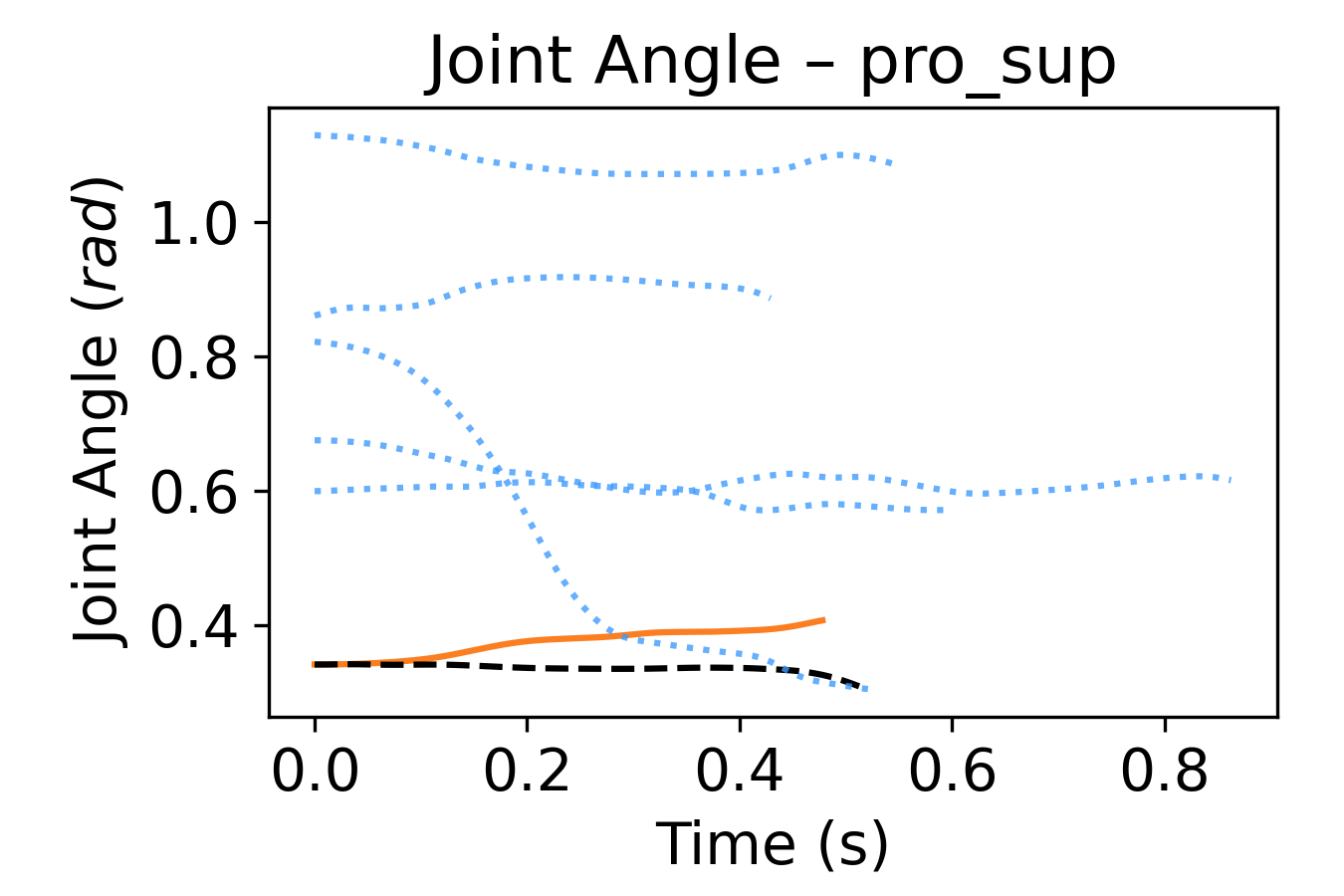}}
	\subfloat{\includegraphics[width=0.25\linewidth, clip]{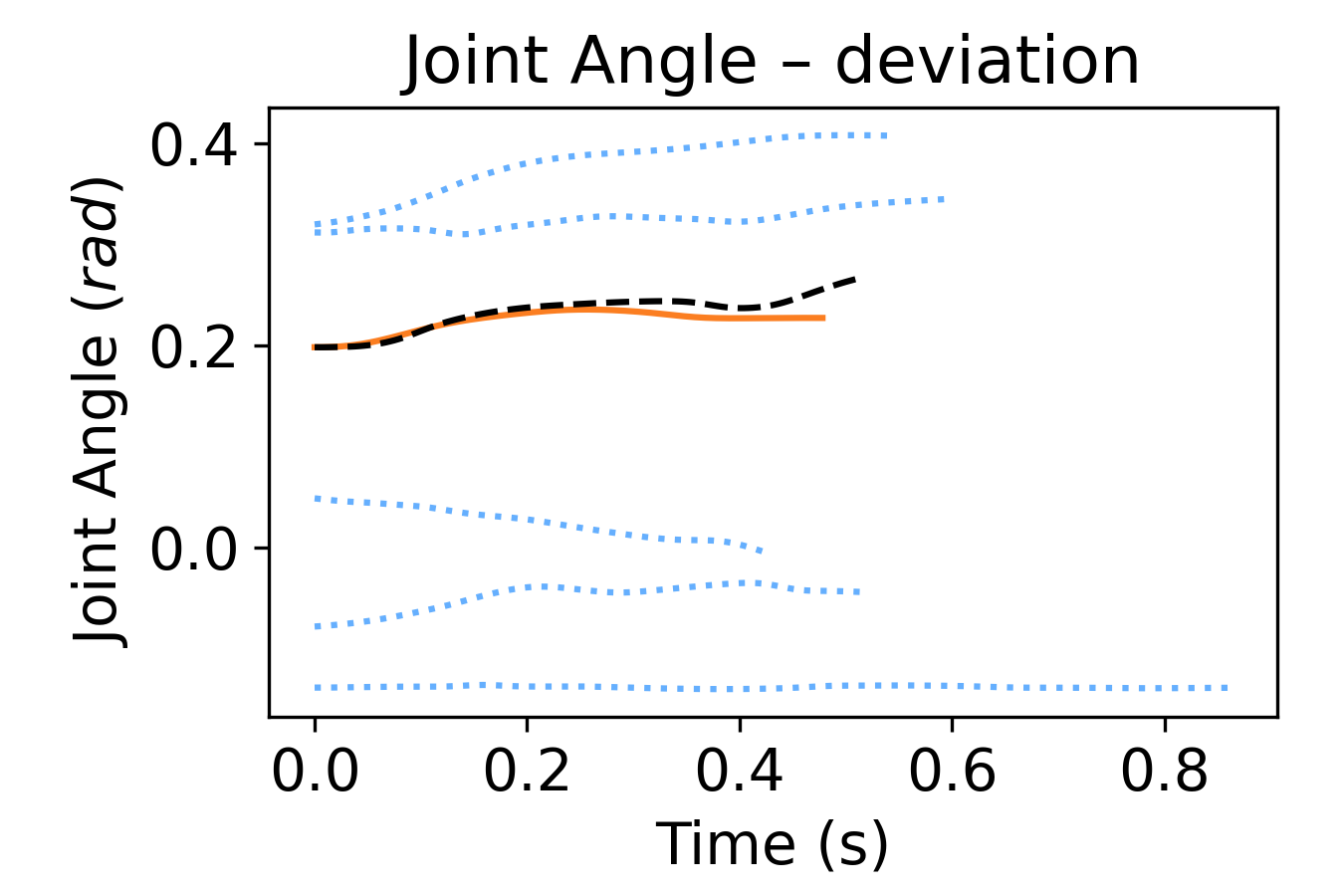}}
	\subfloat{\includegraphics[width=0.25\linewidth, clip]{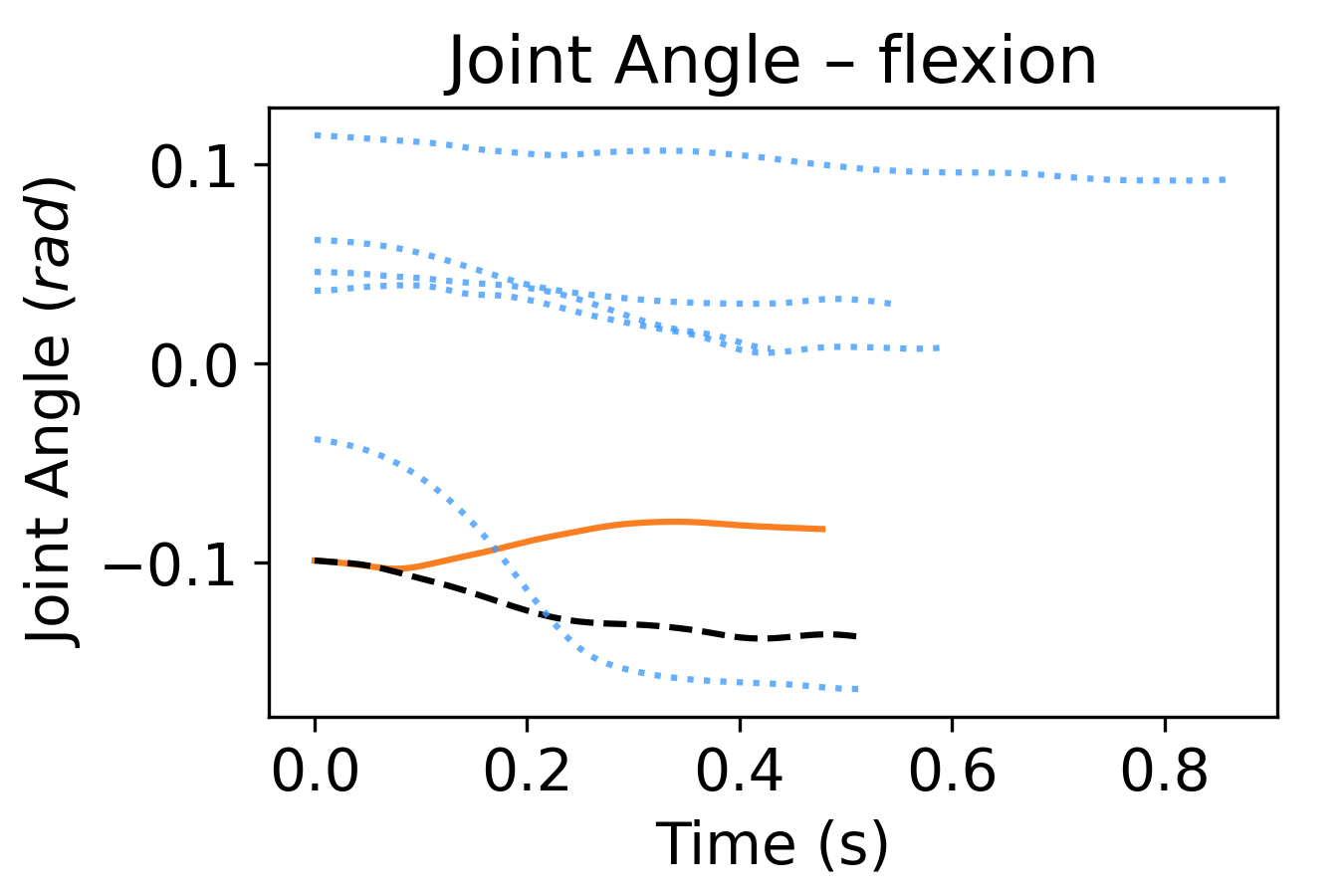}}\\
	
	\subfloat{\includegraphics[width=0.25\linewidth, clip]{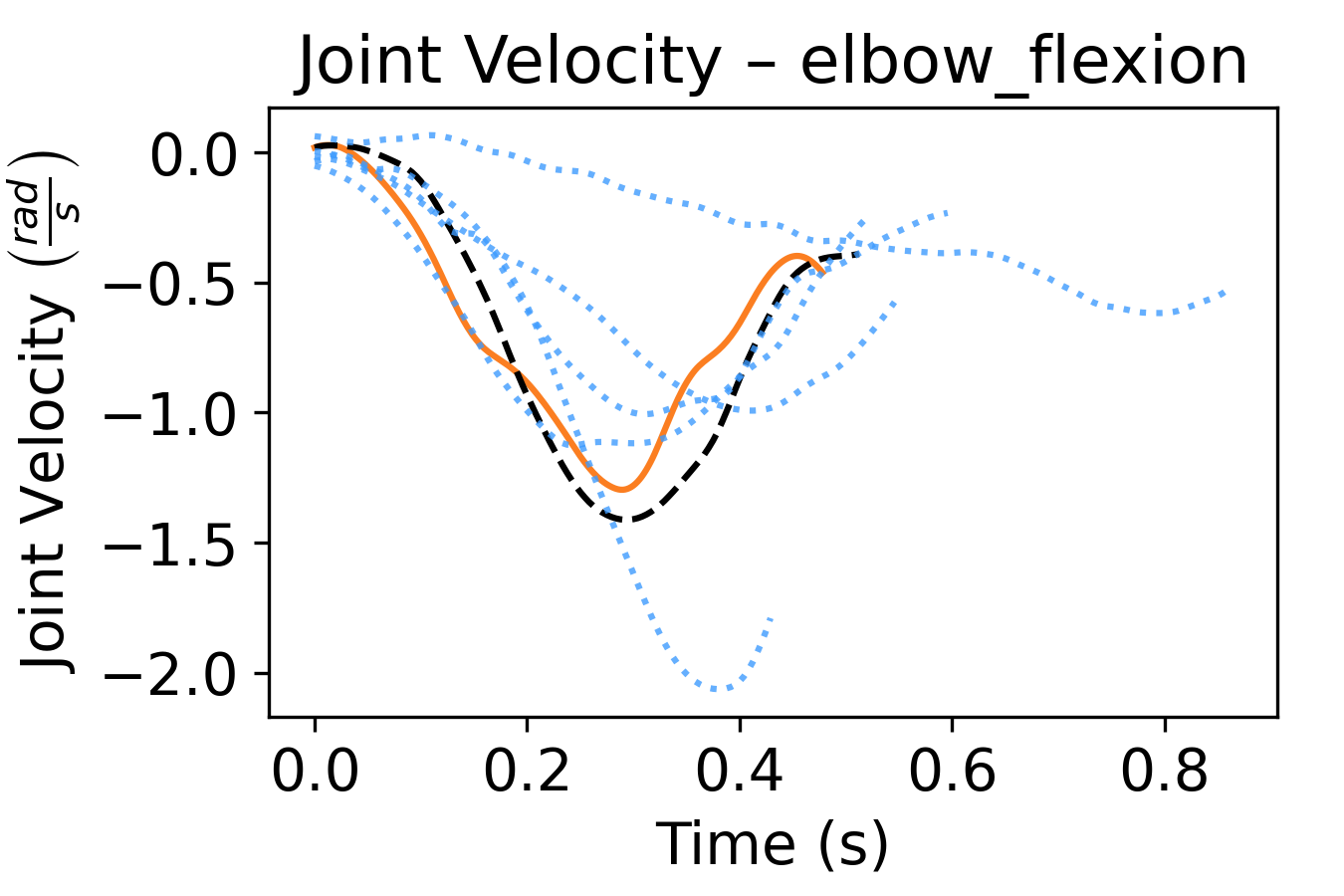}}
	\subfloat{\includegraphics[width=0.25\linewidth, clip]{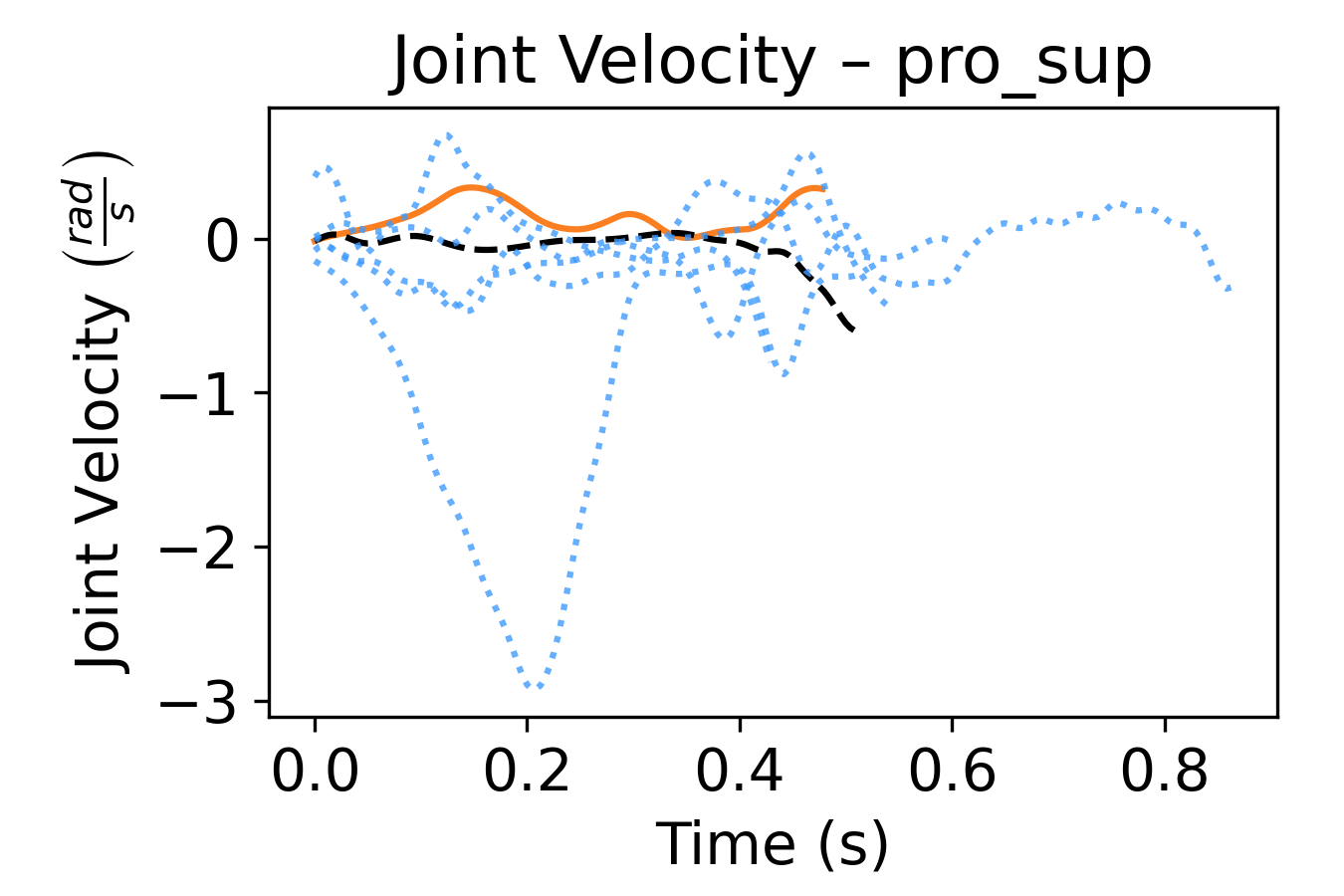}}
	\subfloat{\includegraphics[width=0.25\linewidth, clip]{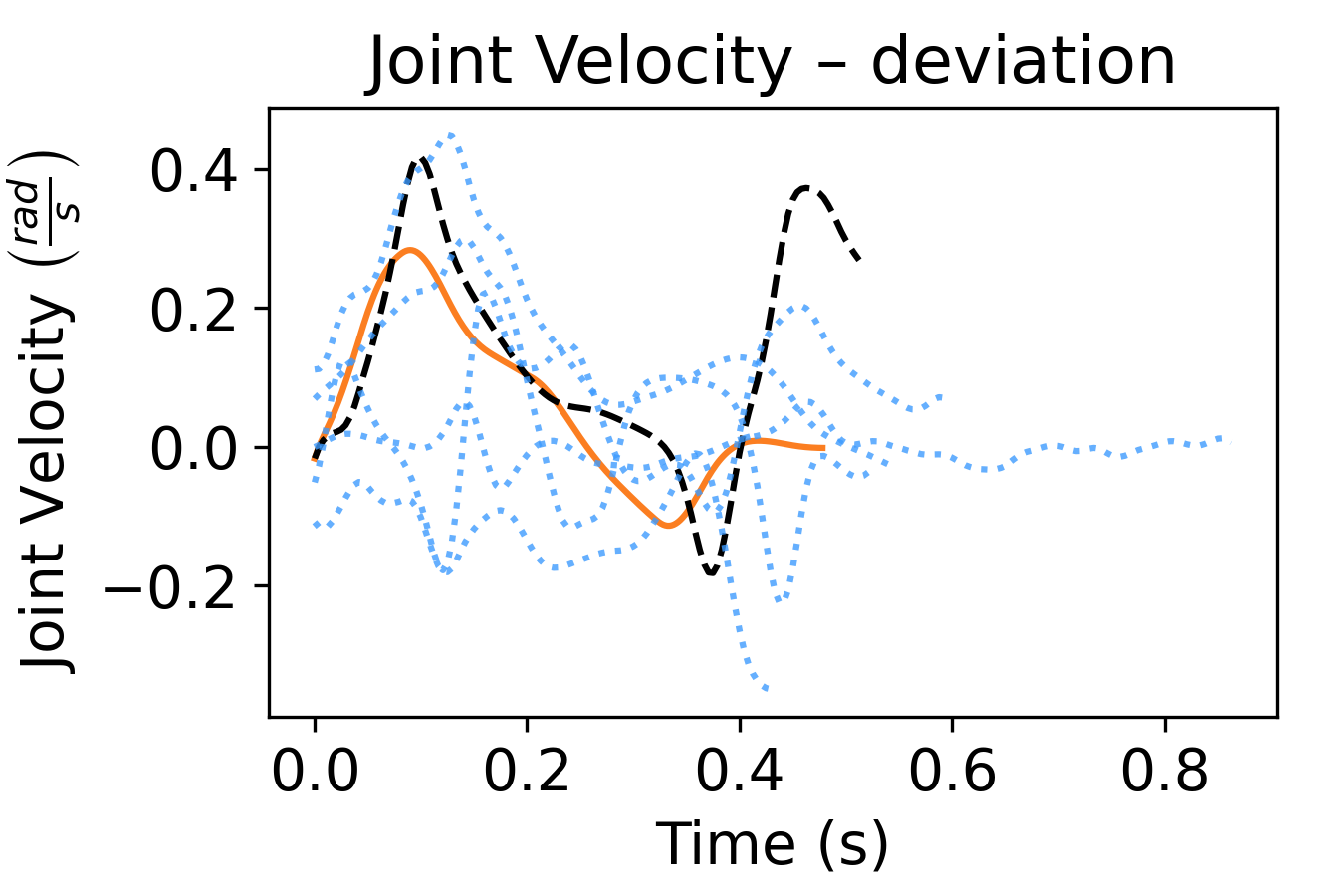}}
	\subfloat{\includegraphics[width=0.25\linewidth, clip]{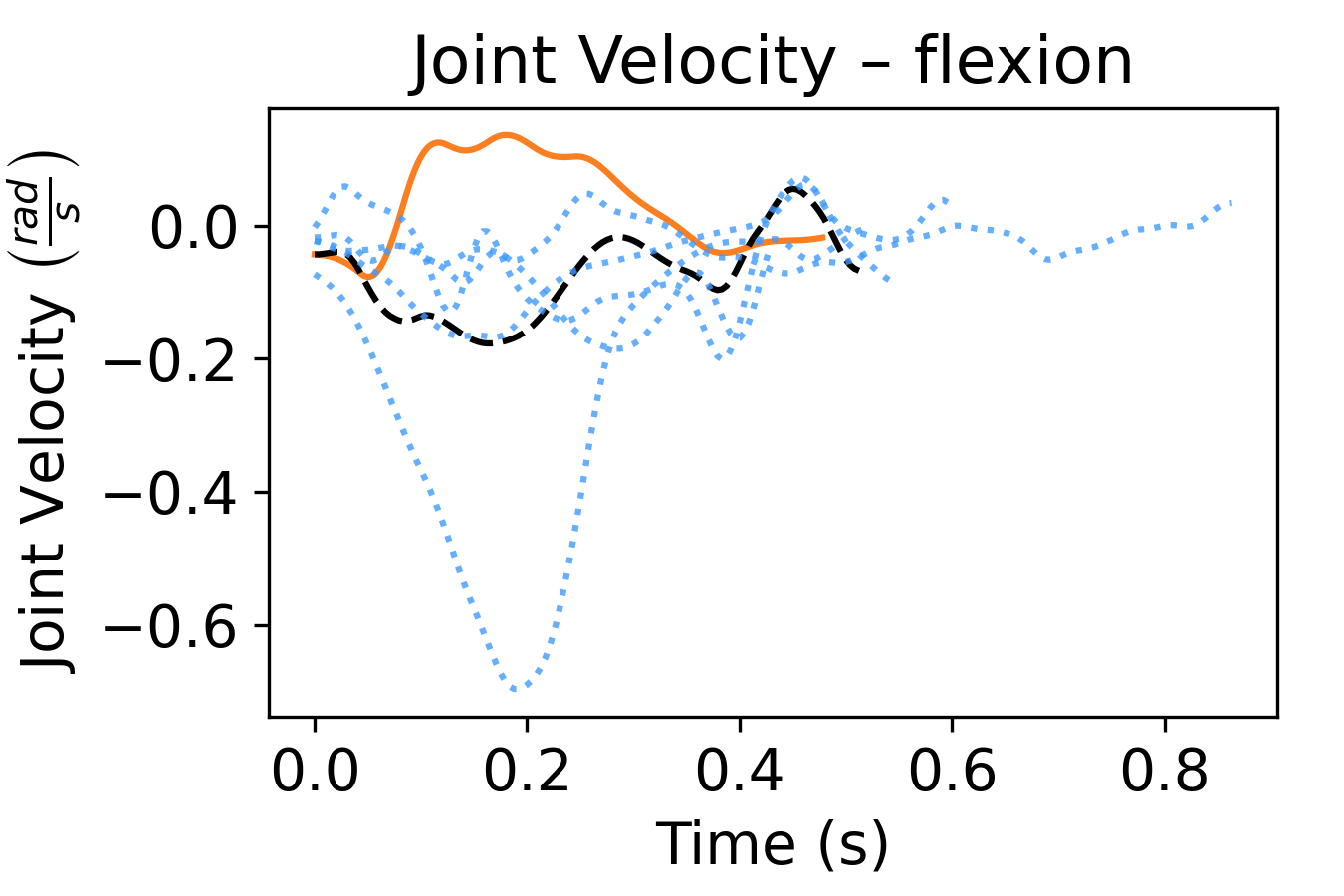}}
	
	\caption{Given an interaction technique (here: \textbf{Virtual Cursor ID}) and a movement direction (here: movements from target 8 to target 9), the characteristic cursor and joint trajectories of an individual user (here: U2, black dashed lines; trajectories of the remaining users are shown as blue dotted lines for comparison) can be predicted by our simulation (orange solid lines).}
	\label{fig:CursorID_qual}
\end{figure}
\newpage

\begin{figure}[h!]
	\centering
	\subfloat{\includegraphics[width=0.25\linewidth, clip]{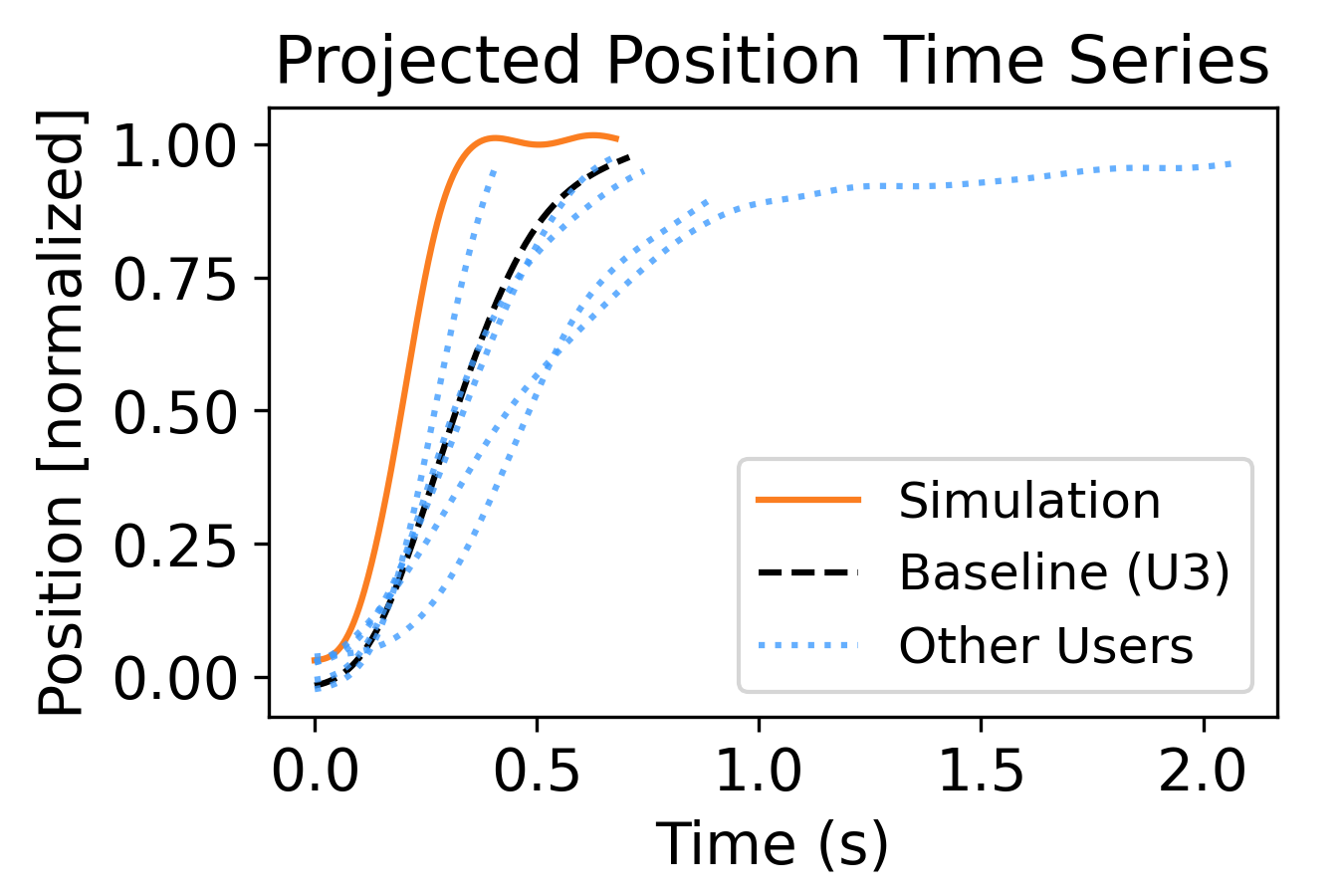}}
	\subfloat{\includegraphics[width=0.25\linewidth, clip]{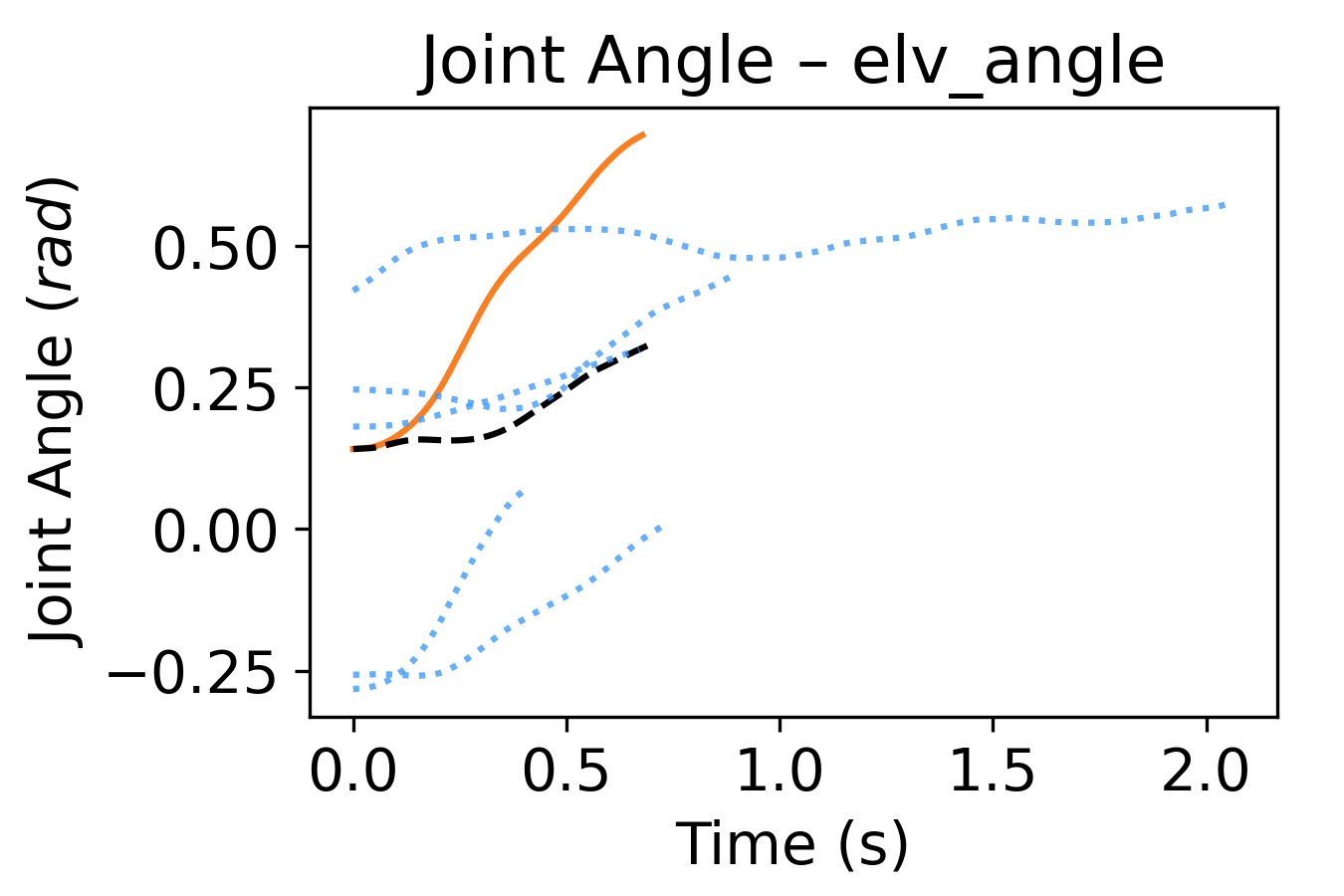}}
	\subfloat{\includegraphics[width=0.25\linewidth, clip]{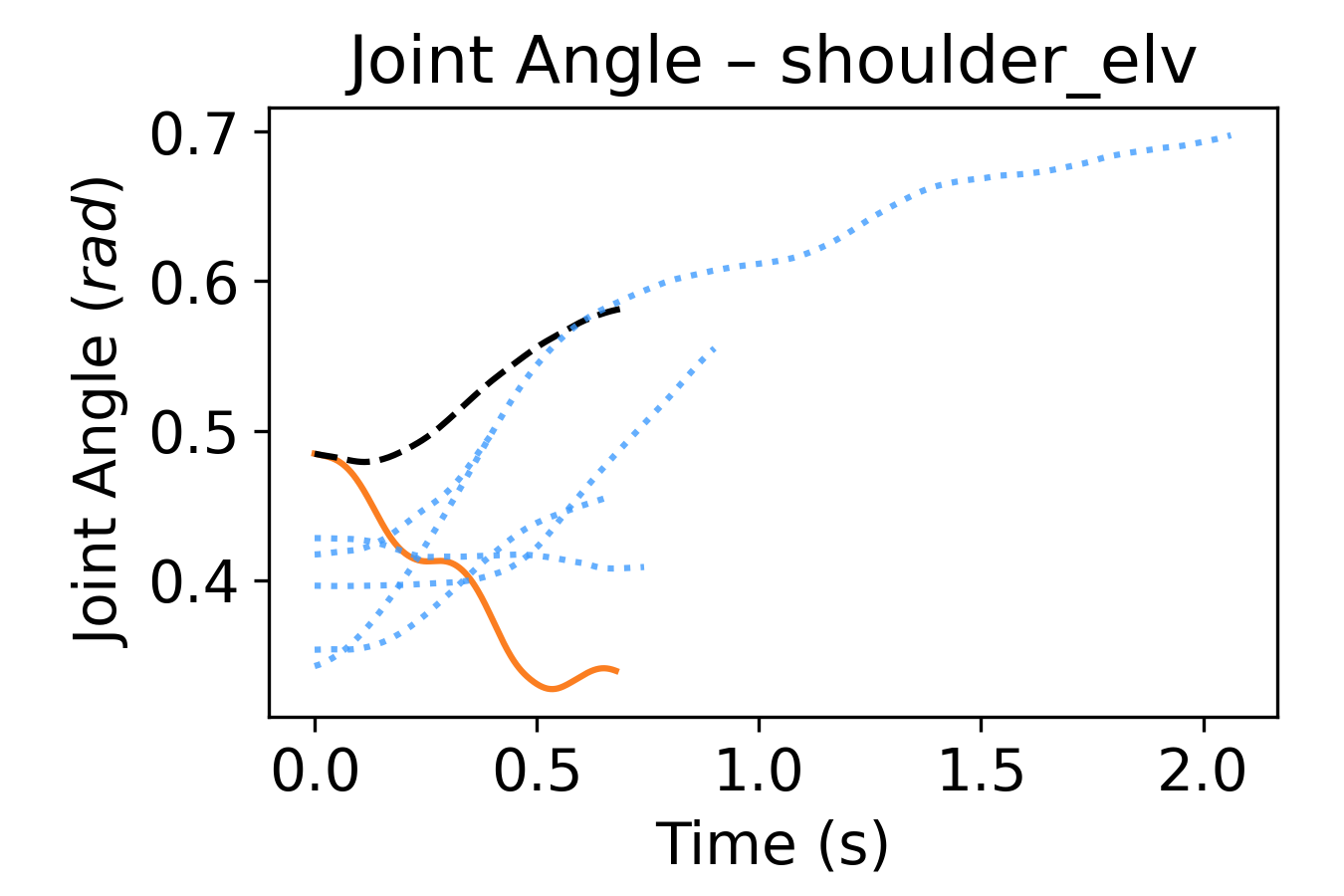}}
	\subfloat{\includegraphics[width=0.25\linewidth, clip]{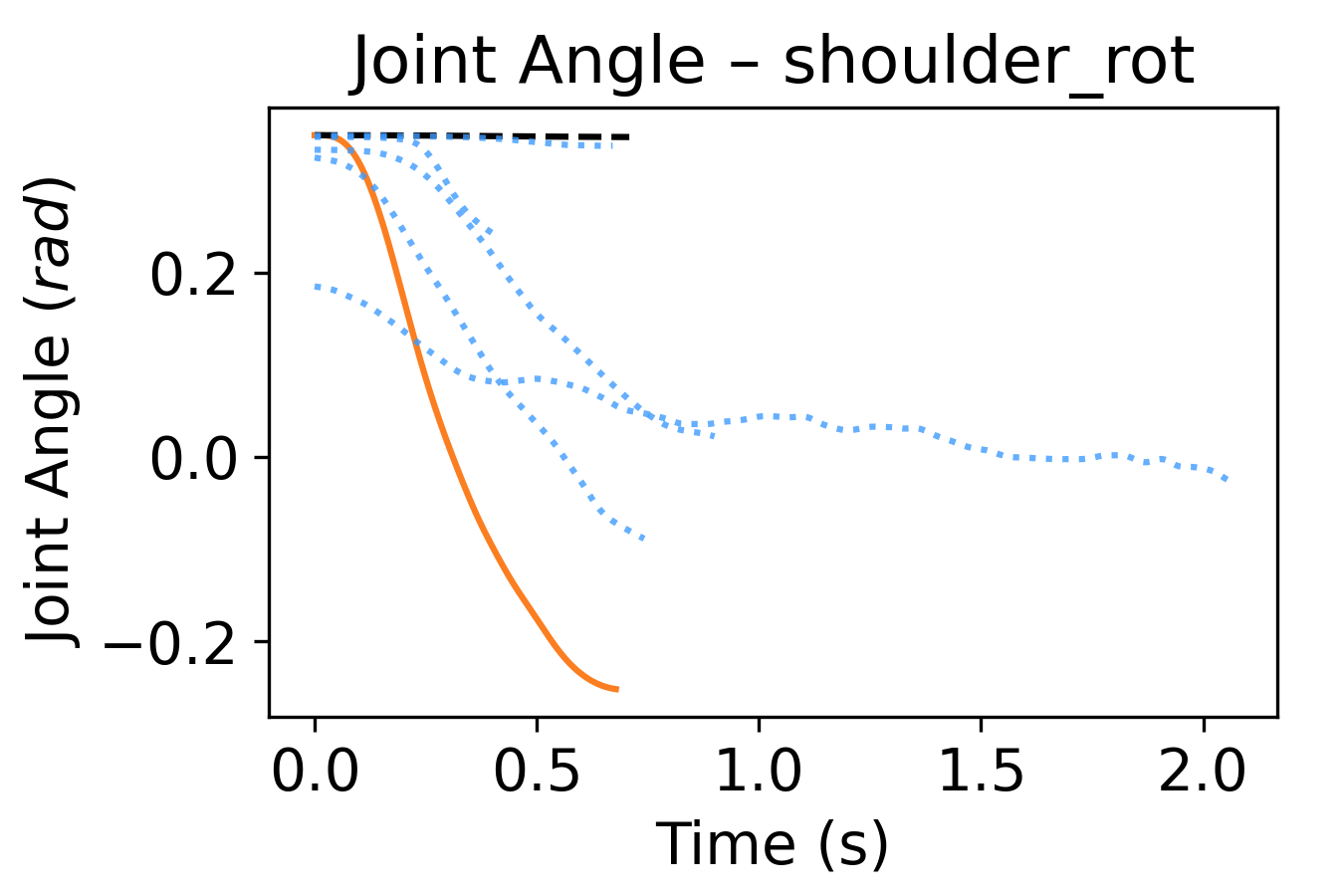}}\\
	
	\subfloat{\includegraphics[width=0.25\linewidth, clip]{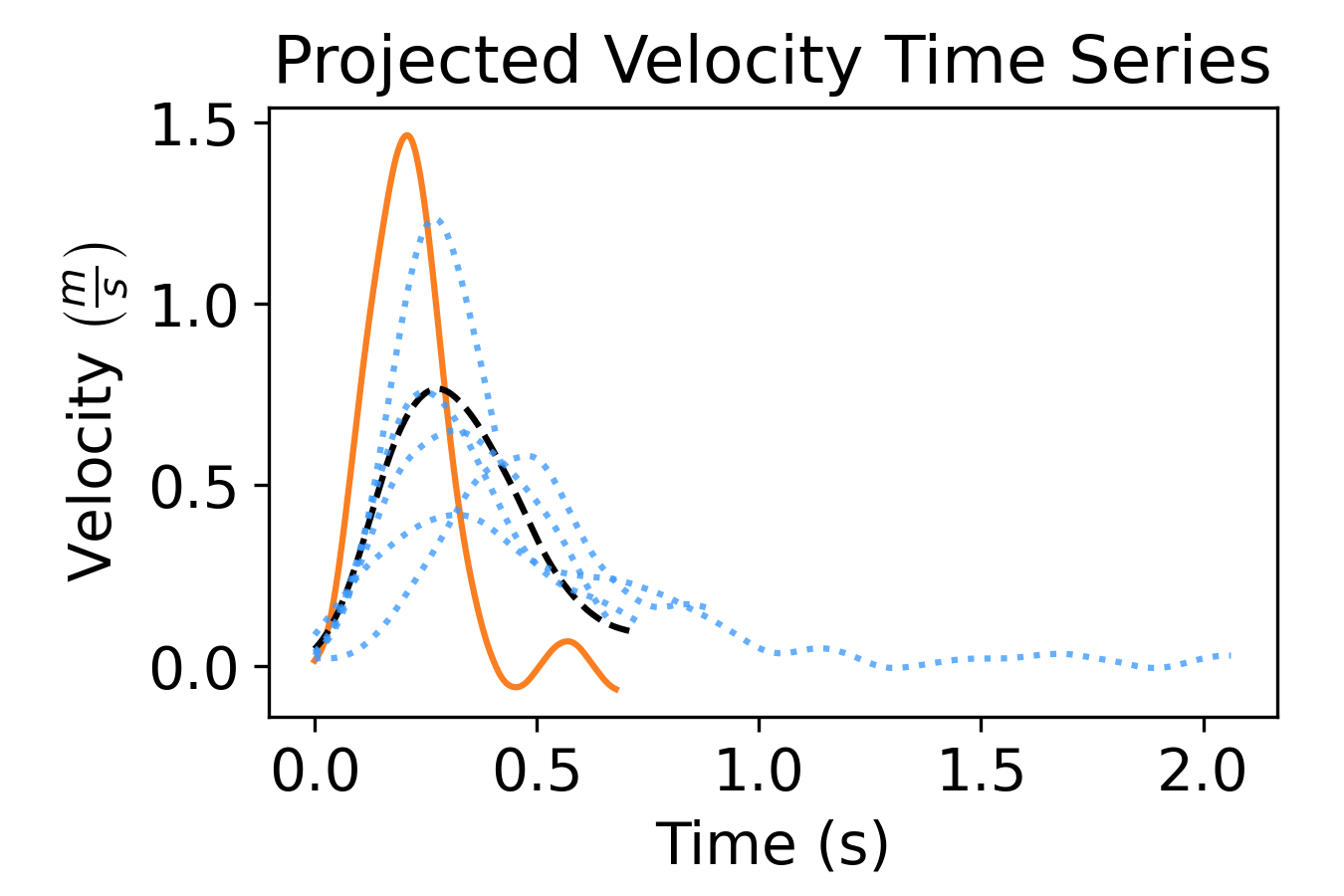}}
	\subfloat{\includegraphics[width=0.25\linewidth, clip]{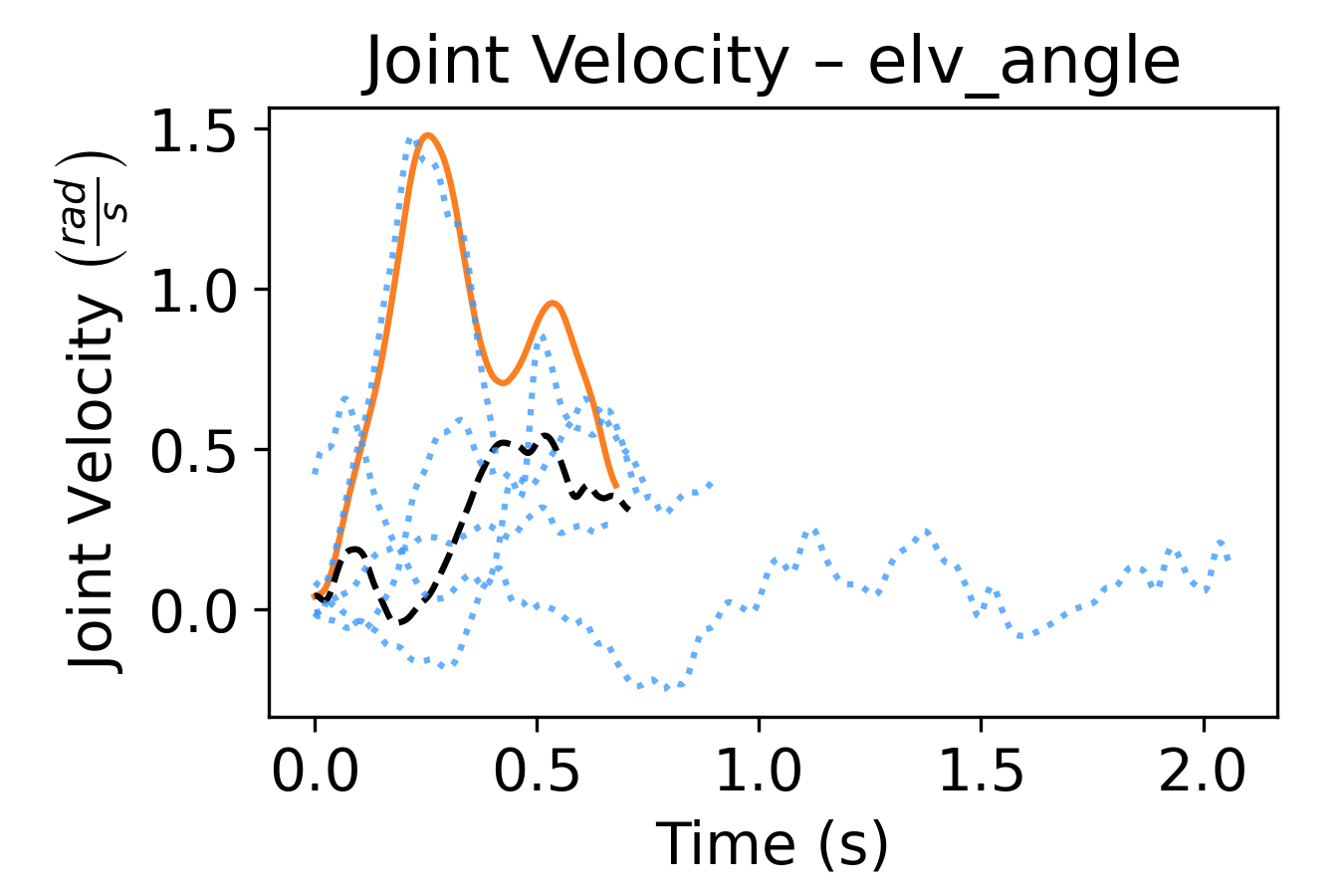}}
	\subfloat{\includegraphics[width=0.25\linewidth, clip]{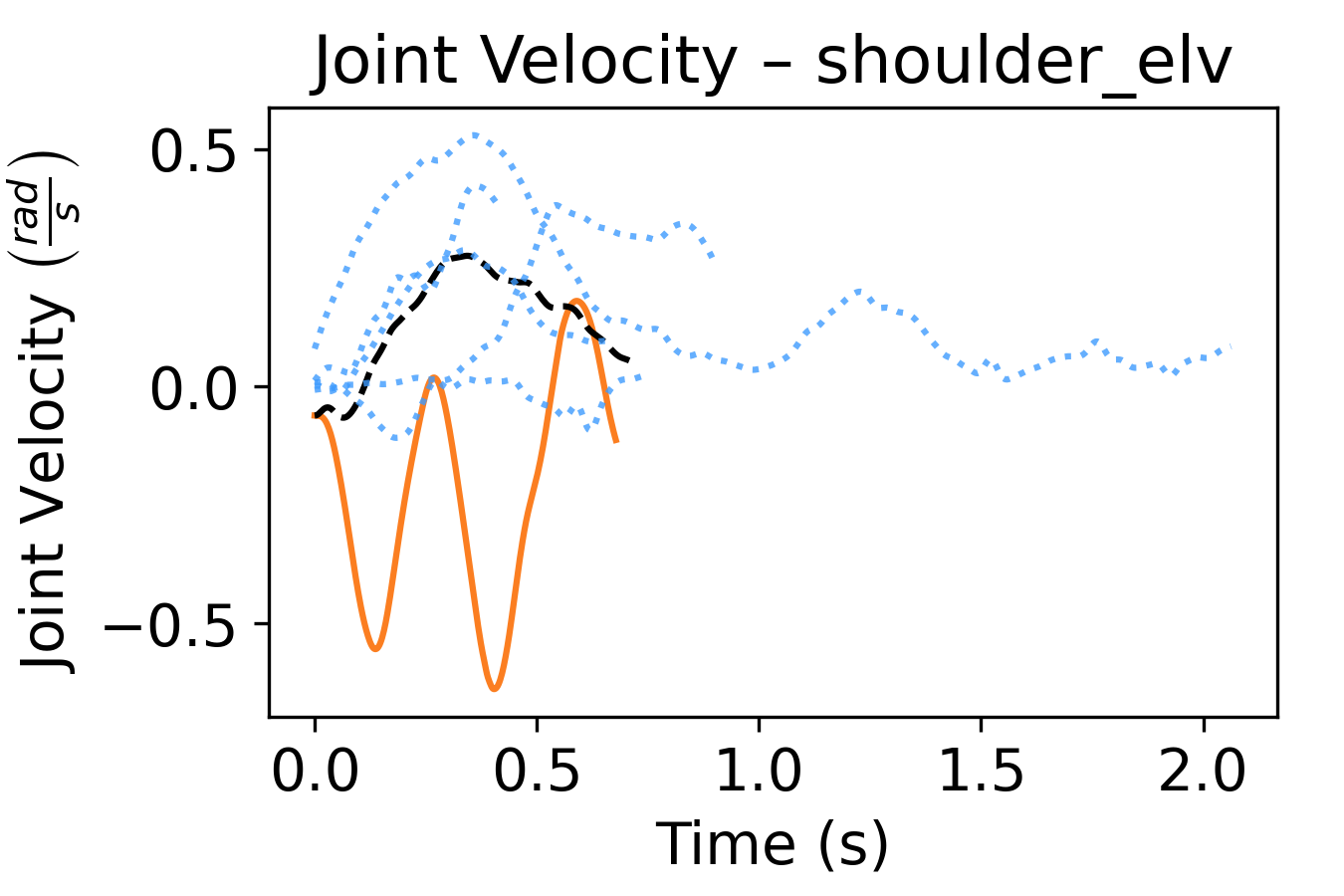}}
	\subfloat{\includegraphics[width=0.25\linewidth, clip]{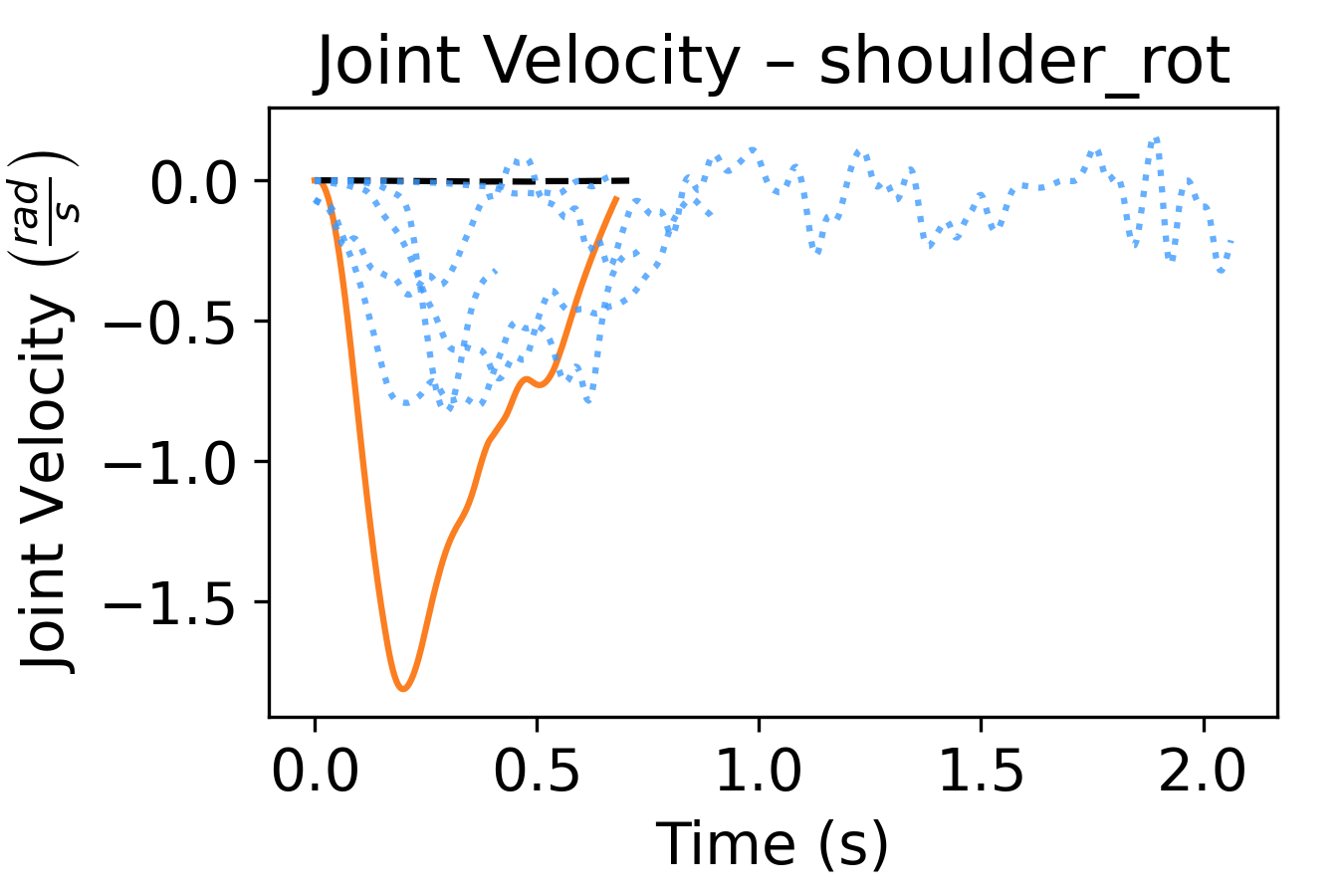}}\\
	
	\subfloat{\includegraphics[width=0.25\linewidth, clip]{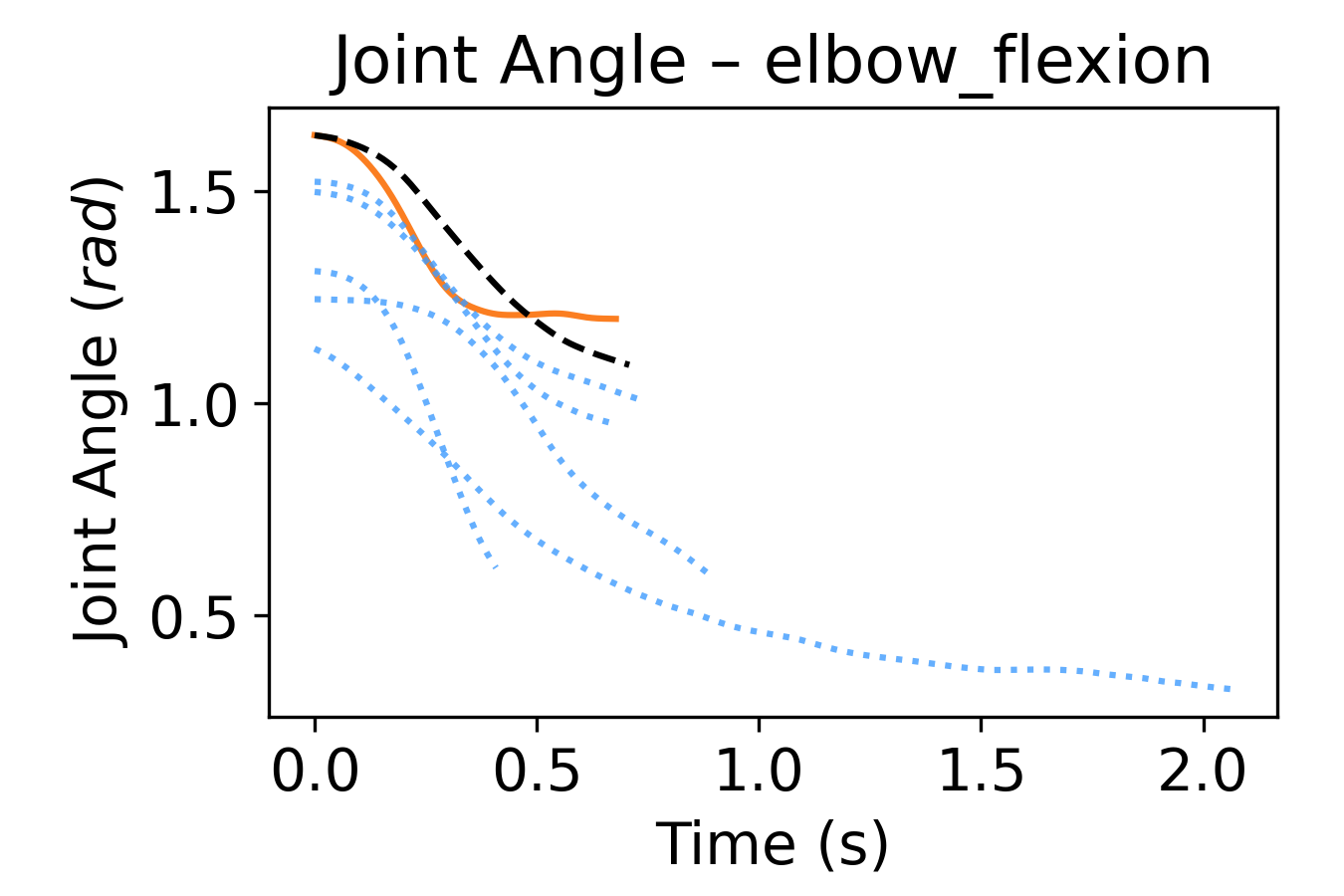}}
	\subfloat{\includegraphics[width=0.25\linewidth, clip]{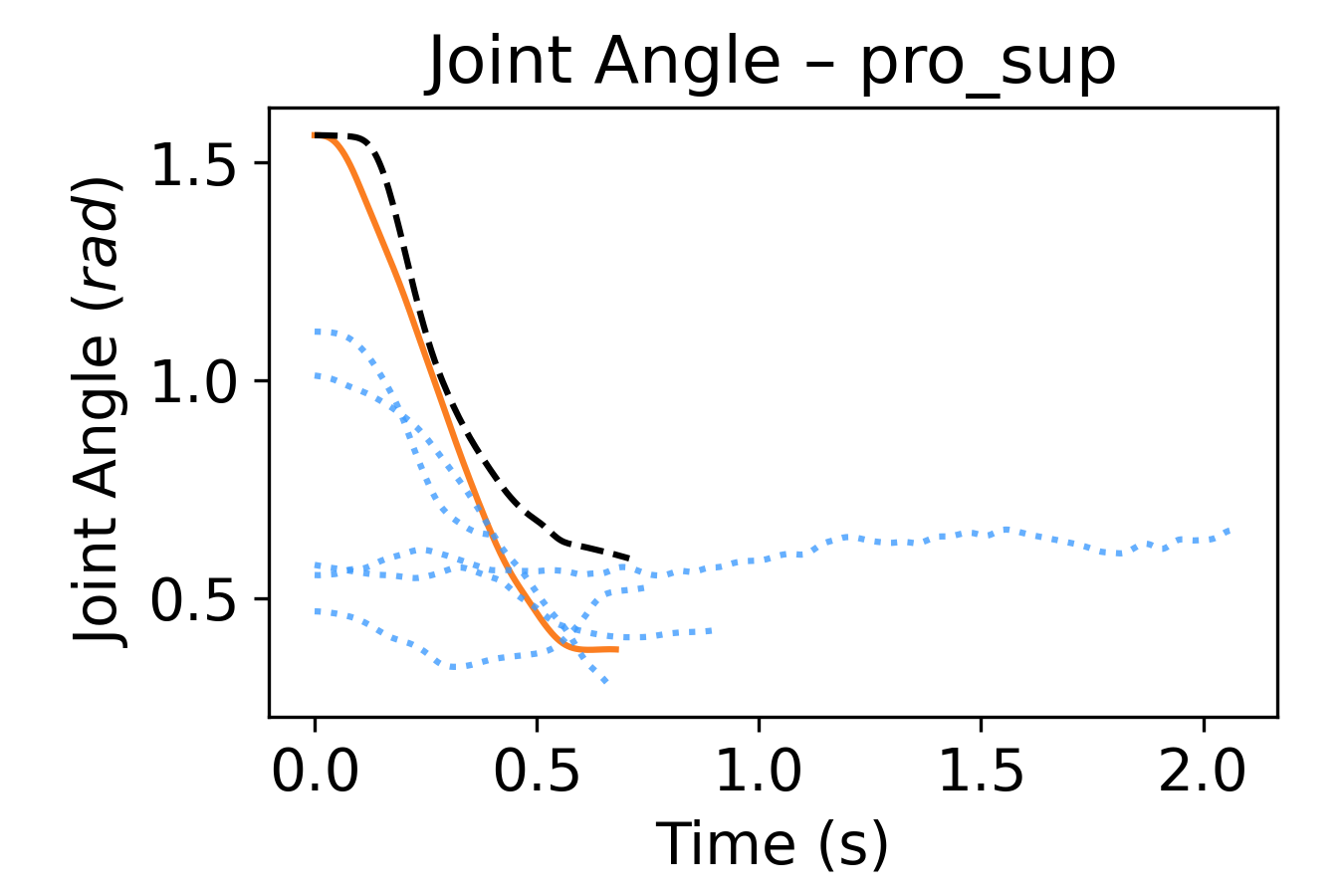}}
	\subfloat{\includegraphics[width=0.25\linewidth, clip]{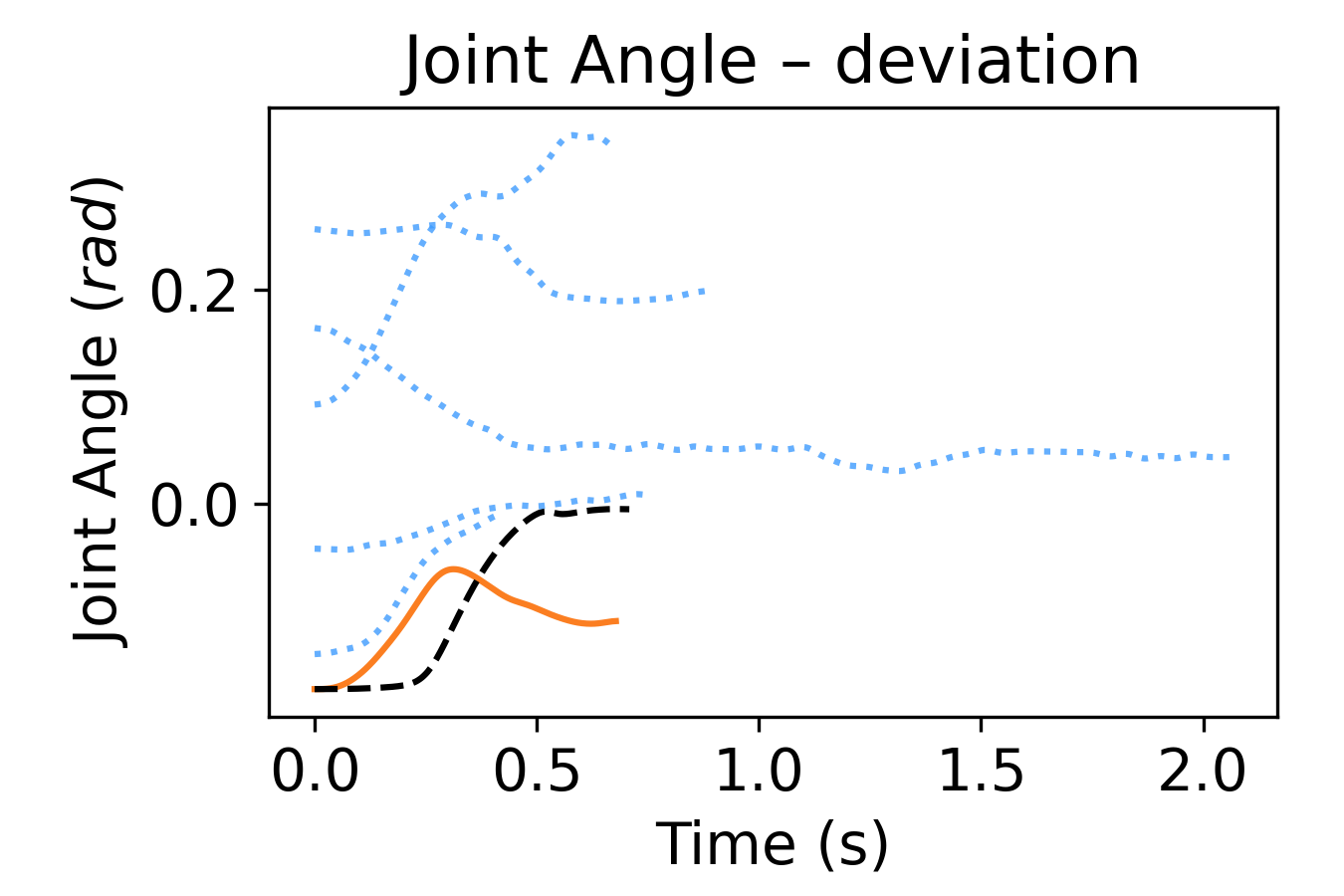}}
	\subfloat{\includegraphics[width=0.25\linewidth, clip]{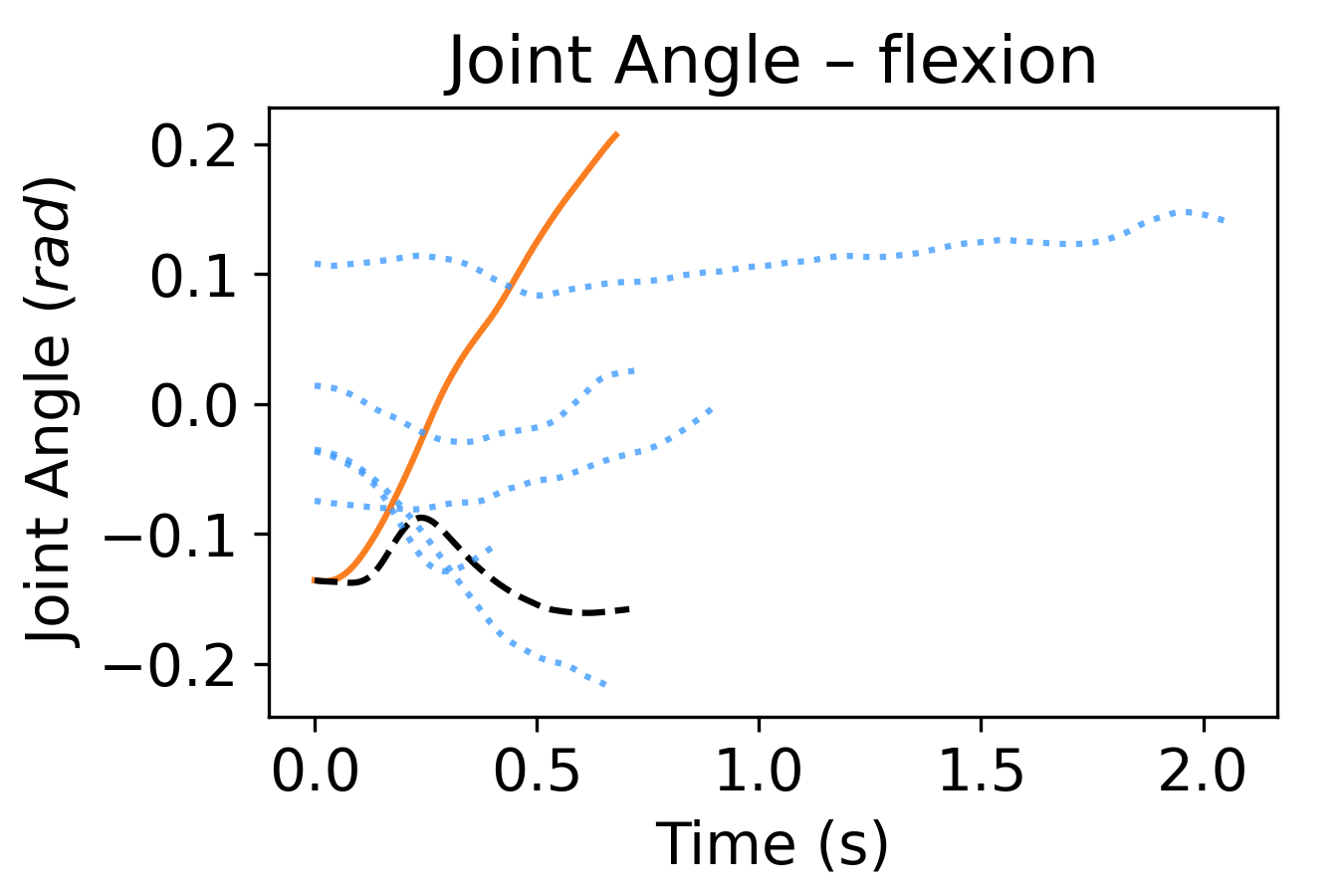}}\\
	
	\subfloat{\includegraphics[width=0.25\linewidth, clip]{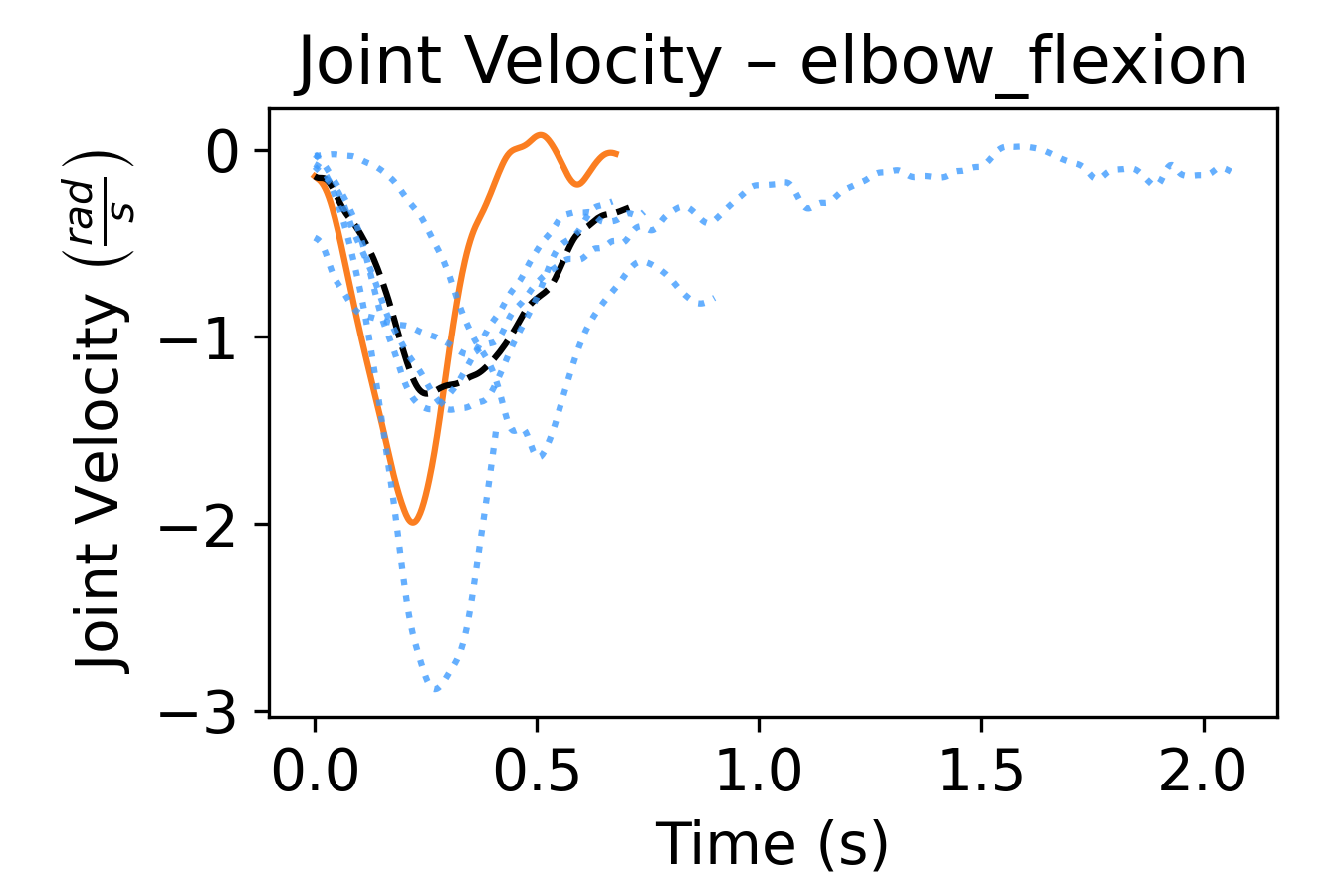}}
	\subfloat{\includegraphics[width=0.25\linewidth, clip]{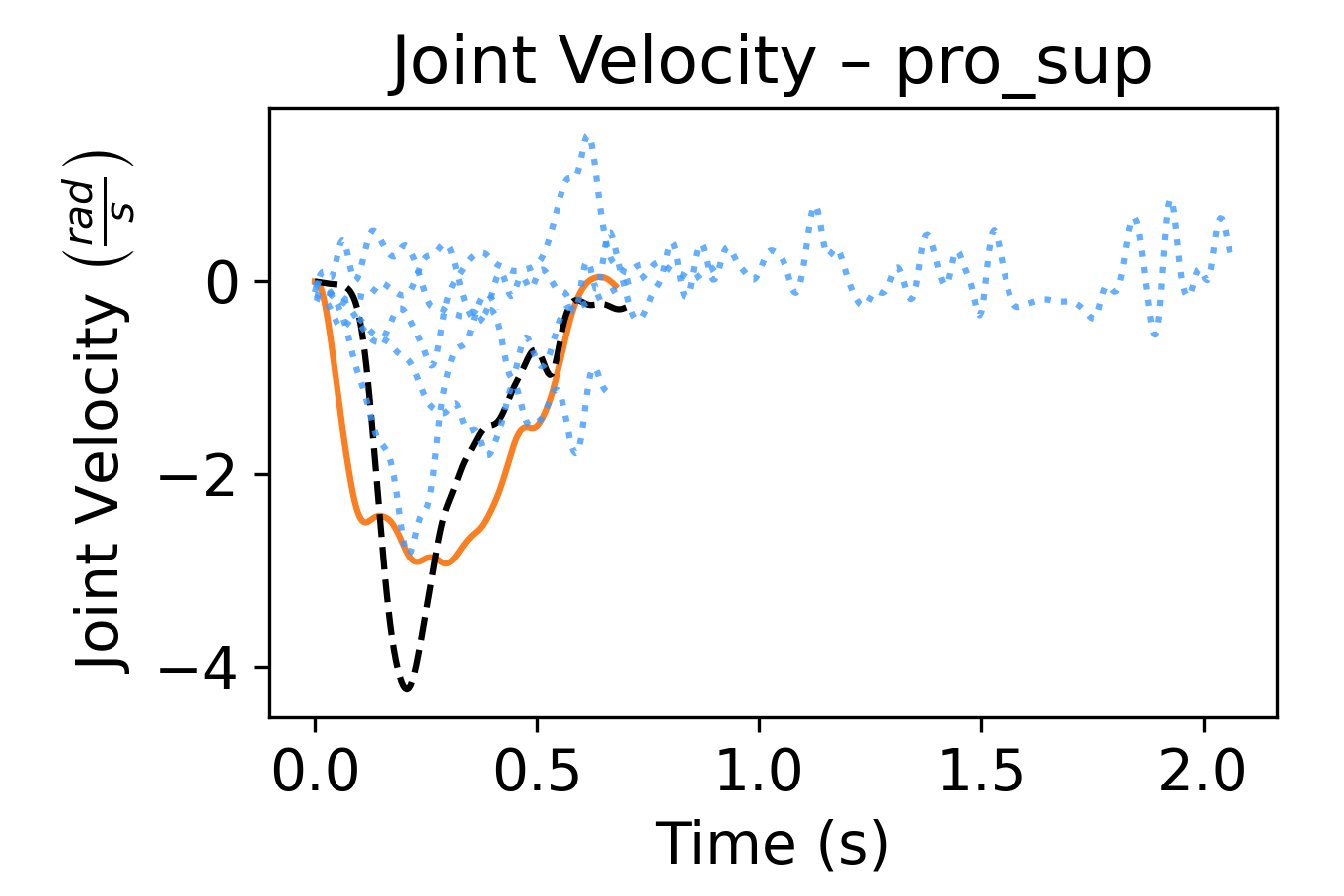}}
	\subfloat{\includegraphics[width=0.25\linewidth, clip]{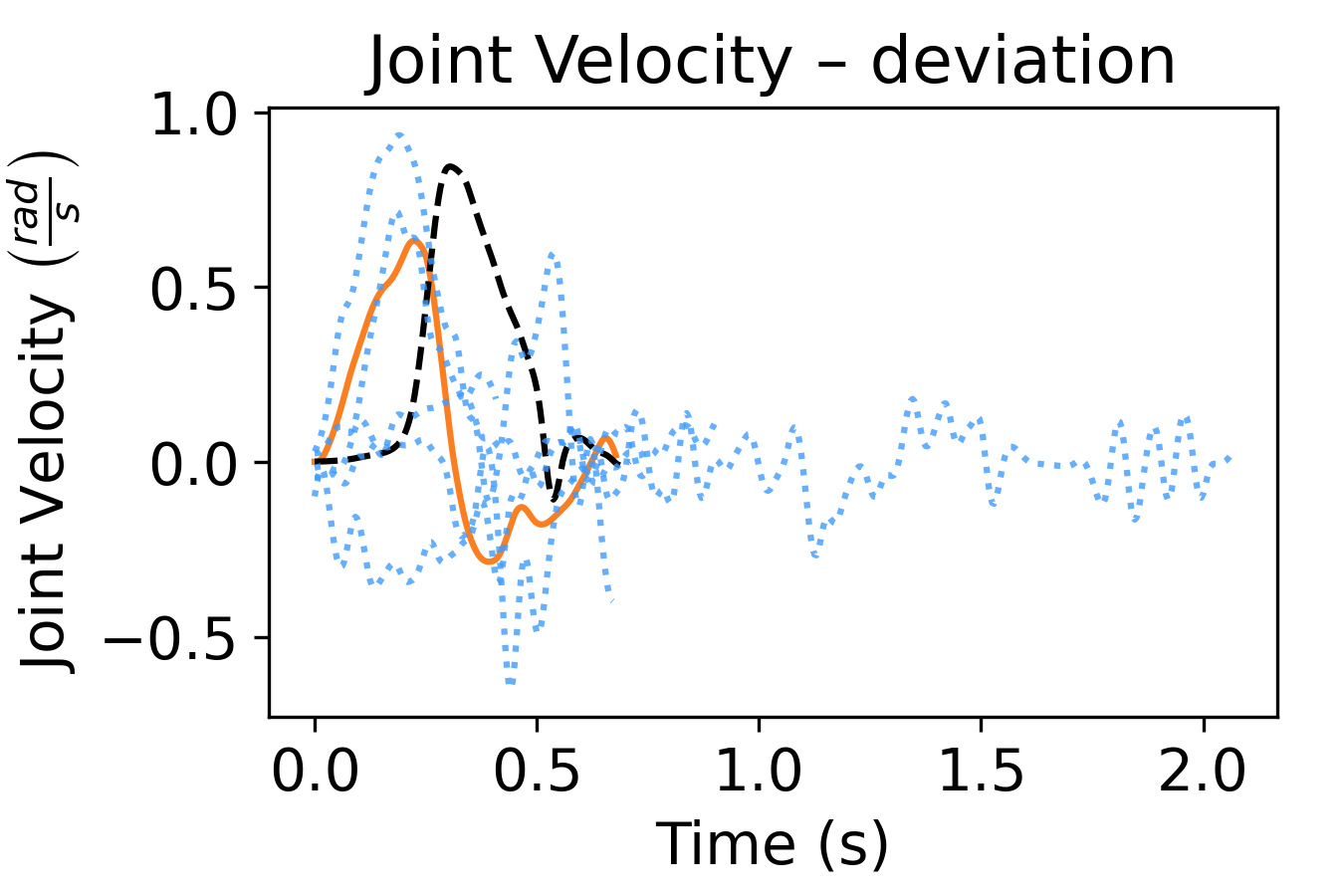}}
	\subfloat{\includegraphics[width=0.25\linewidth, clip]{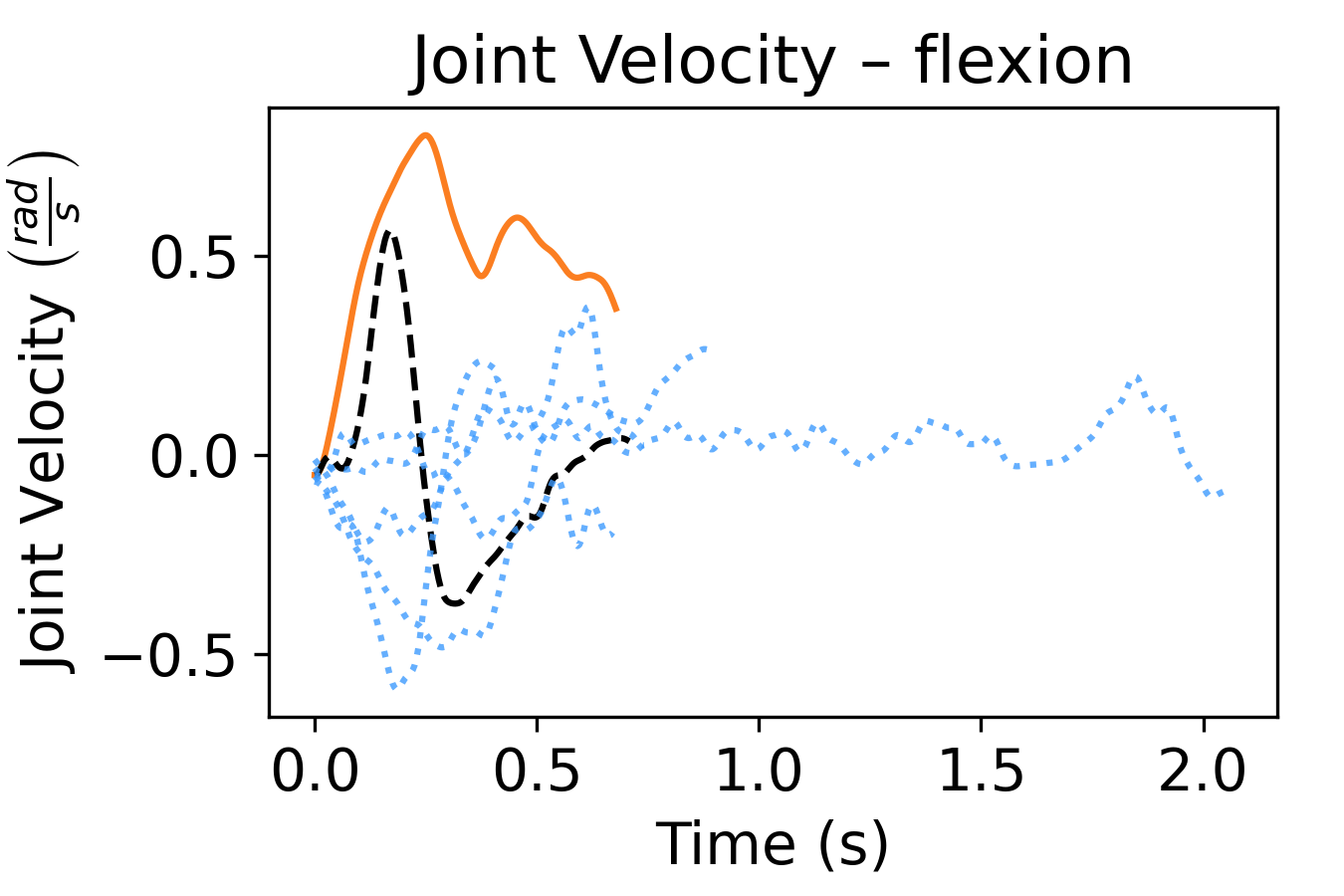}}
	
	\caption{Given an interaction technique (here: \textbf{Virtual Cursor Ergonomic}) and a movement direction (here: movements from target 8 to target 9), the characteristic cursor and joint trajectories of an individual user (here: U3, black dashed lines; trajectories of the remaining users are shown as blue dotted lines for comparison) can be predicted by our simulation (orange solid lines).}
	\label{fig:CursorErgonomic_qual}
\end{figure}
\newpage

\begin{figure}[h!]
	\centering
	\subfloat{\includegraphics[width=0.25\linewidth, clip]{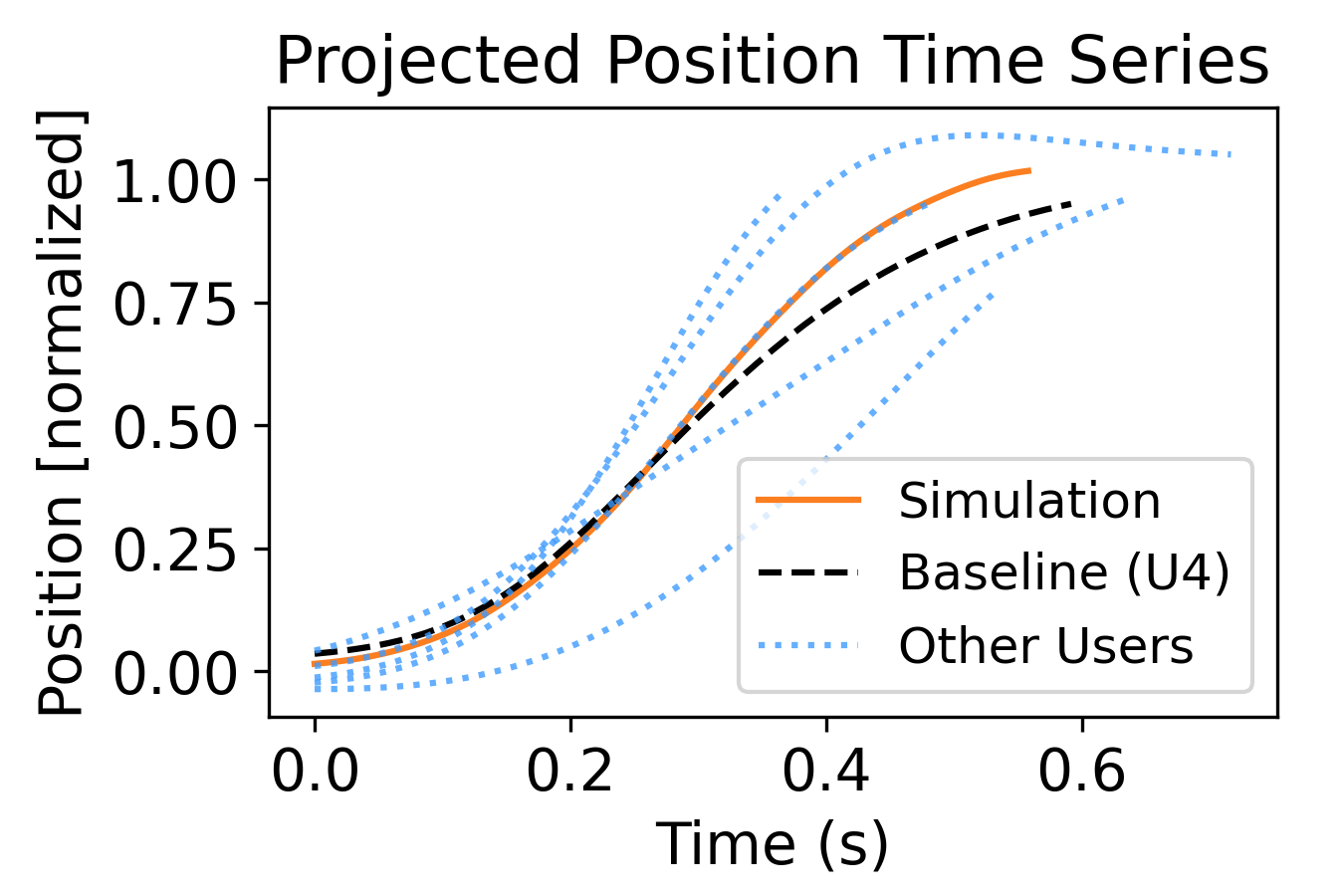}}
	\subfloat{\includegraphics[width=0.25\linewidth, clip]{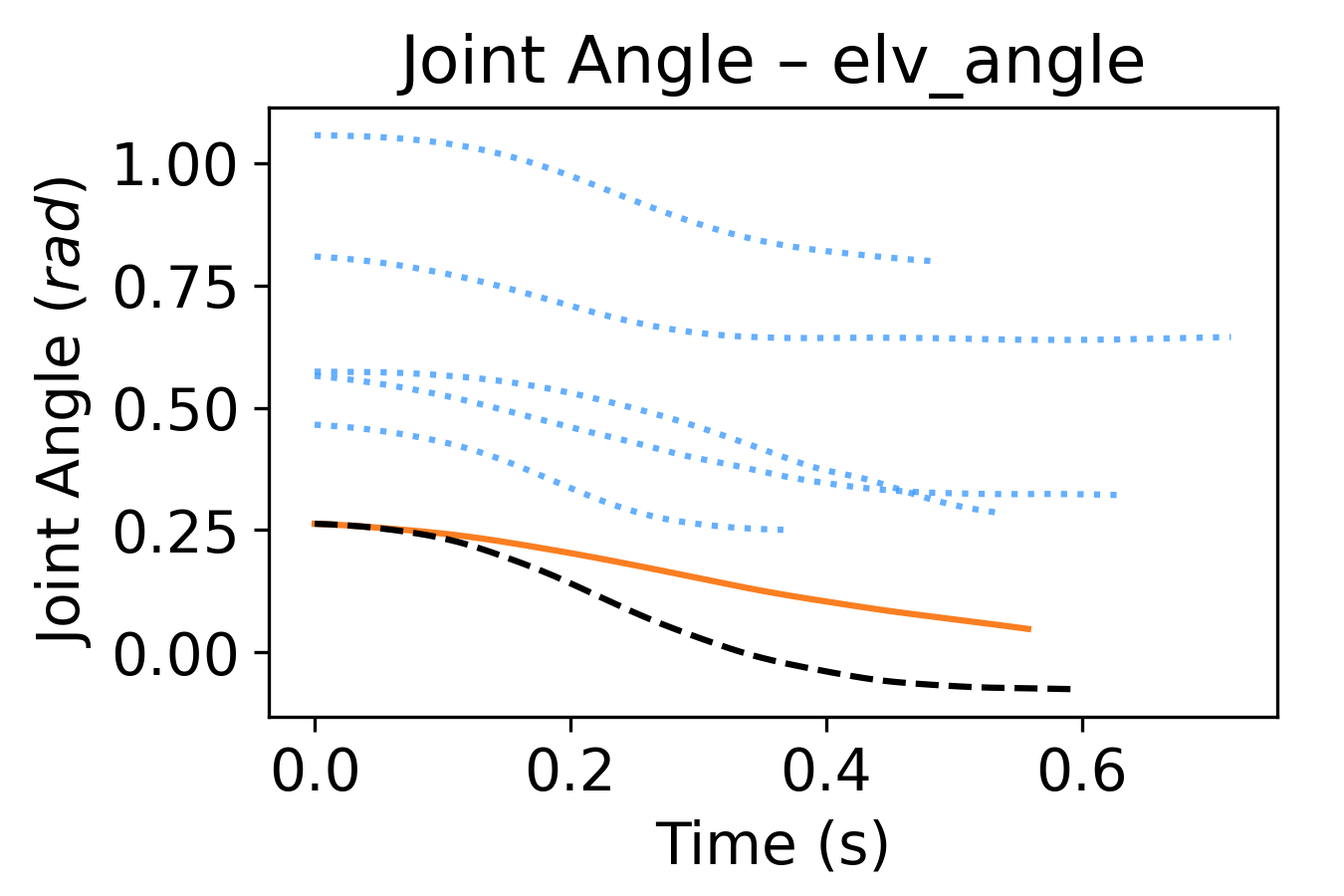}}
	\subfloat{\includegraphics[width=0.25\linewidth, clip]{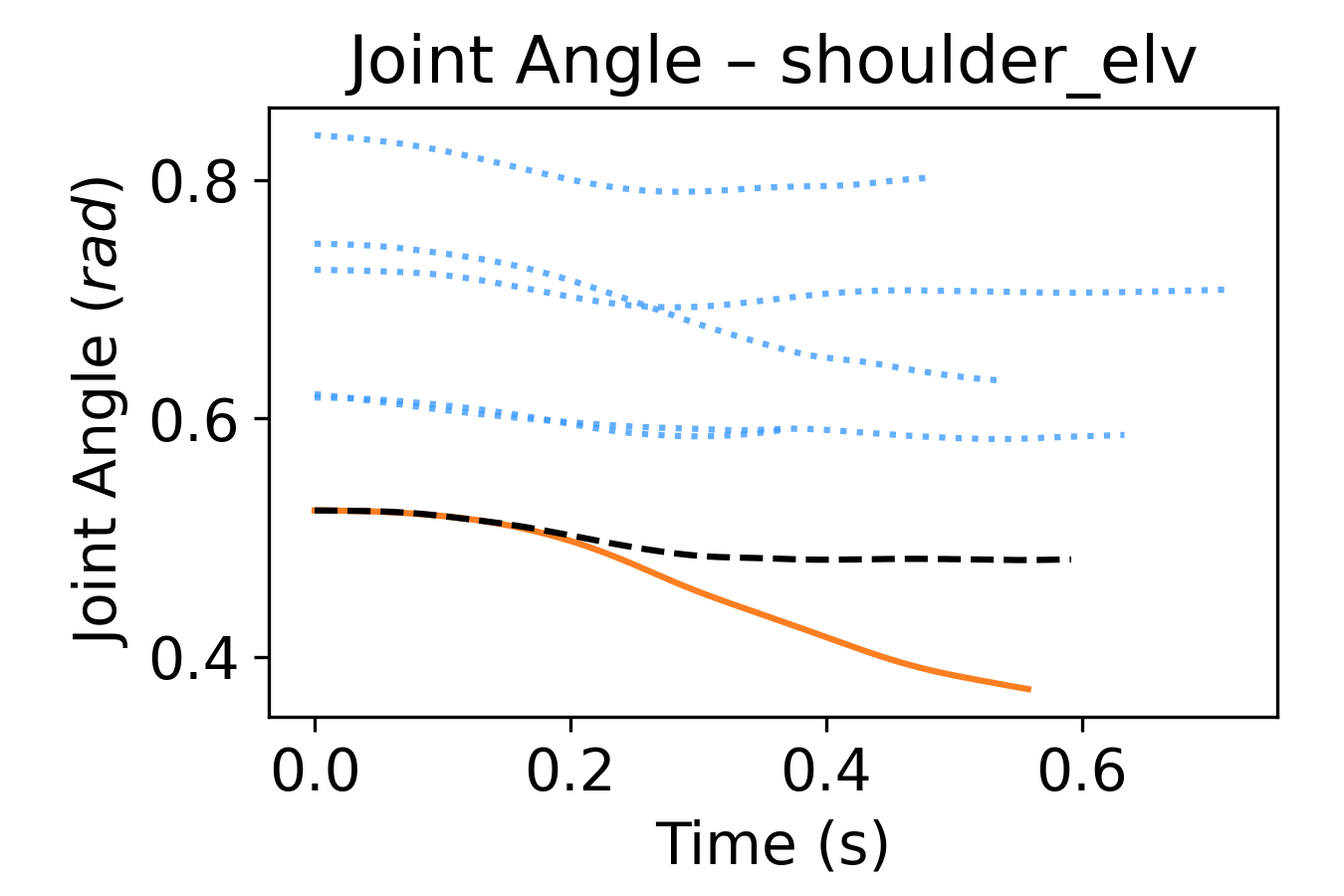}}
	\subfloat{\includegraphics[width=0.25\linewidth, clip]{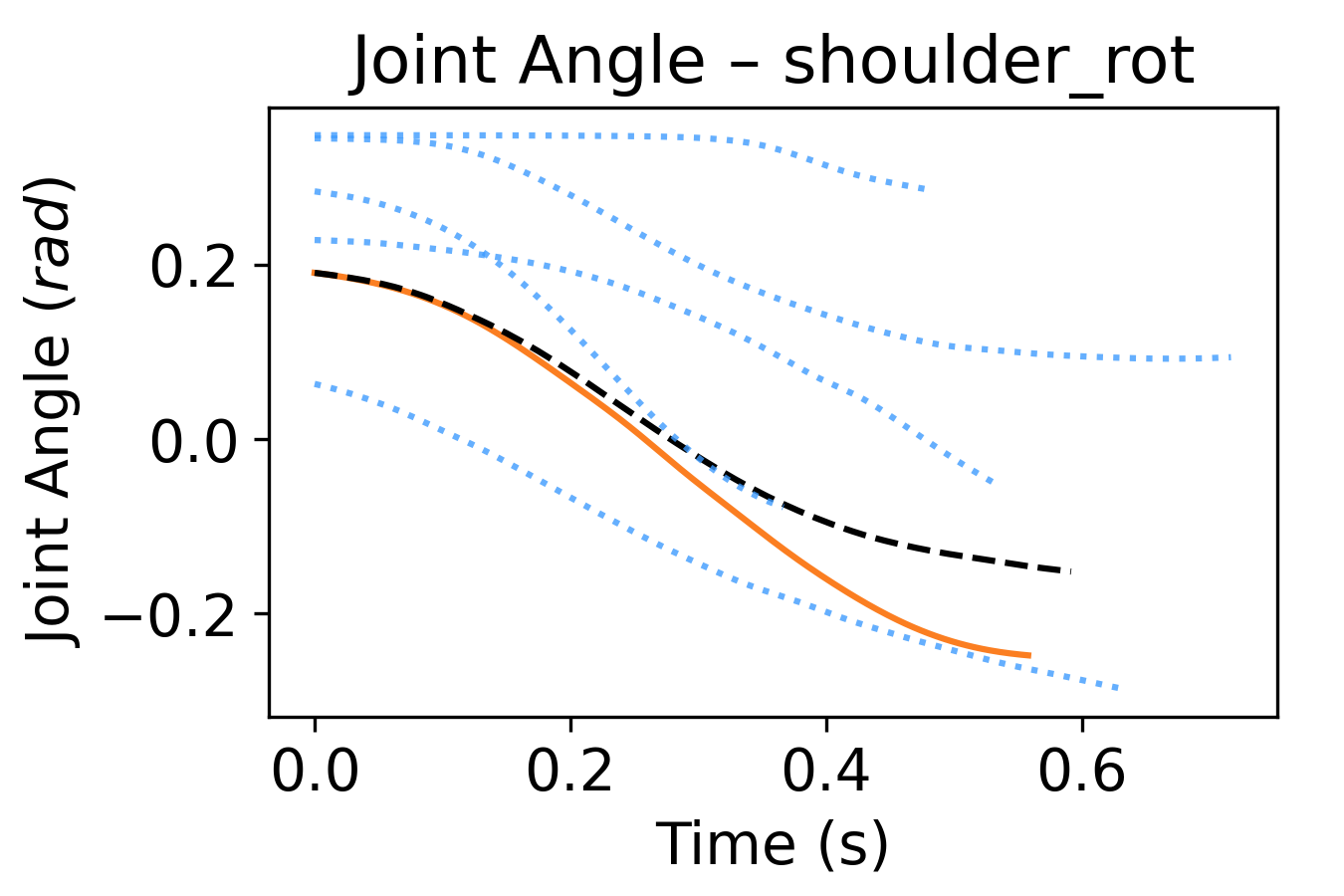}}\\
	
	\subfloat{\includegraphics[width=0.25\linewidth, clip]{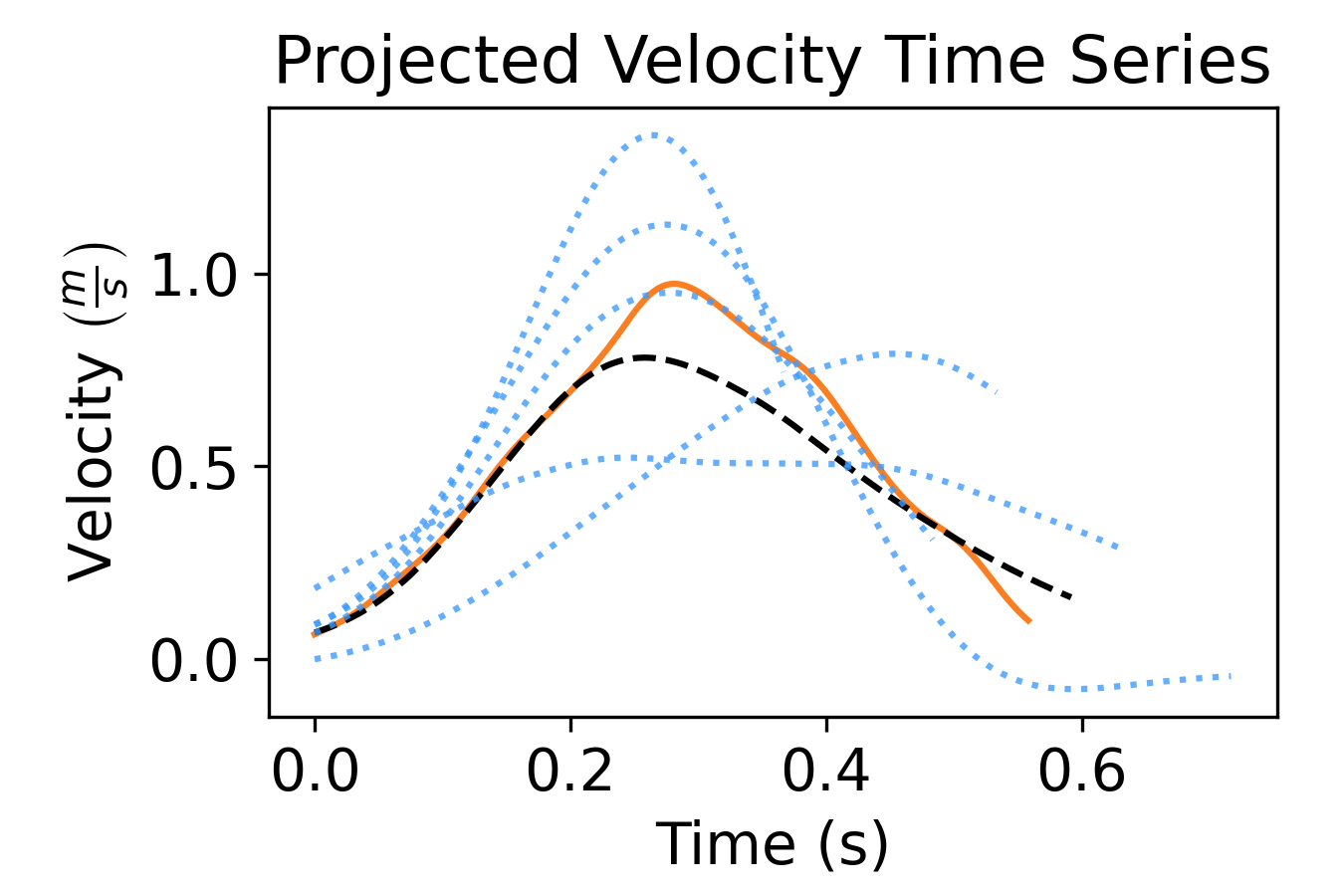}}
	\subfloat{\includegraphics[width=0.25\linewidth, clip]{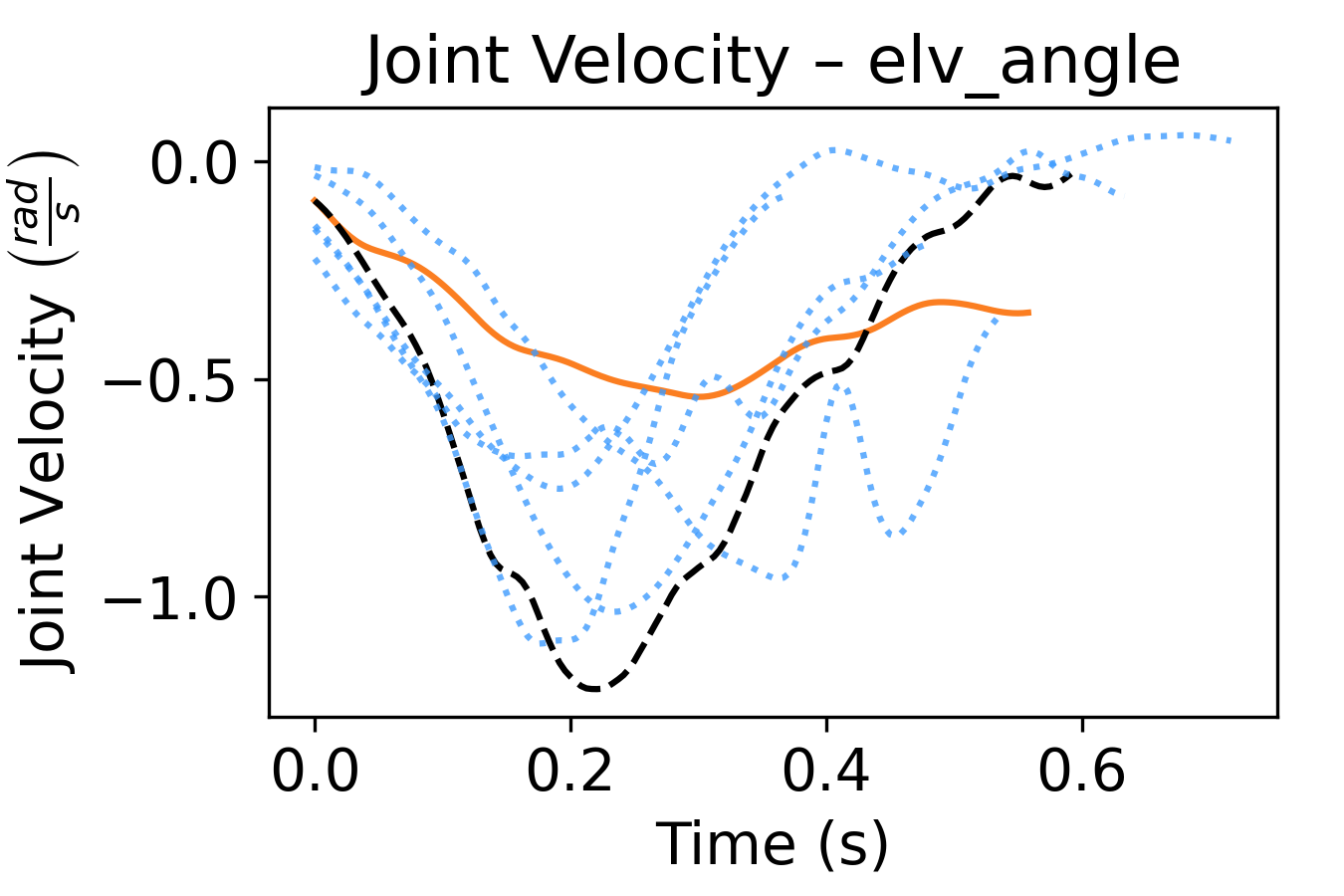}}
	\subfloat{\includegraphics[width=0.25\linewidth, clip]{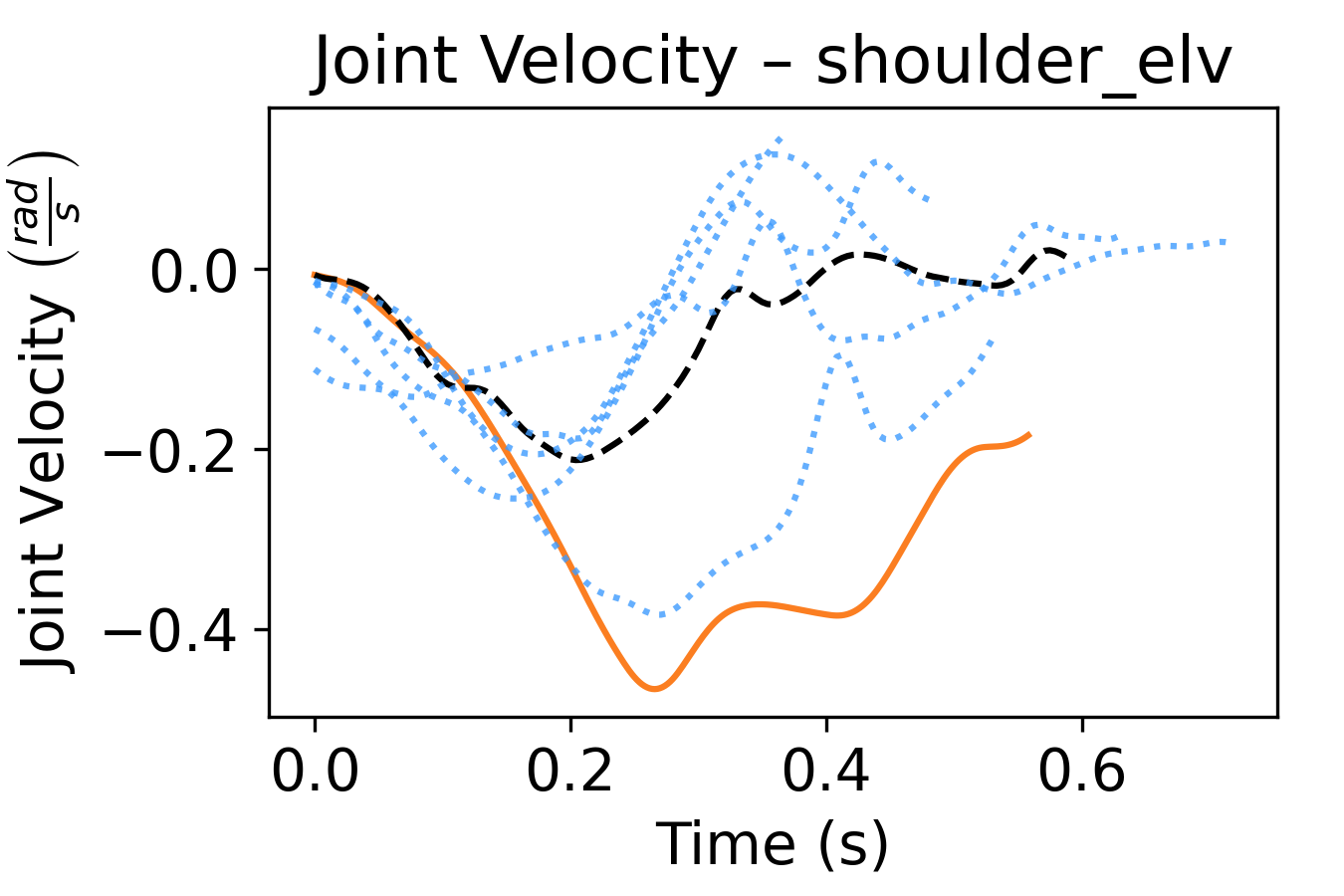}}
	\subfloat{\includegraphics[width=0.25\linewidth, clip]{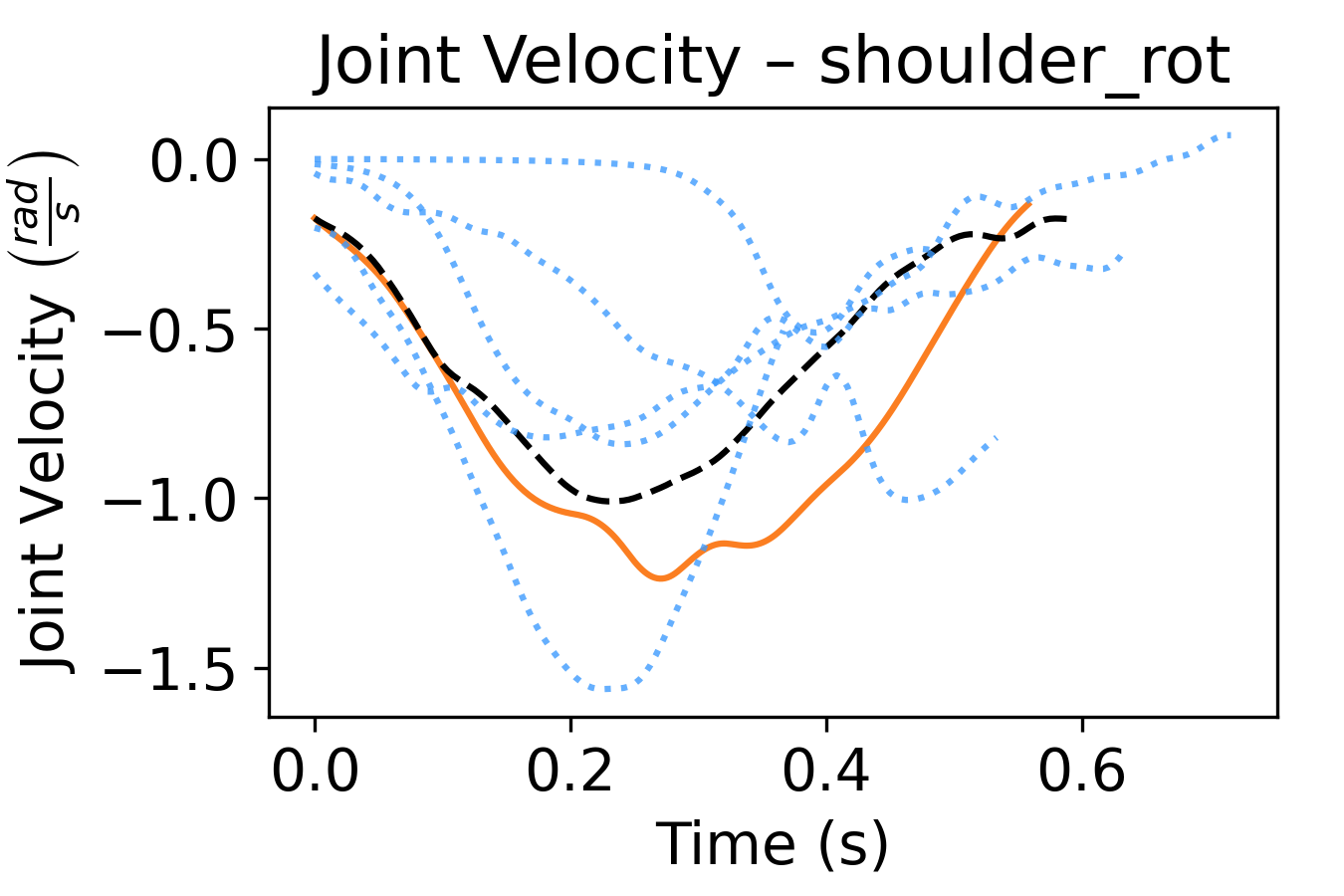}}\\

	\subfloat{\includegraphics[width=0.25\linewidth, clip]{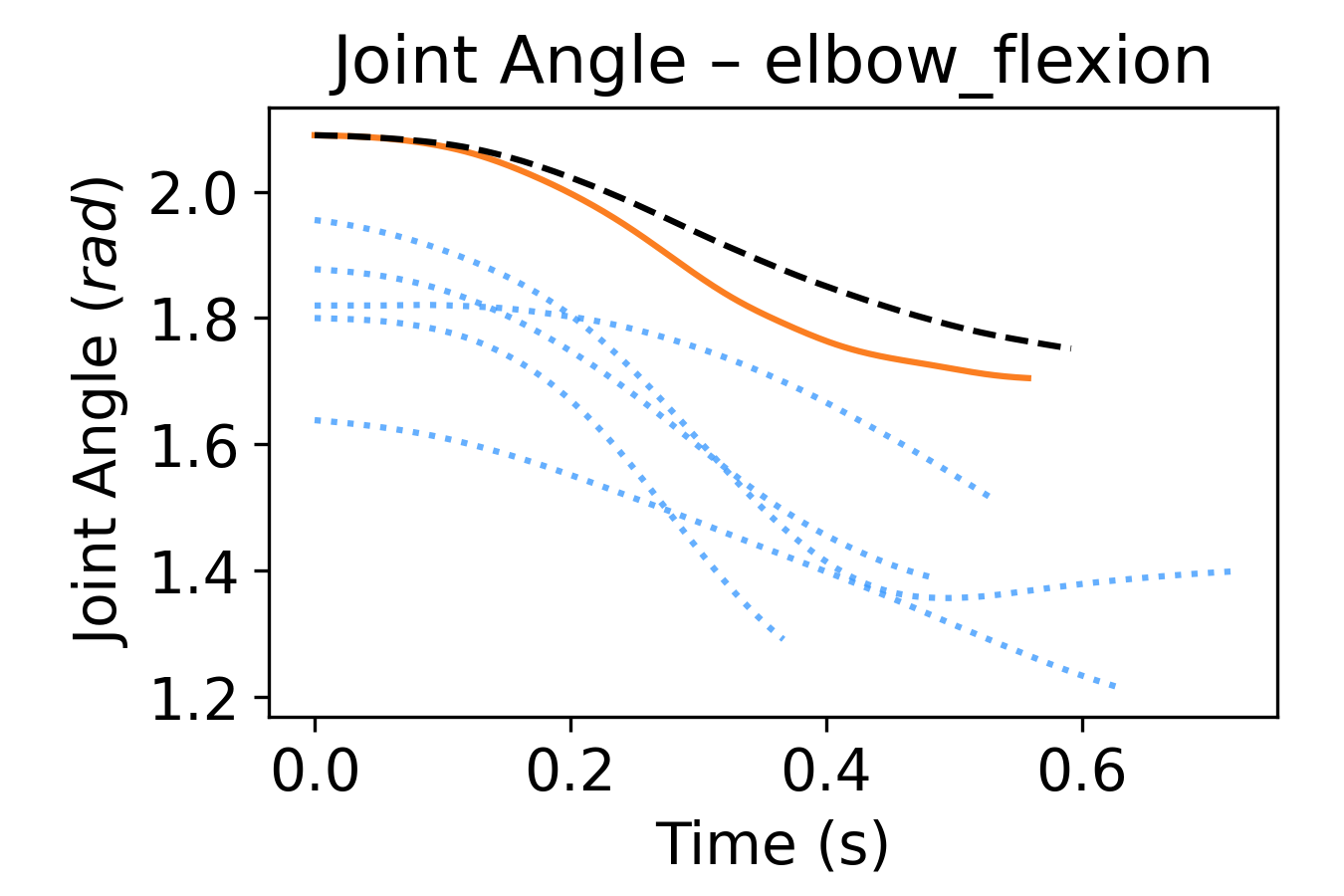}}
	\subfloat{\includegraphics[width=0.25\linewidth, clip]{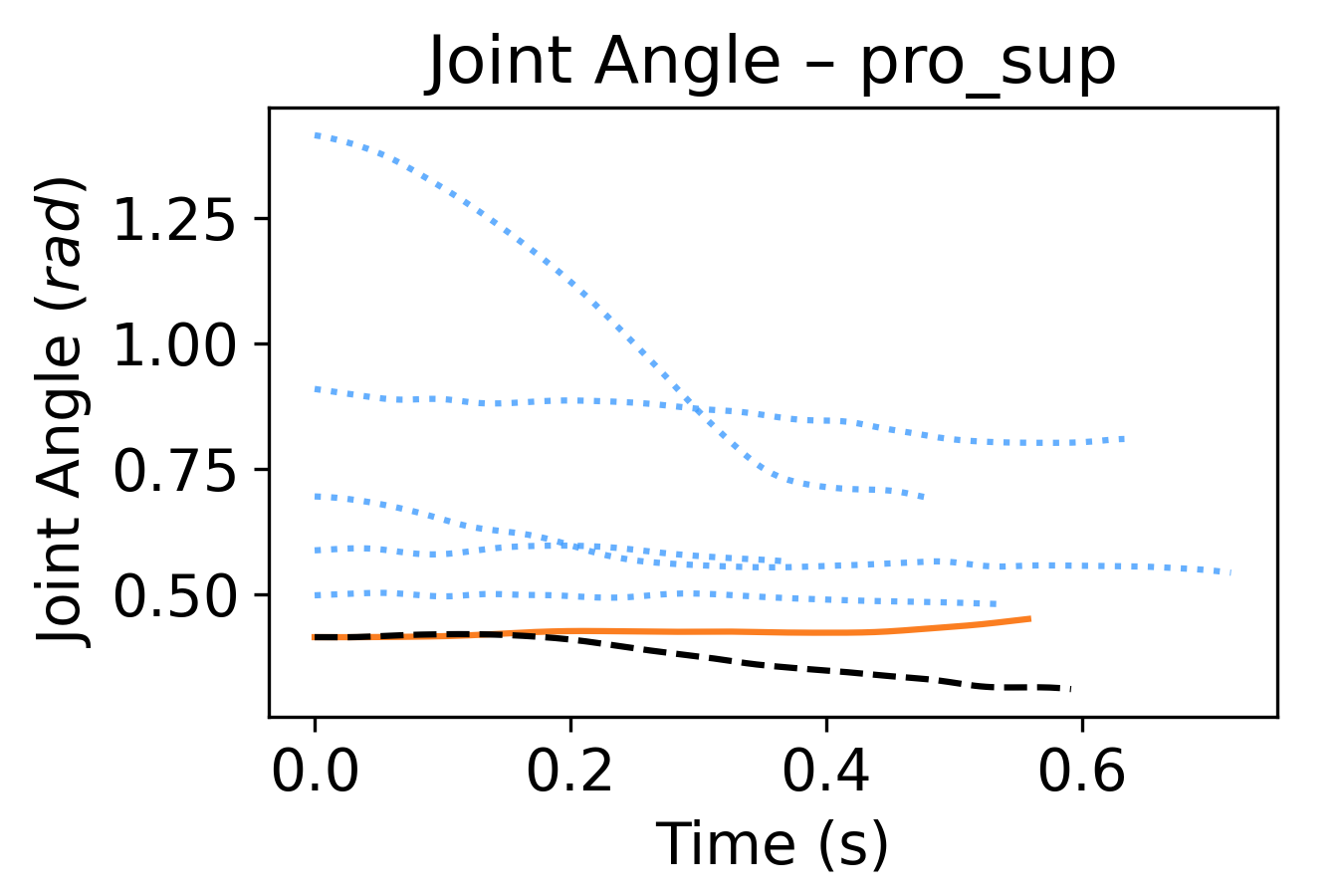}}	
	\subfloat{\includegraphics[width=0.25\linewidth, clip]{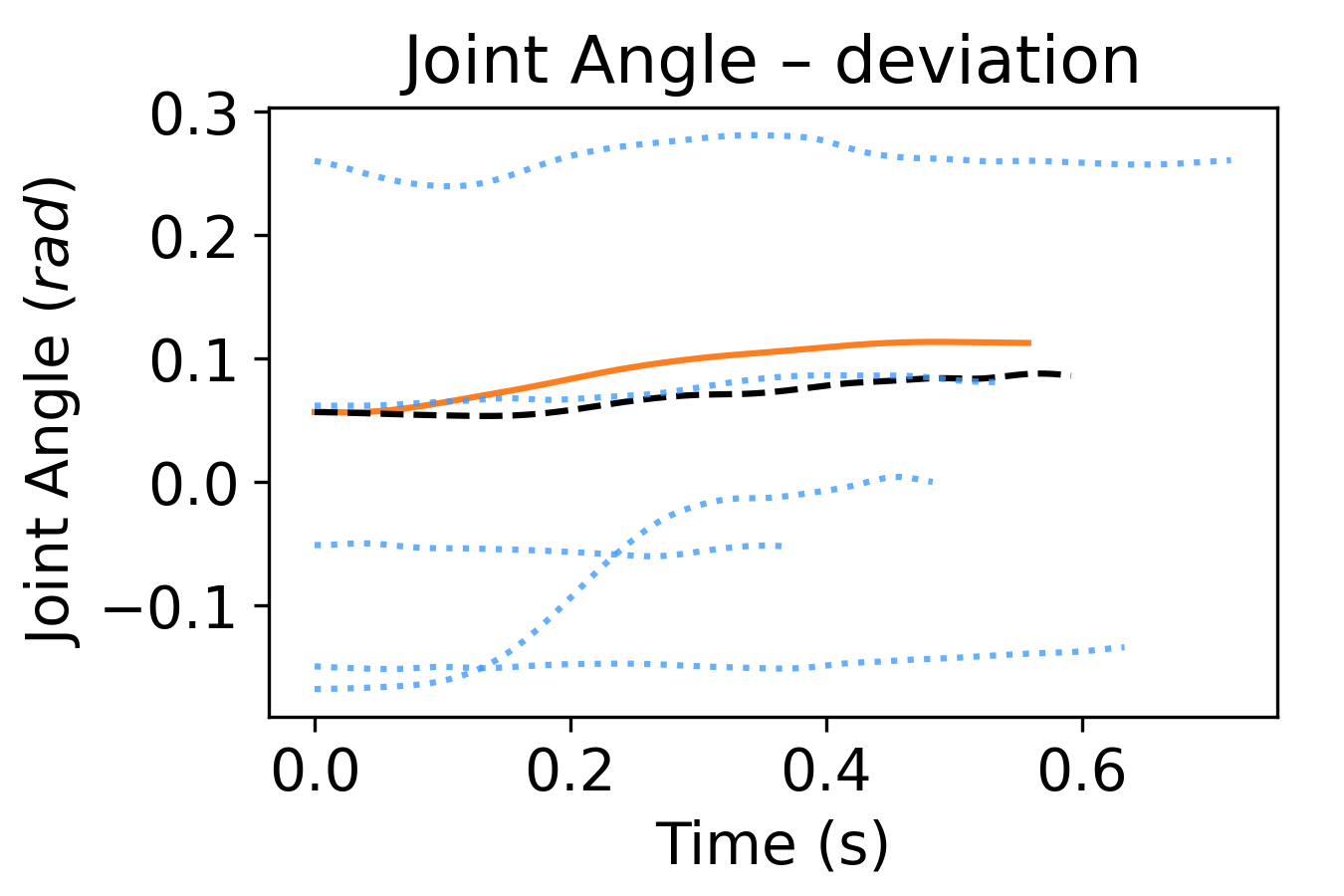}}
	\subfloat{\includegraphics[width=0.25\linewidth, clip]{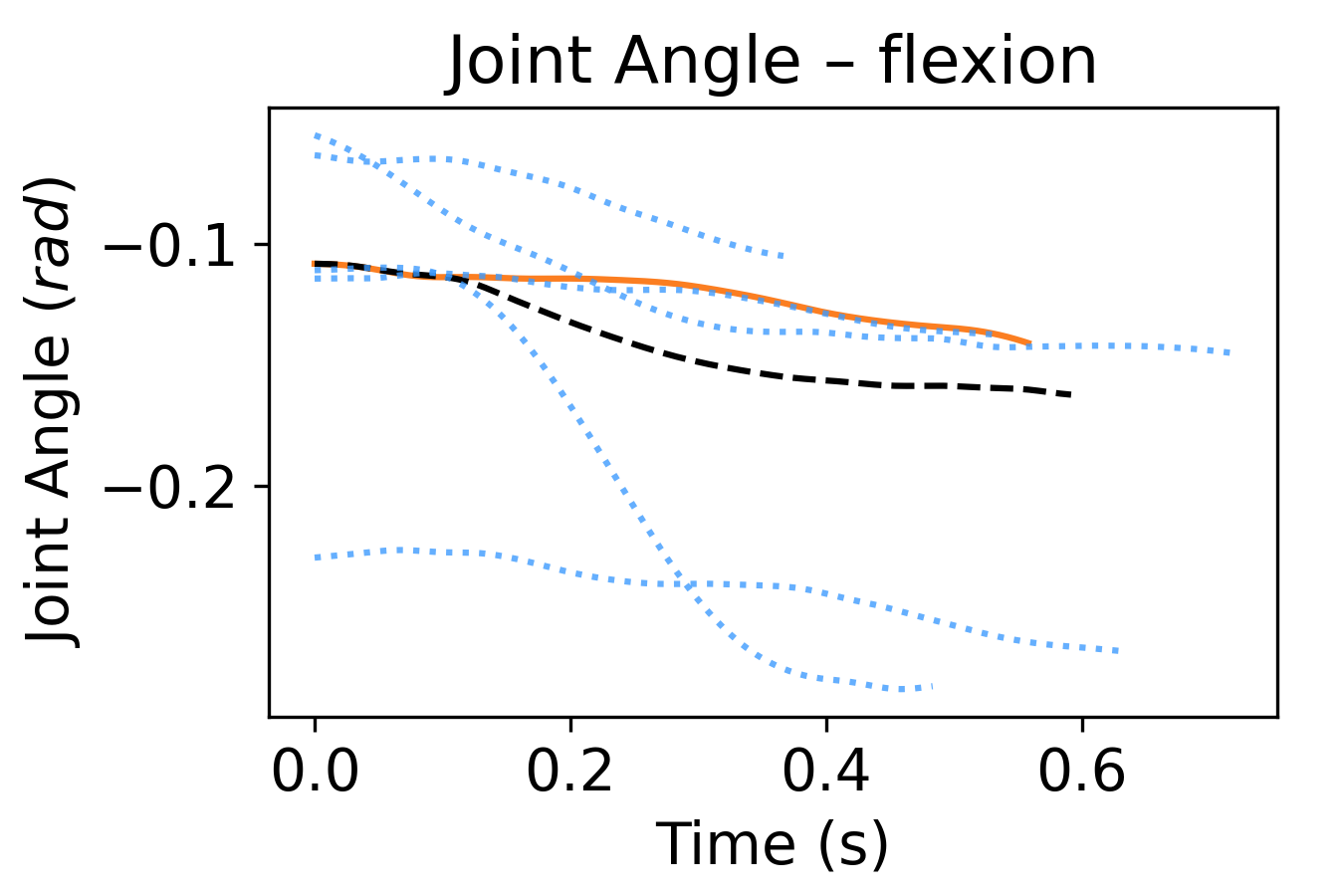}}\\
	
	\subfloat{\includegraphics[width=0.25\linewidth, clip]{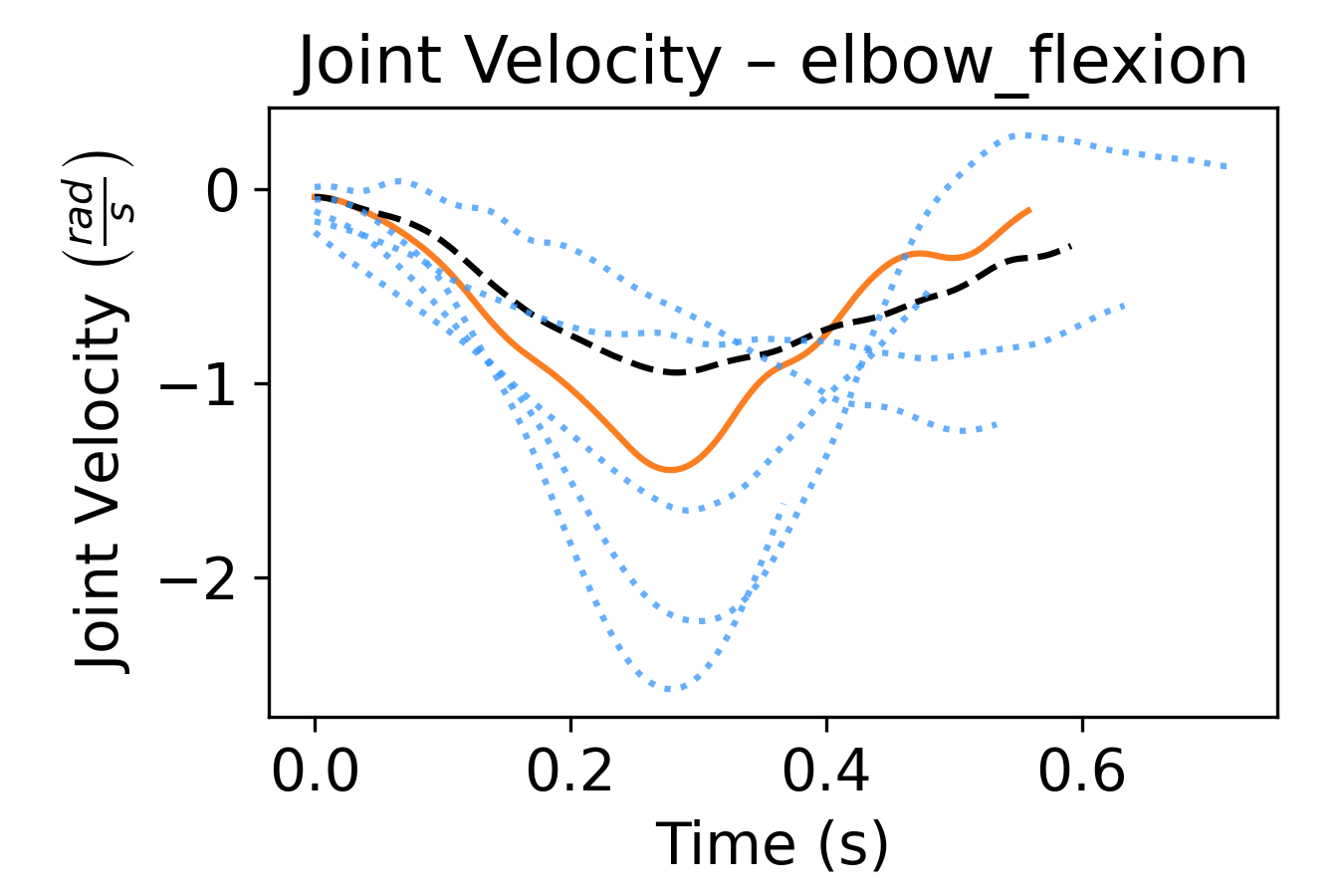}}
	\subfloat{\includegraphics[width=0.25\linewidth, clip]{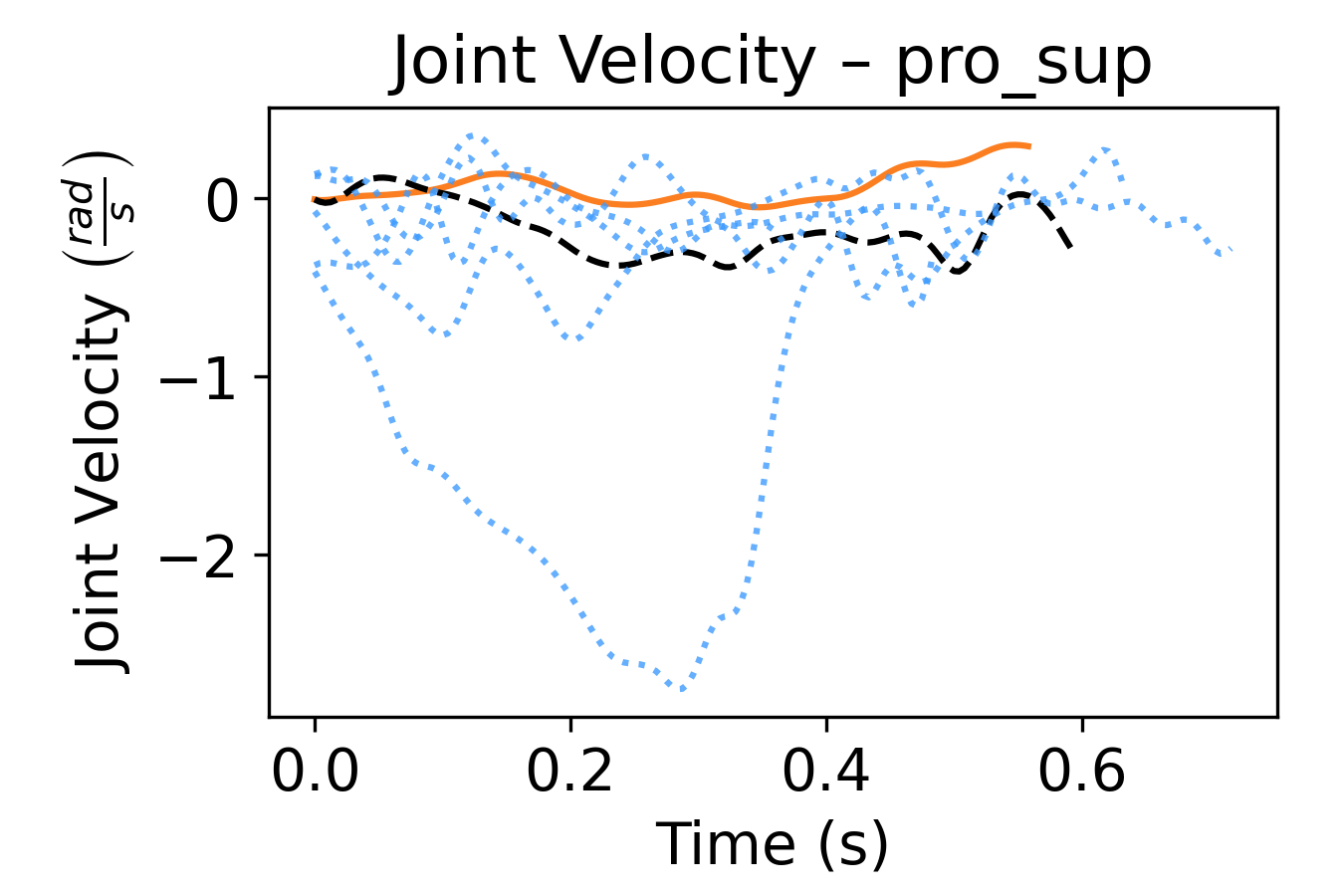}}
	\subfloat{\includegraphics[width=0.25\linewidth, clip]{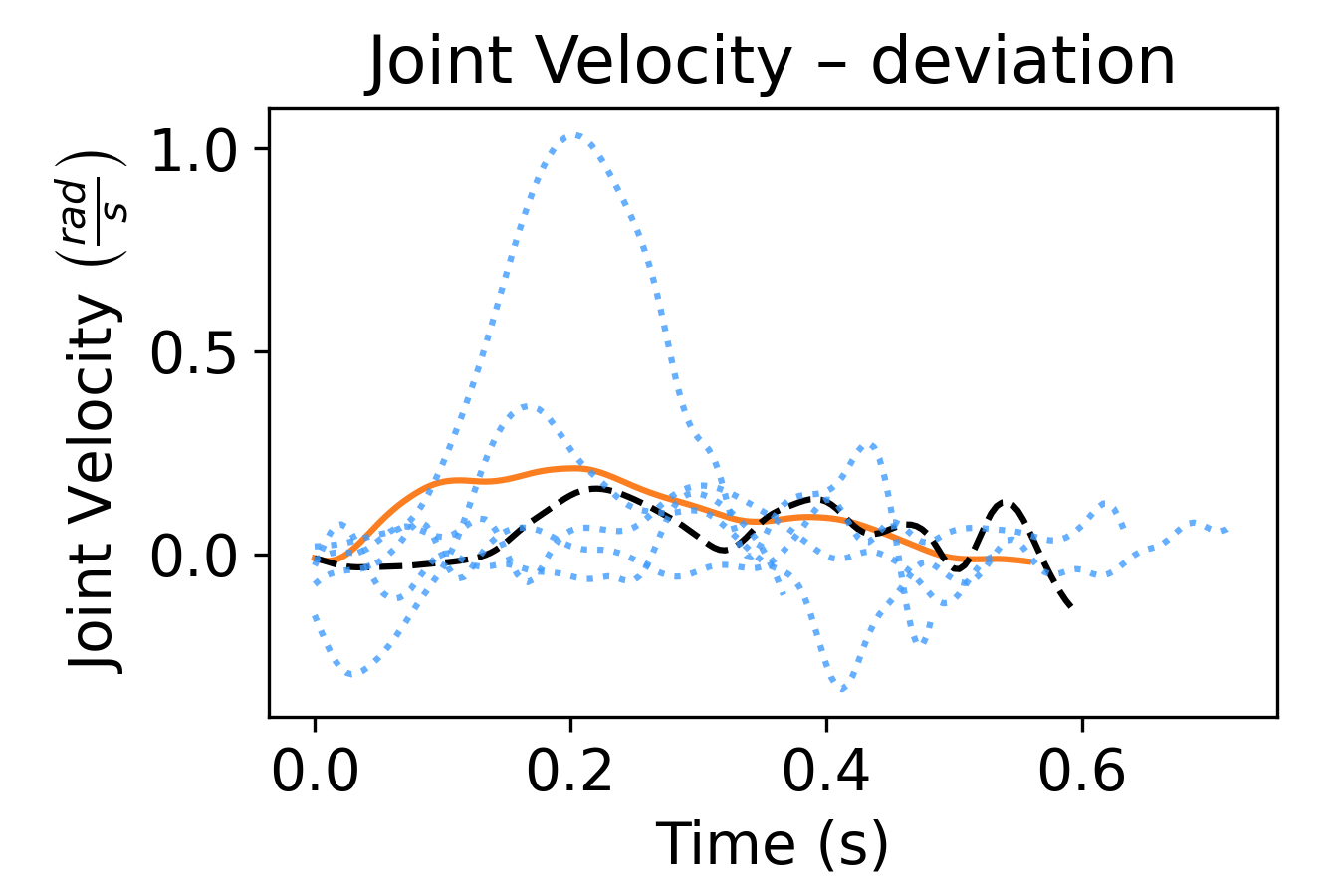}}
	\subfloat{\includegraphics[width=0.25\linewidth, clip]{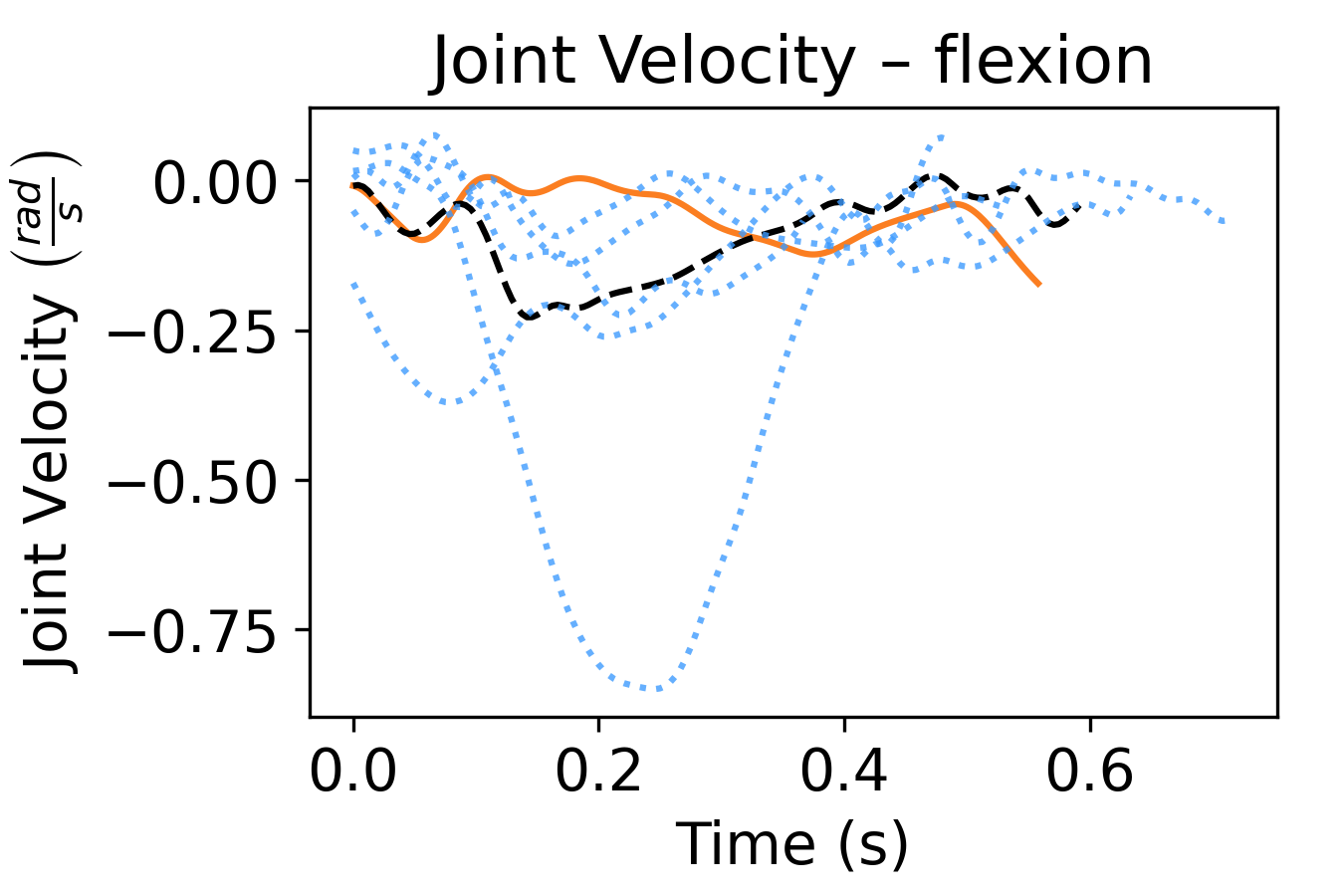}}
	
	\caption{Given an interaction technique (here: \textbf{Virtual Pad ID}) and a movement direction (here: movements from target 8 to target 9), the characteristic cursor and joint trajectories of an individual user (here: U4, black dashed lines; trajectories of the remaining users are shown as blue dotted lines for comparison) can be predicted by our simulation (orange solid lines).}
	\label{fig:PadID_qual}
\end{figure}

\begin{figure}[h!]
	\centering

	\subfloat{\includegraphics[width=0.25\linewidth, clip]{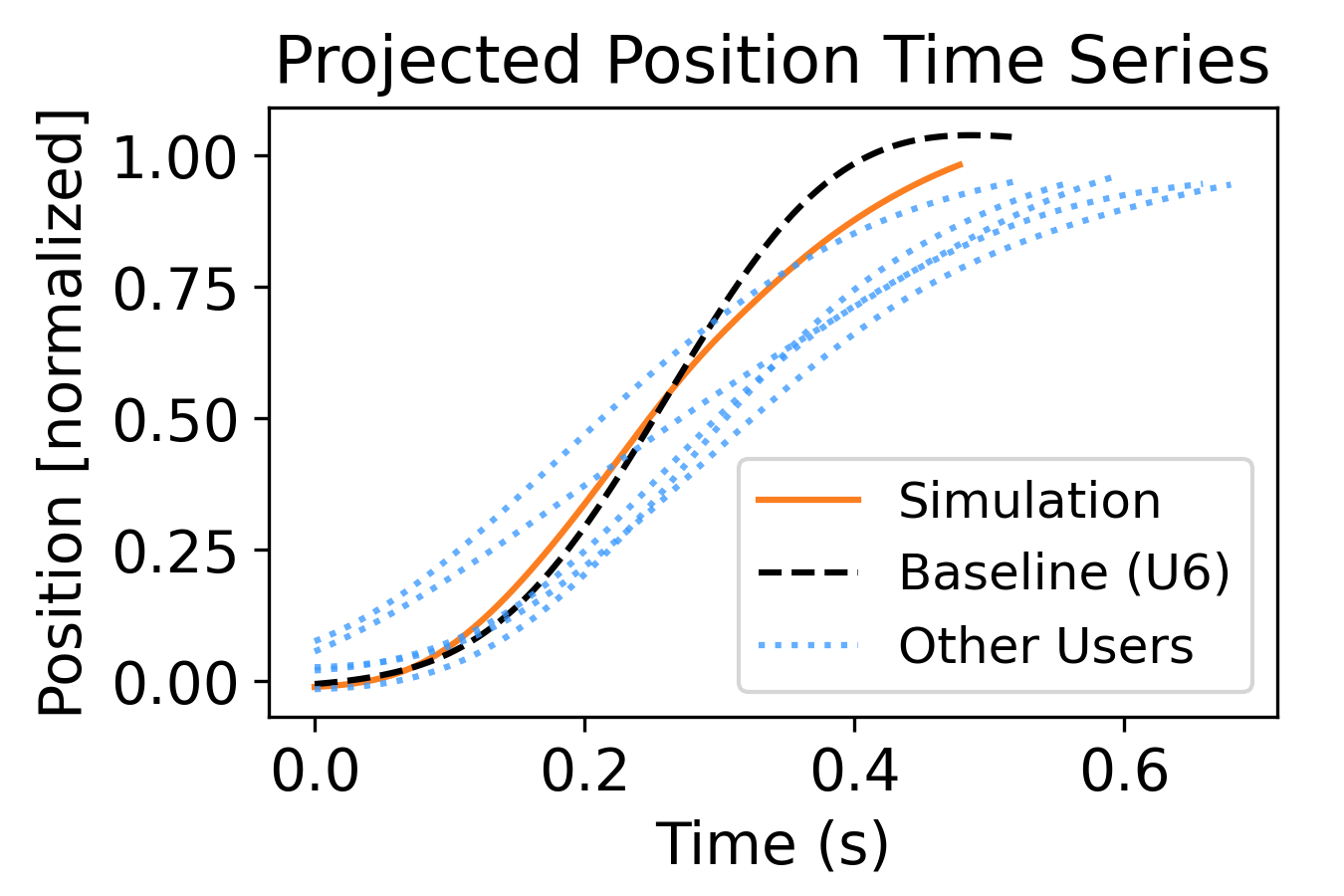}}
	\subfloat{\includegraphics[width=0.25\linewidth, clip]{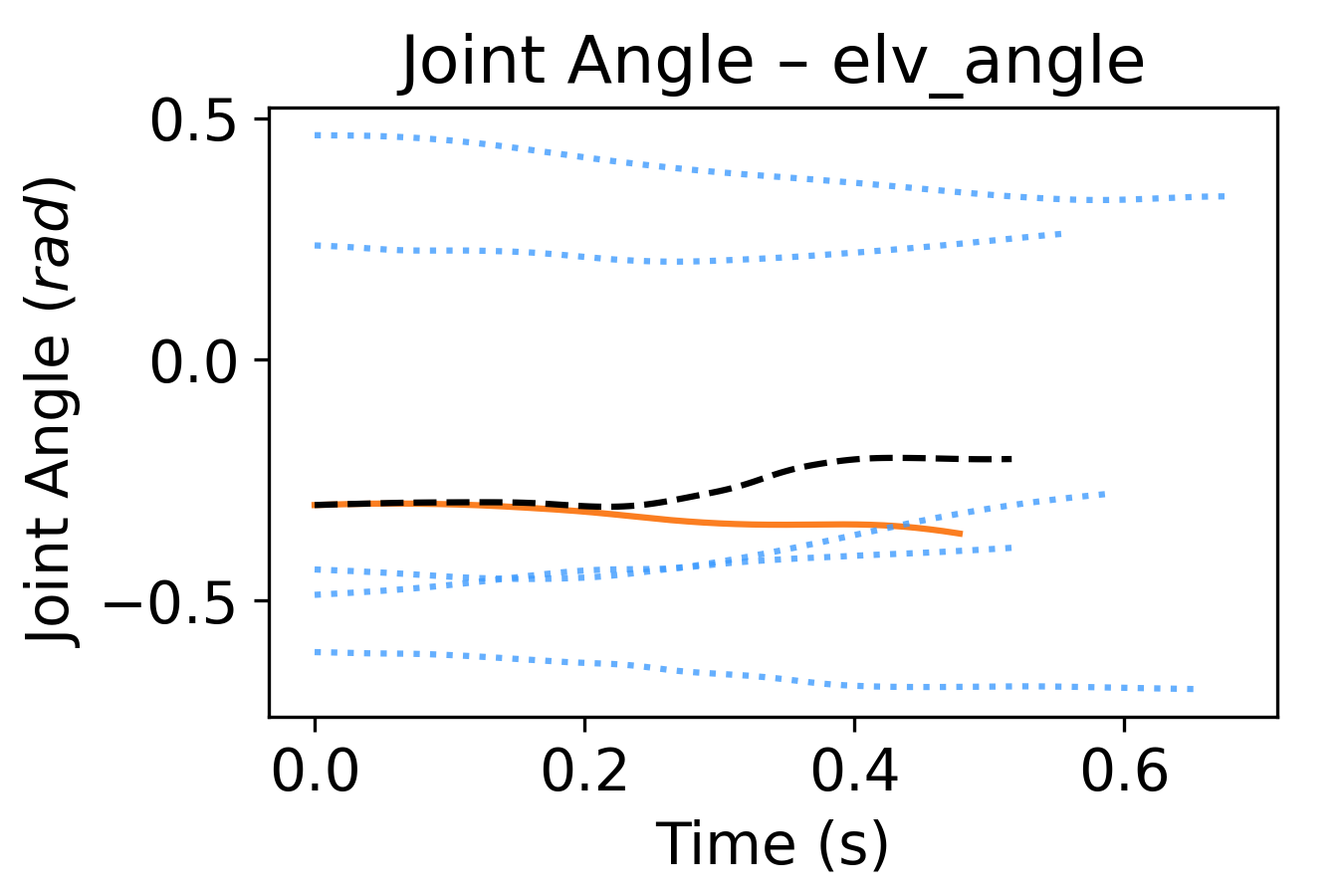}}
	\subfloat{\includegraphics[width=0.25\linewidth, clip]{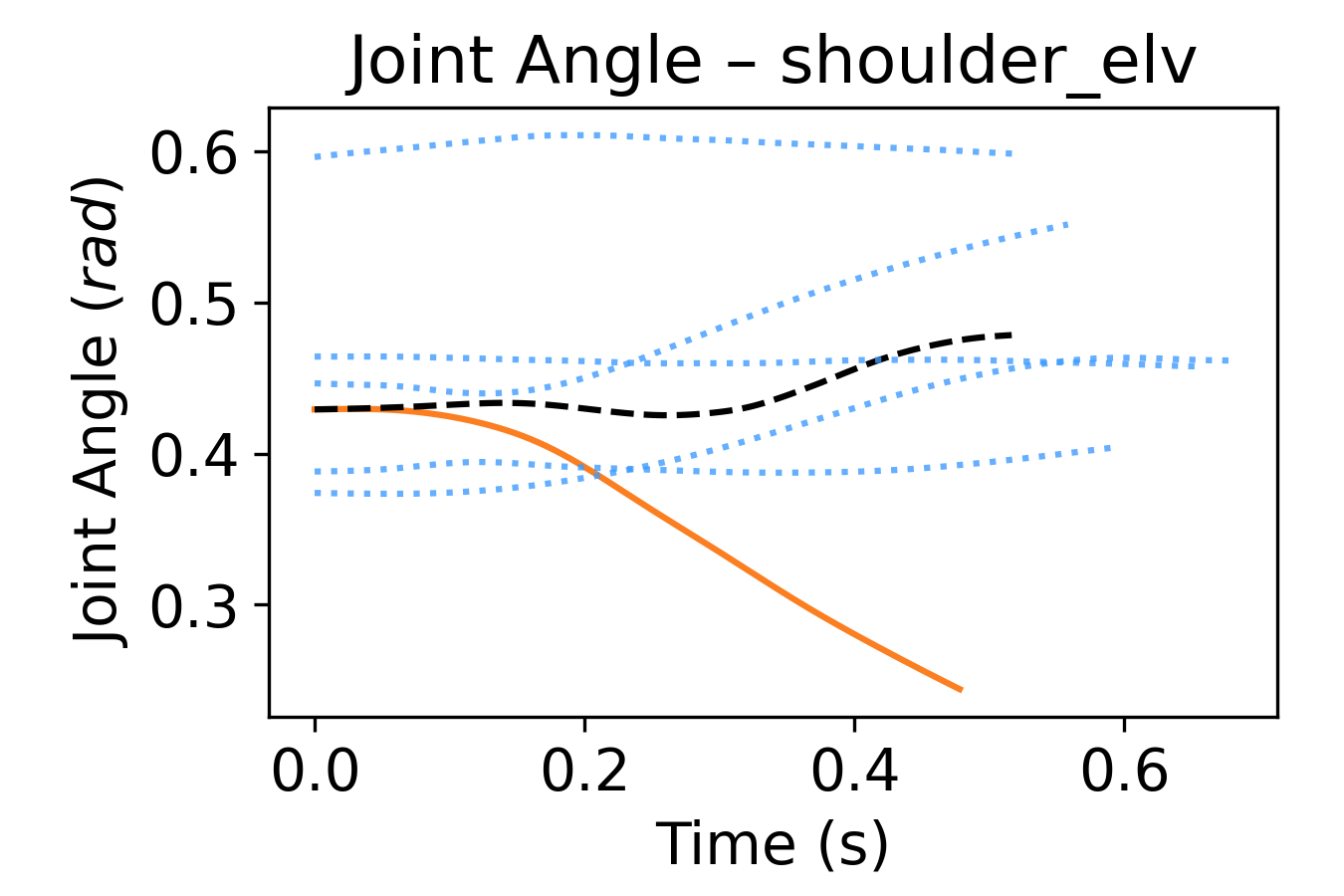}}
	\subfloat{\includegraphics[width=0.25\linewidth, clip]{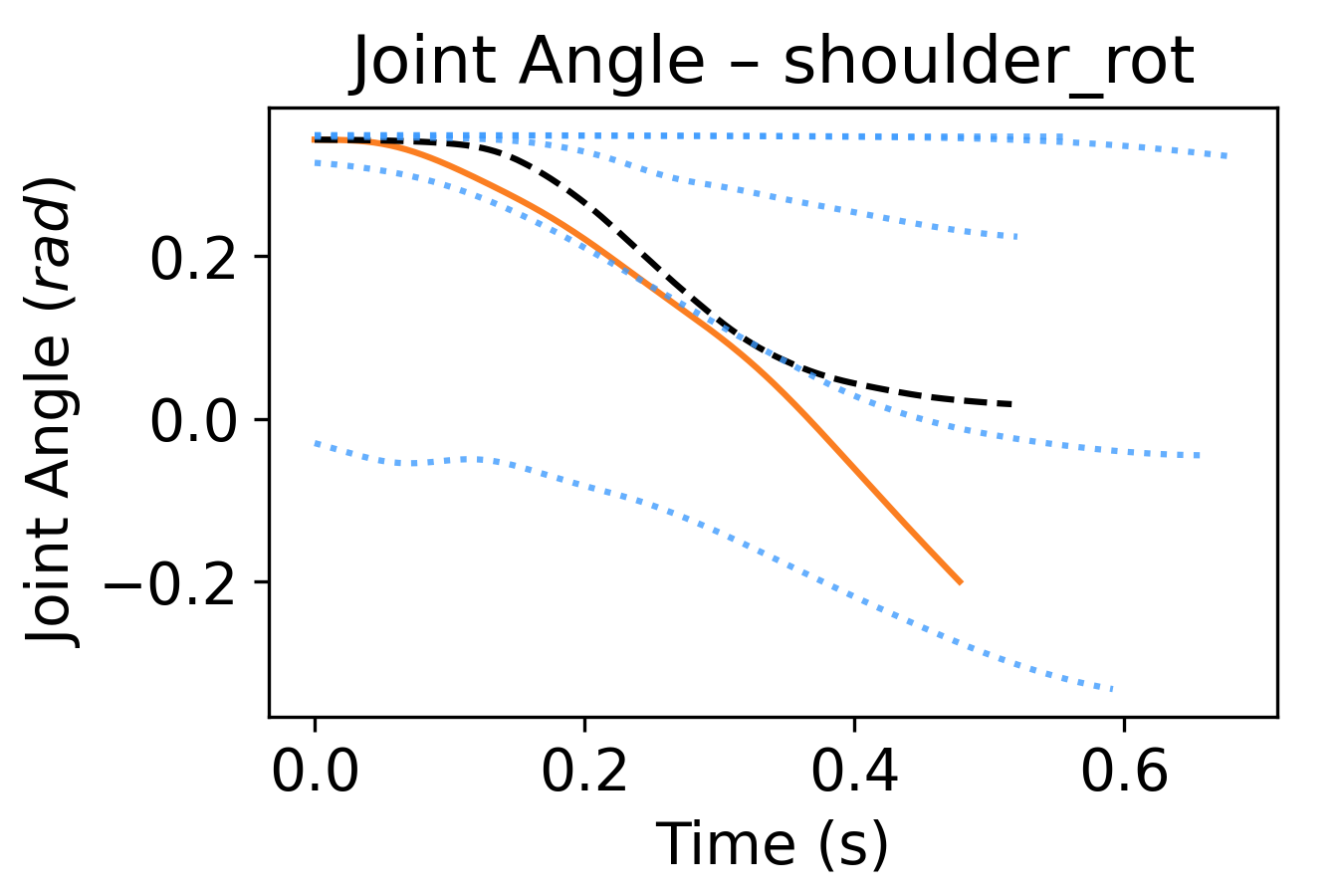}}\\
	
	\subfloat{\includegraphics[width=0.25\linewidth, clip]{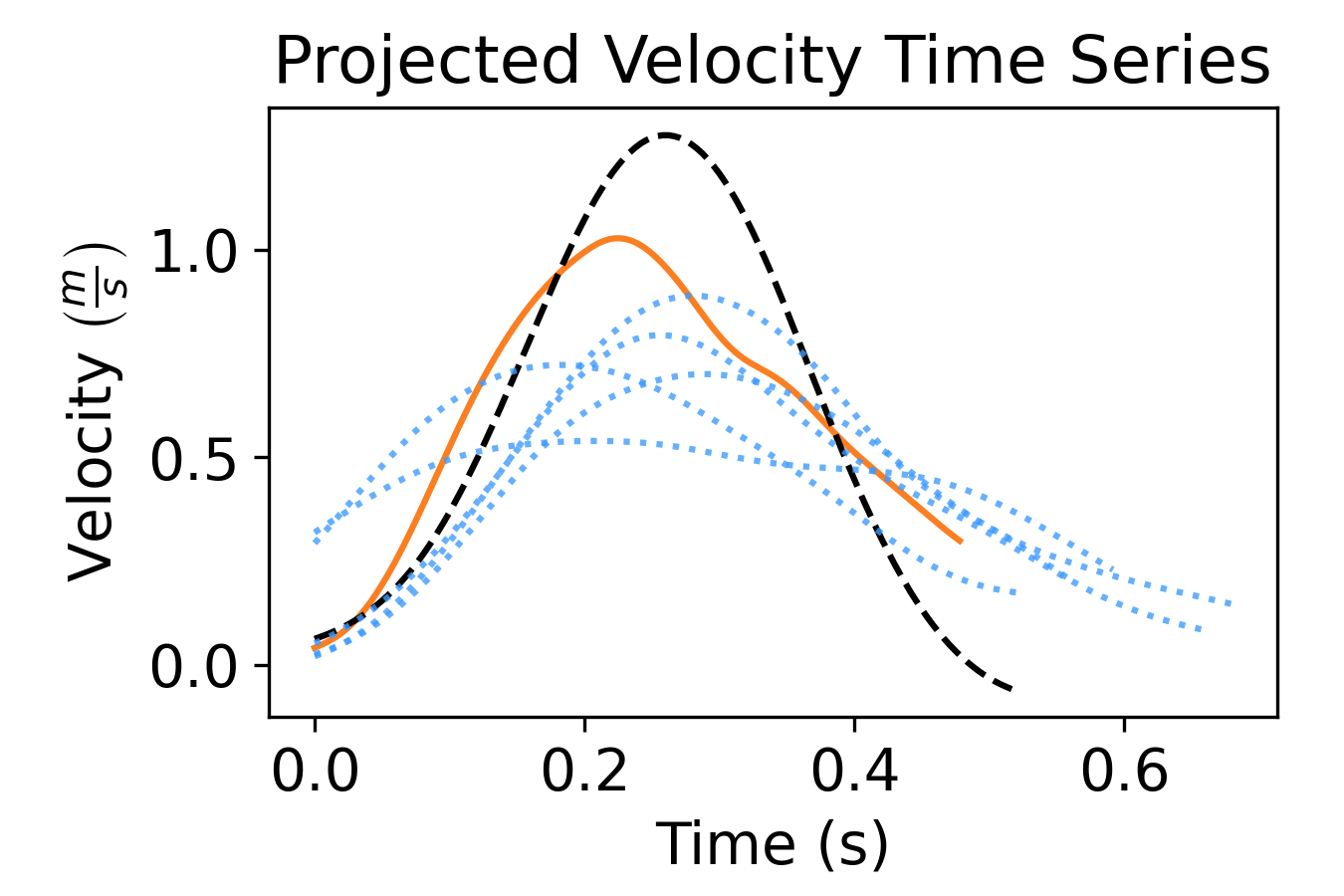}}
	\subfloat{\includegraphics[width=0.25\linewidth, clip]{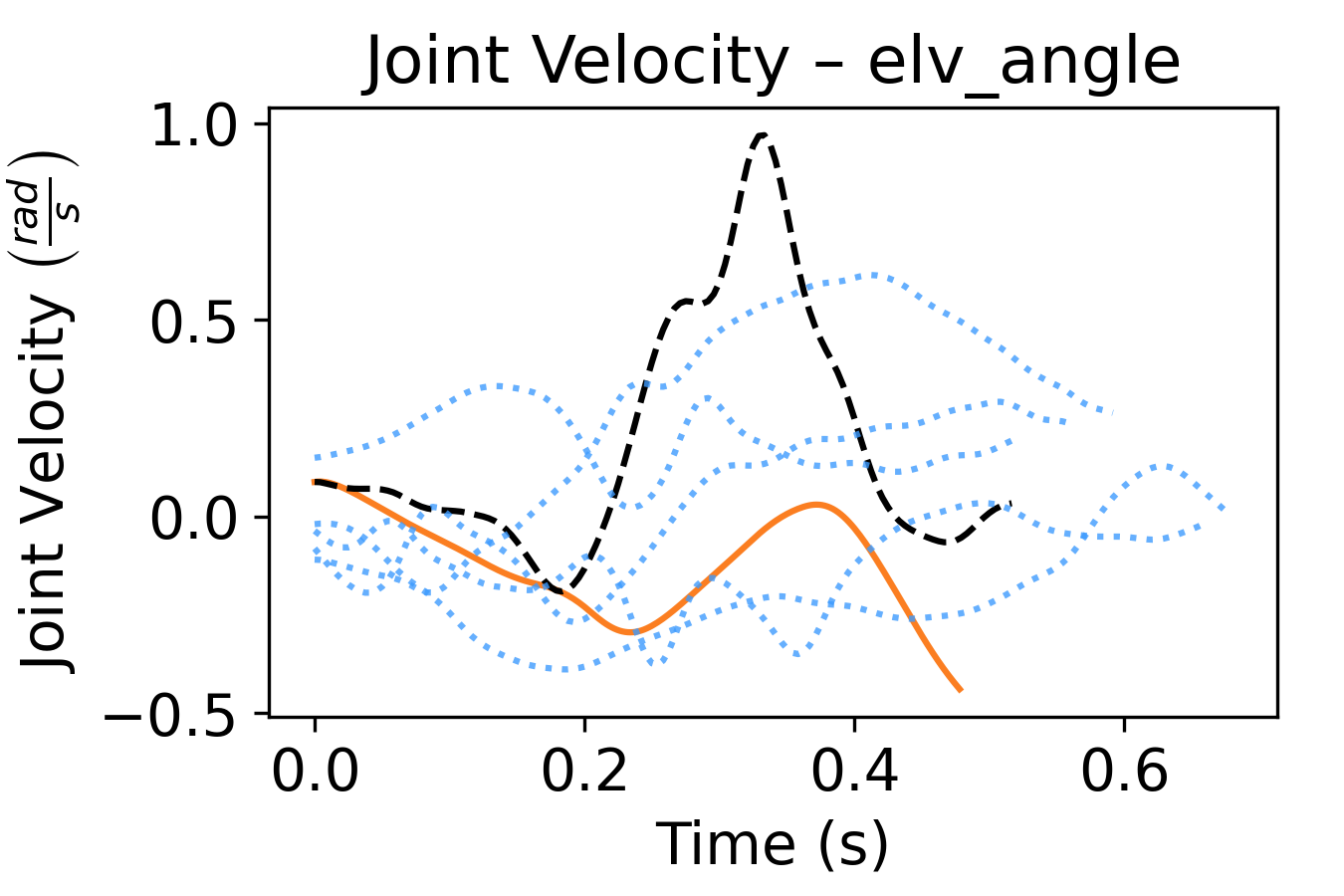}}
	\subfloat{\includegraphics[width=0.25\linewidth, clip]{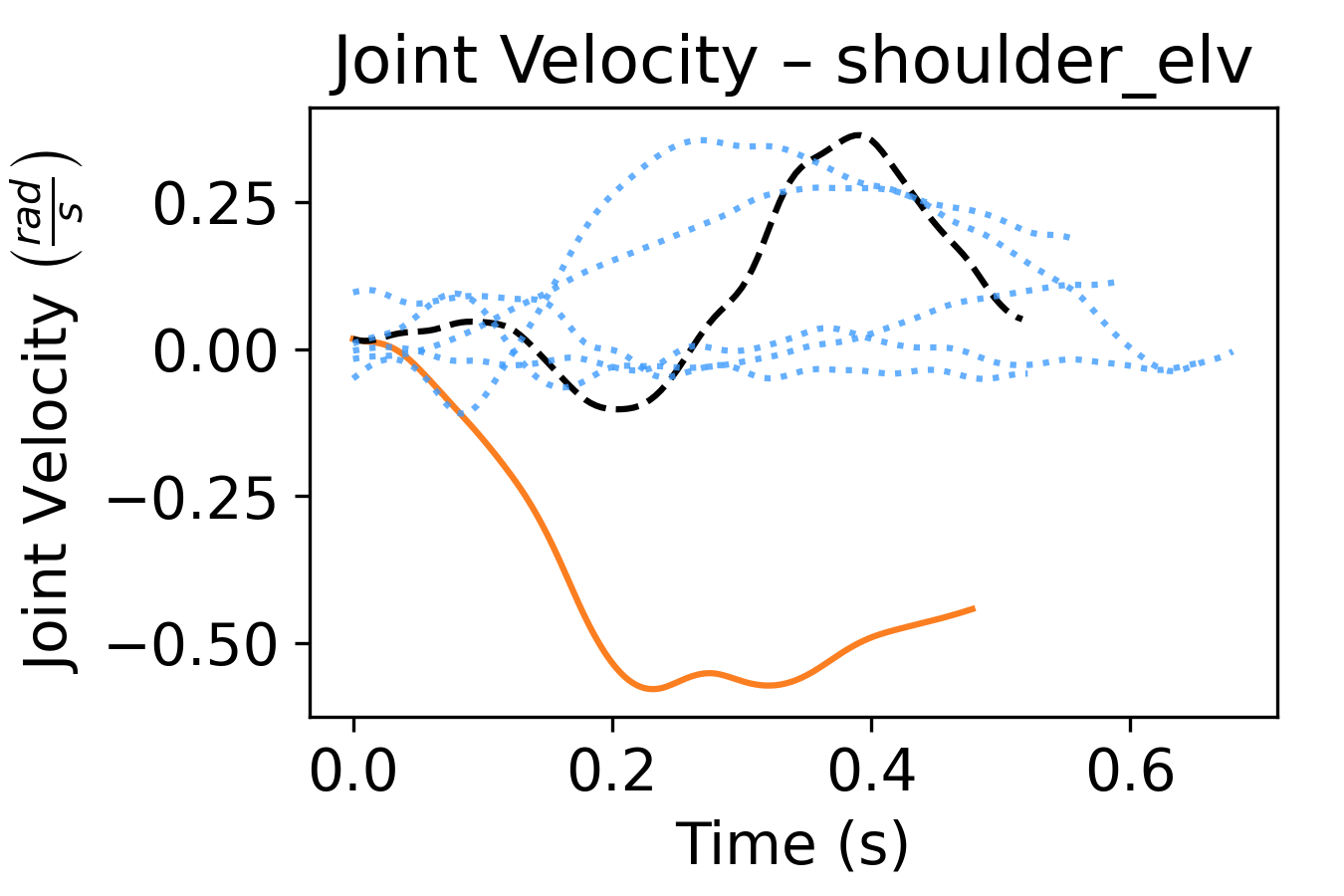}}
	\subfloat{\includegraphics[width=0.25\linewidth, clip]{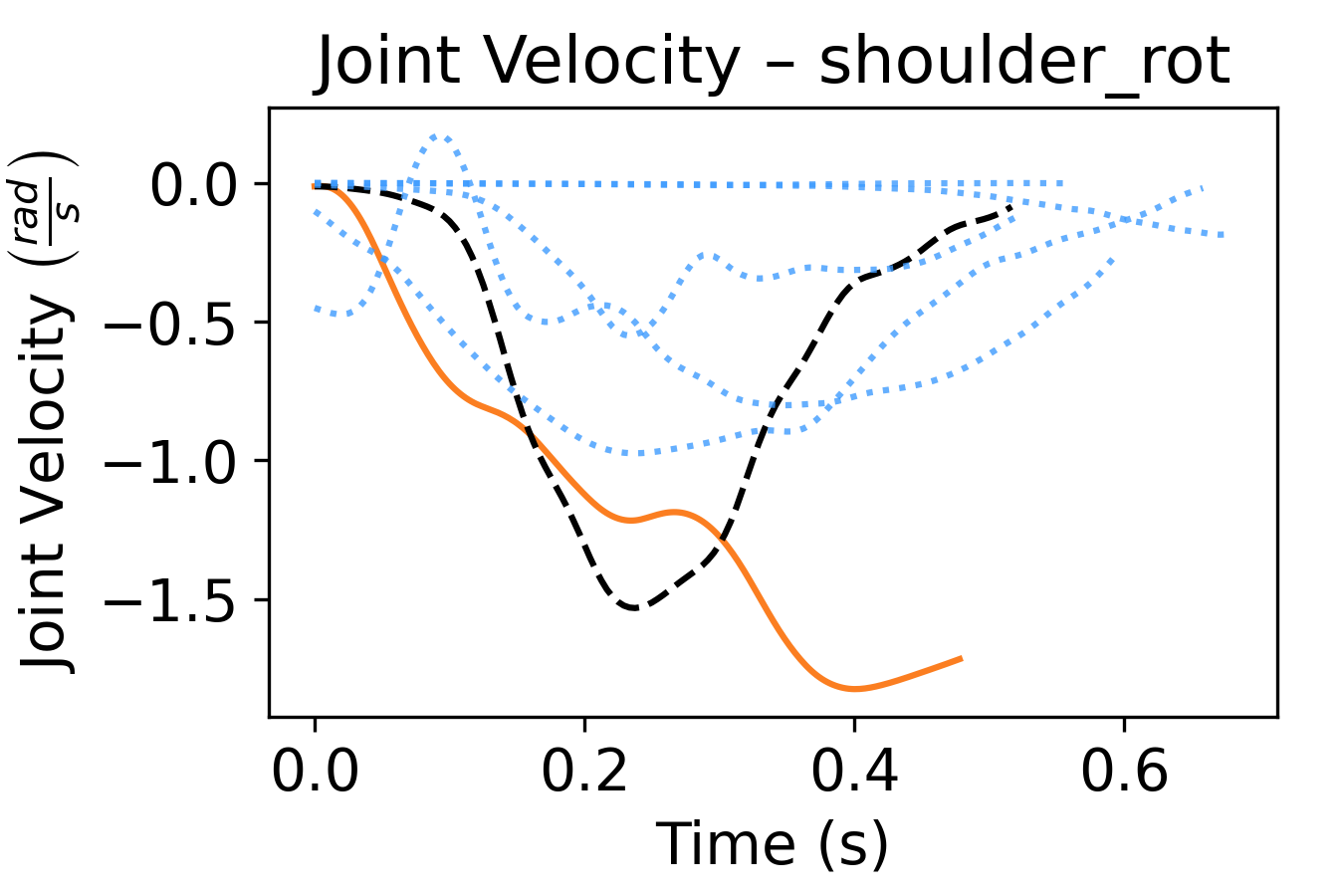}}\\
	
	\subfloat{\includegraphics[width=0.25\linewidth, clip]{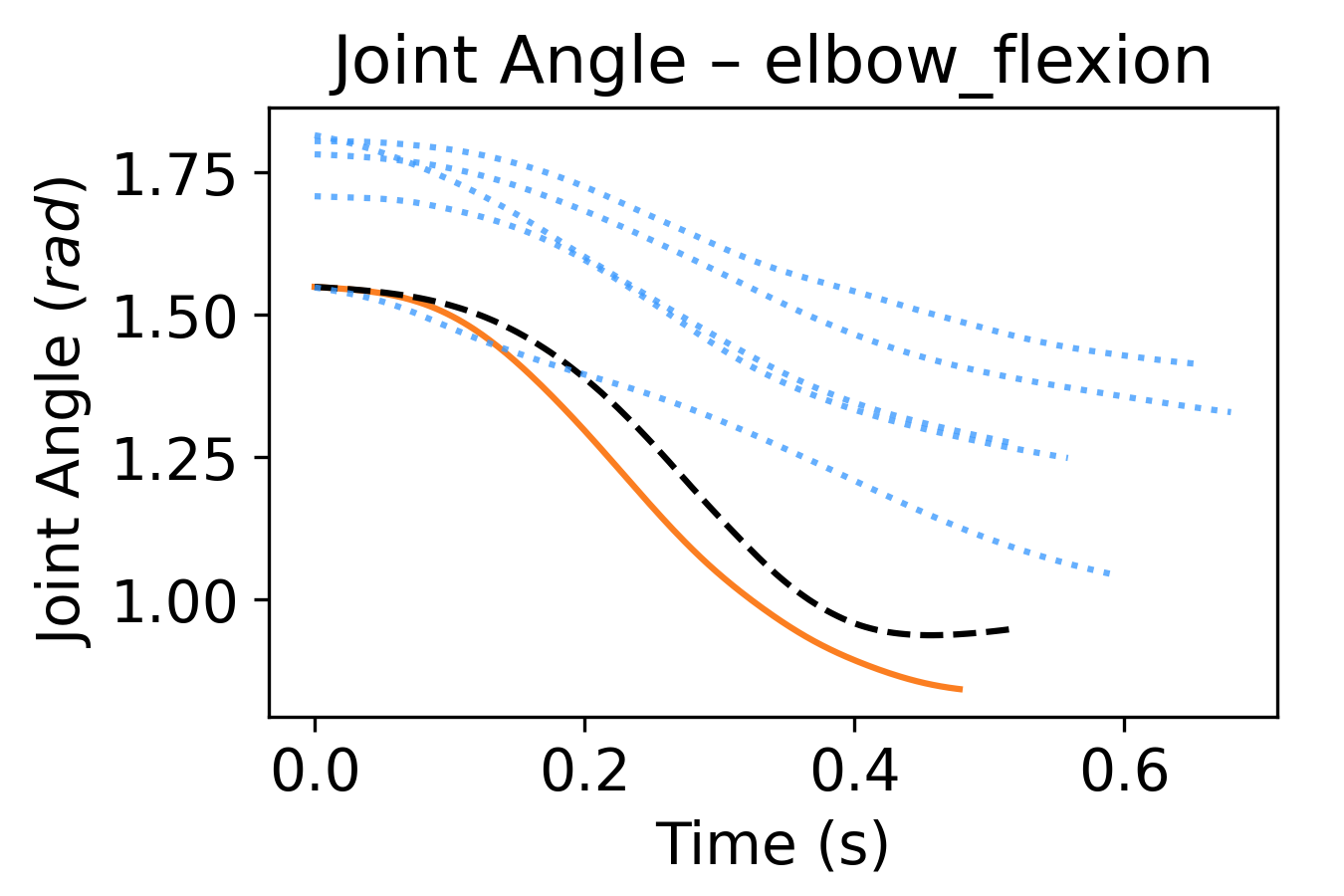}}
	\subfloat{\includegraphics[width=0.25\linewidth, clip]{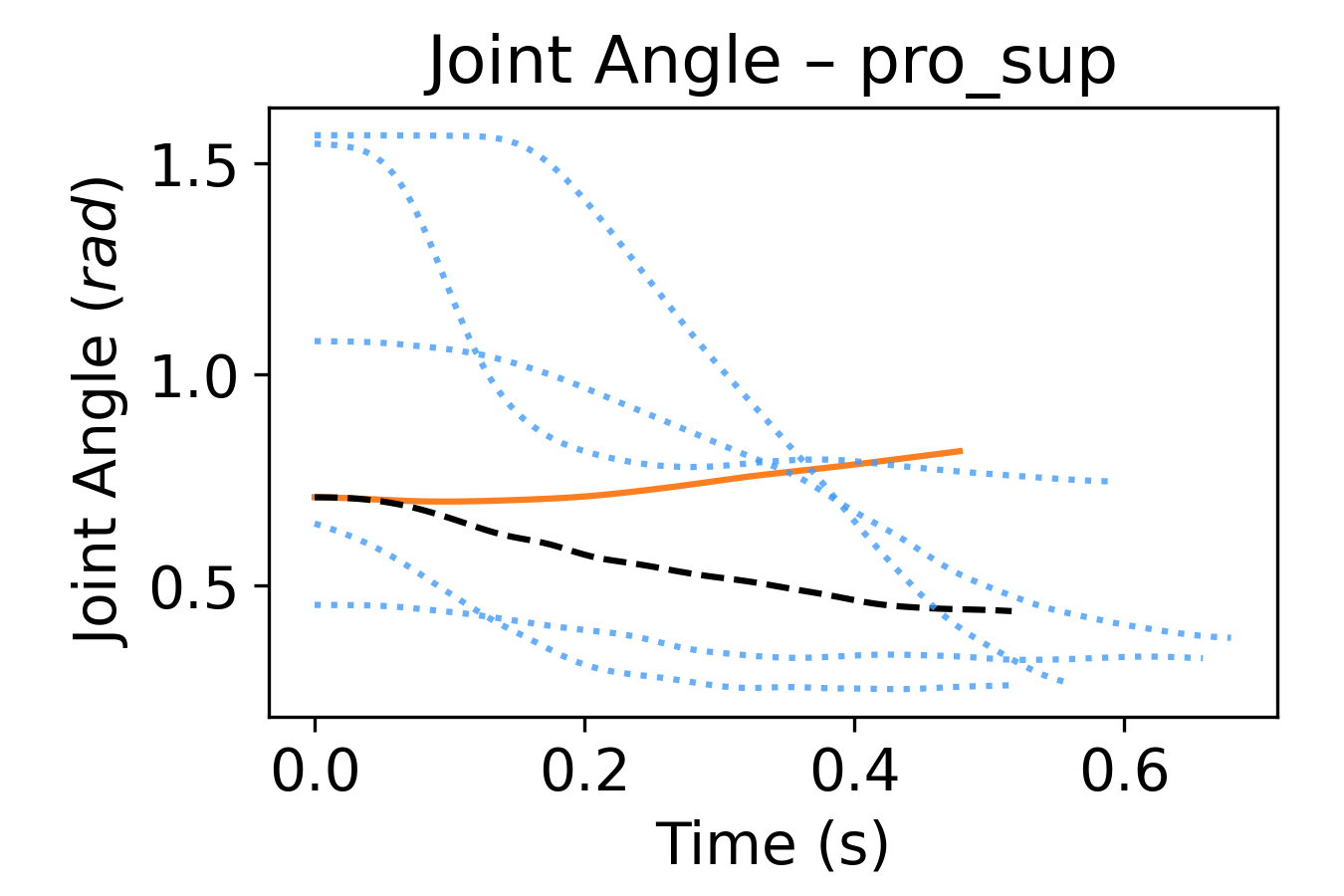}}
	\subfloat{\includegraphics[width=0.25\linewidth, clip]{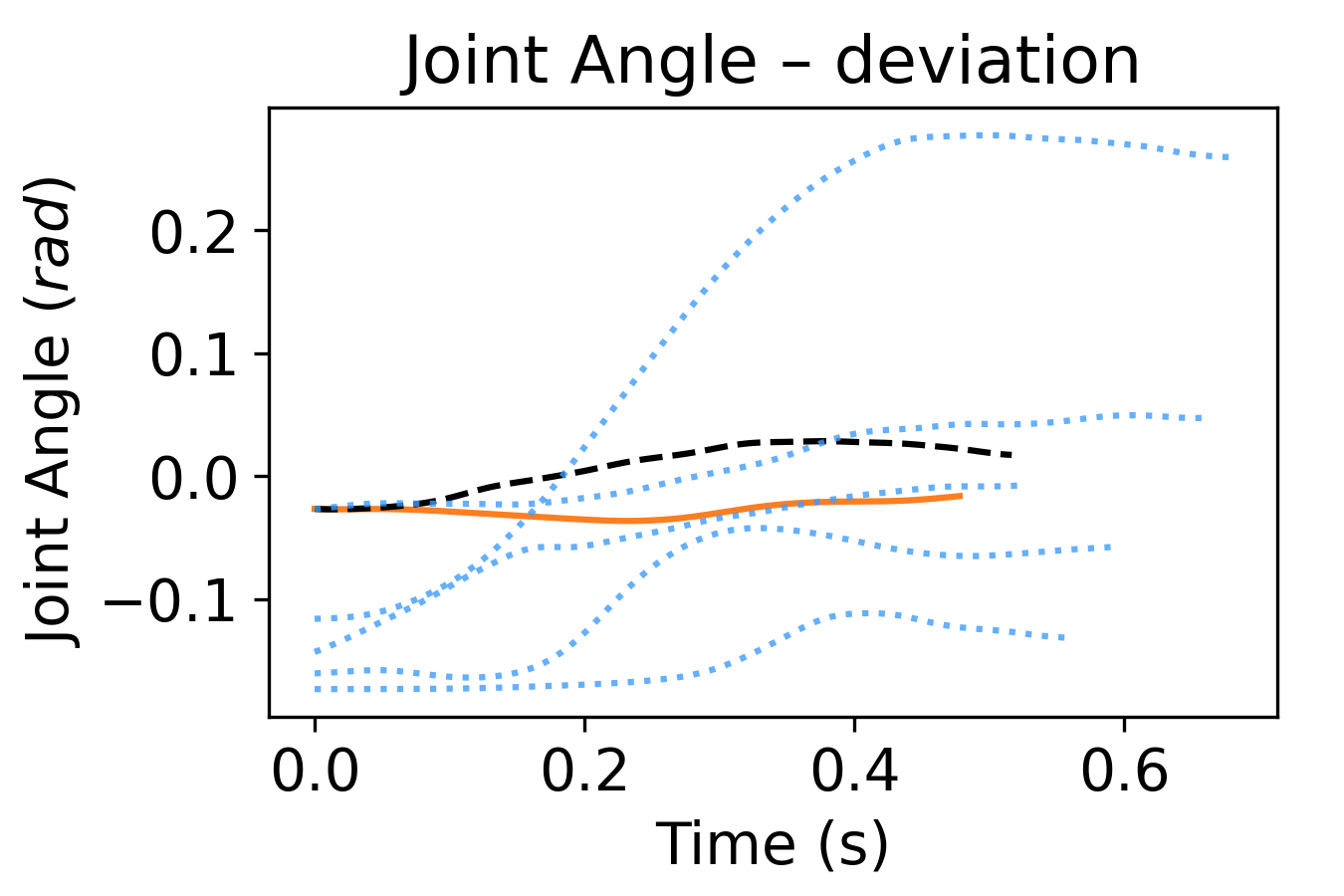}}
	\subfloat{\includegraphics[width=0.25\linewidth, clip]{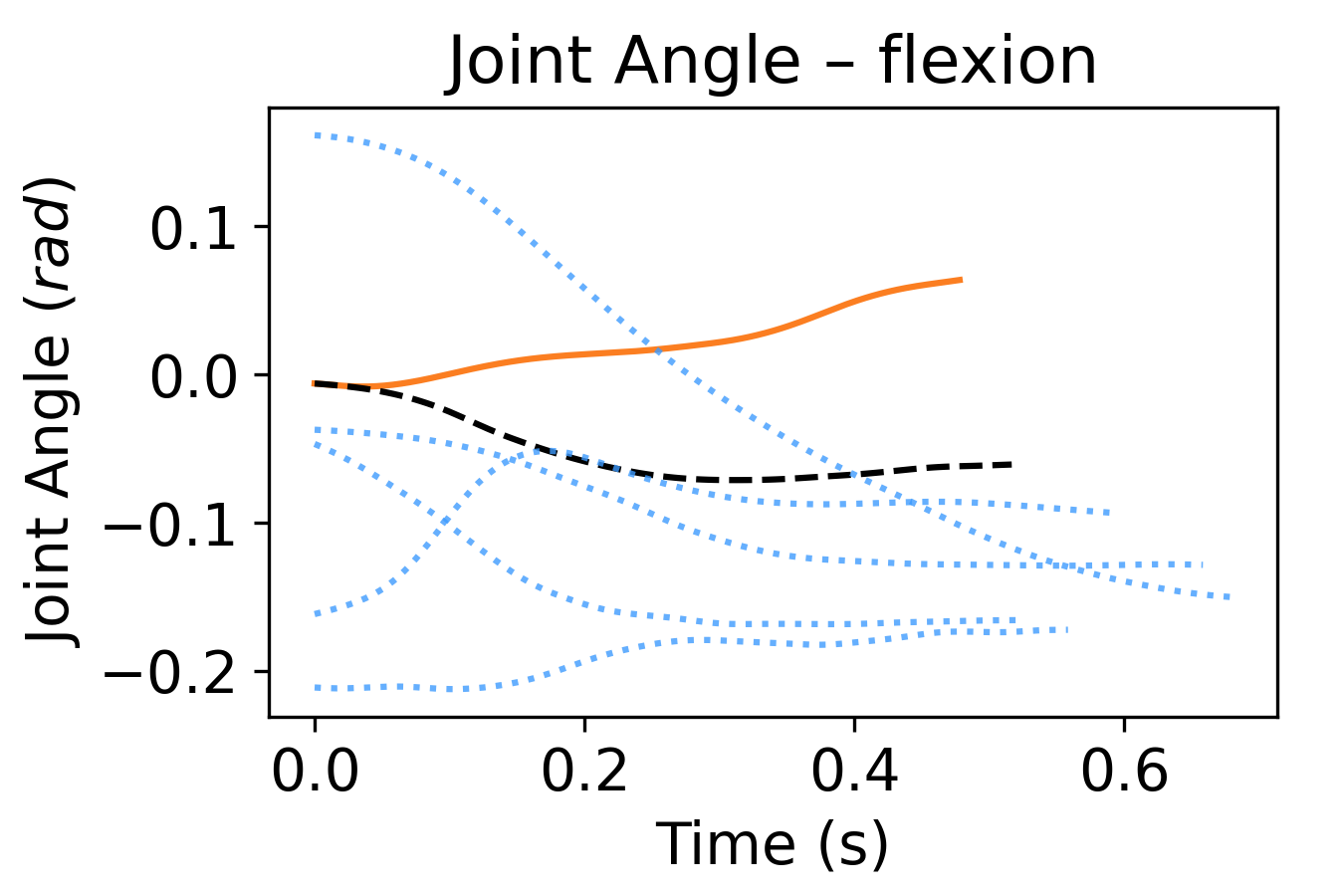}}\\
	
	\subfloat{\includegraphics[width=0.25\linewidth, clip]{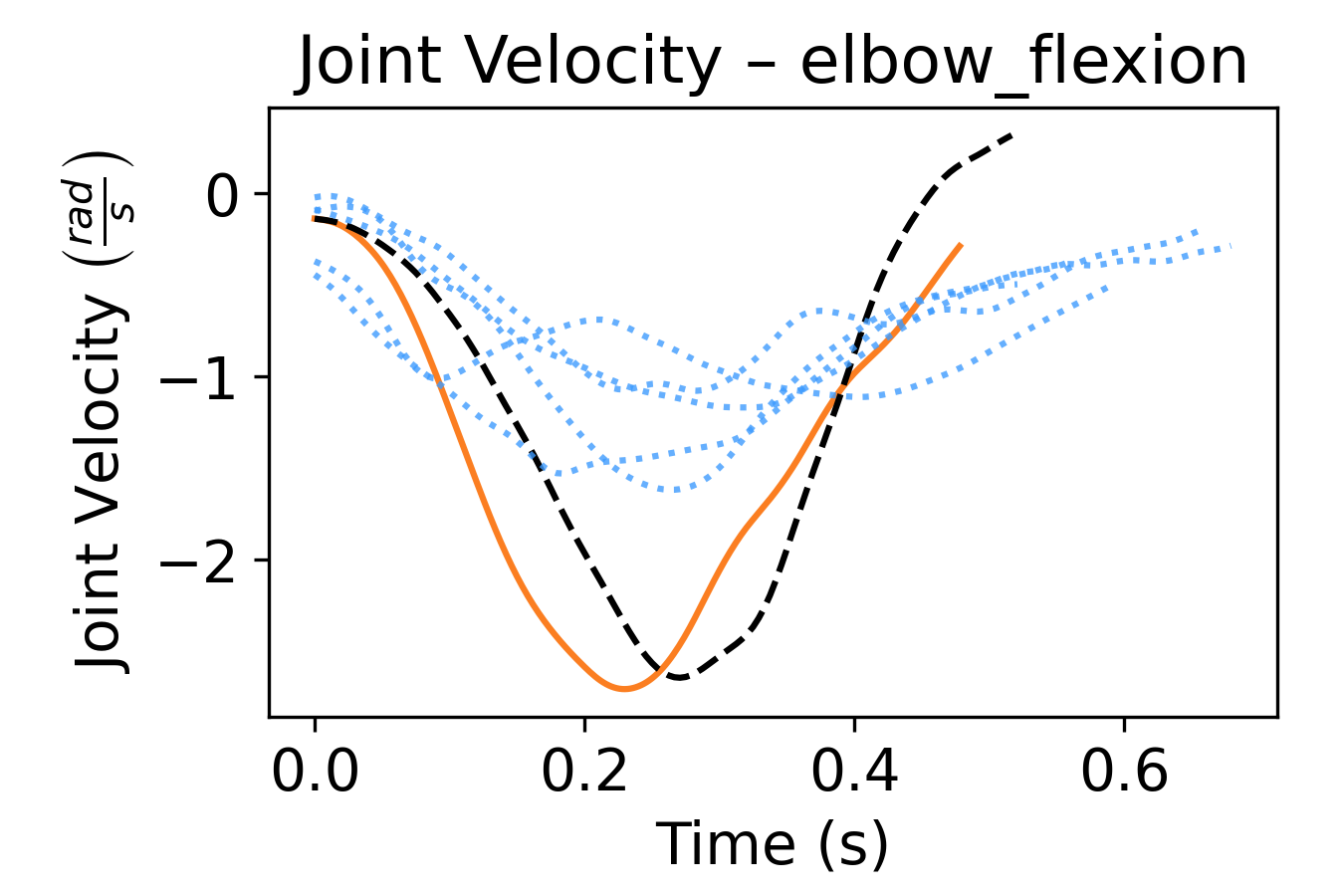}}
	\subfloat{\includegraphics[width=0.25\linewidth, clip]{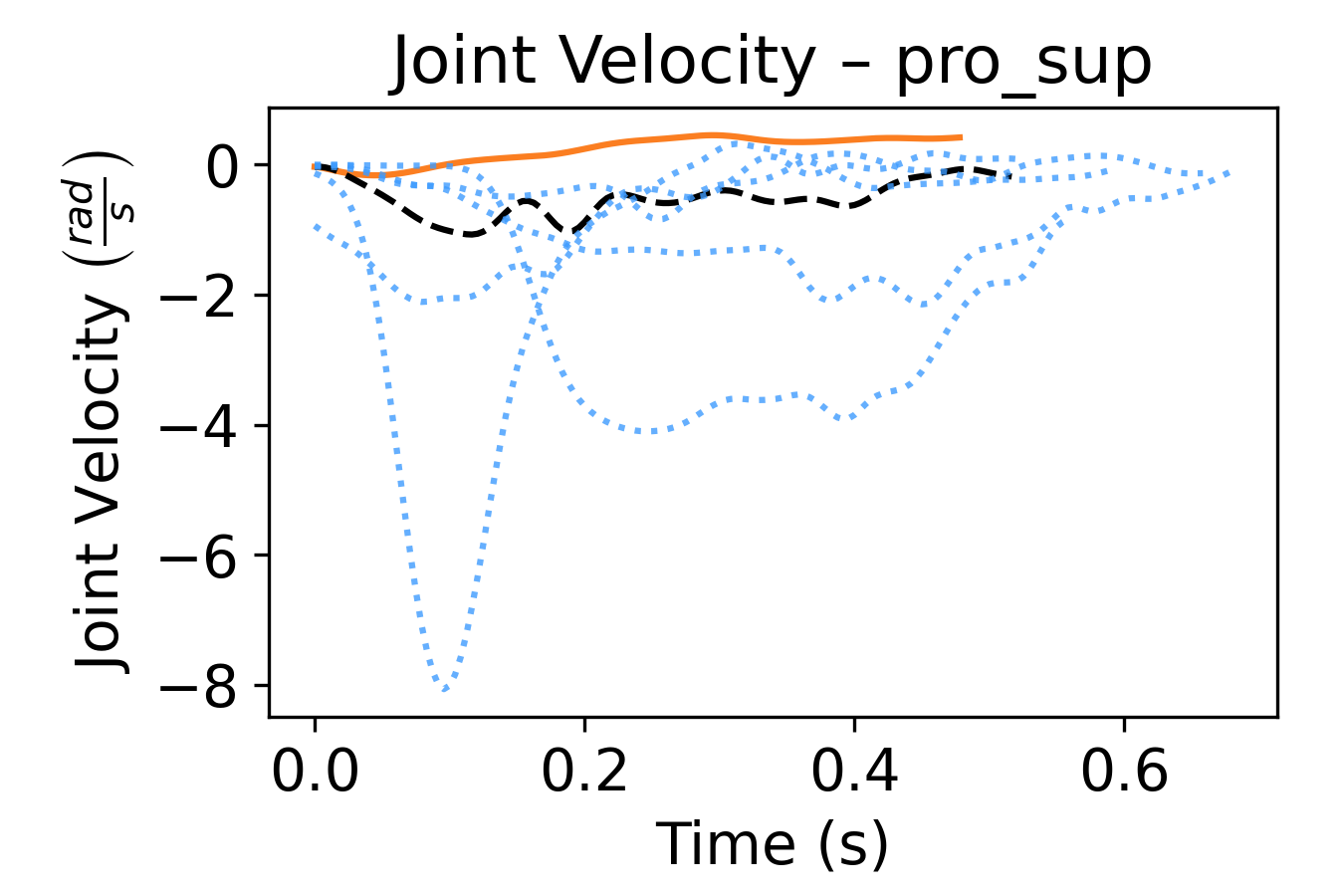}}
	\subfloat{\includegraphics[width=0.25\linewidth, clip]{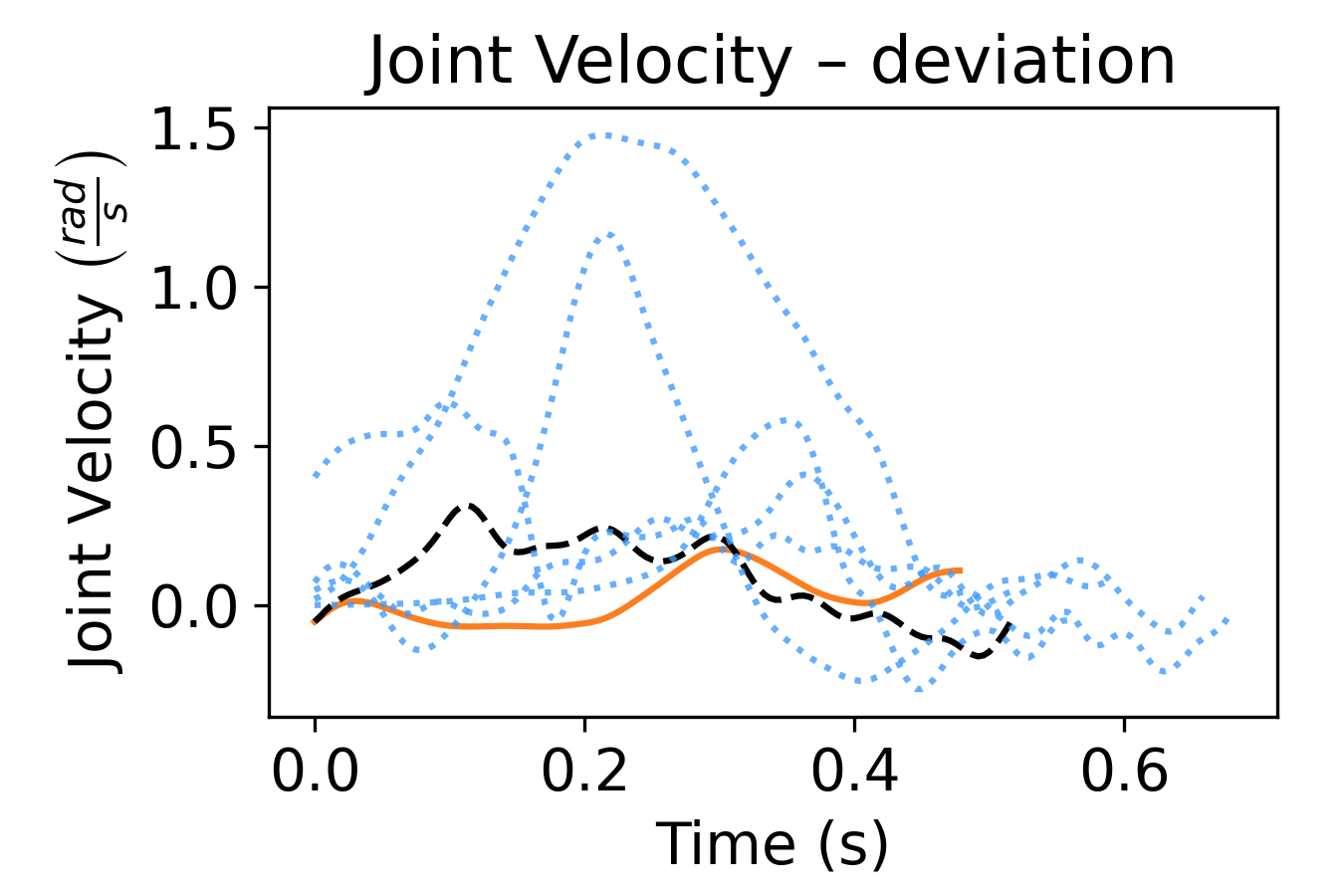}}
	\subfloat{\includegraphics[width=0.25\linewidth, clip]{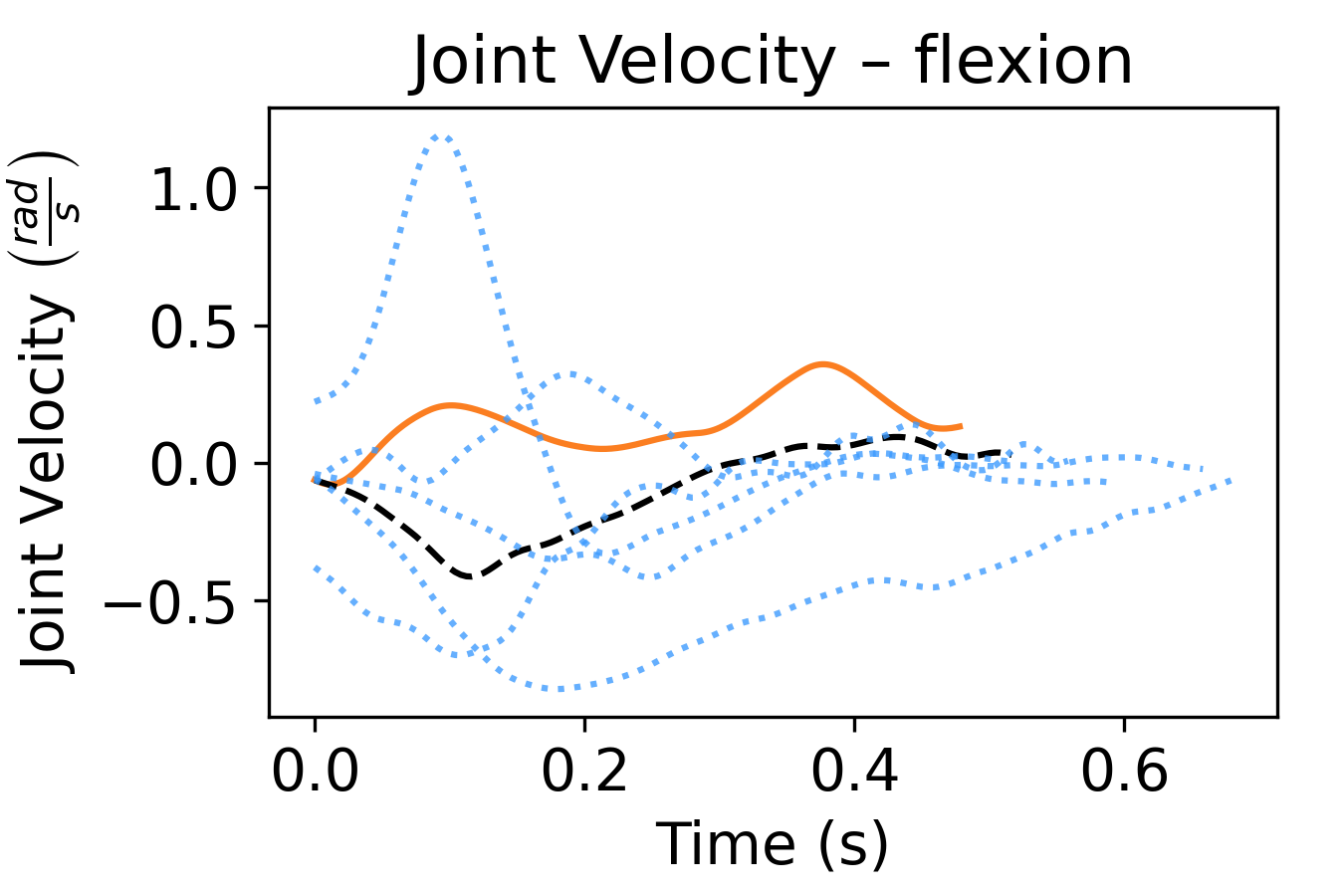}}
	
	\caption{Given an interaction technique (here: \textbf{Virtual Pad Ergonomic}) and a movement direction (here: movements from target 8 to target 9), the characteristic cursor and joint trajectories of an individual user (here: U6, black dashed lines; trajectories of the remaining users are shown as blue dotted lines for comparison) can be predicted by our simulation (orange solid lines).}
	\label{fig:PadErgonomic_qual_2}
\end{figure}

\clearpage

\subsection{Effects of the Cost Weights}

\begin{figure}[!h]
	\subfloat{\includegraphics[width=0.2\linewidth, clip]{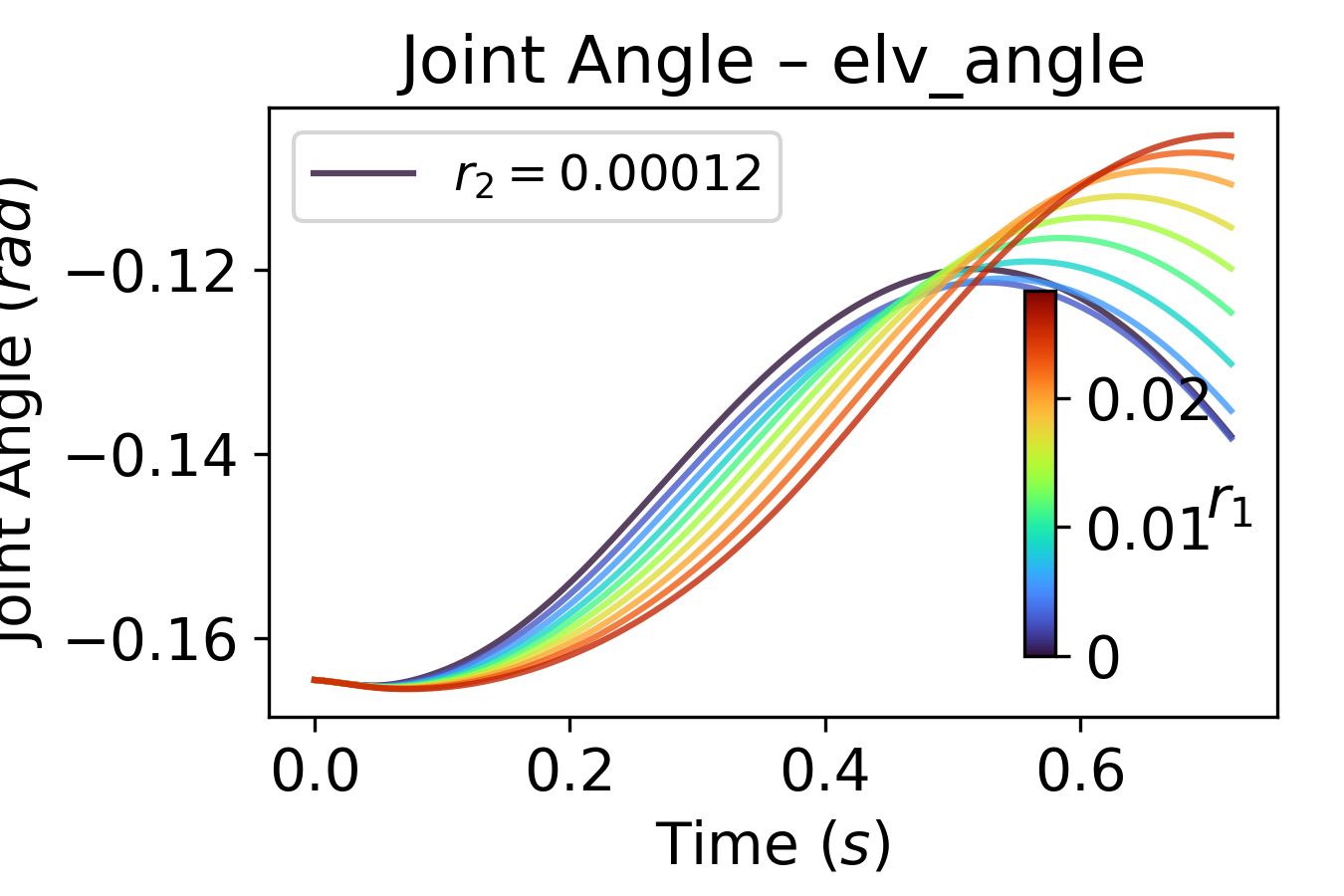}}
	\subfloat{\includegraphics[width=0.2\linewidth, clip]{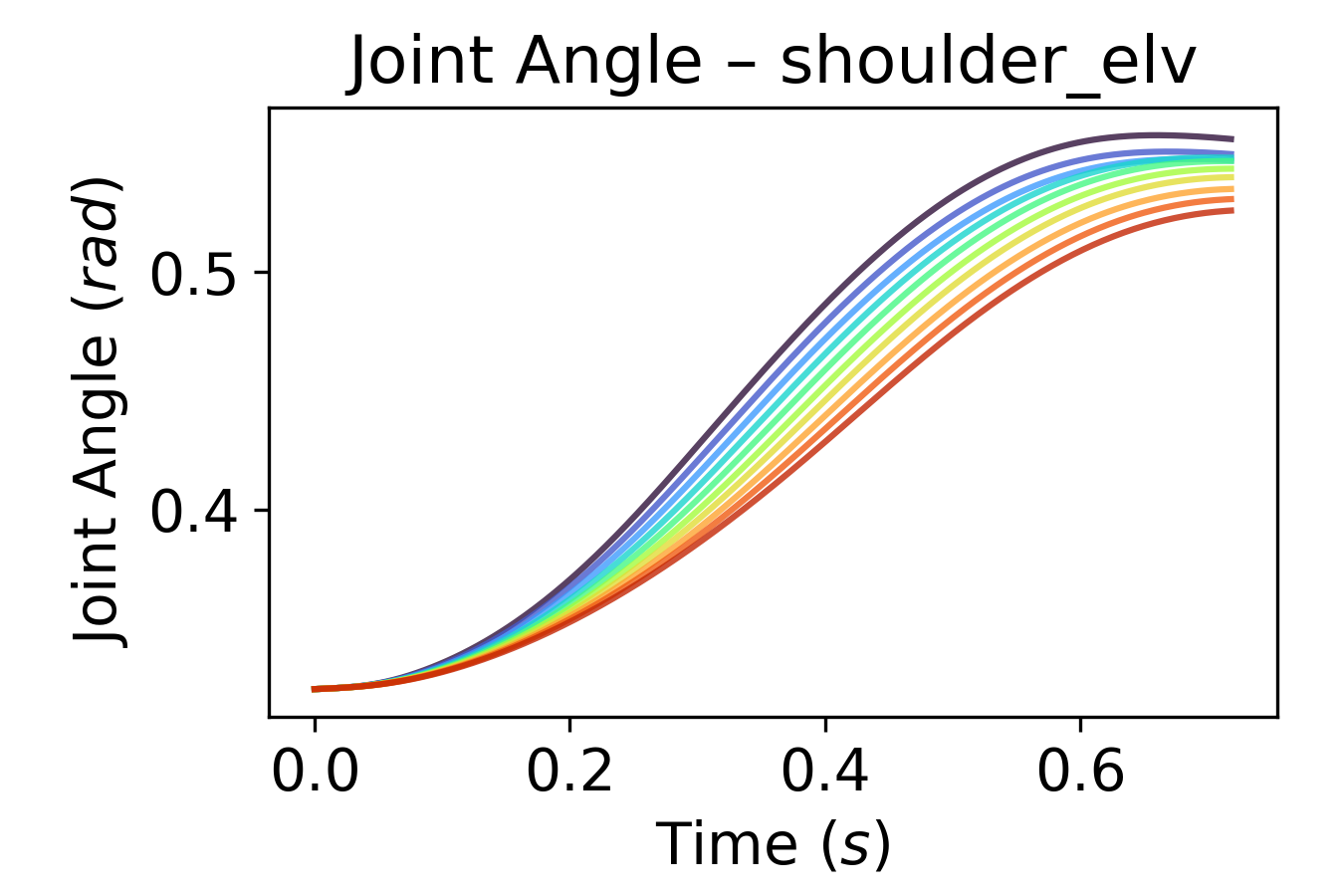}}
	\subfloat{\includegraphics[width=0.2\linewidth, clip]{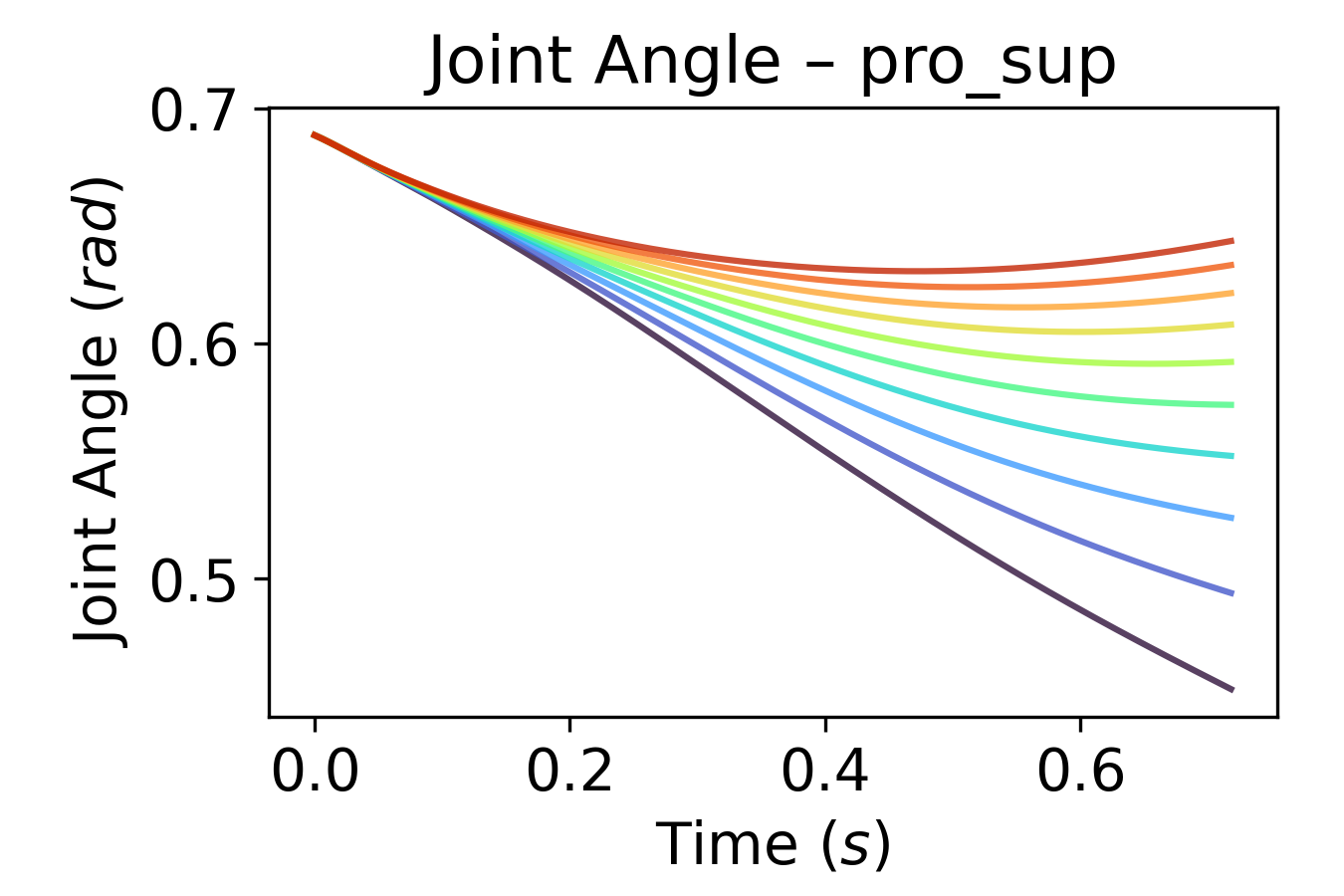}}
	\subfloat{\includegraphics[width=0.2\linewidth, clip]{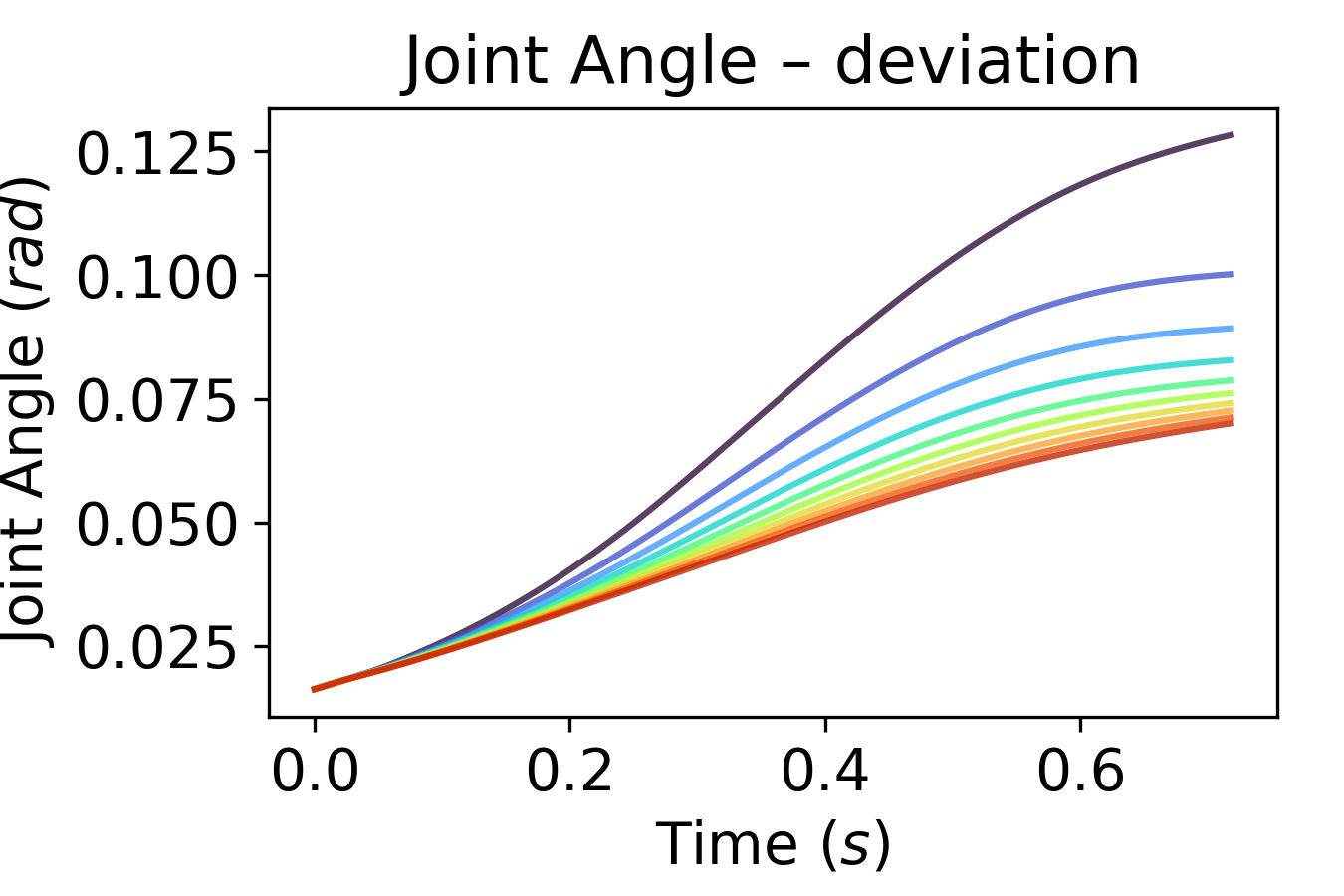}}
	\subfloat{\includegraphics[width=0.2\linewidth, clip]{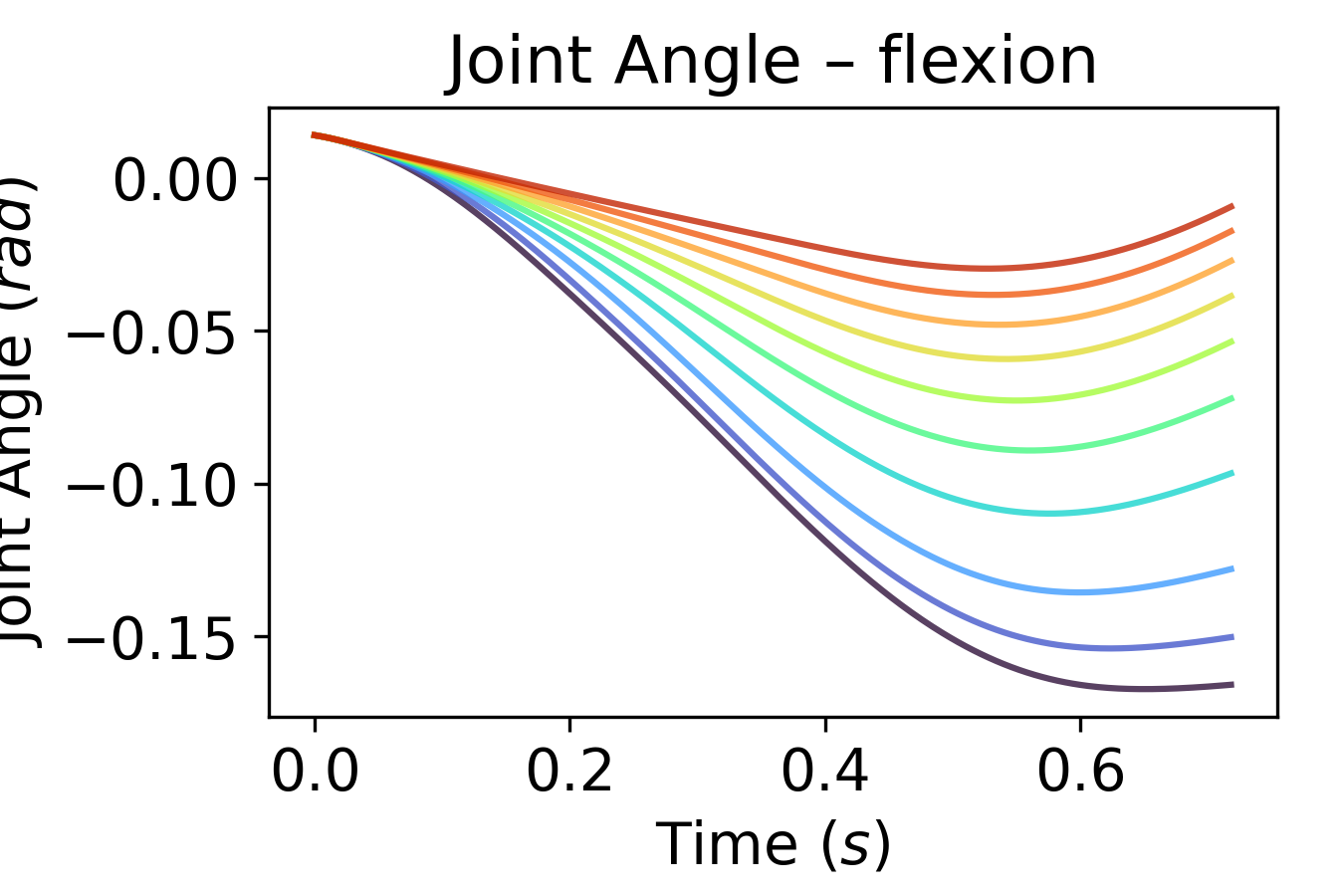}}\\
	\subfloat{\includegraphics[width=0.2\linewidth, clip]{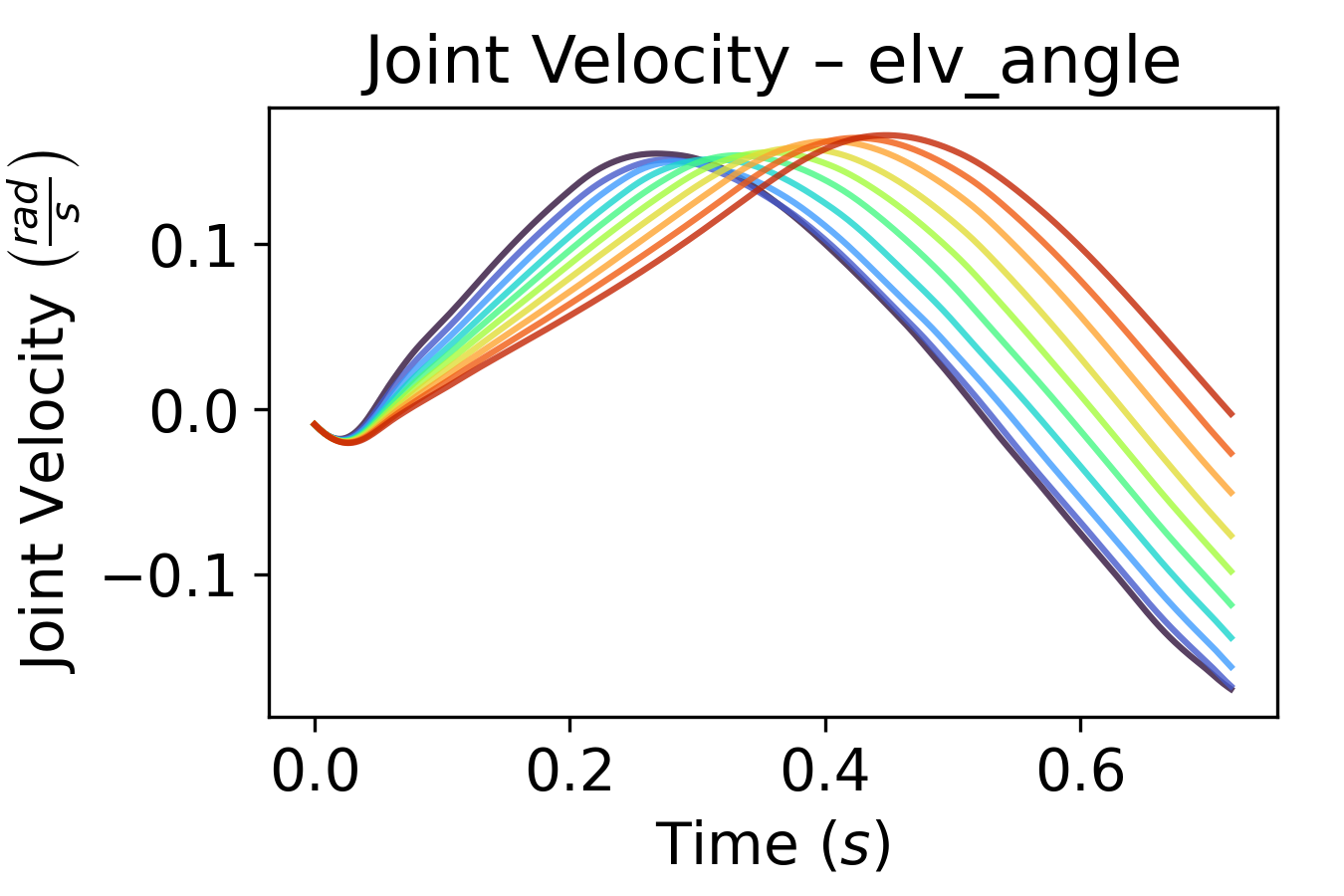}}
	\subfloat{\includegraphics[width=0.2\linewidth, clip]{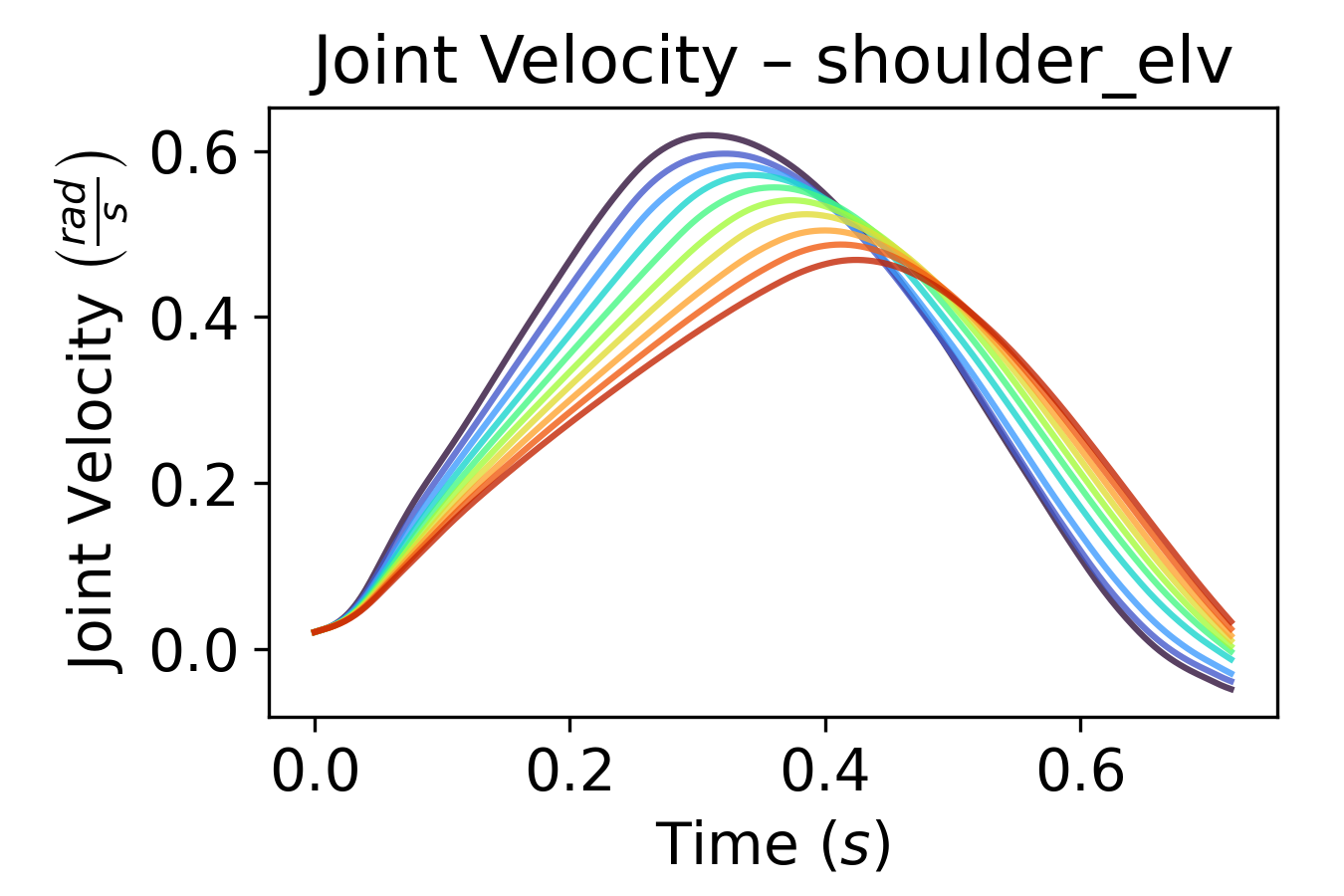}}
	\subfloat{\includegraphics[width=0.2\linewidth, clip]{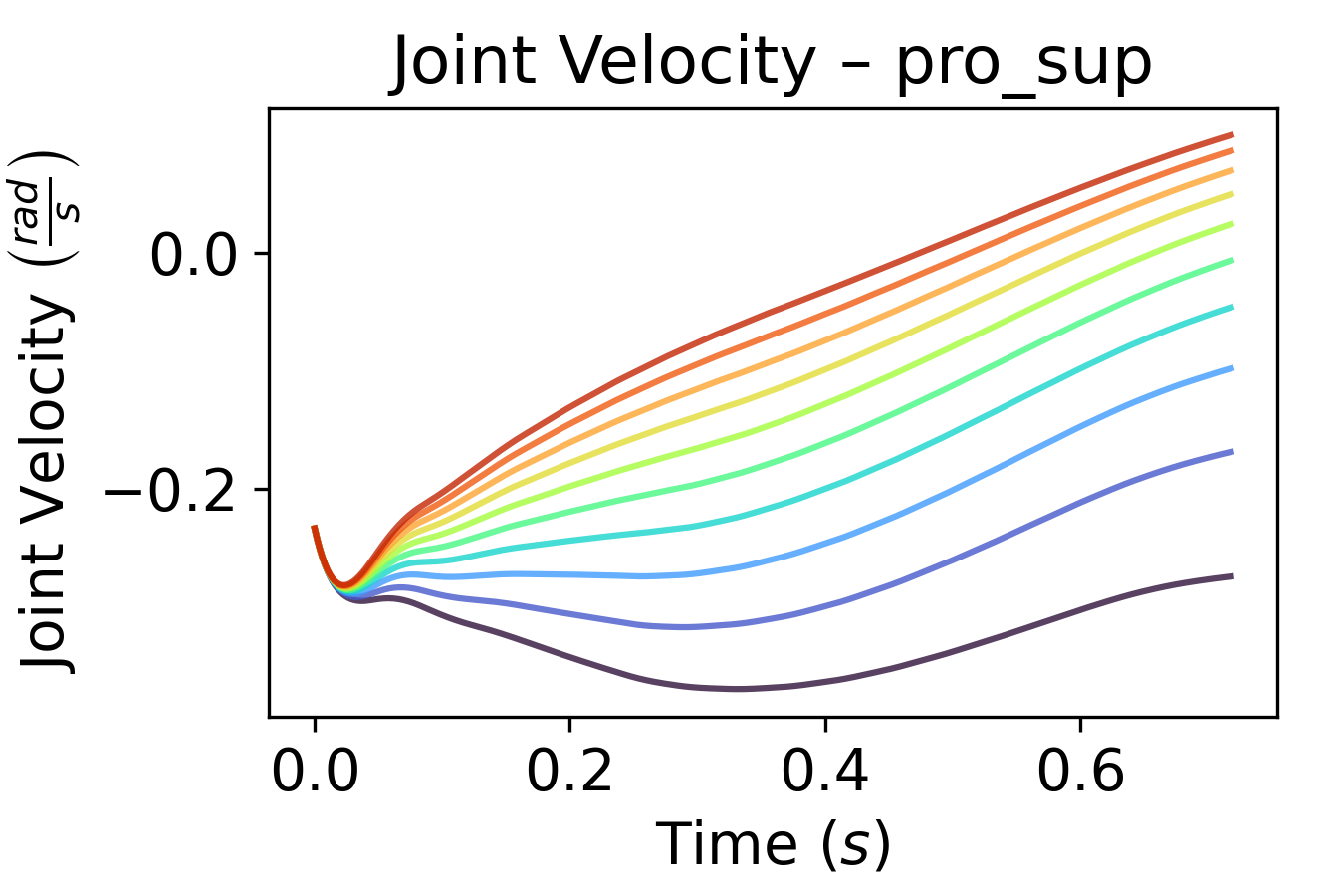}}
	\subfloat{\includegraphics[width=0.2\linewidth, clip]{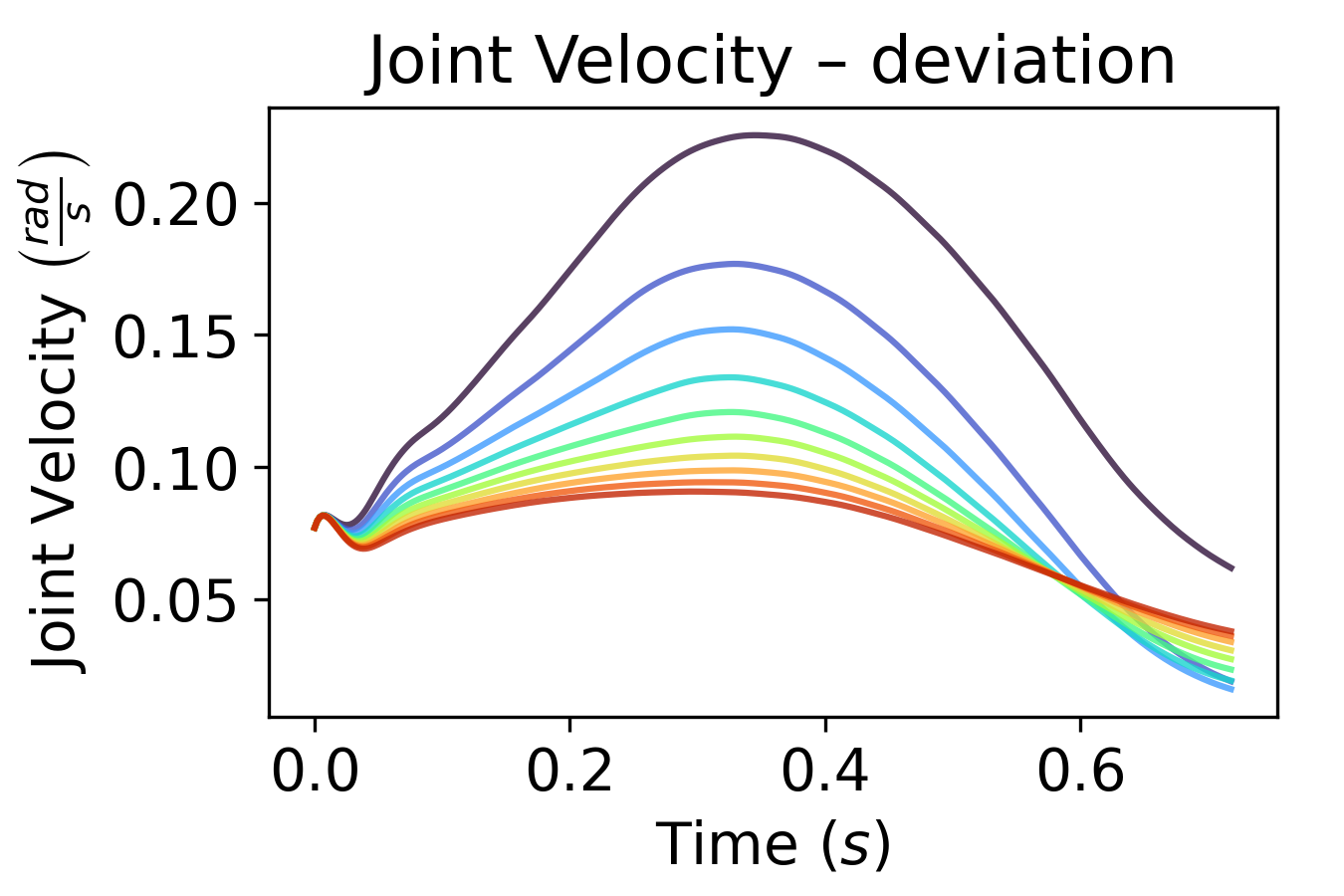}}
	\subfloat{\includegraphics[width=0.2\linewidth, clip]{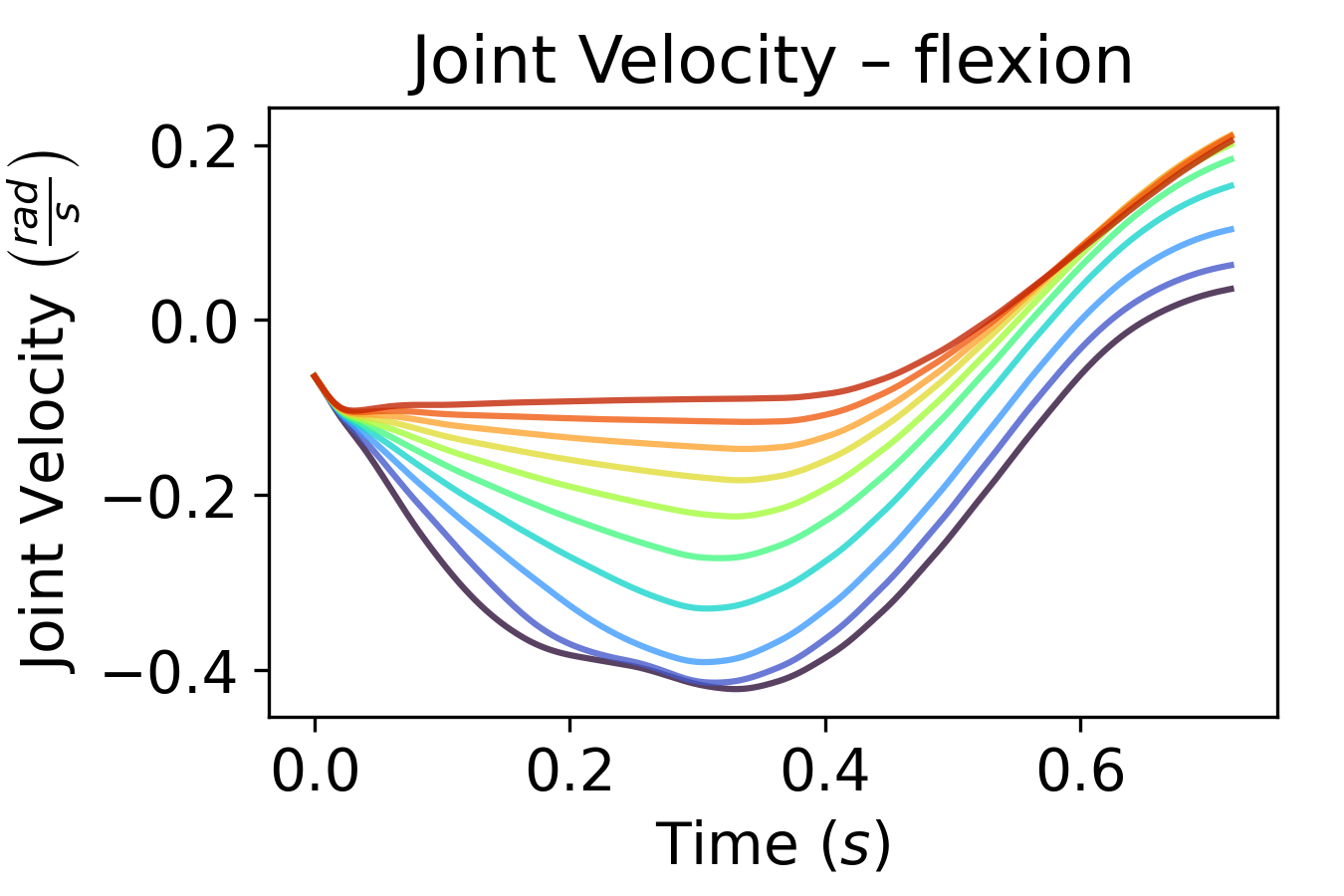}}
	
	\caption{Trajectories of the JAC simulation for different cost weights $r_1$. 
		Remaining joints and cursor trajectories are shown in Figure~\ref{fig:accjoint_r1_qual}.}
	\label{fig:accjoint_r1_qual_otherjoints}
\end{figure}

\begin{figure}[!h]
	\subfloat{\includegraphics[width=0.2\linewidth, clip]{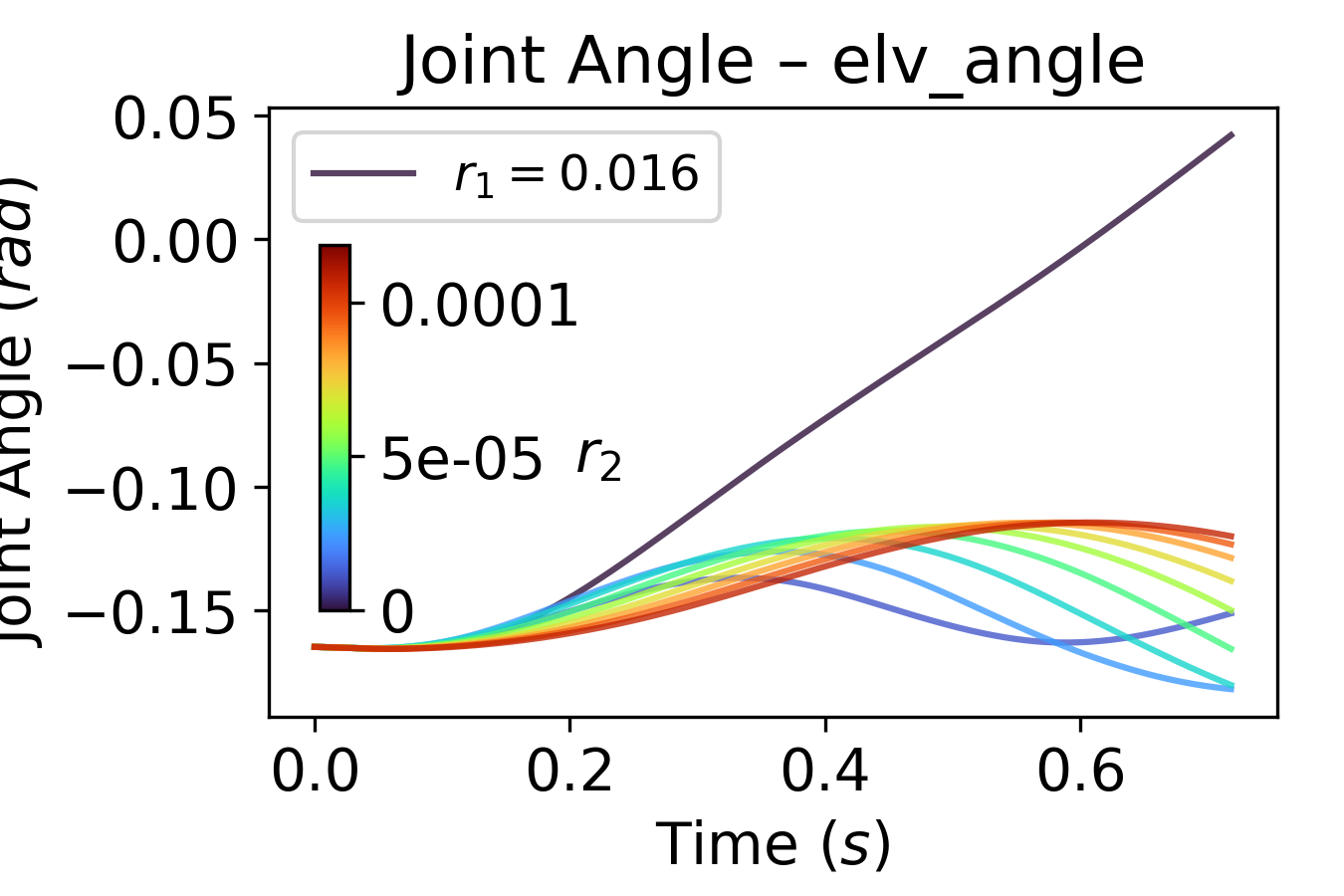}}
	\subfloat{\includegraphics[width=0.2\linewidth, clip]{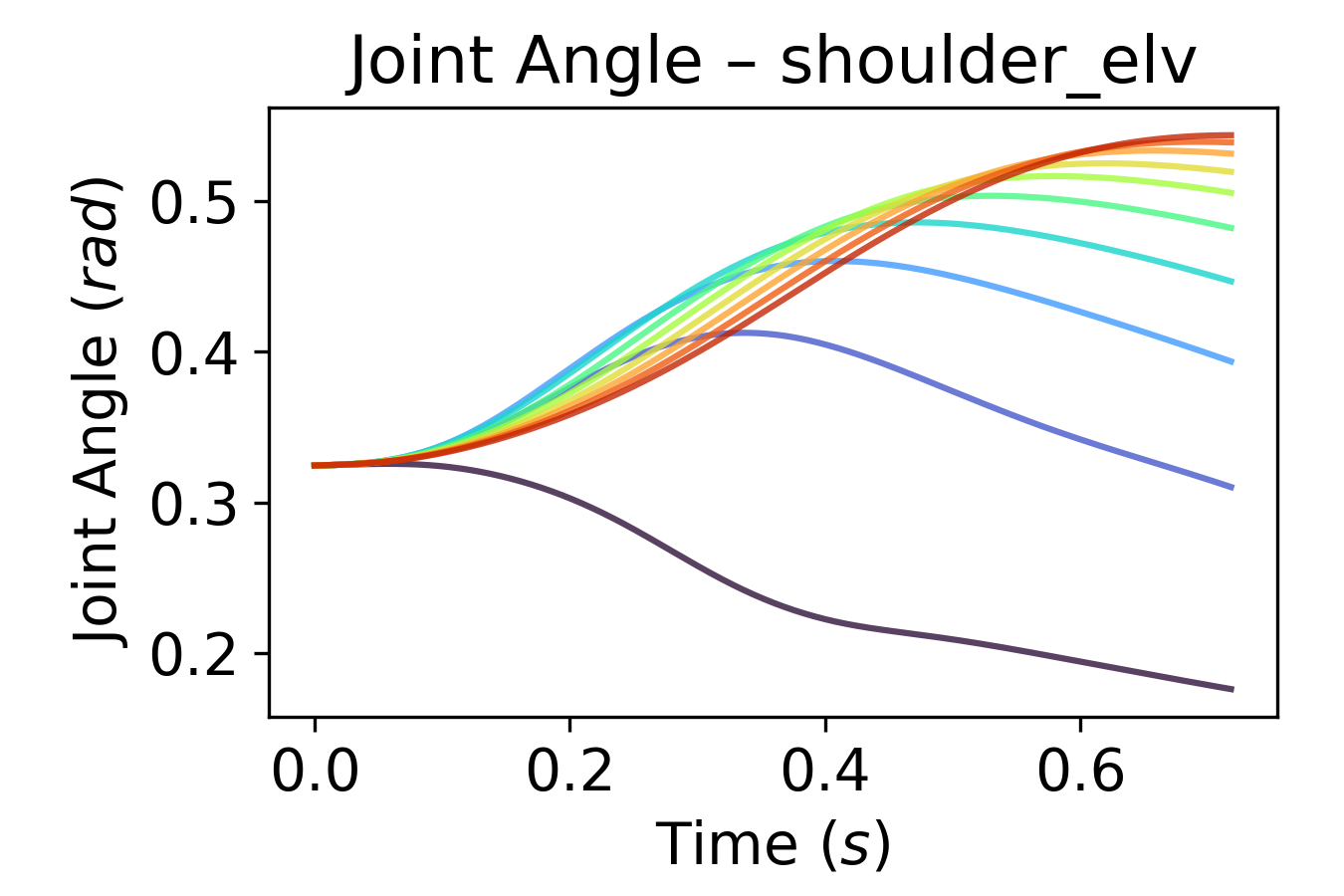}}
	\subfloat{\includegraphics[width=0.2\linewidth, clip]{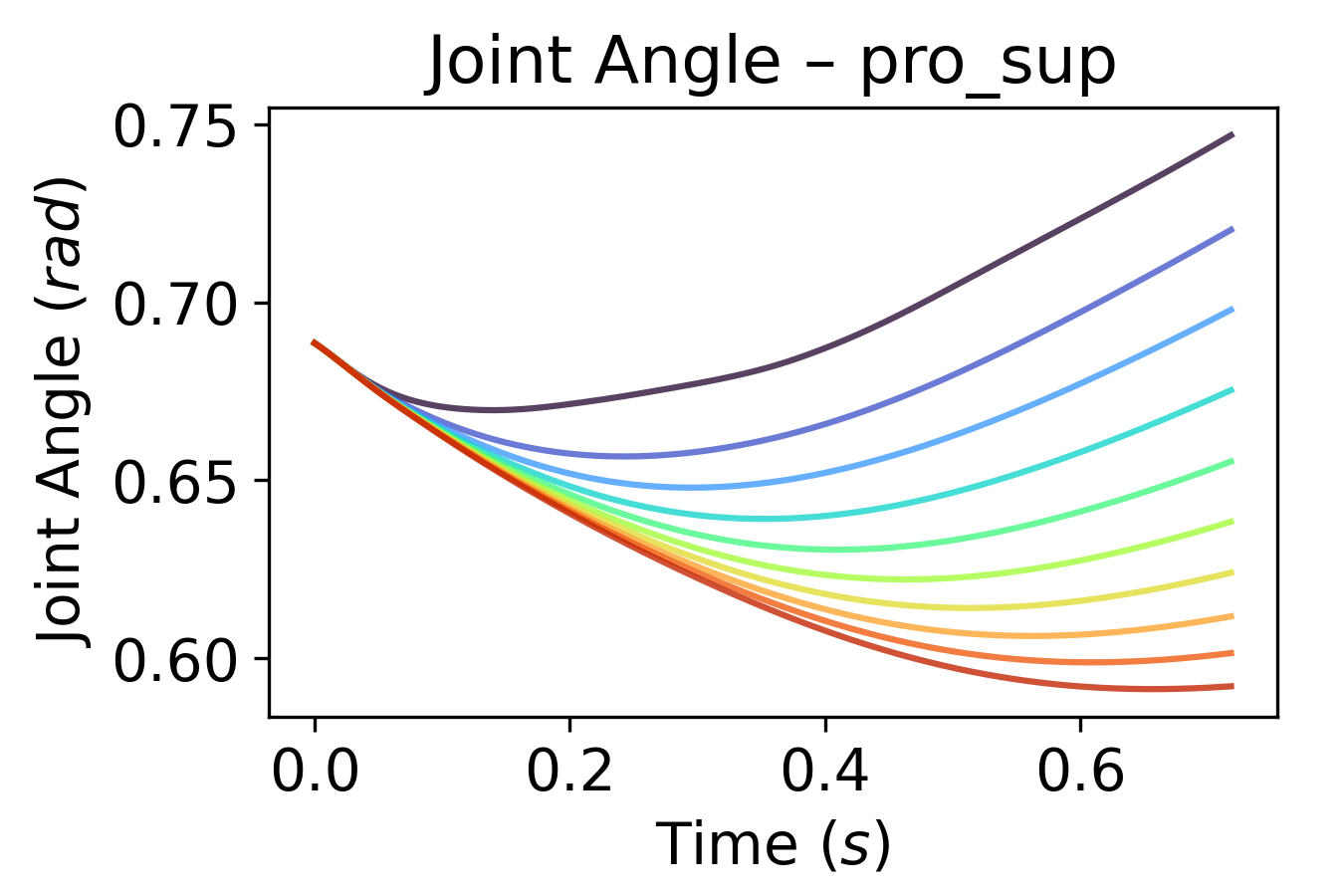}}
	\subfloat{\includegraphics[width=0.2\linewidth, clip]{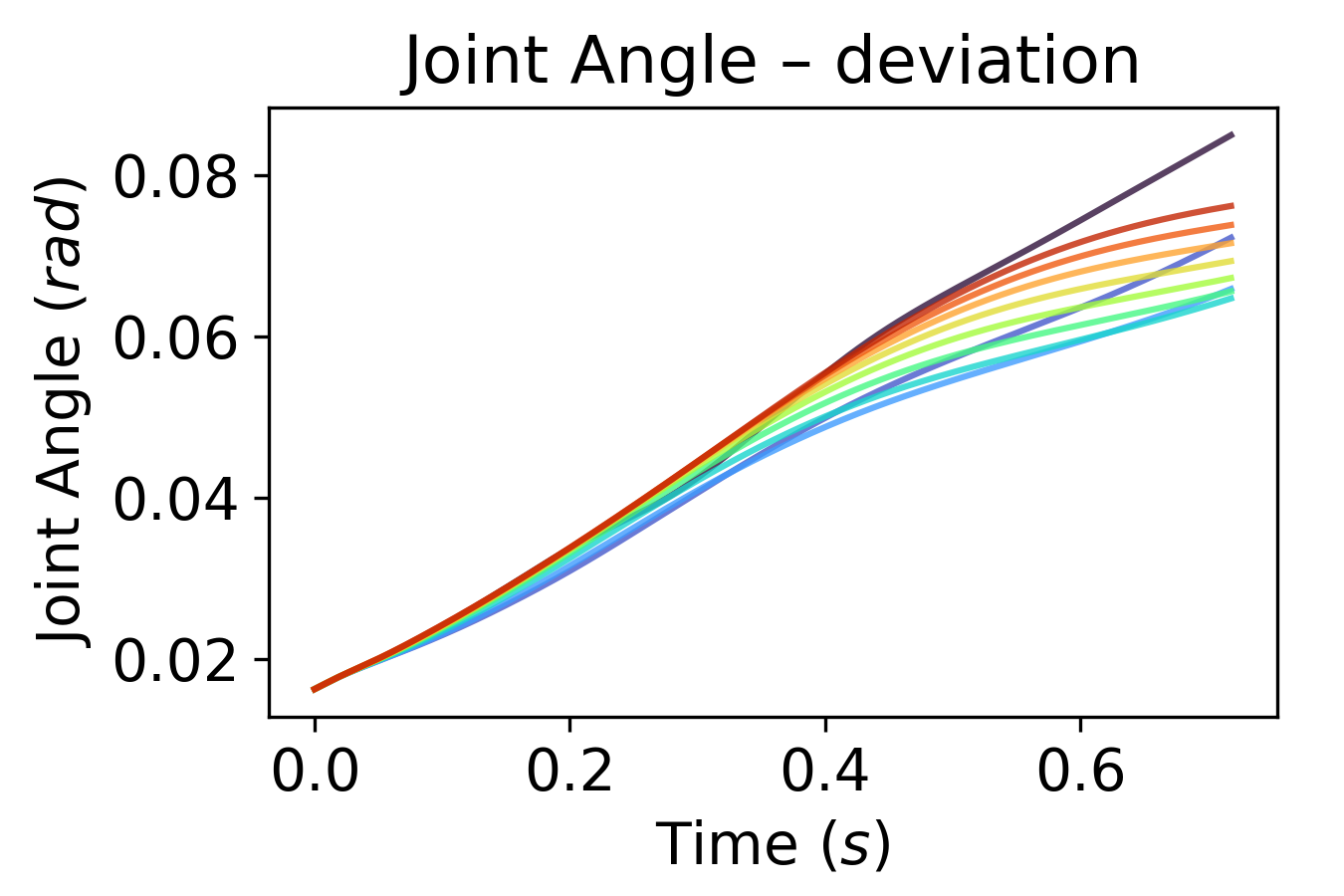}}
	\subfloat{\includegraphics[width=0.2\linewidth, clip]{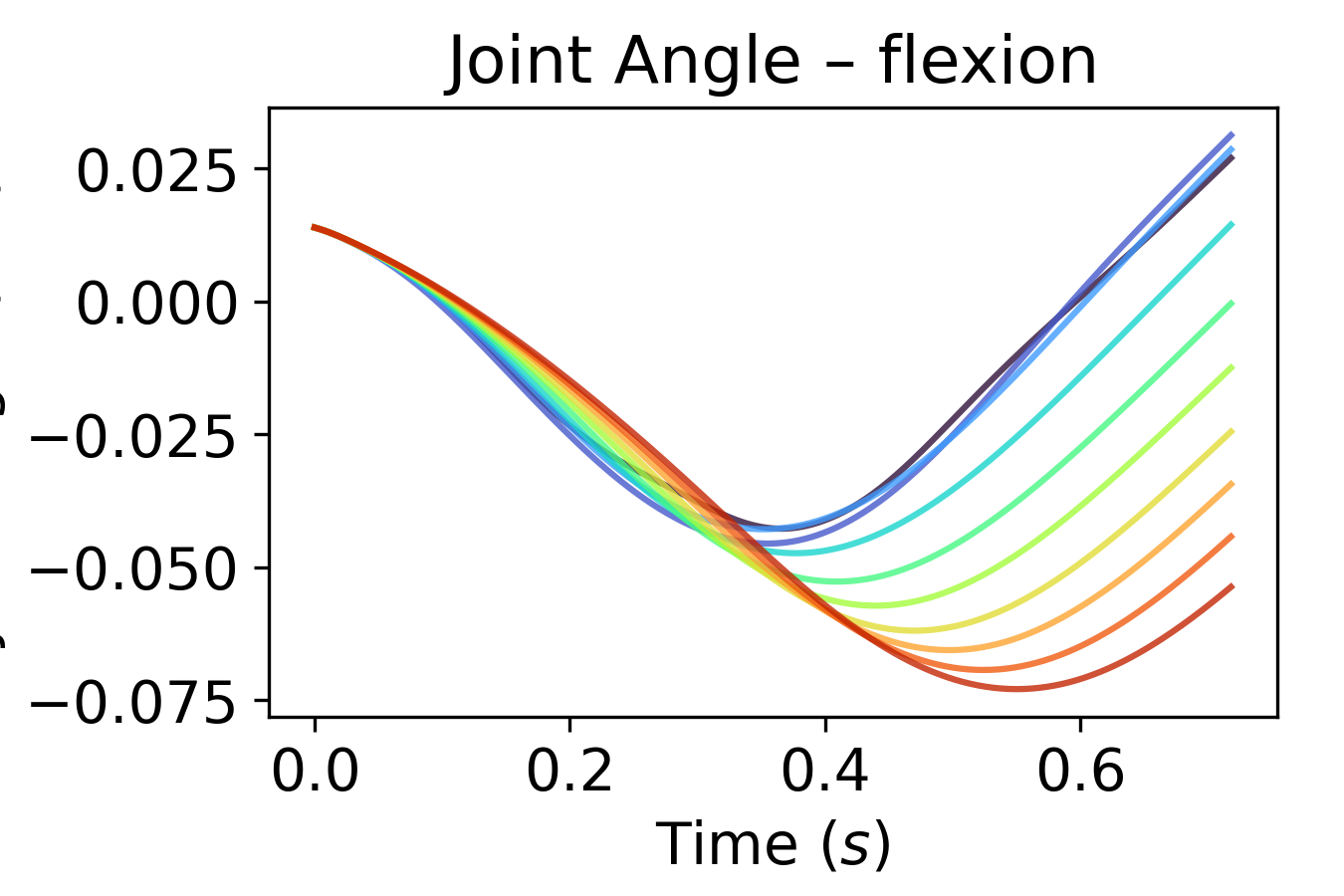}}\\
	\subfloat{\includegraphics[width=0.2\linewidth, clip]{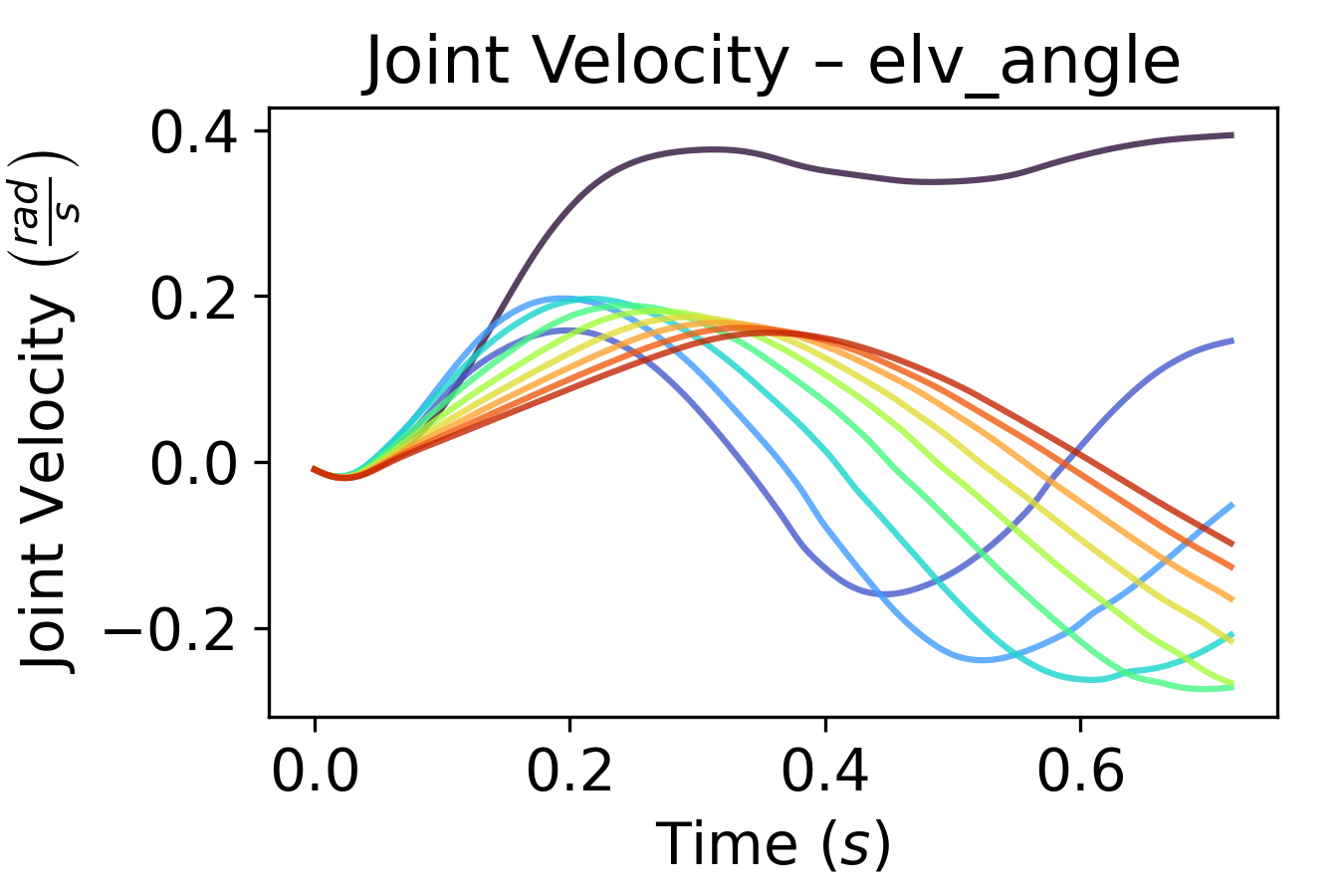}}
	\subfloat{\includegraphics[width=0.2\linewidth, clip]{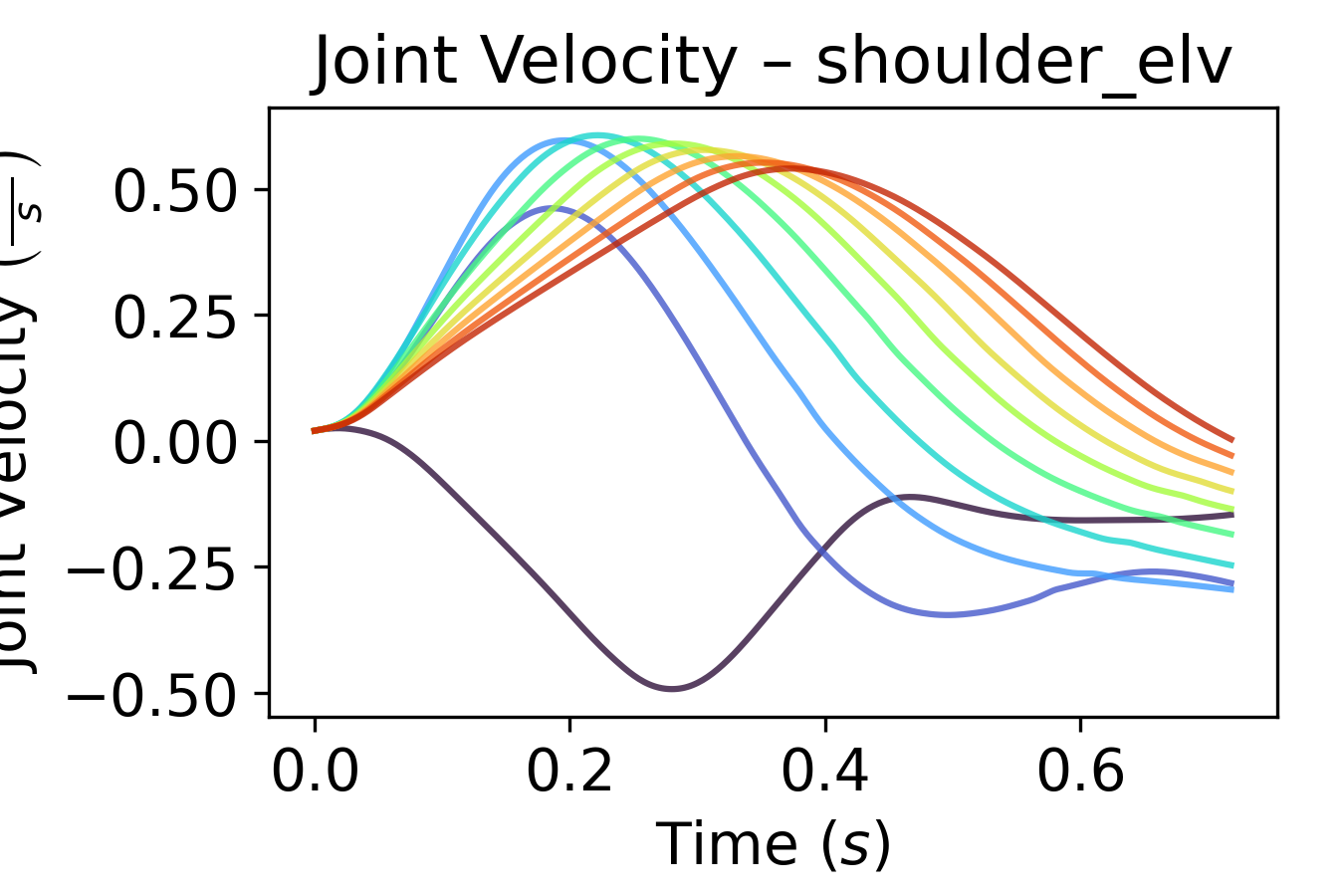}}
	\subfloat{\includegraphics[width=0.2\linewidth, clip]{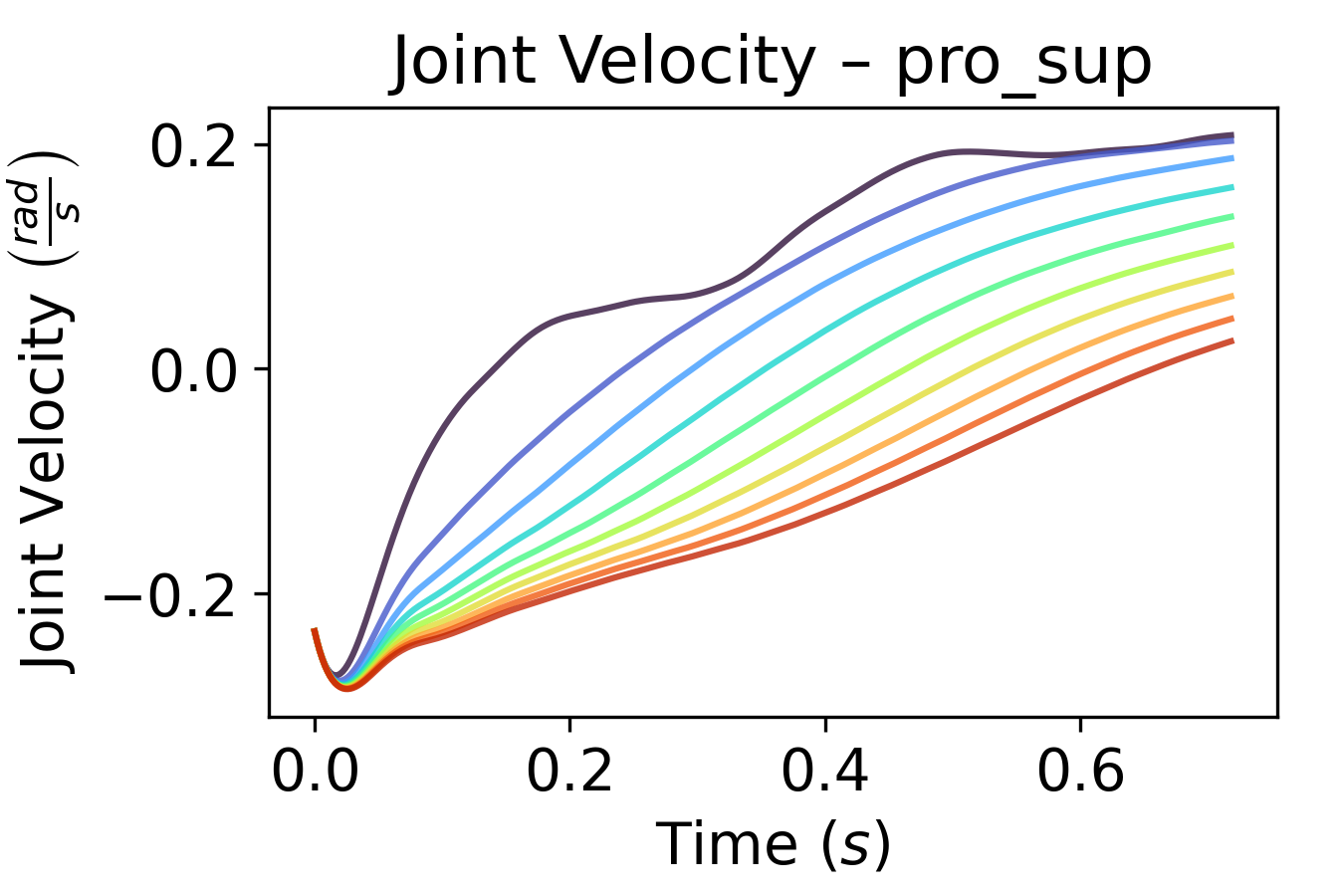}}
	\subfloat{\includegraphics[width=0.2\linewidth, clip]{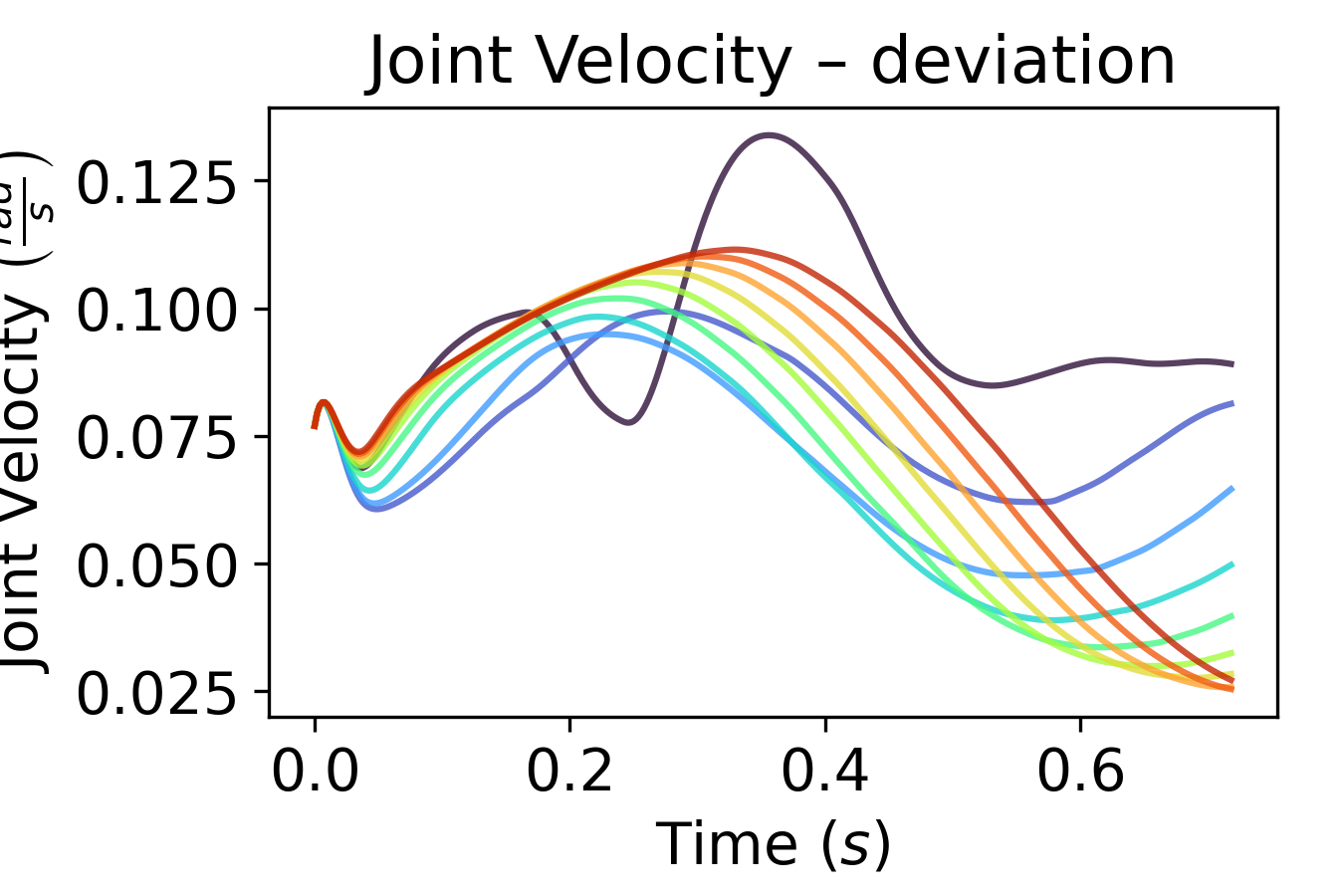}}
	\subfloat{\includegraphics[width=0.2\linewidth, clip]{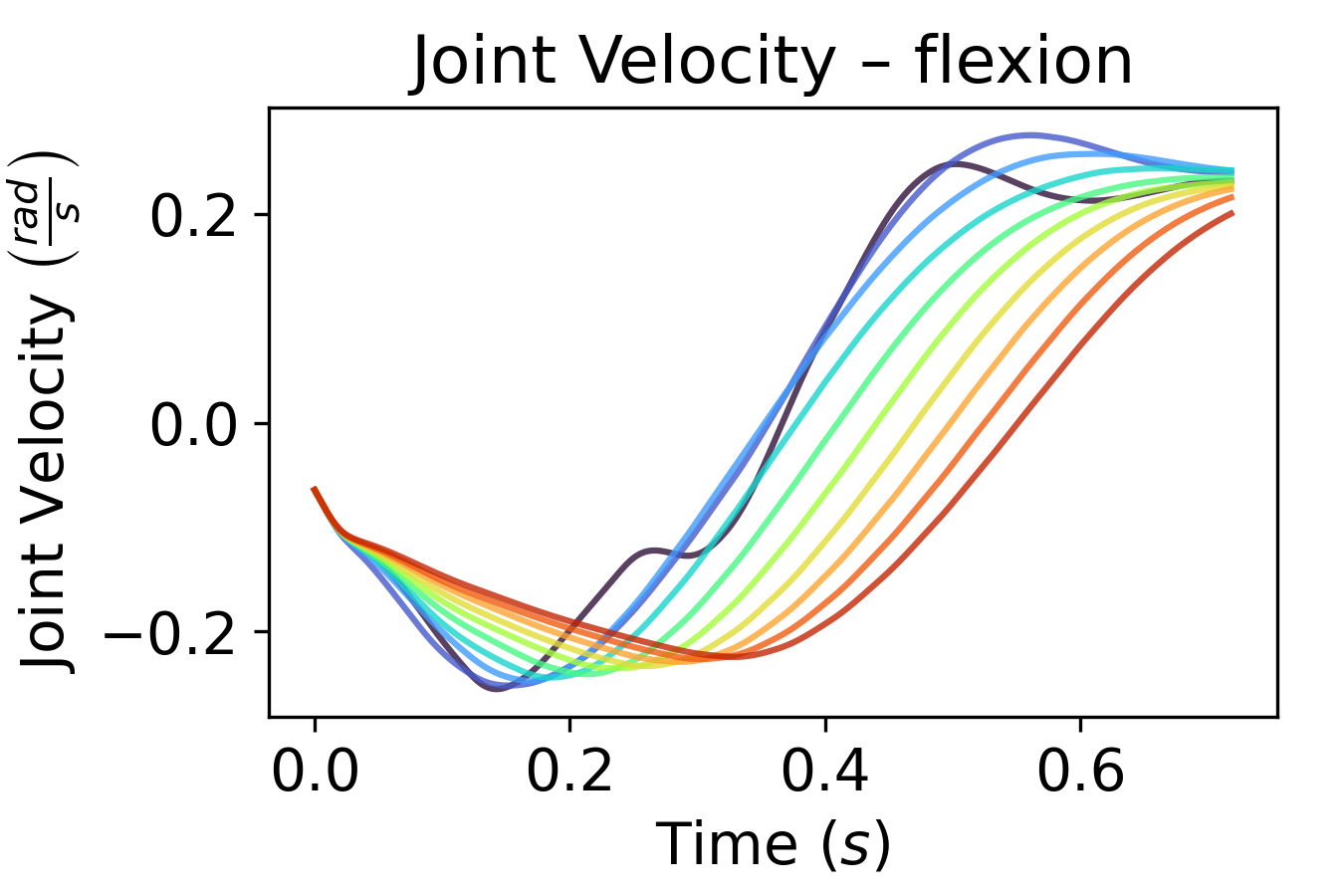}}
	
	\caption{Trajectories of the JAC simulation for different cost weights $r_2$. %
	Remaining joints and cursor trajectories are shown in Figure~\ref{fig:accjoint_r2_qual}.}
	\label{fig:accjoint_r2_qual_otherjoints}
\end{figure}

\clearpage

\subsection{Effects of the MPC Horizon}

\begin{figure}[!h]
	\subfloat{\includegraphics[width=0.2\linewidth, clip]{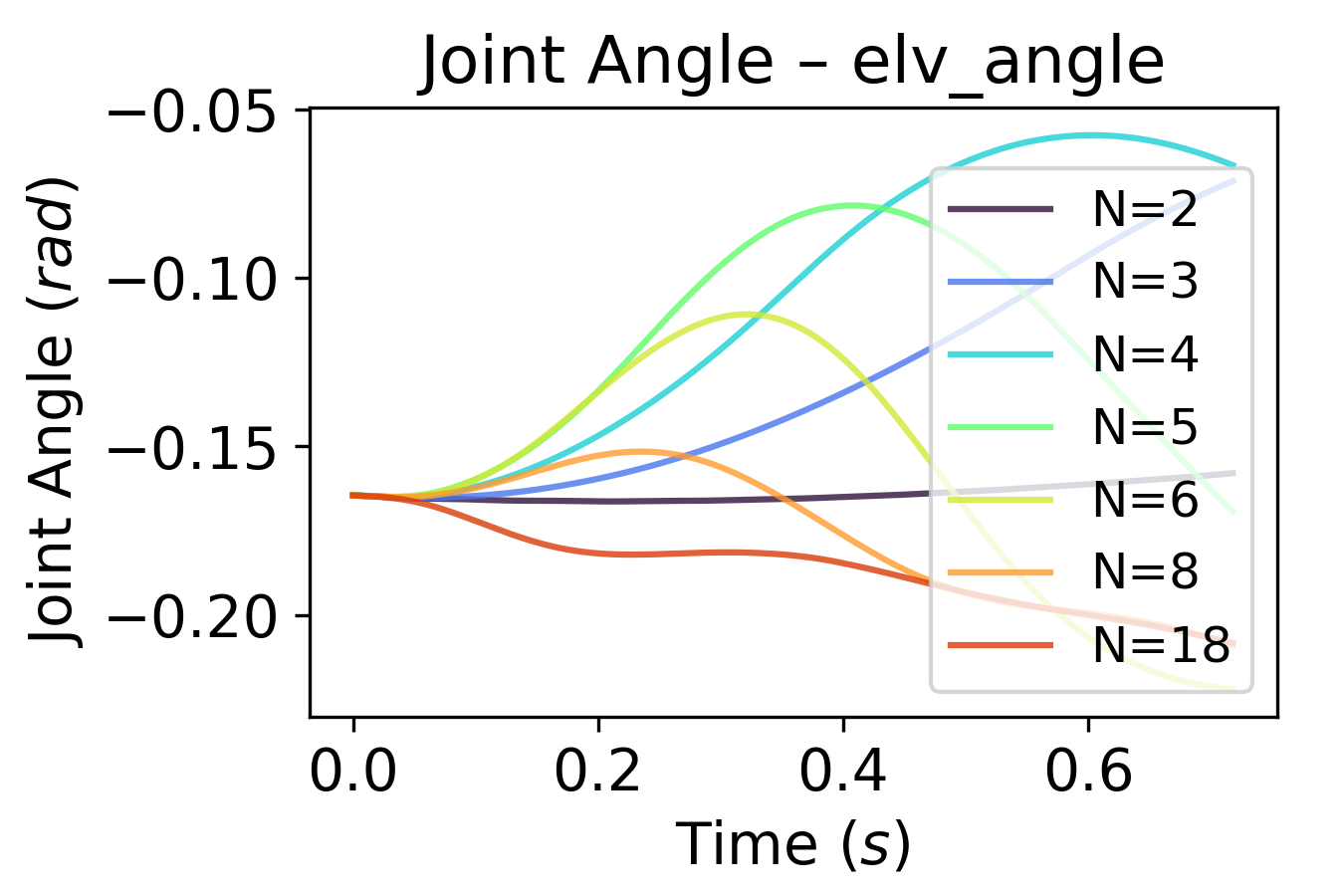}}
	\subfloat{\includegraphics[width=0.2\linewidth, clip]{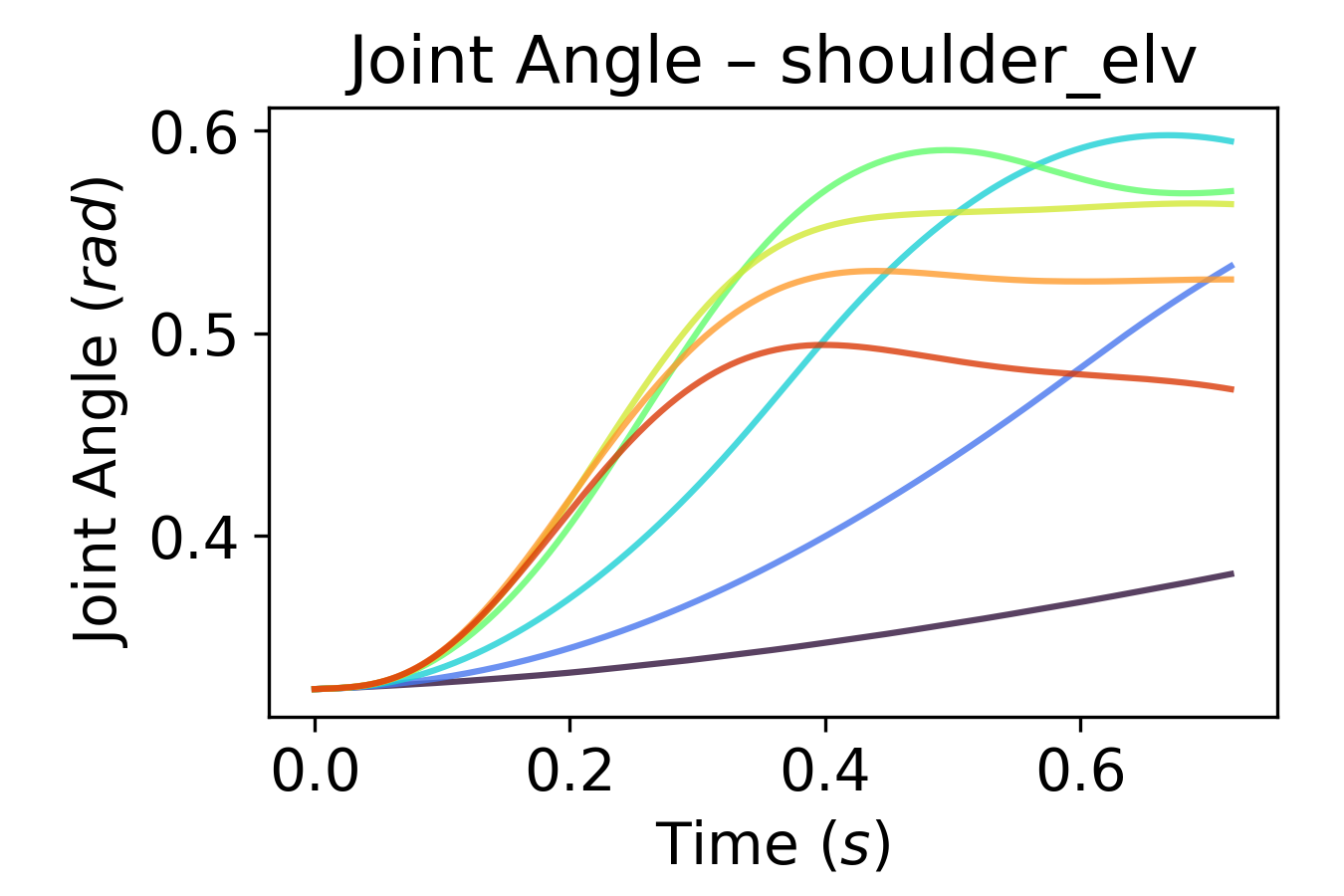}}
	\subfloat{\includegraphics[width=0.2\linewidth, clip]{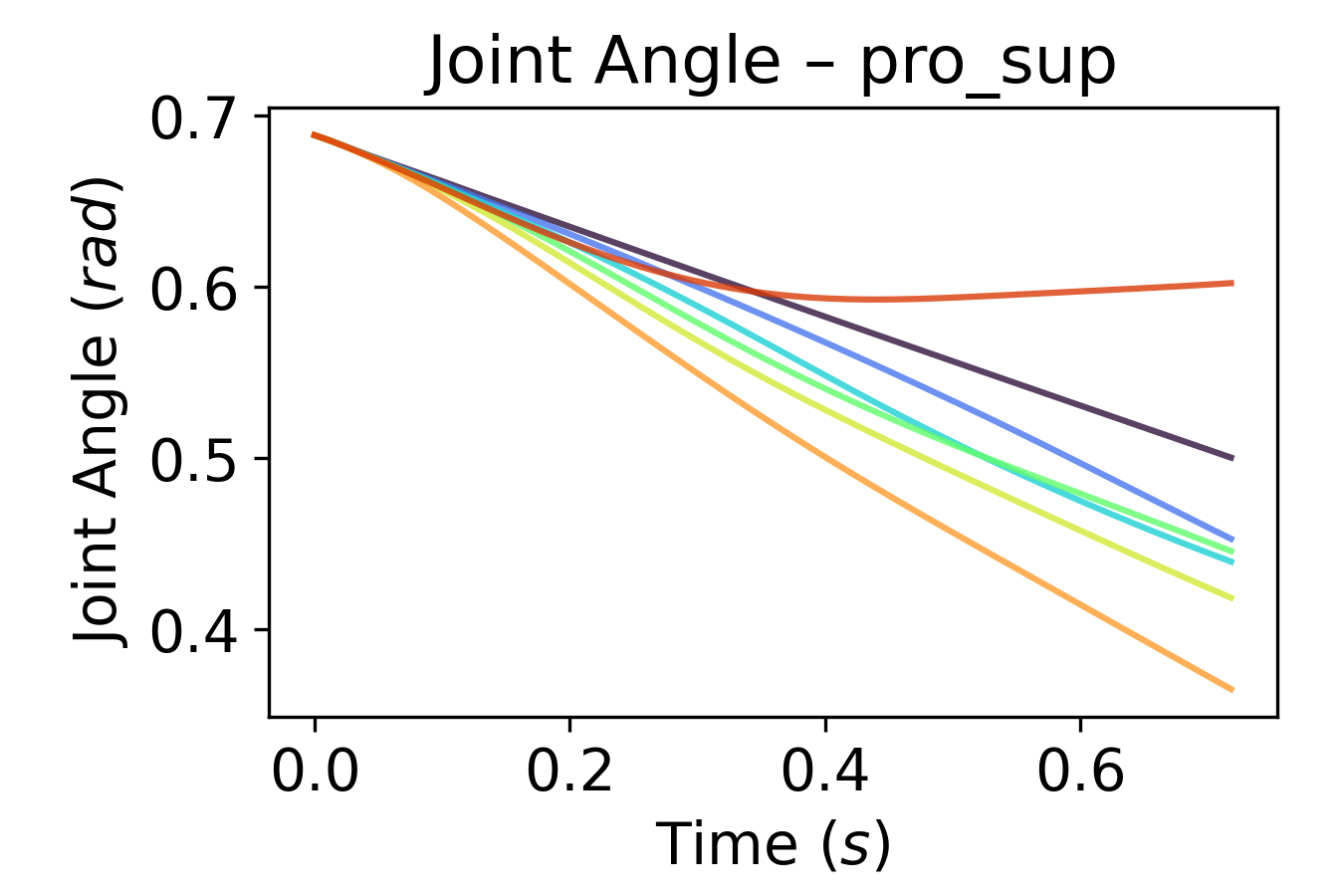}}
	\subfloat{\includegraphics[width=0.2\linewidth, clip]{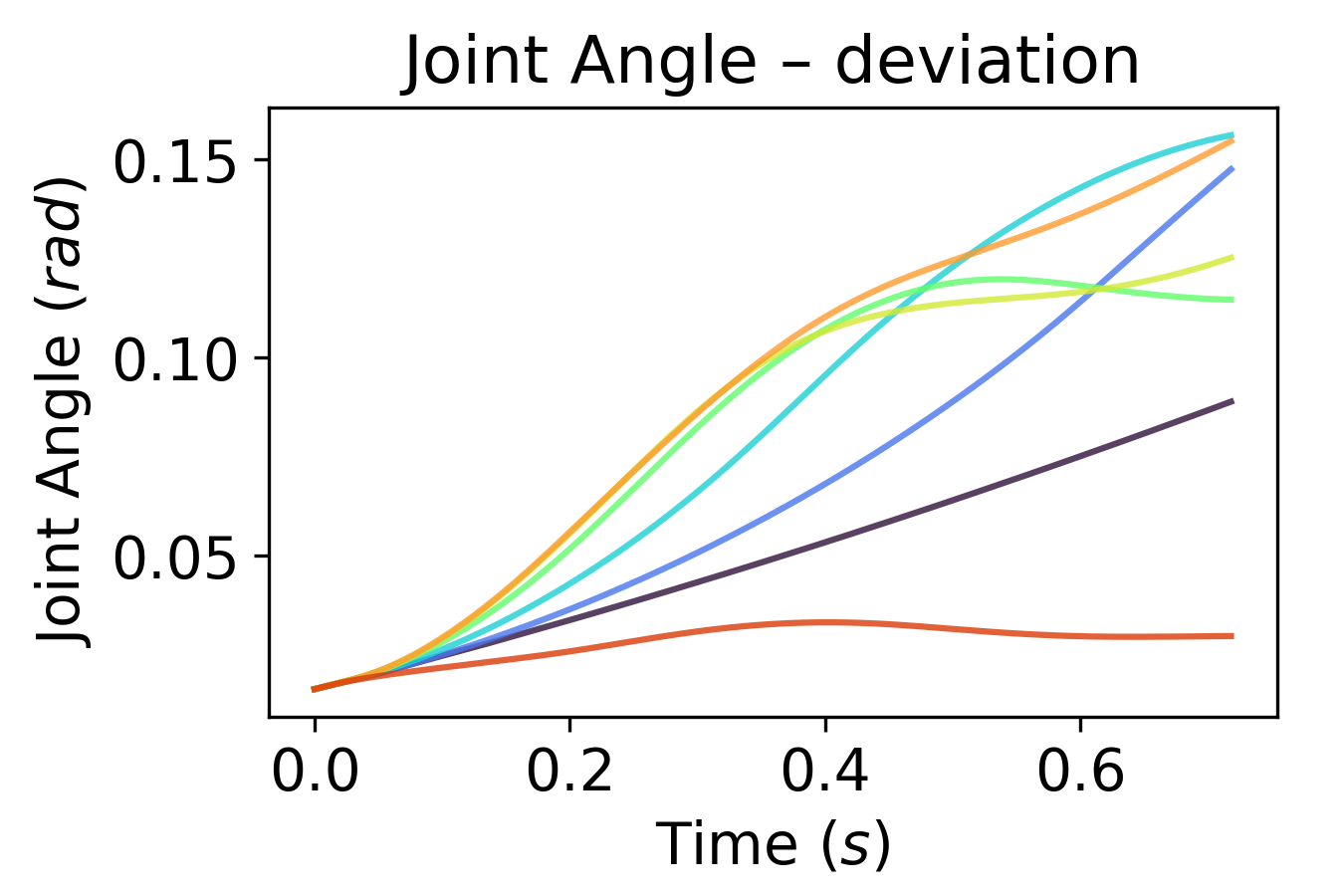}}
	\subfloat{\includegraphics[width=0.2\linewidth, clip]{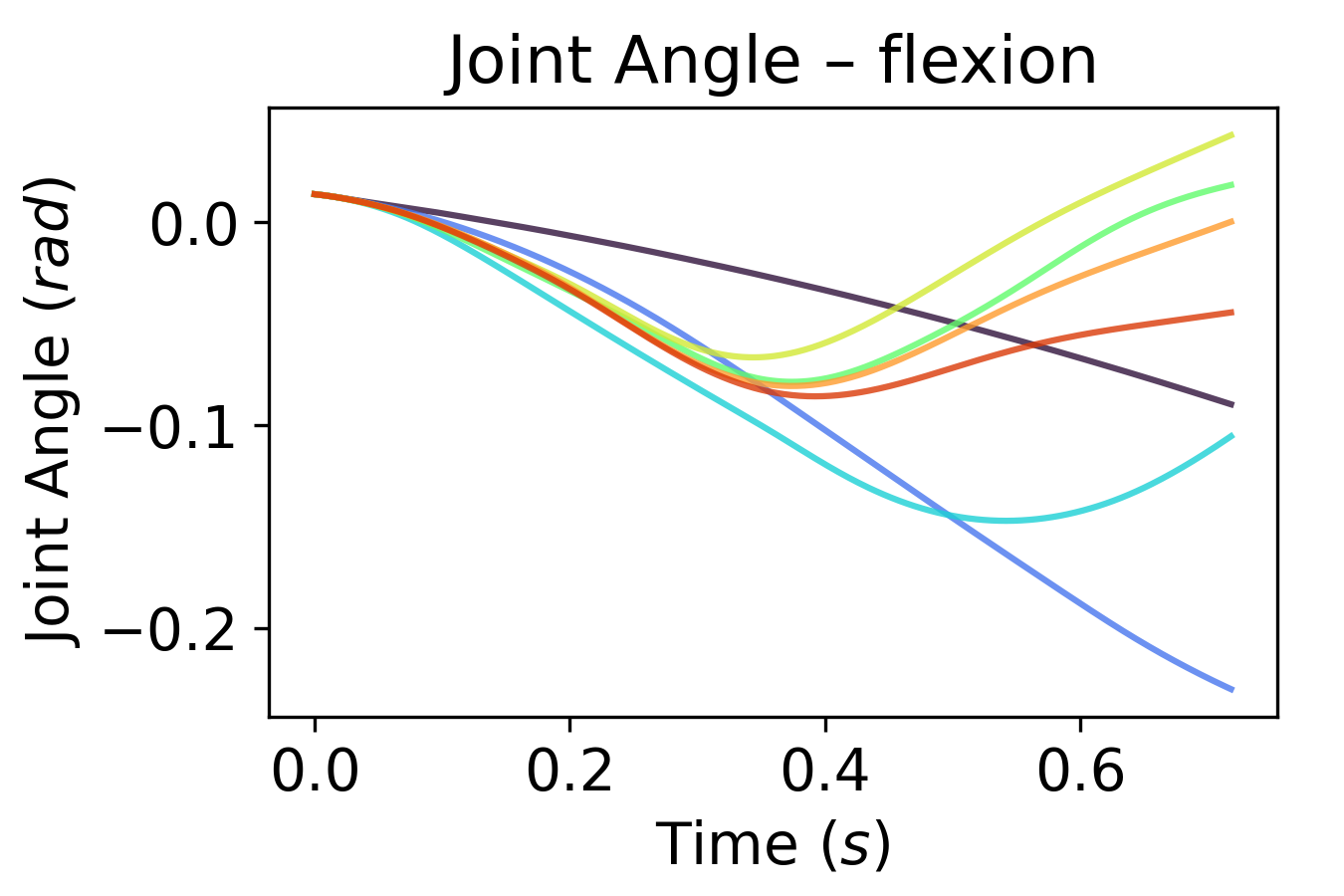}}\\	
	\subfloat{\includegraphics[width=0.2\linewidth, clip]{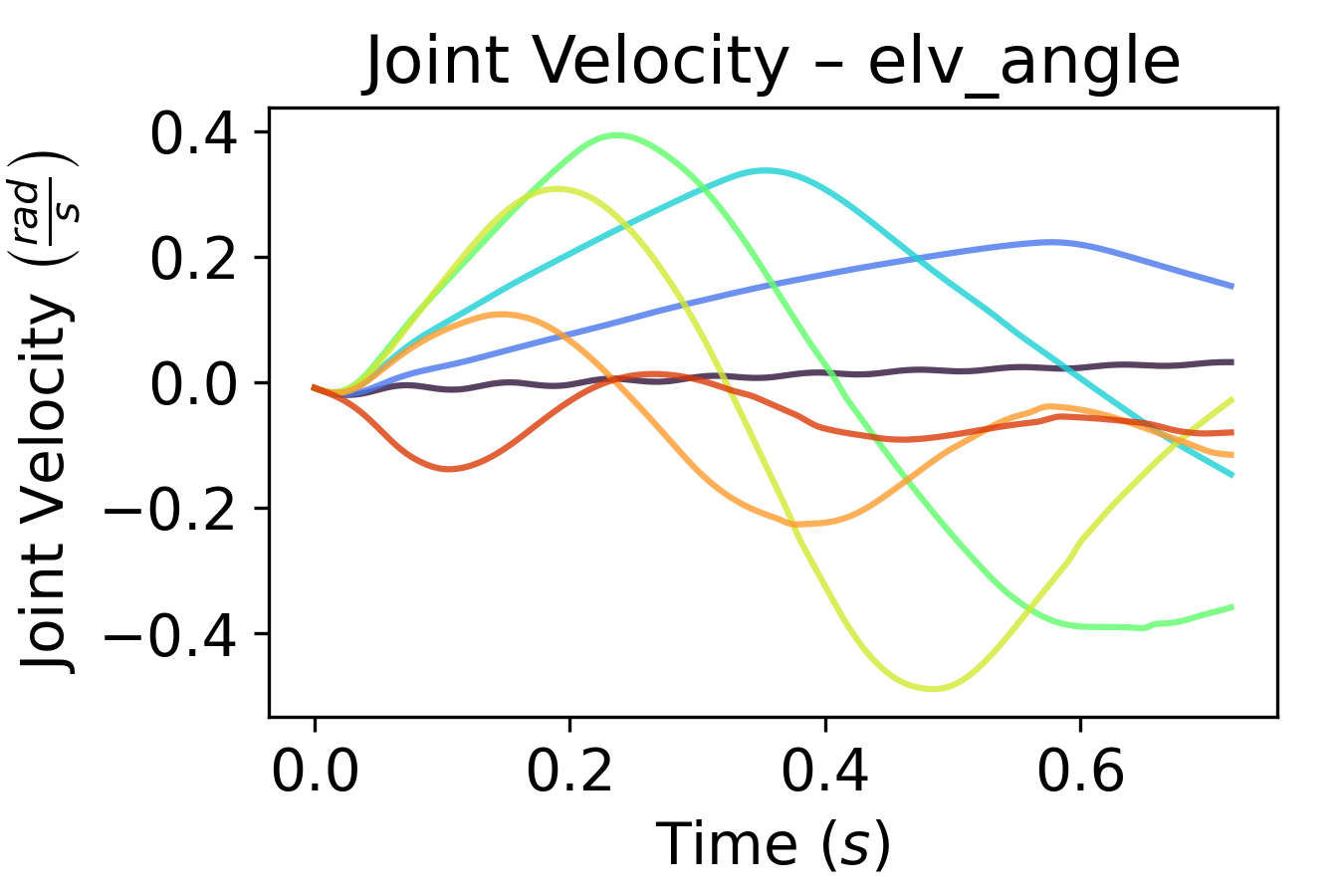}}
	\subfloat{\includegraphics[width=0.2\linewidth, clip]{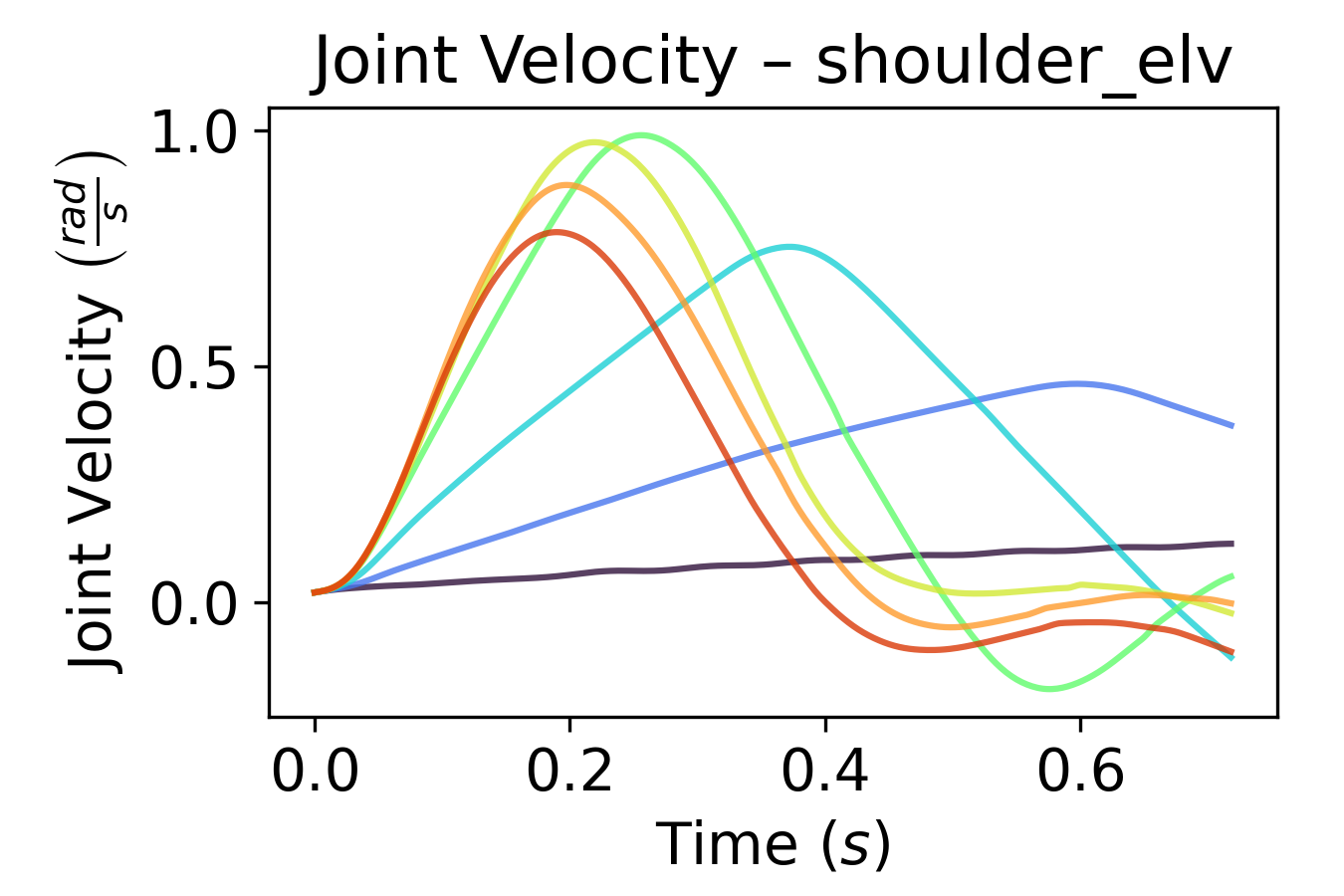}}
	\subfloat{\includegraphics[width=0.2\linewidth, clip]{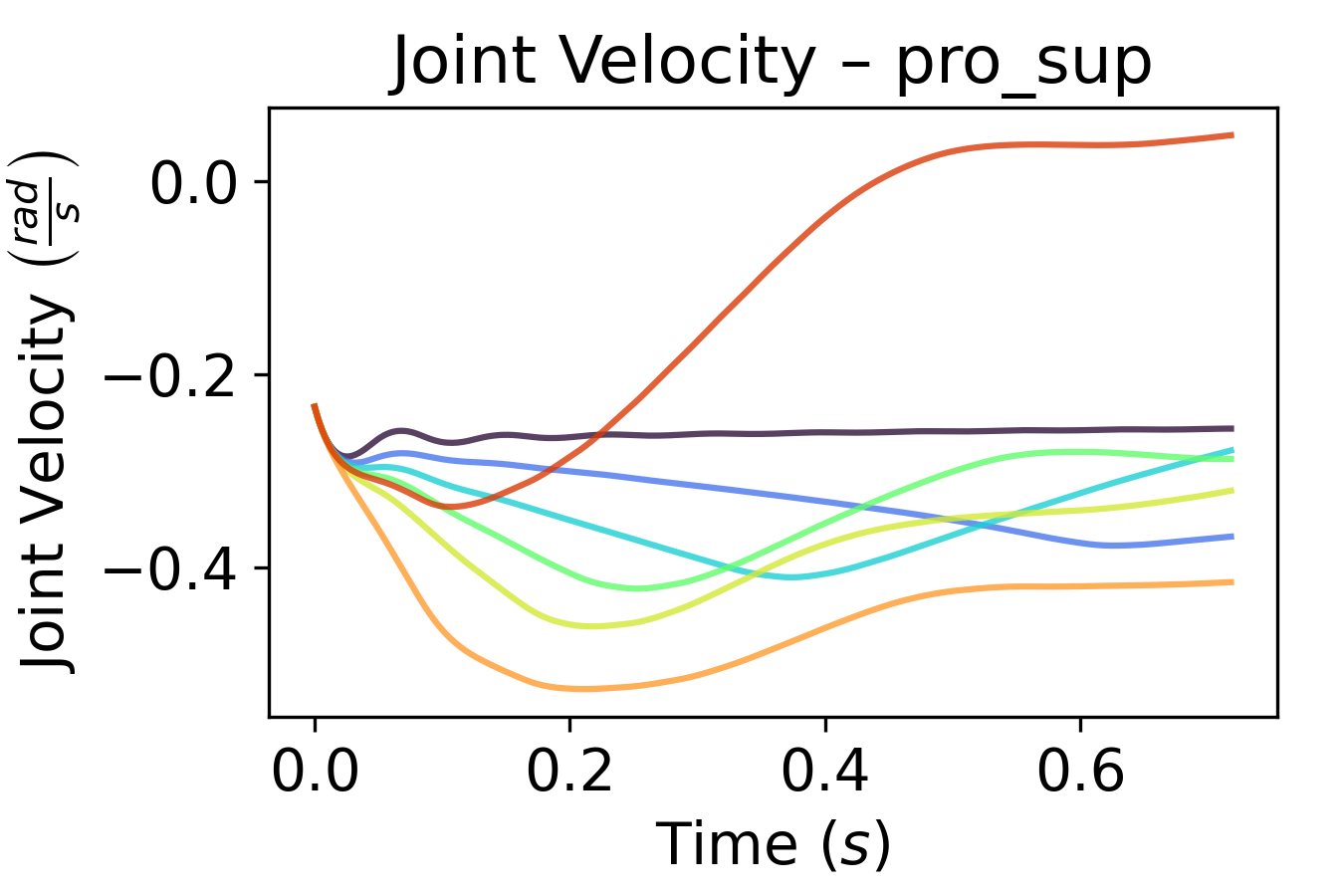}}
	\subfloat{\includegraphics[width=0.2\linewidth, clip]{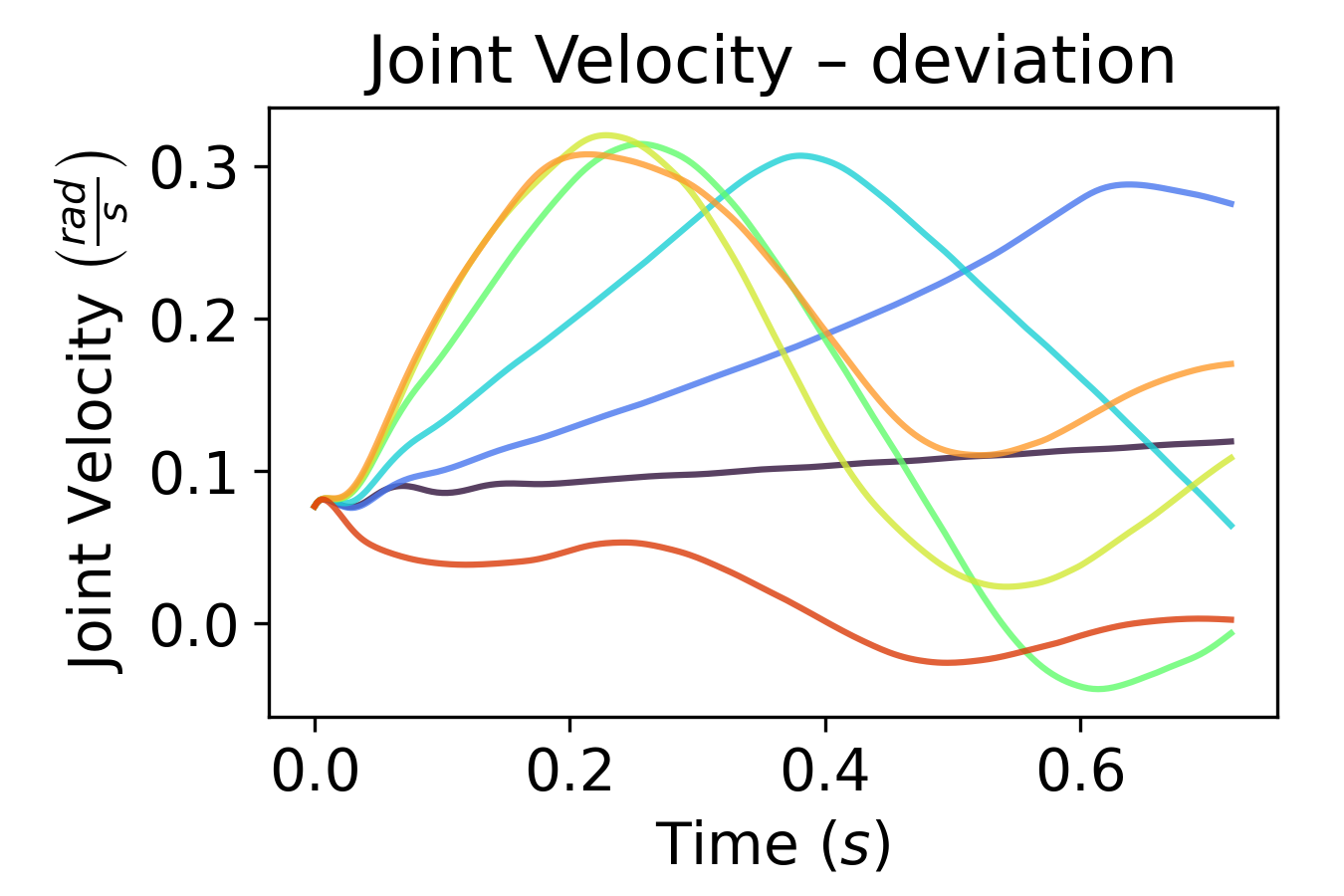}}
	\subfloat{\includegraphics[width=0.2\linewidth, clip]{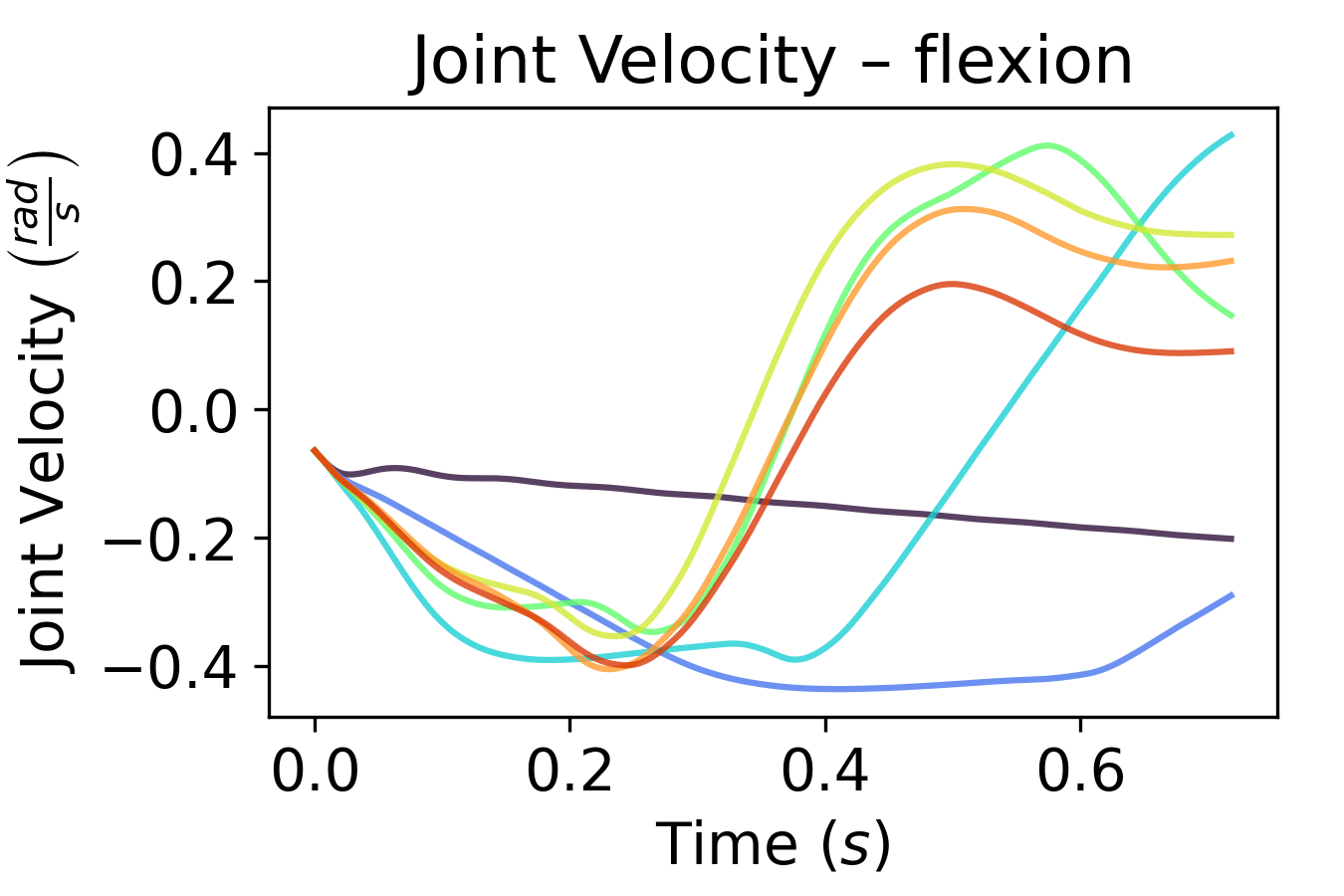}}
	
	\caption{Joint trajectories of one trial for varying MPC horizon $N$, using the Joint Acceleration Costs (JAC), without control noise.
	Remaining joints and cursor trajectories are shown in Figure~\ref{fig:accjoint_N_qual}.}
	\label{fig:accjoint_N_qual_otherjoints}
\end{figure}

\clearpage